\documentclass{pspum-l}

\theoremstyle{definition}

\theoremstyle{remark}

\input{epsf}
\usepackage{graphicx}
\catcode`\@=11
\def\numberbysection{\@addtoreset{equation}{section}
\def\theequation{\thesection.\arabic{equation}}}
\numberwithin{equation}{section}
\numberwithin{table}{section}
\begin{document}

\title{{CONFORMAL FRACTAL GEOMETRY \&~BOUNDARY~QUANTUM GRAVITY}}

\author{\sc Bertrand Duplantier}
\address{Service de Physique Th\'{e}orique, URA CNRS 2306\\
CEA/Saclay\\
F-91191 Gif-sur-Yvette Cedex\\
FRANCE\\
e-mail: bertrand@spht.saclay.cea.fr}


\rightline{\it Am\'erique, \^O ma Norv\`ege !}
\rightline{Dominique Fourcade}
\rightline{{\it Le sujet monotype}}
\rightline{{P.O.L. (1997)}}

\bigskip

\subjclass[2000]{Primary 60D05; Secondary 81T40, 05C80, 60J65, 60J45, 30C85, 60K35, 82B20, 82B41, 82B43}

\begin{abstract}

This article gives a comprehensive description of the fractal geometry of conformally-invariant (CI) scaling curves,
in the plane or half-plane. It focuses on deriving critical exponents associated with interacting random paths,
 by exploiting an underlying quantum gravity (QG) structure, which uses KPZ maps
relating exponents in the plane to those on a random lattice, i.e., in a fluctuating metric. This is accomplished within
the framework of conformal field theory (CFT),
with applications to well-recognized critical models, like $O(N)$ and Potts models, and to the Stochastic L\"owner Evolution (SLE).
Two fundamental ingredients of the QG construction are relating bulk and Dirichlet boundary exponents, and
establishing additivity rules for QG boundary conformal dimensions associated with mutually-avoiding random sets.  From
these we derive
the non-intersection exponents for random walks (RW's) or Brownian paths, self-avoiding walks (SAW's), or arbitrary
mixtures thereof.
The multifractal (MF) function $f(\alpha,\,c)$ of the harmonic measure (i.e.,
electrostatic potential, or diffusion field) near any conformally
invariant fractal boundary, is given as a function
of the central charge $c$ of the associated CFT. 
A Brownian path, a
SAW in the scaling limit, or a critical percolation cluster have identical spectra corresponding
to the same central charge $c=0$, with a Hausdorff dimension $D={\rm sup}_{\alpha} f(\alpha; c=0)=4/3$,
which nicely vindicates Mandelbrot's conjecture
for the Brownian frontier dimension.
The Hausdorff dimensions $D_{\rm H}$ of a non-simple scaling curve or
cluster hull, and $D_{\rm EP}$ of its external perimeter or frontier, are shown to
obey the ``superuniversal'' duality equation $(D_{\rm H}-1)(D_{\rm
EP}-1)=\frac{1}{4}$, valid for any value of the central charge $c$. Higher multifractal
functions, like the double spectrum $f_2(\alpha,\alpha';c)$ of the double-sided harmonic measure, are also considered.
The universal mixed MF spectrum $f(\alpha,\lambda;c)$
describing the local winding rate $\lambda$ and singularity exponent $\alpha$ of the harmonic measure near any CI scaling curve is given. 
 The fundamental duality which exists between simple and non-simple random paths is established via an algebraic
 symmetry  of the KPZ quantum gravity map. An extended dual KPZ relation is then introduced for the SLE,
which commutes with the  $\kappa \to \kappa'=16/\kappa$ duality for ${\rm SLE}_{\kappa}$. This allows us to calculate the SLE exponents
from  simple QG rules. These rules are established from the general structure of
 correlation functions of arbitrary interacting random sets on a random lattice, as derived from random matrix theory.
\end{abstract}


\maketitle

\bigskip

\tableofcontents
\section{\sc{Introduction}}
\label{sec.intro}

\subsection{A Brief Conformal History}

\subsubsection{Brownian Paths, Critical Phenomena, and Quantum Field Theory}

Brownian motion is the archetype of a random process,
hence its great importance in probability
theory. The Brownian path is also the archetype of a scale invariant set, and in two dimensions is
a conformally-invariant one, as shown by P. L\'evy \cite{PLevy}. It is therefore perhaps the most natural random fractal \cite{mandelbrot}.  On the other hand, Brownian paths are intimately linked with quantum field theory (QFT).
 Intersections of Brownian paths indeed provide the random geometrical mechanism underlying QFT
  \cite{symanzyk}. In a Feynman diagram, any propagator line can be represented by a Brownian path,
  and the vertices are
  intersection points of the Brownian paths.
  This equivalence is widely used in polymer theory
  \cite{PGG,desC} and in rigorous studies of second-order phase
transitions and field theories  \cite{aizenman1}. Families of
universal critical exponents are in particular associated with {\it non-
intersection} probabilities of collections of random
walks (RW's) or Brownian paths, and these play an important role both in probability theory and quantum field theory
\cite{lawler1,fisher,aizenman2,duplantier1}.

A perhaps less known fact is the existence of highly non-trivial geometrical, actually {\it fractal}
(or {\it multifractal}), properties of Brownian paths or their subsets \cite{mandelbrot}. These types of geometrical
fractal properties generalize to all universality classes of, e.g., random walks (RW's), loop-erased random walks
(LERW's), self-avoiding walks (SAW's) or
polymers, Ising, percolation
and Potts models, $O(N)$ models, which are related in an essential way to standard critical phenomena and
field theory. The random fractal geometry is particularly rich in two dimensions.


\subsubsection{Conformal Invariance and Coulomb Gas}
 In {\it two dimensions} (2D), the notion of {\it conformal invariance} \cite{BPZ,friedan,cardylebowitz}, and the
 introduction of the so-called
 ``Coulomb gas techniques''
 have brought a wealth of exact results (see,
e.g.,  \cite{dennijs,nien,cardy,S1,duplantier4,DS2,SD,cardy3}).
Conformal field theory (CFT) has
lent strong support to the conjecture that statistical systems at
their critical point, in their scaling (continuum) limit, produce
{\it conformally-invariant} (CI) fractal structures, examples of
which are the continuum scaling limits of RW's, LERW's, SAW's, critical
Ising or Potts clusters.  A prominent role was played by Cardy's equation for the crossing probabilities in 2D percolation  \cite{cardy3}.
 To understand conformal invariance in a rigorous way  presented a mathematical challenge
(see, e.g.,  \cite{langlands,ai1,ben}).

In the particular case of planar Brownian paths,  Beno\^{\i}t Mandelbrot  \cite{mandelbrot}
made the following famous conjecture in 1982: {\it in two dimensions, the external frontier of a planar
Brownian path has a Hausdorff dimension
\begin{equation}
D_{\rm Brown.\, fr.}=\frac{4}{3}, \label{Mand}
\end{equation}
identical to that
of a planar self-avoiding walk} \cite{nien}. This identity has played an important role
in probability theory and theoretical physics in recent years, and will be a central theme in this article.
We shall understand this identity in the light of ``quantum gravity'', to which we turn now.

\subsubsection{Quantum Gravity and the KPZ Relation}
Another breaktrough, not wide\-ly noticed at the time, was the introduction of ``2D quantum gravity'' (QG) in the
statistical mechanics of 2D critical systems.  V. A. Kazakov gave the solution of the Ising model on a random planar lattice \cite{kazakov}.
The  astounding discovery by Knizhnik, Polyakov, and Zamolodchikov of the ``KPZ'' map between critical
exponents in the standard plane and in a random 2D metric \cite{KPZ} led to the relation of the exponents found in \cite{kazakov} to those of Onsager
(see also \cite{david2}).
The first other explicit solutions and checks of KPZ were obtained for SAW's \cite{DK} and for the $O(N)$ model \cite{kostovgaudin}.

\subsubsection{Multifractality}
The concepts of generalized dimensions and associated {\it multifractal} (MF)
 measures were developed in parallel two decades ago \cite
{mandelbrot2,hentschel,frisch,halsey}. It was later realized that
multifractals and field theory have deep connections, since
the algebras of their respective correlation functions reveal
striking similarities  \cite{cates}.

A particular example is given by classical potential theory, i.e., that of the electrostatic or
diffusion field near critical fractal boundaries,
or near diffusion
limited aggregates (DLA). The self-similarity of the fractal boundary is indeed reflected
in a multifractal behavior of the moments of the potential. In DLA,
the potential, also called harmonic measure, actually determines
the growth process \cite{halmeakproc,halseyerice,meak}.
For equilibrium statistical fractals, a first analytical example of multifractality was studied in Ref.
  \cite{cates et witten}, where the fractal boundary was chosen to be a simple RW, or a
SAW, both accessible to renormalization group
methods near four dimensions. In two dimensions, the existence of a multifractal spectrum for the Brownian path frontier
was established rigorously \cite{lawler97}.

In 2D, in analogy to the simplicity of the classical
method of conformal transforms to solve electrostatics of {
Euclidean} domains, a {\it universal}
solution could be expected  for the distribution of potential near any CI fractal in the plane.
It was clear that these multifractal spectra
should be linked with the conformal invariance classification, but outside the realm of the
usual {\it rational} exponents. That presented a second challenge to the theory.

\subsection{Elaborating Conformal Geometrical Structures}
\subsubsection{Brownian Intersection Exponents}
It was already envisioned in the
mid-eigh\-ties that the critical properties of planar Brownian paths, whose conformal invariance was
 well established \cite{PLevy}, could be the opening gate to rigorous studies
of two-dimensional critical phenomena
\footnote{It is perhaps interesting to note that P.-G. de Gennes originally studied polymer
theory with the same hope of understanding from that perspective the broader class of critical phenomena. It turned out to be historically the converse:
the Wilson-Fisher renormalization group approach to spin models with $O(N)$ symmetry yielded the polymer critical exponents
as the special case of the $N \to 0$ limit \cite{PGG}.}.
Michael Aizenman, in a seminar in the Probability Laboratory of University of Paris VI
in 1984, promised a good bottle of Bordeaux wine for the resolution of the $\zeta_2$ exponent governing in two dimensions
the non-intersection
probability up to time $t$, $P_2(t) \approx t^{-\zeta_2}$, of two Brownian paths 
\footnote{The Ch\^ateau Margaux 1982 bottle was finally drunk in 2001 Chez Panisse, Berkeley CA.}. The precise values of the family $\zeta_L$
governing the similar non-intersection
properties of $L$ Brownian paths were later conjectured from conformal invariance and numerical studies in \cite{duplantier2}
(see also \cite{sokal,burdzy}). They correspond to a CFT with central charge $c=0$. Interestingly
enough, however, their analytic derivation  resisted  attempts by standard ``Coulomb-gas'' techniques.

\subsubsection{Spanning Trees and LERW}
The related random process, the ``loop-erased random walk'', introduced in \cite{duke}, in which the loops of a simple RW are erased
sequentially, could also be expected to be accessible
to a rigorous approach. Indeed, it can be seen as the backbone of a spanning tree, and
the Coulomb gas predictions for the associated exponents \cite{D6,majumdar} were
obtained rigorously by  determinantal or Pfaffian techniques by R. Kenyon \cite{kenyon1}, in addition to the
conformal invariance of crossing probabilities \cite{kenyon2}. They correspond to a CFT with central charge $c=-2$.

\subsubsection{Conformal Invariance and Brownian Cascade Relations}
The other route was followed by W. Werner \cite{werner}, joined later by G.F. Lawler, who concentrated
on  Brownian path intersections, and on their general conformal invariance properties. They derived in particular important
``cascade relations'' between Brownian intersection exponents of packets of Brownian paths \cite{lawler2}, but still without a
derivation of the conjectured values of the latter.

\subsection{Quantum Gravity}
\subsubsection{QG and Brownian Paths, SAW's and Percolation}
In the Brownian cascade structure the author recognized an underlying quantum gravity structure.
This led to an analytical derivation of
the (non-)intersection exponents for Brownian paths \cite{duplantier7}. The same QG structure, properly understood, also
gave access to exponents
of mixtures of RW's and SAW's, to the harmonic measure multifractal spectra of the latter two \cite{duplantier8},
 of a percolation cluster \cite{duplantier9}, and to the rederivation of
path-crossing exponents in percolation of ref. \cite{ADA}. Mandelbrot's conjecture (\ref{Mand})
also follows from
\begin{equation}
D_{\rm Brown.\, fr.}=2-2{\zeta}_{\frac{3}{2}}=\frac{4}{3}. \label{mandel}
\end{equation}
It was also observed there that the whole
class of Brownian paths, self-avoiding walks, and percolation clusters,
 possesses the same harmonic MF spectrum in two dimensions, corresponding to a unique central charge $c=0$.
 Higher MF spectra were also calculated \cite{duplantier10}.
 Related results were obtained in \cite{lawler3,cardy2}.

\subsubsection{General CI Curves and Multifractality}
The general solution for the potential distribution near any conformal
fractal in 2D  was finally obtained from the same quantum gravity structure \cite{duplantier11}.
The exact multifractal spectra describing the singularities of the harmonic measure
 along the fractal boundary depend only on
the so-called
{\it central charge c}, the parameter which labels the universality class of the underlying CFT.

\subsubsection{Duality}
A corollary  is the existence of a subtle geometrical {\it duality} structure
in boundaries of
random paths or clusters \cite{duplantier11}. For instance, in the Potts model, the {\it external perimeter}
(EP) of a Fortuin-Kasteleyn cluster, which bears the electrostatic charge and is a
{\it simple} (i.e., double point free) curve, differs from the full
cluster's  hull, which bounces onto itself in the scaling limit.  The EP's Hausdorff dimension $D_{\rm EP}$,
 and the hull's Hausdorff dimension $D_{\rm H}$
obey a duality relation:
\begin{equation}
(D_{\rm EP}-1)(D_{\rm H}-1)=\frac{1}{4}\, , \label{D-D}
\end{equation}
where $ D_{\rm EP} \leq D_{\rm H} $.
This generalizes the case of percolation hulls \cite{GA},
elucidated in \cite{ADA}, for which: $D_{\rm EP}=4/3, D_{\rm H}=7/4$. Notice
that the symmetric point of (\ref{D-D}), $D_{\rm EP}=D_{\rm H}=3/2$, gives the
maximum dimension of a  simple
conformally-invariant random curve in the plane.


\subsection{Stochastic L\"owner Evolution}
\subsubsection{SLE and Brownian Paths}
Meanwhile, O. Schramm, trying to reproduce by a continuum stochastic process both the conformal invariance and
Markov properties
of the scaling limit of loop-erased random walks, invented in 1999 the so-called ``Stochastic L\"owner Evolution'' (SLE)
\cite{schramm1},
a process parametrized by an auxiliary one-dimensional Brownian motion of speed $\kappa$. It became quickly recognized as
a breakthrough since
it provided a powerful analytical tool to describe conformally-invariant scaling curves for various values of $\kappa$. It in particular describes
 LERW for $\kappa=2$, and
hulls of critical percolation
clusters for $\kappa=6$. More generally, it was clear that it described the continuum limit of
 hulls of critical clusters, and that the
 $\kappa$ parameter is actually in
one-to-one correspondance to the usual Coulomb gas
coupling constant $g$, $g=4/\kappa$ (see, e.g., \cite{BDjsp}).

Lawler, Schramm and Werner were then able to rigorously derive the Brownian intersection exponents \cite{lawler4}, as well
as Mandelbrot's conjecture \cite{lawler5} by relating them to the properties of ${\rm SLE}_{\kappa=6}$. S. Smirnov related the continuum limit of site percolation on the triangular lattice to
 the ${\rm SLE}_{\kappa=6}$ process \cite{smirnov1}, and derived Cardy's equation \cite{cardy3} from it. Other well-known percolation scaling behaviors
 follow from this \cite{lawler6,smirnov2}. The scaling limit of the LERW has also been rigorously shown to be the ${\rm SLE}_{\kappa=2}$
 \cite{LSWLERW}, as anticipated in \cite{schramm1}, while that of SAW's is expected to correspond to $\kappa=8/3$ \cite{BDjsp,SAWLSW}.

\subsubsection{$\kappa\kappa'=16$ Duality for the ${\rm SLE}_{\kappa}$}
The ${\rm SLE}_{\kappa}$ trace essentially describes boundaries of conformally-invariant random clusters.
For $\kappa \leq 4$, it is a simple path, while for $\kappa > 4$ it bounces onto itself. One can establish a dictionnary between
the results obtained by quantum gravity and Coulomb gas techniques for Potts and $O(N)$ models \cite{duplantier11},
and those concerning the SLE \cite{BDjsp} (see below). The duality equation
(\ref{D-D}) then brings in a $\kappa\kappa'=16$ {\it duality} property \cite{duplantier11,BDjsp}
between Hausdorff dimensions:
\begin{equation}
\left[D_{\rm EP}(\kappa)-1\right] \left[ D_{\rm
H}(\kappa)-1\right]=\frac{1}{4},\ \kappa \geq 4\ ,
\label{dualione}
\end{equation}
where  $$D_{\rm EP}(\kappa)=D_{\rm H}(\kappa'=16/\kappa),\quad
\kappa'\leq 4 $$
gives the dimension of the frontier of
a non-simple ${\rm SLE}_{\kappa \geq 4}$ trace as the Hausdorff dimension of the simple
${\rm SLE}_{16/{\kappa}}$ trace. Actually, this extends to the whole multifractal spectrum of the harmonic measure near the
${\rm SLE}_{\kappa}$, which is identical to that of the ${\rm SLE}_{16/{\kappa}}$ \cite{duplantier11,BDjsp}. From that result was originally stated the duality prediction that
 the frontier of the non-simple ${\rm SLE}_{\kappa \geq 4}$ path
is locally a simple ${\rm SLE}_{16/{\kappa}}$ path \cite{duplantier11,BDjsp,duplantierdual}.

The SLE geometrical properties per se are an active subject of investigations \cite{RS}.  The value of the
 Hausdorff dimension of the SLE trace,
$D_{\rm H}(\kappa)=1+{\kappa}/{8}$, has been obtained rigorously by V. Beffara \cite{beffara}, in agreement with the value predicted by the Coulomb gas approach
\cite{nien,SD,duplantier11, BDjsp}. The duality
(\ref{dualione}) predicts $D_{\rm EP}(\kappa)=1+({\kappa}/{8})\vartheta(4-\kappa)+({2}/{\kappa})\vartheta (\kappa -4)$ for the dimension of the SLE frontier \cite{duplantier11,BDjsp}.

\subsubsection{Recent Developments} The mixed multifractal spectrum
\cite{binder} describing the local rotations and singularities of the harmonic measure near the SLE boundary
has been obtained \cite{DB}.
A two-parameter family of Stochastic L\"owner Evolution processes, the $\rm{SLE}(\kappa,\rho)$ processes,
has been introduced recently \cite{CRLSW}.
The relationship of $\rm{SLE}_{\kappa}$ to standard conformal field theory has been pointed out and developed 
 recently \cite{CRLSW,BB,LF1}. Boundary correlators in 2D
quantum gravity, which were originally calculated via the Liouville
field theory \cite{FZZ,PT}, and are related to our quantum gravity approach, have been recovered from discrete models
on a random lattice \cite{KKK}. A description of collections of SLE's in terms
of Dyson's circular ensembles has been proposed \cite{cardy2003}. The two-parameter 
family $\rm{SLE}(\kappa,\rho)$ has been studied further \cite{WW}, in
particular in relation to the duality property mentioned above \cite{dubedat}.

\subsection{Synopsis}\footnote{The first part of this paper (sections \ref{sec.inter}-\ref{sec.higher})
is a slightly expanded version of ref. \cite{BDjsp}. The second part (sections \ref{sec.winding}-\ref{sec.multifSLE}
and appendices) gives a detailed description
of local windings and singularities along CI scaling curves. It then focuses on SLE and quantum gravity, and on their various dualities,
mirrored in that relating simple and non-simple paths. Finally, it offers detailed arguments
leading to the quantum gravity approach
to interacting random paths.}
The aim of the present article is to give a comprehensive description of conformally-invariant fractal geometry,
and of its underlying quantum gravity structure. In particular, we show how the repeated use of KPZ maps between
the critical exponents in the complex plane $\mathbb C$ and those in quantum gravity
allows the determination of a very large class of critical exponents arising in planar critical statistical systems, including the
multifractal ones, and their reducing to simple irreducible elements. Two key elements are relating the bulk exponents
in quantum gravity to the Dirichlet boundary ones, and establishing simple additivity rules for the latter. Within this
unifying perspective, we cover many well-recognized geometrical models, like RW's or SAW's and their intersection properties,
Potts and $O(N)$ models, and the multifractal properties thereof. 

We also adapt the quantum gravity
formalism to the ${\rm SLE}_{\kappa}$ process, revealing there a hidden algebraic duality in the KPZ map itself, which in turn
translates into the geometrical $\kappa \to \kappa'=16/\kappa$ duality between simple and non-simple SLE traces.
This KPZ algebraic duality also explains the duality which exists within the class of Potts and $O(N)$ models
between hulls and external frontiers.

In section \ref{sec.inter} we  first establish the values of the intersection exponents of random walks
or Brownian paths from quantum gravity. The combinatorial details are given in appendix \ref{Brownapp}.
In section \ref{sec.mixing} we then move to the critical properties of arbitrary sets mixing simple random walks
or Brownian paths and self-avoiding walks, with arbitrary interactions thereof. The  multifractal spectrum of the harmonic measure
near Brownian paths or self-avoiding ones is studied in section \ref{sec.harmonic}, including the case of the double-sided potential.
Section \ref{sec.perco} yields the related multifractal spectrum for percolation clusters. This completes the description of the universality 
class of central charge $c=0$.

 We address in section \ref{sec.conform} the general
 solution for the multifractal potential distribution near any conformal
 fractal in 2D, which allows determination of the Hausdorff dimension of the frontier.
 The multifractal spectra depend only on the central charge c, which labels the universality class of the underlying CFT.

 Another feature is the consideration in section \ref{sec.higher} of  higher multifractality, which occurs in a
 natural way in the joint distribution
 of potential on both sides of a random CI scaling path, or more generally, in the distribution of
 potential between the
 branches of a {\it star} made of an arbitrary number of CI paths. The associated universal multifractal spectrum
 will depend on two variables, or more generally, on $m$ variables
 in the case of an $m$-arm star.

 Section \ref{sec.winding} describes the more subtle mixed multifractal spectrum associated with the local rotations
 and singularities along a conformally-invariant curve, as seen by the harmonic measure \cite{binder,DB}.
 Here quantum gravity and Coulomb gas techniques must be fused.

Section \ref{sec.geodual} focuses on the $O(N)$ and Potts models, on the ${\rm SLE}_{\kappa}$, and on the correspondence between
them. This is exemplified for the geometric duality existing between their frontiers and full boundaries or hulls. The various Hausdorff
dimensions of $O(N)$ lines, Potts cluster boundaries, and SLE's traces are given. 

Conformally invariant paths have quite different critical properties and obey different quantum gravity rules, depending on
whether they are
{\it simple paths or not}. The next sections are devoted to the elucidation of this difference, and its treatment
 within a unified framework.

A fundamental algebraic duality which exists in the KPZ map is studied in section \ref{sec.duality}, and applied to the
construction rules for critical exponents associated with non-simple paths versus simple ones. These dual rules
are established in appendices  \ref{ONapp} and \ref{BBapp} from considerations of quantum gravity.

In section \ref{sec.SLEKPZ}, we construct an extended KPZ formalism for the ${\rm SLE}_{\kappa}$ process,
which is valid for all values of the parameter $\kappa$.  It corresponds to the usual KPZ formalism for $\kappa \leq 4$ (simple paths), 
and to the algebraic dual one for $\kappa > 4$ (non-simple paths). The composition rules for calculating critical exponents 
involving several random paths in the SLE process are given, as well as some short-distance expansion results.  The multi-line 
exponents for the SLE, and the equivalent ones for $O(N)$ and Potts models are listed.

Finally, in section \ref{sec.multifSLE} the extended SLE quantum gravity formalism is applied to the calculation of all harmonic measure
exponents near multiple SLE traces, near a boundary or in open space.

Appendix \ref{Brownapp} details the calculation of Brownian non-intersection exponents in quantum gravity.
Appendix \ref{ONapp} describes the calculation of $O(N)$ model exponents from quantum gravity, and establishes
the relation between boundary and bulk exponents and the additivity rules for Dirichlet boundary ones. These QG relations are actually
sufficient to determine all exponents without further calculations. Finally, appendix \ref{BBapp} establishes the
general relation between boundary and bulk exponents in quantum gravity, as well as the boundary additivity rules. They follow from a fairly
universal structure of correlation functions in quantum gravity.


The  quantum gravity techniques used here are perhaps not widely
known in the statistical mechanics community at-large, since they originally belonged to
string or random matrix theory. These techniques, moreover, are
not yet within the realm of rigorous mathematics. However, the
correspondence extensively used here, which exists between scaling
laws in the plane, and on a random Riemann surface appears to be 
fundamental, and, in my opinion, illuminates many of the geometrical properties of conformally-invariant random curves in the plane.\\

\section{\sc{Intersections of Random Walks}}
\label{sec.inter}

\subsection{Non-Intersection Probabilities}
\subsubsection{Planar Case}
\begin{figure}[tb]
\begin{center}
\includegraphics[angle=0,width=.5\linewidth]{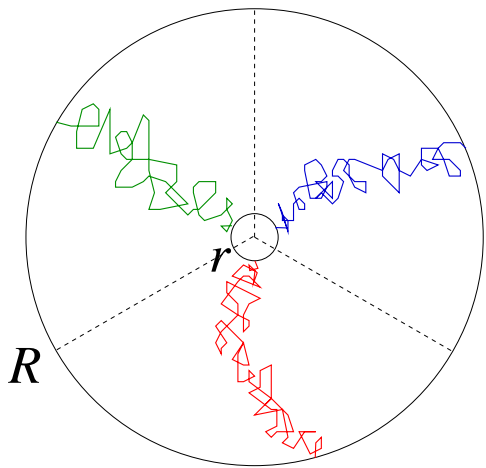}
\end{center}
\caption{Three non-intersecting planar random walks crossing an annulus from $r$ to $R$.}
\label{fig.rw111}
\end{figure}
Let us first define the so-called (non-){\it intersection exponents} for random walks or
Brownian motions. While simpler than the multifractal
exponents considered above, in fact they generate the latter. Consider
a number $L$ of independent random walks
$B^{(l)},l=1,\cdots,L$ in $\mathbb Z ^{2}$ (or Brownian paths in $\mathbb R^{2}=\mathbb C),$
starting at fixed neighboring points, and the probability
\begin{equation}
{ P}_{L}\left( t\right) =P\left\{ \cup^{L}_{l, l'=1} (B^{(l)}\lbrack
0,t\rbrack \cap B^{(l')}\lbrack 0,t\rbrack) =\emptyset \right\},
\label{pl}
\end{equation}
that the intersection of their paths up to time $t$ is
empty \cite{lawler1,duplantier1}. At large times one
expects this probability to decay as
\begin{equation}
{P}_{L}\left( t\right) \approx t^{-\zeta_{L}}, \label{zeta}
\end{equation}
where $\zeta _{L} $ is a {\it universal} exponent
depending only on $L$.
Similarly, the probability that the Brownian paths altogether traverse the
annulus ${\mathcal %
D}\left( r, R\right)$ in $\mathbb C$ from the inner boundary circle of radius $r$ to the outer
one at distance $R$ (Fig. \ref{fig.rw111}) scales as
\begin{equation}
{P}_{L}\left( R\right) \approx \left({r}/{R}\right)^{2\zeta_{L}},
\label{zetaR}
\end{equation}
These exponents can be generalized to $d$ dimensions. Above the upper critical  dimension
$d=4$, RW's  almost surely do not intersect and $\zeta _{L}\left( d \geq 4\right)=0 $. The existence of
exponents $\zeta _{L}$ in $d=2, 3$ and their universality
have been proven \cite{burdzy}, and they can be calculated near $
d=4$ by renormalization theory  \cite{duplantier1}.\\

\subsubsection{Boundary Case}

\begin{figure}[tb]
\begin{center}
\includegraphics[angle=0,width=.5\linewidth]{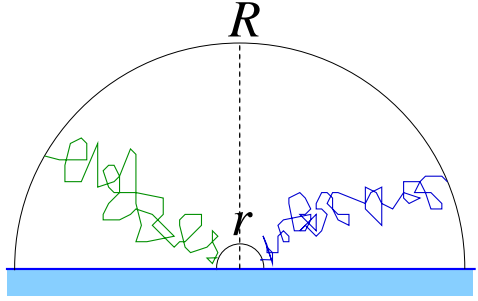}
\end{center}
\caption{Two mutually-avoiding random walks crossing a half-annulus from $r$ to $R$ 
in the half-plane $\mathbb H$.}
\label{fig.rwhalf111}
\end{figure}

A generalization was introduced  \cite{duplantier2} for $L$ walks
constrained to stay in the half-plane $\mathbb H$ with Dirichlet boundary conditions on $\partial \mathbb H$
, and starting at neighboring
points near the boundary. The non-intersection probability
$\tilde{{P}_{L}}\left( t\right) $ of their paths is governed by a
boundary critical exponent $\tilde{\zeta}_{L}$ such that
\begin{equation}
\tilde{P} _{L}\left( t\right) \approx t^{-\tilde{\zeta}_{L}}.
\label{zetat}
\end{equation}
One can also consider the probability that the Brownian paths altogether traverse the
half-annulus ${\mathcal D}\left( r, R\right)$ in $\mathbb H$, centered on the boundary line $\partial H$,
 from the inner boundary circle of radius $r$ to the outer
one at distance $R$ (Fig. \ref{fig.rwhalf111}). It scales as
\begin{equation}
{\tilde {P}}_{L}\left( R\right) \approx \left({r}/{R}\right)^{2\tilde \zeta_{L}}. \label{zetaRt}
\end{equation}

\subsubsection{Conformal Invariance and Weights}
It was first conjectured from conformal invariance arguments and
numerical simulations that in two dimensions  \cite{duplantier2}
\begin{equation}
\zeta _{L}=h_{0, L}^{\left( c=0\right) }=\frac{1}{24}\left(
4L^{2}-1\right), \label{Zeta}
\end{equation}
and for the half-plane
\begin{equation}
2\tilde{\zeta}_{L}=h_{1, 2L+2}^{\left( c=0\right)
}=\frac{1}{3}L\left( 1+2L\right), \label{zC2}
\end{equation}
where $h_{p,q}^{(c)}$ denotes the Ka\v {c} conformal weight
\begin{equation}
h_{p,q}^{(c)}=\frac{\left[ (m+1)p-mq\right] ^{2}-1}{4m\left(
m+1\right) }, \label{K}
\end{equation}
of a minimal conformal field theory of central charge
$c=1-6/[m\left(
m+1\right)] ,$ $m\in \mathbb N ^{*}$  \cite{friedan}. For Brownian motions
$c=0,$ and $m=2.$\\

\subsubsection{Disconnection Exponent}
A discussion of the intersection exponents of random walks a priori requires a number $L \geq 2$ of them. Nonetheless, for $L=1$,
the exponent has a meaning: the non-trivial value $\zeta _{1}=1/8$ actually gives 
the {\it disconnection exponent} governing the probability that
the origin of a single Brownian path remains accessible from infinity
without the path being crossed, hence stays connected to infinity. On a Dirichlet boundary, $\tilde \zeta_1$ retains its
standard value $\tilde \zeta_1=1$,
which can be derived directly, e.g., from the Green function formalism. It corresponds to
a path extremity located on the  boundary, which always stays accessible
due to Dirichlet boundary conditions.\\

\subsection{Quantum Gravity}
\subsubsection{Introduction}
To derive the intersection exponents above, the idea
 \cite{duplantier7} is to map the original random walk problem in
the plane onto a random lattice with planar geometry, or, in other
words, in presence of two-dimensional {\it quantum gravity}
 \cite{KPZ}. The key point is that the random walk
intersection exponents on the random lattice are related to those
in the plane. Furthermore, the RW intersection problem can be
solved in quantum gravity. Thus, the exponents $\zeta_{L}$ (eq.
(\ref{Zeta})) and $\tilde{\zeta}_{L}$ (eq. (\ref{zC2})) in the
standard complex plane or half-plane are derived from this mapping to a random
lattice or Riemann surface with fluctuating metric.

Random surfaces, in relation to string theory  \cite{see2}, have
been the subject and source of important developments in
statistical mechanics in two dimensions. In particular, the
discretization of string models led to the consideration of
abstract random lattices $G$, the connectivity fluctuations of
which represent those of the metric, i.e., pure 2D quantum gravity
 \cite{boulatov}.\\

\subsubsection{KPZ Relation}
One can put any 2D statistical model (e.g.,
Ising model  \cite{kazakov}, self-avoiding walks  \cite{DK}) on the
random planar graph $G$, thereby obtaining a new critical
behavior, corresponding to the confluence of the criticality of
the infinite random surface $G$ with the critical point of the original
model. The critical system ``dressed by gravity'' has a larger
 symmetry under diffeomorphisms. This allowed Knizhnik, Polyakov, and
Zamolodchikov (KPZ)  \cite{KPZ} (see also  \cite{david2}) to establish the
existence of a fundamental relation between the conformal dimensions
$\Delta ^{\left( 0\right) }$ of scaling operators in the
plane and those in presence of gravity, $\Delta$:
\begin{equation}
\label{KPZg}
\Delta ^{\left( 0\right)}=U_{\gamma}(\Delta)=\Delta \frac{
\Delta -\gamma  }{1-\gamma},
\end{equation}
where $\gamma$, the {\it string susceptibility exponent}, is related to the central charge of the
statistical model in the plane:
\begin{equation}
\label{c(g)}
c=1-6\gamma^{2}/\left(1-\gamma\right),\;\; \gamma \leq 0.
\end{equation}
The same relation applies between conformal weights $\tilde \Delta ^{\left( 0\right)}$ in the half-plane $\mathbb H$ and
$\tilde \Delta $ near the boundary of a disk with fluctuating metric:
\begin{equation}
\label{KPZgb}
\tilde \Delta ^{\left( 0\right) }=U_{\gamma}\left(\tilde \Delta\right )=\tilde \Delta \frac{
\tilde \Delta -\gamma  }{1-\gamma}.
\end{equation}

For a minimal model of the series~(\ref{K}), $\gamma=-1/m$,
and the conformal weights in the plane $\mathbb C$ or half-plane $\mathbb H$
are $\Delta _{p,q}^{\left( 0\right) }:=
h_{p,q}^{\left( c\right) }.$\\

\subsubsection{Random Walks in Quantum Gravity}
Let us now consider as a statistical model {\it random walks} on a
{\it random graph}. We know  \cite{duplantier2} that the central
charge $c=0$, whence $m=2$, $\gamma=-1/2.$ Thus the KPZ relation
becomes
\begin{equation}
\Delta ^{\left( 0\right) }={U}_{\gamma=-1/2}\left( \Delta
\right)=\frac{1}{3}\Delta \left( 1+2\Delta
\right):= U(\Delta), \label{KPZ}
\end{equation}
which has exactly the same analytical form as equation
(\ref{zC2})! Thus, from this KPZ equation one infers that the conjectured planar
Brownian intersection exponents  in the complex plane $\mathbb C$ (\ref{Zeta}) and  in $\mathbb H$ (\ref{zC2}) must
  be equivalent to the following Brownian intersection exponents in quantum gravity:
\begin{eqnarray}
 \label{delta}
{\Delta }_{L}&=&\frac{1}{2}\left(L-\frac{1}{2}\right),\\
\label{deltatilde}
\tilde{\Delta }_{L}&=&L.
\end{eqnarray}
Let us now sketch the derivation of these quantum gravity
exponents \cite{duplantier7}. A more detailed proof is given in appendix \ref{Brownapp}.

\subsection{Random Walks on a Random Lattice}
\subsubsection{Random Graph Partition Function}
Consider the set of planar random graphs $G$, built up with,
e.g., trivalent vertices tied together in a {\it random way}. The topology is fixed here to be that of a sphere $
\left( {\mathcal S}\right) $ or a disk $\left( \mathcal {D}\right)
$. The partition function is defined as
\begin{equation}
Z_{}(\beta,\chi )=\sum _{G(\chi)}{\frac{1}{S(G)}}e^{-\beta \left| G\right| },
\label{Zchi1}
\end{equation}
where $\chi $ denotes the fixed Euler characteristic of graph $G$; $\chi=2\;\left(
{\mathcal S}\right) ,1\;\left( {\mathcal D}\right)$; $\left| G\right|$
is the number of vertices of $G$, $S\left( G\right) $ its symmetry
factor. The partition sum converges for all values of the
parameter $\beta $ larger than some critical $\beta_c$. At $\beta
\rightarrow \beta_c^{+},$ a singularity appears due to the
presence of infinite graphs in (\ref{Zchi1})
\begin{equation}
Z_{}\left( \beta , \chi \right) \sim \left( \beta -\beta_c\right)
^{2-\gamma _{\rm str}(\chi)}, \label{Zchi2}
\end{equation}
where $\gamma _{\rm str}(\chi)$ is the string susceptibility
exponent, which depends on the topology of $G$ through the Euler characteristic.
 For pure gravity as described in (\ref{Zchi1}), the
embedding dimension $d=0$ coincides with the central charge $c=0,$
and  \cite{kostov}
\begin{equation}\gamma _{\rm str}(\chi)=2-\frac{5}{4} \chi , (c=0).
\label{gamchi}
\end{equation}
 In particular $\gamma_{\rm
str}(2)=-\frac{1}{2}$ for the spherical topology,  and $\gamma_{\rm
str}(1)=\frac{3}{4}$. The string susceptibility exponent appearing in KPZ formula (\ref{KPZg}) is the planar one
$$\gamma=\gamma _{\rm str}(\chi=2).$$

A particular partition function will play an important role later, that of the doubly punctured sphere. It is defined as
\begin{equation}
 Z\lbrack \hbox to 8.5mm{\hskip 0.5mm 
$\vcenter{\epsfysize=.45truecm\epsfbox{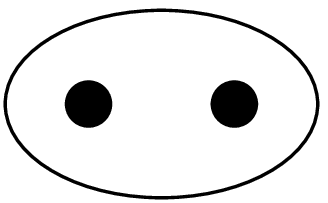}}$
              \hskip 10mm}
 \rbrack \
:= \frac{\partial^{2}}{\partial \beta ^{2}}
Z_{}(\beta,\chi =2)=\sum _{G(\chi=2)}{\frac{1}{S(G)}}\left| G\right|^2 e^{-\beta \left| G\right| }.
\label{Zdotdot}
\end{equation}
Owing to (\ref{Zchi2}) it scales as
\begin{equation}
 Z\lbrack \hbox to 8.5mm{\hskip 0.5mm
$\vcenter{\epsfysize=.45truecm\epsfbox{fig2.eps}}$
              \hskip 10mm}
 \rbrack \
\sim \left( \beta -\beta_c\right)
^{-\gamma _{\rm str}(\chi=2)}.
\label{Zdotdots}
\end{equation}
\\

\subsubsection{Random Walk Partition Functions}
\begin{figure}[tb]
\begin{center}
\includegraphics[angle=0,width=.5\linewidth]{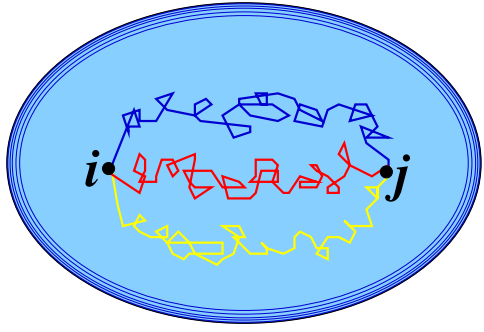}
\end{center}
\caption{$L=3$ mutually-avoiding random walks on a random sphere.}
\label{Fig.wmrw1}
\end{figure}

 Let us now consider a set of $L$ random walks ${\mathcal B}=\{{B}_{ij}^{\left( l\right)
},l=1,...,L\}$ on the {\it random graph} $G$ with the special
constraint that they start at the same vertex $i\in G,$ end at the
same vertex $j \in G$, and have no intersections in between. We
introduce the $L-$walk partition function on the random lattice
 \cite{duplantier7}:
\begin{equation}
Z_L(\beta ,z)=\sum _{{\rm planar}\ G}\frac{1}{ S(G)} e^{-\beta
\left| G\right| }\sum _{i,j\in G} \sum_{\scriptstyle
B^{(l)}_{ij}\atop\scriptstyle l=1,...,L} z^{\left| {\mathcal
B}\right| }, \label{Zl}
\end{equation}
where a fugacity $z$ is associated with the total number $\left|
{\mathcal B}\right| =\left| \cup^{L}_{l=1} B^{(l)}\right| $ of
vertices visited by the walks (Fig. \ref{Fig.wmrw1}).\\

\subsubsection{RW Boundary Partition Functions}
\begin{figure}[tb]
\begin{center}
\includegraphics[angle=0,width=.5\linewidth]{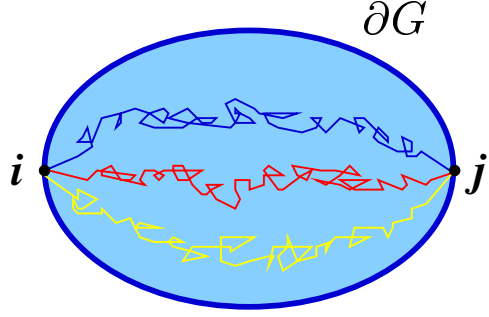}
\end{center}
\caption{$L=3$ mutually-avoiding random walks traversing a random disk.}
\label{Fig.wmrw2}
\end{figure}

We generalize this to the {\it boundary} case where $G$ now has
the topology of a disk and where the random walks connect two
sites $i$ and $j$ on the
boundary $\partial G:$%
\begin{equation}
\tilde Z_{L}(\beta , {\beta}^{\prime}, z)=\sum _{ {\rm disk}\ G}
e^{-\beta \left| G\right| }e^{-{\tilde \beta}{\left| \partial
G\right|}} \sum _{{i,j} \in \partial G}\sum_{{\scriptstyle
B^{(l)}_{ij}}\atop{\scriptstyle l=1,...,L}} z^{\left|{\mathcal
B}\right| },
\label{Ztilde}
\end{equation}
where $e^{-{\tilde \beta}}$ is the fugacity associated with the
boundary's length (Fig. \ref{Fig.wmrw2}).
Again, a particular boundary partition function will play a central role, that
of the disk with two punctures on the boundary. It corresponds to the $L=0$ case of the $\tilde Z_L$'s,
and is defined as
\begin{equation}
Z( \hbox to 9.5mm{\hskip 0.5mm
$\vcenter{\epsfysize=.45truecm\epsfbox{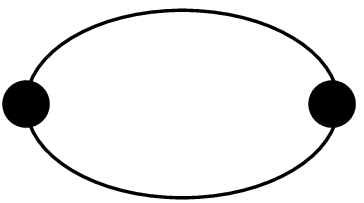}}$
              \hskip -80mm}
)=\tilde Z_{L=0}(\beta , {\beta}^{\prime})=\sum _{ {\rm disk}\ G}
e^{-\beta \left| G\right| }e^{-{\tilde \beta}{\left| \partial
G\right|}} \left| \partial
G\right|^2.
\label{Ztilde0}
\end{equation}

The double grand canonical partition functions (\ref{Zl}) and (\ref{Ztilde})
associated with non-intersecting RW's on a random lattice can be
calculated exactly  \cite{duplantier7}. The detailed calculations are given in appendix \ref{Brownapp}.
One in particular uses an equivalent representation of the random walks by their
{\it forward (or backward) trees}, which are trees uniformly spanning the sets of visited sites. This turns the
RW's problem into the solvable one of random trees on random graphs (see, e.g.,  \cite{DK}).\\

\subsubsection{Scaling Laws for Partition Functions}
The critical behavior of
$Z_{L}$ is characterized by the existence of two critical values of the parameters,
$\beta_c$ where the random lattice size diverges, and $z_c$ where the set
of sites visited by the random walks also diverges. The critical behavior
of $Z_{L}\left( \beta ,z\right)$ is then obtained by taking the
double
scaling limit $\beta \rightarrow \beta_c^{+}$ (infinite random surface) and $
z\rightarrow z_c^{-}$ (infinite RW's), such that the average lattice and RW's sizes respectively scale as
\begin{equation}
|G| \sim (\beta-\beta_c)^{-1},  \left| {\mathcal B}\right| \sim (z_c-z)^{-1}.
\label{sizes}
\end{equation}
The analysis of the singular
behavior in terms of conformal weights is performed by using
{\it finite-size scaling} (FSS)  \cite{DK}, where one must have (see appendix \ref{Brownapp})
$$\left| {\mathcal B}\right| \sim \left| G\right| ^\frac{1}{2} \iff
 z_c-z \sim (\beta-\beta_c)^{1/2}.$$
One obtains in this regime the global scaling of the full partition function
 \cite{duplantier7}:
\begin{equation}
Z_{L}\left( \beta ,z\right) \sim \left( \beta -\beta_c\right) ^{L}
\sim {\left| G\right|} ^{-L}.
\label{Zll}
\end{equation}

The interpretation of partition function $Z_{L}$ (\ref{Zl}) is the following:
It represents a random surface with two {\it
punctures} where two conformal operators, of conformal weight
$\Delta _{L}$, are located (here two vertices of $L$
non-intersecting RW's), and, using a graphical notation, scales as
\begin{equation}
Z_{L}\sim Z\lbrack \hbox to 8.5mm{\hskip 0.5mm
$\vcenter{\epsfysize=.45truecm\epsfbox{fig2.eps}}$
              \hskip 10mm}
 \rbrack \ \times \left| G\right|
^{-2\Delta _{L}},\
\label{Zls}
\end{equation}
where the partition function of the doubly punctured surface is the
second derivative of $Z_{ }(\beta,\chi=2)$ (\ref{Zdotdot}):
\begin{equation}
 Z\lbrack \hbox to 8.5mm{\hskip 0.5mm
$\vcenter{\epsfysize=.45truecm\epsfbox{fig2.eps}}$
              \hskip 10mm}
 \rbrack \
= \frac{\partial^{2}}{\partial \beta ^{2}}
Z_{}(\beta,\chi =2).
\label{Zdots}
\end{equation}
From (\ref{Zdotdots}) we find
\begin{equation}
Z_{L}\sim  \left| G\right|
^{\gamma _{\rm str}(\chi =2)-2\Delta _{L}}.\
\label{Zlds}
\end{equation}
Comparing the latter to (\ref{Zll}) yields
\begin{equation}
2\Delta _{L}-\gamma _{\rm str}(\chi =2)=L, \label{deltal}
\end{equation}
where we recall that $\gamma _{\rm str}(\chi =2)=-1/2$. We thus get the first previously-announced
result
\begin{equation}
\Delta _{L}=\frac{1}{2}\left(L-\frac{1}{2}\right).
\label{deltaL}
\end{equation}
\\

\subsubsection{Boundary Scaling}
For the boundary partition function $\tilde Z_{L}$ (\ref{Ztilde})
a similar analysis can be performed near the triple critical point $(\beta_c,\tilde\beta_c,z_c)$,
where the boundary length also diverges. One finds that the average boundary length $|\partial G|$
must scale with the area $|G|$ in a natural way (see appendix \ref{Brownapp})
\begin{equation}
|\partial G|\sim |G|^{1/2}.
\label{FSSdGG}
\end{equation}
The boundary partition
function $ \tilde Z_{L}$ corresponds to two boundary operators of
conformal weights ${\tilde \Delta}_{L},$ integrated over $\partial G,$
on a random surface with the topology of a disk, or in terms of scaling behavior
\begin{equation}
\tilde Z_{L}\sim Z( \hbox to 9.5mm{\hskip 0.5mm
$\vcenter{\epsfysize=.45truecm\epsfbox{fig3.eps}}$
              \hskip -80mm}
) \times \left|\partial G\right| ^{-2\tilde{\Delta }_{L}},
\label{Zlts}
\end{equation}
using the graphical representation of the two-puncture partition function (\ref{Ztilde0}).
\\

\subsubsection{Bulk-Boundary Relation}
From the exact calculation of the {\it boundary} partition
function (\ref{Ztilde}), and of the boundary puncture {\it disk} one (\ref{Ztilde0}),
one gets the further scaling equivalence to the
{\it bulk} partition function (\ref{Zl}):
\begin{equation}
Z_{L} \sim \frac{\tilde Z_{L}}{Z( \hbox to 9.5mm{\hskip 0.5mm
$\vcenter{\epsfysize=.45truecm\epsfbox{fig3.eps}}$
              \hskip -80mm}
)},
\label{ratio}
\end{equation}
where the equivalence holds true in terms of scaling behavior. It intuitively means that carving away from
the $L$-walk boundary partition function the contribution of one connected domain with two
boundary punctures brings us back to the $L$-walk bulk partition function. This fundamental scaling equivalence
is explained in appendix \ref{Brownapp}.
Comparing eqs.~(\ref{Zlts}), (\ref{ratio}), and~(\ref{Zlds}), and
using the FSS (\ref{FSSdGG}) gives
\begin{equation} {\tilde \Delta}_{L}=2\Delta_L-\gamma_{\rm str} (\chi=2).
\label{deltat}
\end{equation}
This relation between bulk and Dirichlet boundary behaviors in quantum gravity is quite general and will also
play a
fundamental role
in the study of other critical systems in two dimensions. It is studied in full detail in
appendices  \ref{ONapp} and \ref{BBapp}. From (\ref{deltal}) we finally find the second annonced result:
\begin{equation} {\tilde \Delta}_{L}=L.
\label{tdeltal}
\end{equation}

Applying the quadratic KPZ relation (\ref{KPZ}) to $\Delta
_{L}$ (\ref{deltaL})
and ${\tilde \Delta}_{L}$ (\ref{tdeltal}) above finally yields the values in
the plane ${\mathbb C}$ or half-plane $\mathbb H$
\begin{eqnarray}
\nonumber
\zeta _{L}&=&{U}_{\gamma=-1/2}\left( \Delta_L
\right)=\frac{1}{24} \left( 4L^2-1\right)\\
\nonumber
2\tilde \zeta _{L}&=&{U}_{\gamma=-1/2}\left( \tilde \Delta_L
\right)=\frac{1}{3}L \left( 1+2L \right),
\end{eqnarray}
\\
as previously annonced.

\vfill\eject
\subsection{Non-Intersections of Packets of Walks}
\subsubsection{Definition}
\begin{figure}[tb]
\begin{center}
\includegraphics[angle=0,width=.5\linewidth]{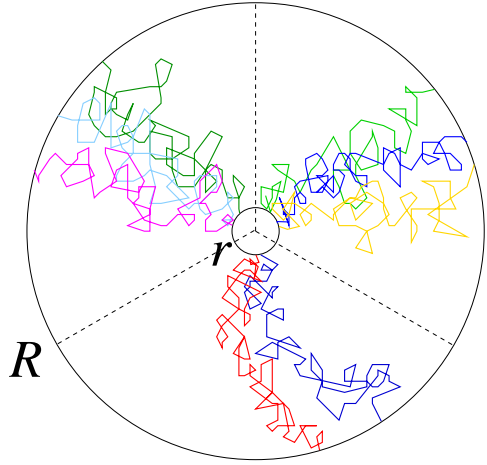}
\end{center}
\caption{Packets of $n_1=3, n_2=3$, and $n_3=2$ independent planar random walks, in a mutually-avoiding 
star configuration, and crossing an annulus from $r$ to $R$.}
\label{fig.rw332}
\end{figure}
\begin{figure}[tb]
\begin{center}
\includegraphics[angle=0,width=.5\linewidth]{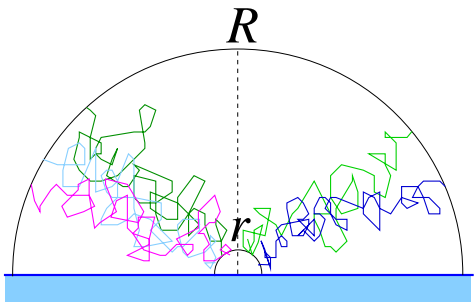}
\end{center}
\caption{Packets of $n_1=3$, and $n_2=2$ independent random walks, in a mutually-avoiding 
star configuration, and crossing the half-annulus from $r$ to $R$ in the half-plane $\mathbb H$.}
\label{fig.rwhalf332}
\end{figure}

Consider configurations made of $L$ mutually-avoiding bunches $l=1,\cdots,L$, each of them
made of $n_{l}$ walks {\it transparent} to each
other, i.e., $n_l$ independent RW's  \cite{werner}. All of them start at neighboring points (Fig. \ref{fig.rw332}).
The probability of non-intersection of the $L$ packets up to time $t$
scales as
\begin{equation}
P_{n_{1},\cdots,n_{L}}(t) \approx t^{-\zeta
(n_{1},\cdots,n_{L})},
\label{Pbunch}
\end{equation}
and near a Dirichlet boundary (Fig. \ref{fig.rwhalf332})
\begin{equation}
\tilde P_{n_{1},\cdots,n_{L}}(t) \approx t^{-\tilde \zeta
(n_{1},\cdots,n_{L})}.
\label{Ptbunch}
\end{equation}
The original case of $L$ mutually-avoiding RW's corresponds to $n_{1}=..=n_{L}=1$.
The probability for the same $L$ Brownian path packets to cross the annulus $\mathcal D(r,R)$ in $\mathbb C$
(Fig. \ref{fig.rw332})
scales accordingly as
\begin{equation}
P_{n_{1},\cdots,n_{L}}(r) \approx \left(r/R\right)^{-2\zeta
(n_{1},\cdots,n_{L})},
\label{Pbunchr}
\end{equation}
and, near a Dirichlet boundary in $\mathbb H$ (Fig. \ref{fig.rwhalf332}), as
\begin{equation}
\tilde P_{n_{1},\cdots,n_{L}}(r) \approx \left(r/R\right)^{-2\tilde \zeta
(n_{1},\cdots,n_{L})}.
\label{Ptbunchr}
\end{equation}

The generalizations of former exponents $\zeta_L$, as well as ${\tilde \zeta}_{L}$, describing these $L$ packets
can be written as
conformal weights
$$\zeta (n_{1},\cdots,n_{L})=\Delta
^{(0)}
\left\{ n_{l}\right\}$$ in the plane ${\mathbb C}$, and $$2{\tilde \zeta}(n_{1},\cdots,n_{L}) ={\tilde %
\Delta}^{(0)}\left\{ n_{l}\right\} $$ in the half-plane $\mathbb H$. They can be calculated from quantum gravity, via
their conterparts $\Delta\left\{ n_{l}\right\}$ and $\tilde \Delta \left\{ n_{l}\right\}$. The details are given in
appendix \ref{Brownapp}. We here sketch the main steps.

\subsubsection{Boundary Case}
One introduces the analogue
${\tilde Z}\left\{ n_{1},\cdots, n_{L}\right\}$ of partition function (\ref{Ztilde}) for the $L$ packets of walks.
In presence of
gravity each bunch contributes its own {\it normalized
boundary partition function} as a factor, and this yields a natural
generalization of the scaling equation~(\ref{ratio}) (see appendix \ref{Brownapp})
\begin{equation}
\frac{{\tilde Z}\left\{ n_{1},\cdots, n_{L}\right\}}{
Z( \hbox to 9.5mm{\hskip 0.5mm
$\vcenter{\epsfysize=.45truecm\epsfbox{fig3.eps}}$
              \hskip -80mm}
)} \sim \prod _{l=1}^{L} \star \left\{\frac{{\tilde Z}\left( n_{l}\right)}{Z(
\hbox to 9.5mm{\hskip 0.5mm
$\vcenter{\epsfysize=.45truecm\epsfbox{fig3.eps}}$
              \hskip -80mm}
)}\right\}, \label{Zn}
\end{equation}
where the star product is to be understood as a scaling equivalence. Given the definition of boundary conformal weights
(see (\ref{Zlts})), the normalized left-hand fraction is to be identified
with $\left|\partial G\right| ^{-2{\tilde \Delta} \left\{n_{1},\cdots,n_{L}\right\}}$, while each normalized factor
${\tilde Z}\left( n_{l}\right)/{Z(
\hbox to 9.5mm{\hskip 0.5mm
$\vcenter{\epsfysize=.45truecm\epsfbox{fig3.eps}}$
              \hskip -80mm}
)}$ is to be identified
with $\left|\partial G\right| ^{-2{\tilde \Delta} \left( n_{l}\right)}$.
 Here ${\tilde \Delta}(n)$ is the boundary dimension of a
{\it single} packet of $n$ mutually transparent walks on the random surface.
 The {\it factorization} property
(\ref{Zn}) therefore immediately implies the {\it additivity of boundary
conformal dimensions in presence of gravity}
\begin{equation}
{\tilde \Delta}\left\{n_{1},\cdots,n_{L}\right\} =\sum ^{L}_{l=1}
{\tilde \Delta}(n_{l}). \label{deltan}
\end{equation}
In the standard plane $\mathbb C$, a packet of $n$ independent random walks has a trivial boundary conformal dimension
${\tilde \Delta}^{(0)}(n)=n{\tilde \Delta}^{(0)}(1)=n,$ since for a single walk ${\tilde \Delta}^{(0)}(1)=1,$
as can be seen using the Green function formalism. We therefore know
${\tilde \Delta}(n)$ exactly, since it suffices to take the positive {\it inverse} of the KPZ map
(\ref{KPZ}) to get
\begin{equation}
{\tilde \Delta}(n)={U}_{\gamma=-1/2}^{-1}(n)=\frac{1}{4}(\sqrt{24n+1}-1).
\label{deltatn}
\end{equation}
One therefore finds:
\begin{equation}
{\tilde \Delta}\left\{n_{1},\cdots,n_{L}\right\} =
\sum ^{L}_{l=1}{U}_{\gamma=-1/2}^{-1}(n_l)=\sum ^{L}_{l=1}\frac{1}{4}(\sqrt{24n_l+1}-1).
\label{deltan'}
\end{equation}

\subsubsection{Relation to the Bulk}
One similarly defines for $L$ mutually-avoiding packets of $n_{1},\cdots,n_{L}$ independent walks
the generalization $Z\left\{  n_{1},\cdots, n_{L}\right\}$ of the bulk
partition function (\ref{Zl}) for $L$ walks on a random sphere. One then establishes on a random surface
the identification, similar to (\ref{ratio}), of
this bulk partition function with the normalized boundary one (see appendix \ref{Brownapp}):
\begin{equation}
Z\left\{  n_{1},\cdots, n_{L}\right\} \sim \frac{{\tilde Z}\left\{ n_{1},\cdots, n_{L}\right\}}{
Z( \hbox to 9.5mm{\hskip 0.5mm
$\vcenter{\epsfysize=.45truecm\epsfbox{fig3.eps}}$
              \hskip -80mm}
)} . \label{ZnZntilde}
\end{equation}
By definition of quantum conformal weights (appendix \ref{Brownapp}), the left-hand term of (\ref{ZnZntilde})
scales as $|G|^{-2\Delta \left\{n_{1},\cdots,n_{L}\right\}+\gamma _{\rm str}(\chi =2)}$, while the
right-hand term scales, as written above, as $\left|\partial G\right| ^{-2{\tilde \Delta}
\left\{n_{1},\cdots,n_{L}\right\}}$.
Using the area to perimeter scaling relation (\ref{FSSdGG}), we thus get the identity
existing in quantum gravity between bulk and boundary conformal weights, similar to (\ref{deltal}):
\begin{equation}
2\Delta \left\{  n_{1},\cdots, n_{L}\right\}-\gamma _{\rm str}(\chi =2)=
{\tilde \Delta}\left\{n_{1},\cdots,n_{L}\right\}, \label{deltanL}
\end{equation}
with   $\gamma _{\rm str}(\chi =2)=-\frac{1}{2}$ for pure gravity.\\

\subsubsection{Back to the Complex Plane}
In the plane, using once again the KPZ relation (\ref{KPZ}) for
${\tilde \Delta}\left\{ n_{l}\right\}$  and ${\Delta}\left\{ n_{l}\right\}$, we obtain the general results
 \cite{duplantier7}
\begin{eqnarray}
\nonumber
2{\tilde
\zeta}(n_{1},\cdots,n_{L})&=&\tilde \Delta^{(0)}\{n_{1},\cdots,n_{L}\}=U\left(\tilde \Delta\left\{n_{1},\cdots,n_{L}
\right\}\right)\\
\nonumber
\zeta (n_{1},\cdots,n_{L})  &=&\Delta^{(0)}\{n_{1},\cdots,n_{L}\}=U\left(\Delta\left\{n_{1},\cdots,n_{L}\right\}\right),
\label{'ZetaL}
\end{eqnarray}
where we set $U:={U}_{\gamma=-1/2}$. One can finally write, using (\ref{deltatn}) and (\ref{deltan'})
\begin{eqnarray}
\label{Zetall}
2{\tilde \zeta}(n_{1},\cdots,n_{L})&=&U(x)=\frac{1}{3} x(1+2x)\\
\label{Zetal}
\zeta (n_{1},\cdots,n_{L})&=& V(x):=U\left[\frac{1}{2}\left(x-\frac{1}{2}\right)\right]=\frac{1}{24} (4x^{2}-1),\\
x&=&\sum_{l=1}^{L}{U}^{-1}(n_{l})=\sum_{l=1}^{L}\frac{1}{4}(\sqrt{24n_{l}+1}-1).
\label{ZetaL}
\end{eqnarray}
Lawler and Werner  \cite{lawler2} established
the existence of two functions $U$ and $V$ satisfying the
structure (\ref{Zetall}-\ref{ZetaL}) by purely probabilistic means,
using the geometrical conformal invariance of Brownian motions. The quantum gravity approach
here explains this structure in terms of linearity of boundary quantum gravity
(\ref{deltan},\ref{deltan'}), and yields the explicit functions
\begin{eqnarray}
U(x)&=&{U}_{\gamma=-1/2}(x)\\
V(x)&=&U\left[\frac{1}{2}\left(x-\frac{1}{2}\right)\right],
\end{eqnarray}
as KPZ maps (\ref{Zetall}--\ref{Zetal}). The same expressions for these
functions has also been derived in probability
theory from the equivalence to ${\rm SLE}_6$  \cite{lawler4}.

\subsubsection{Particular Values and Mandelbrot's Conjecture}
The first few values are:
\begin{eqnarray}
\nonumber
{\tilde \Delta}(n=1)&=&{U}_{\gamma=-1/2}^{-1}(1)=1\\
\nonumber
{\tilde \Delta}(n=2)&=&{U}_{\gamma=-1/2}^{-1}(2)=\frac{3}{2}\\
\nonumber
{\tilde \Delta}(n=3)&=&{U}_{\gamma=-1/2}^{-1}(3)=\frac{1}{4}(\sqrt{73}-1).
\end{eqnarray}
Let us introduce the notation $1^{(
L)}=\stackrel{L}{\overbrace{1,1,\cdots,1}}$ for $L$ mutually-avoiding walks in a star
configuration. Then the exponent $\zeta (2,1^{(
L)})$ describing a two-sided walk and $L$ one-sided walks, all
mutually-avoiding, has the value
\begin{eqnarray}
\nonumber
\zeta (2,1^{(L)})&=&V\left[{L {U}^{-1}(1)+{U}^{-1}(2)}\right]=V(L +\frac{3}{2})\\
\nonumber
&=&\zeta_{L +\frac{3}{2}}=\frac{1}{6}(L+1)(L+2).
\end{eqnarray}
For $L=1$, $\zeta (2,1)=\zeta_{L=5/2}=1$ correctly gives
the exponent governing the escape probability of a RW from a given origin near another RW  \cite{lawlerisrael}.
(By construction the second one indeed appears as made of two independent RW's diffusing away from the origin.)

For $L=0$ one finds the non-trivial result
$$\zeta
(2,1^{(0)})=\zeta_{L=3/2}=1/3,$$
which describes the accessible points along a RW. It is formally related to the Hausdorff
dimension of the Brownian frontier by $D=2-2\zeta$ \cite{lawler}. Thus we
obtain for the dimension of the Brownian frontier  \cite{duplantier7}
\begin{equation}
D_{\rm Br.\, fr.}=2-2{\zeta}_{\frac{3}{2}}=\frac{4}{3}, \label{mand}
\end{equation}
i.e., the famous {\it Mandelbrot conjecture}. Notice that the accessibility of a point on a
Brownian path is a statistical constraint equivalent to the non-intersection of $L=3/2$ paths.
(The relation of this Hausdorff dimension
to the exponent $\zeta_{3/2}=1/3$ was actually made in December 1997, after a discussion in the
Institute for Advanced Study at Princeton with M. Aizenman and R. Langlands about the meaning of half-integer
indices in critical percolation exponents.) The Mandelbrot conjecture was later
established in probability theory  \cite{lawler5}, using the
analytic properties of the non-intersection exponents derived from the
stochastic L\"owner evolution ${\rm SLE}_6$  \cite{schramm1}.

The quantum geometric structure made explicit here allows
generalizations to self-avoiding walks and percolation, which we now describe.\\

\section{\sc{Mixing Random \& Self-Avoiding Walks}}
\label{sec.mixing}
We now generalize the scaling structure obtained in the preceding
section to arbitrary sets of random or self-avoiding walks interacting together
 \cite{duplantier8} (see also  \cite{lawler2,lawler3})

\subsection{General Star Configurations}

\subsubsection{Star Algebra}
Consider a
general copolymer ${\mathcal S}$ in the plane ${\mathbb C}$ (or
in ${\mathbb Z }^{2}$), made of an arbitrary mixture of RW's or Brownian paths $%
\left( \makebox{set}{\rm \;}{\mathcal B}\right) ,$ and 
SAW's or polymers $\left( \makebox{set}{\rm \;}{\mathcal P}\right)$, all
starting at neighboring points, and diffusing away, i.e., in a {\it star} configuration.
In the plane, any successive pair $\left( A,B\right) $ of
such paths, $A,B\in {\mathcal B}$ or ${\mathcal P},$ can be
constrained in a specific way: either they avoid each other
$\left( A\cap B=\emptyset ,\makebox{ denoted }A\wedge B\right) ,$ or
they are independent, i.e., ``transparent'' and can cross each
other (denoted $A\vee B)$ \cite{duplantier8,ferber}. This notation
allows any {\it nested} interaction structure  \cite{duplantier8};
one can decide for instance that the
branches $%
\left\{ {\mathcal P}_{\ell }\in {\mathcal P}\right\} _{\ell =1,...,L}$ of an
$L$-star
polymer, all mutually-avoiding, further avoid a collection of Brownian paths $%
\left\{ \mathcal B_{k}\in {\mathcal B}\right\} _{k=1,...,n},$ all
transparent to each other, which structure is represented by:
\begin{equation}
{\mathcal S}=\left( \bigwedge\nolimits_{\ell =1}^{L}{\mathcal P}_{\ell
}\right) \wedge \left( \bigvee\nolimits_{k=1}^{n}\mathcal B_{k}\right) .
\label{vw}
\end{equation}
A priori in 2D the order of the branches of the star polymer 
 matters and is intrinsic to the $\left( \wedge ,\vee \right)$ notation.

 \subsubsection{Conformal Operators and Scaling Dimensions}
To each {\it specific} star copo\-lymer center ${\mathcal S}$ is
attached a local conformal scaling operator $\Phi_{\mathcal S}$, which represents the presence of the
star vertex, with a scaling dimension
$x\left( {\mathcal S}\right)$  \cite{duplantier4,DS2,duplantier8}. When the star is constrained to
stay in a {\it half-plane} $\mathbb H$, with Dirichlet boundary conditions, and
 its core placed near the {\it
boundary} $\partial \mathbb H$, a new boundary scaling operator ${\tilde \Phi}_{\mathcal S}$ appears, with a boundary
scaling dimension $\tilde{x}\left( {\mathcal S}\right)$  \cite
{duplantier4,DS2}. To obtain proper scaling, one has to construct
 the partition functions of Brownian paths and polymers
having the same mean size $R$  \cite{duplantier4}. These partition functions then scale as powers of $R$,
with an exponent which mixes the
scaling dimension of the star core ($x\left( {\mathcal S}\right)$ or $\tilde{x}\left( {\mathcal S}\right)$),
with those of star dangling ends.

\subsubsection{Partition Functions}
It is convenient to define for each star $%
{\mathcal  S}$ a grand canonical partition function  \cite
{duplantier4,DS2,ferber}, with fugacities $z$ and $z^{\prime }$ for
the total lengths $\left| {\mathcal  B}\right| $ and $\left| {\mathcal  P}\right| $ of
RW or SAW paths:
\begin{equation}
{\mathcal  Z}_{R}\left( {\mathcal  S}\right) =\sum_{{\mathcal  B},{\mathcal  P}\subset {\mathcal  S}%
}z^{\left| {\mathcal  B}\right| }z^{\prime \left| {\mathcal  P}\right| }\;{\bf 1}%
_{R}\left( {\mathcal  S}\right) ,  \label{zr}
\end{equation}
where one sums over all RW and SAW configurations respecting the mutual-avoi\-dance constraints built in star ${\mathcal  S}$ 
(as in (\ref{vw})),
further
constrained by the indicatrix $
\;{\bf 1}_{R}\left( {\mathcal  S}\right) $ to stay within a disk of radius $R$
centered on the star. At the critical values $z_{c}=\mu
_{B}^{-1},z_{c}^{\prime }=\mu _{P}^{-1},$ where  $\mu _{B}$ is
the coordination number of the underlying lattice for the RW's, and $\mu _{P}$ the
effective one for the SAW's, ${\mathcal  Z}_{R}$ has a power law decay  \cite
{duplantier4}
\begin{equation}
{\mathcal  Z}_{R}\left( {\mathcal  S}\right) \sim R^{-x\left( {\mathcal  S}\right) -x^{\bullet
}}.  \label{zrs}
\end{equation}
Here $x\left( {\mathcal  S}\right) $ is the scaling dimension of the operator $\Phi_{\mathcal S}$,
 associated only with the singularity
occurring at the center of the star where all critical paths meet, while $x^{\bullet}$ 
is the contribution of the independent dangling ends. 
It reads $x^{\bullet   }=\left\| {\mathcal  B}\right\| x_{B,1}+\left\| {\mathcal  P}%
\right\| x_{P,1}-2{\mathcal  V},$ where $\left\| {\mathcal  B}\right\| $ and $\left\|
{\mathcal  P}\right\| $ are respectively the total numbers of Brownian or polymer
paths of the star; $x_{B,1}$ or $x_{P,1}$ are the scaling dimensions of the
extremities of a {\it single} RW ($x_{B,1}=0$) or SAW ($x_{P,1}=\frac{5}{48}$)%
 \cite{duplantier4,nien}. The last term in (\ref{zrs}), in which ${\mathcal  V}=\left\| {\mathcal  B}
\right\| +\left\| {\mathcal  P}\right\| $ is the number of dangling vertices, corresponds  to the
integration over extremity positions in the disk of radius $R$.

When the star is constrained to stay in a {\it half-plane} with its core
placed near the {\it boundary}, its partition function scales as  \cite
{duplantier4,duplantier2}
\begin{equation}
\tilde{{\mathcal  Z}}_{R}\left( {\mathcal  S}\right) \sim R^{-\tilde{x}\left( {\mathcal  S}%
\right) -x^{\bullet   }},  \label{zz}
\end{equation}
where $\tilde{x}\left( {\mathcal  S}\right) $ is the boundary scaling dimension, $%
x^{\bullet  }$ staying the same for star extremities in the bulk.

\vfill\eject
\subsection{Quantum Gravity for SAW's \& RW's}

\subsubsection{Scaling Dimensions and Conformal Weights}
Any scaling dimension $x$ in the plane is twice the {\it conformal
weight} $\Delta ^{(0)}$ of the corresponding operator,
while near a boundary they are identical  \cite{BPZ,cardylebowitz}
\begin{equation}
x=2\Delta ^{\left( 0\right) },\quad
\tilde{x}=\tilde{\Delta}^{\left( 0\right) }.  \label{xdelta}
\end{equation}

\subsubsection{KPZ map}
As in section \ref{sec.inter}, the idea is to use the representation where the RW's or
SAW's are on a 2D random lattice, or a random Riemann surface,
i.e., in the presence of 2D {\it quantum gravity} (QG)  \cite
{KPZ}. The general relation (\ref{KPZ}) for Brownian paths depends only on the
central charge $c=0$, which also applies to self-avoiding walks or polymers.  For a critical
system with central charge $c=0$, the two universal functions:
\begin{eqnarray}
U\left( x\right) =U_{\gamma=-\frac{1}{2}}\left( x\right)=\frac{1}{3}x\left( 1+2x\right) , \hskip2mm
V\left( x\right) =\frac{1}{24}\left( 4x^{2}-1\right) ,  \label{U}
\end{eqnarray}
with $V\left( x\right) := U\left(\frac{1}{2}\left(x-\frac{1}{2}\right) \right)$,
generate all the scaling exponents. They transform the conformal weights in bulk quantum gravity,
${\Delta}$, or in boundary QG, $\tilde{\Delta}$, into the plane and half-plane ones (\ref{xdelta}):
\begin{eqnarray}
{\Delta}^{\left( 0\right)}=U({\Delta}),\;
\tilde{\Delta}^{\left( 0\right)}=U(\tilde{\Delta}),\;
{\Delta}^{\left( 0\right)}=V(\tilde{\Delta}).
\label{KPZSAW}
\end{eqnarray}

\subsubsection{Composition Rules}
Consider two  stars $A,B$ joined at their centers, and in a random {\it mutually-avoiding} star-configuration
$ A\wedge B$. Each star is made
of an arbitrary collection of Brownian paths and self-avoiding paths with arbitrary interactions of type (\ref{vw}).
Their respective bulk partition functions (\ref{zr}), (\ref{zrs}), or boundary partition functions
(\ref{zz}) have associated
planar scaling exponents $x\left( A\right) ,x\left( B\right)
,$ or planar boundary exponents $\tilde{x}\left( A\right)
,\tilde{x}\left( B\right)$. The corresponding scaling dimensions in {\it quantum gravity} are then,
for instance for $A$:
\begin{eqnarray}
\label{deltaA}
\tilde {\Delta}\left( A\right)=U^{-1}\left( \tilde{x}\left( A\right)\right),\;\;\;\;
{\Delta}\left( A\right)=U^{-1}\left[\frac{1}{2}{x}\left( A\right)\right],
\end{eqnarray}
where $U^{-1}\left( x\right) $ is the positive inverse of the KPZ map $U$
\begin{equation}
U^{-1}\left( x\right) =\frac{1}{4}\left( \sqrt{24x+1}-1\right) .
\label{u1}
\end{equation}
The key properties are given by the following propositions: \\
$\bullet$ {\it In $c=0$ quantum gravity the  boundary and bulk scaling
dimensions of a given random path set are related by:}
\begin{eqnarray}
\tilde {\Delta}( A)=2{\Delta}\left( A\right) -\gamma_{\rm str}(c=0)=2{\Delta}\left( A\right) +\frac{1}{2}.
\label{tdeltaA=deltaA}
\end{eqnarray}
This generalizes the relation (\ref{deltat}) for non-intersecting Brownian paths.\\
$\bullet$ {\it In quantum gravity the  boundary scaling
dimensions of two mutually-avoiding sets is the sum of their respective boundary scaling
dimensions:}
\begin{eqnarray}
\tilde {\Delta}\left( A\wedge B\right)=\tilde {\Delta}\left( A\right)
+\tilde {\Delta}\left(  B\right).
\label{deltaA+deltaB}
\end{eqnarray}
It generalizes identity (\ref{deltan}) for mutually-avoiding packets of Brownian paths.
The  boundary-bulk relation (\ref{tdeltaA=deltaA}) and the fusion rule (\ref{deltaA+deltaB})
 come from simple convolution  properties of partition functions on a random lattice
 \cite{duplantier7,duplantier8}. They are studied in detail in appendices  \ref{Brownapp} and \ref{BBapp}.

The planar scaling exponents $x\left( A\wedge B\right) $ in $\mathbb C$, and $\tilde{x}\left(
A\wedge B\right)$ in $\mathbb H$ of the two mutually-avoiding stars $A\wedge B$
are then given by the KPZ map (\ref{KPZSAW}) in terms of (\ref{deltaA+deltaB})
\begin{eqnarray}
x\left( A\wedge B\right) &=&2V\left[\tilde {\Delta}\left( A\wedge B\right) \right]
= 2V\left[\tilde {\Delta}\left( A\right)
+\tilde {\Delta}\left(  B\right)\right] \\
\tilde{x}\left( A\wedge B\right) &=&U\left[ \tilde {\Delta}\left( A\wedge B\right) \right]
=U\left[\tilde {\Delta}\left( A\right)
+\tilde {\Delta}\left(  B\right) \right]. \label{xprep}
\end{eqnarray}
Owing to (\ref{deltaA}), these scaling
exponents thus obey the {\it star algebra}
 \cite{duplantier7,duplantier8}
\begin{eqnarray}
x\left( A\wedge B\right) &=&2V\left[ U^{-1}\left( \tilde{x}\left(
A\right)
\right) +U^{-1}\left( \tilde{x}\left( B\right) \right) \right]  \label{x} \\
\tilde{x}\left( A\wedge B\right) &=&U\left[ U^{-1}\left(
\tilde{x}\left( A\right) \right) +U^{-1}\left( \tilde{x}\left(
B\right) \right) \right] . \label{xx}
\end{eqnarray}
On a random surface, $U^{-1}\left( \tilde{x} \right)$ is the
boundary conformal weight corresponding to the value $\tilde{x}$ in the upper-half plane
${\mathbb H}$, and the sum of $U^{-1}$
functions in eq. (\ref{x})  linearly represents the mutually-avoiding juxtaposition
$A \wedge B$ of two sets of random paths near the random
frontier, i.e., the operator product of two ``boundary operators'' on the
random surface. The latter sum is mapped by the functions $U$,
$V$, into the scaling dimensions in ${\mathbb H}$ or ${\mathbb C}$
 \cite{duplantier8}.

These fusion rules (\ref{deltaA+deltaB}), (\ref{x}) and (\ref{xx}), which mix bulk and boundary exponents are already
apparent in the derivation of non-intersection exponents for Brownian paths given in section \ref{sec.inter}
and appendix \ref{Brownapp}. They also apply to the $O(N)$ model, as shown in appendix \ref{ONapp},
and are established in all generality in appendix \ref{BBapp}. They can also be seen as recurrence ``cascade'' relations
in ${\mathbb C}$ between successive conformal
Riemann maps of the frontiers of mutually-avoiding paths onto the half-plane boundary
 ${\partial \mathbb H}$, as in the original work  \cite{lawler2} on Brownian paths.

When the random sets $A$ and $B$ are {\it independent} and can
overlap, their scaling dimensions in the standard plane or half-plane are additive by trivial factorization of
partition functions or probabilities \cite{duplantier8}
\begin{eqnarray}
x\left( A\vee B\right)
=x\left( A\right) +x\left( B\right) ,\;\;\;\;
\tilde{x}\left( A\vee B\right) =
\tilde{x}\left( A\right) +\tilde{x}\left( B\right). \label{add}
\end{eqnarray}
This additivity no longer applies in quantum gravity, since overlapping 
paths get coupled by the fluctuations of the metric, and are no longer independent. In contrast, it is replaced by
 the additivity rule (\ref{deltaA+deltaB}) for mutually-avoiding paths (see appendix \ref{BBapp} for a thorough discussion
 of this additivity property).

It is clear at this stage that the set of equations above is {\it
complete.} It allows for the calculation of any conformal
dimensions associated with a star
structure ${\mathcal S}$ of the most general type, as in (\ref{vw}),
involving $\left( \wedge ,\vee \right) $ operations
separated by nested{\it \ }parentheses  \cite{duplantier8}. Here follow some examples.

\subsection{RW-SAW Exponents }
The single extremity scaling
dimensions are for a RW or a SAW near a Dirichlet boundary 
$\partial {\mathbb H}$  \cite {cardy}
\begin{equation}
\tilde{x}_{B}\left( 1\right)=\tilde{\Delta}_{B}^{\left( 0\right)
}\left( 1\right)
=1,\;\tilde{x}_{P}\left( 1\right)=\tilde{\Delta}%
_{P}^{\left( 0\right) }\left( 1\right) =%
{{\frac{5}{8}}}%
,  \label{num}
\end{equation}
or in quantum gravity
\begin{equation}
\label{numbis}
\tilde{\Delta}_{B}\left( 1\right) =U^{-1}\left(
1\right) =1,\;\tilde{\Delta}_{P}\left( 1\right) =U^{-1}\left(\frac{5}{8}\right)=\frac{3}{4}.
\end{equation} 
Because of the star algebra described above these are the only
numerical seeds, i.e., generators, we need.

Consider  packets of $n$ copies of transparent RW's or $m$
transparent SAW's. Their boundary conformal dimensions in ${\mathbb H}$ are
respectively, by using (\ref{add}) and (\ref{num}),
$\tilde{\Delta}_{B}^{\left( 0\right) }\left( n\right) =n$ and
$\tilde{\Delta}_{P}^{\left( 0\right) }\left( m\right)
=\frac{5}{8}m$.  The inverse mapping to the random
surface yields the quantum gravity conformal weights $\tilde{\Delta}_{B}\left( n\right) =U^{-1}\left(
n\right) $ and $\tilde{\Delta}_{P}\left( m\right) =U^{-1}\left(\frac{5}{8}m\right).$ 
The star made of $L$ packets $\ell \in \left\{1,...,L\right\} $, each of them made of $n_{\ell }$ transparent
RW's and of $m_{\ell }$ transparent SAW's, with the $L$ packets
 mutually-avoiding, has planar scaling dimensions
\begin{eqnarray}
\tilde{\Delta}^{\left( 0\right) }\left\{ n_{\ell
},m_{\ell}\right\} &=&U\left( \tilde{\Delta}\left\{ n_{\ell },m_{\ell}\right\}\right) \\
\Delta
^{\left( 0\right)
}\left\{ n_{\ell },m_{\ell }\right\} &=&V\left( \tilde{\Delta}\left\{ n_{\ell },m_{\ell}\right\}%
\right) , \\
\label{gtDelta}
\tilde{\Delta}\left\{ n_{\ell },m_{\ell}\right\}
&=&\sum\nolimits_{\ell =1}^{L}U^{-1}\left( n_{\ell }+
{ {\frac{5}{8}}}%
m_{\ell }\right)\\
\nonumber
&=&\sum\nolimits_{\ell =1}^{L}\frac{1}{4}\left( \sqrt{24\left( n_{\ell}+ {\frac{5}{8}}m_{\ell}\right)+1}-1\right).
\end{eqnarray}
Take, as an example,  a copolymer star ${\mathcal S}_{L,L^{\prime }}$ made of $L$ RW's and $
L^{\prime }$ SAW's, all mutually-avoiding
$\left( \forall \ell=1,\cdots,L ,n_{\ell }=1, m_{\ell}=0;\;\;\forall \ell
^{\prime }=1,\cdots,L',n_{\ell^{\prime } }=0, m_{\ell ^{\prime }}=1\right)$. In quantum gravity the
linear boundary conformal weight (\ref{gtDelta}) is simply\-
$\tilde{\Delta}\left({\mathcal S}_{L,L^{\prime }}\right) =L+\frac{3}{4}L^{\prime }$. By the $U$ and $V$ maps, it gives the scaling dimensions in
  $\partial {\mathbb H}$ and ${\mathbb C}$
\begin{eqnarray*}
\tilde{\Delta}^{\left( 0\right) }\left( {\mathcal S}_{L,L^{\prime }}\right)&=&{\frac {1}{3}}
\left( L+{\frac {3 }{ 4}}L^{\prime }\right) \left( 1+2L+{\frac {3 }{ 2}} L^{\prime }\right)  \\
\Delta ^{\left( 0\right)}\left( {\mathcal S}_{L,L^{\prime }}\right)
&=&{\frac {1 }{ 24}}\left[ 4\left( L+{\frac {3 }{ 4}} L^{\prime }\right) ^{2}-1\right],
\end{eqnarray*}
recovering for $L=0$ the SAW star-exponents  \cite{DS2} and for $L^{\prime}=0$ the RW non-intersection exponents 
in $\partial {\mathbb H}$ and ${\mathbb C}$ obtained in section \ref{sec.inter}
 \begin{eqnarray*}
2\tilde \zeta_L&=&\tilde{\Delta}^{\left( 0\right) }\left( {\mathcal S}_{L,L^{\prime }=0}\right)={\frac {1}{3}}
L \left( 1+2L\right)  \\
\zeta_L&=&\Delta ^{\left( 0\right)}\left( {\mathcal S}_{L,L^{\prime }=0}\right)
={\frac {1 }{ 24}}\left( 4 L ^{2}-1\right).
\end{eqnarray*}

This encompasses all previously known exponents for RW's and SAW's
 \cite{duplantier2,duplantier4,DS2}. In particular we arrive at a
striking {\it scaling equivalence: a self-avoiding walk is exactly
equivalent to $5/8$ of a Brownian motion} \cite{duplantier8}. Similar results were 
later obtained in probability theory, based on the general
structure of
 ``completely conformally-invariant processes'', which correspond
 exactly to  $c=0$ central charge conformal field theories  \cite {lawler3,lawler4}.
 The construction of the scaling limit of SAW's still eludes a rigorous approach, though it is
 predicted to correspond to ``stochastic L\"owner evolution'' ${\rm SLE}_{\kappa}$ with
 $\kappa=8/3$, equivalent to a Coulomb gas with $g=4/\kappa=3/2$ (see  section \ref{sec.geodual} below).\\

\vfill\eject
\section{\sc{Harmonic Measure of Brownian and Self-Avoiding Paths}}
\label{sec.harmonic}

\subsection{Harmonic Measure and Potential}
\subsubsection{Introduction}
The {\it harmonic measure}, i.e., the diffusion or electrostatic potential
field near an equipotential fractal boundary \cite{BBE}, or,
equivalently, the electric charge appearing on the frontier of a
perfectly conducting fractal, possesses a self-similarity
property, which is reflected in a {\it multifractal}
behavior. Cates and Witten  \cite{cates et witten} considered the
case of the Laplacian diffusion field near a simple random walk,
or near a self-avoiding walk. The associated exponents can be
recast as those of star copolymers made of a bunch of independent
RW's diffusing away from a generic point of the absorber. The
exact solution to this problem in two dimensions is as follows
 \cite{duplantier8}. From a mathematical point of view, it can
also be derived from the results of refs
 \cite{lawler2,lawler3,lawler4,lawler5} taken altogether.

\subsubsection{Harmonic Measure}
Consider a two-dimensional very large ``absorber'' ${\mathcal S}$.
One  defines the harmonic measure $H\left( w\right) $ as the
probability that a random walker launched from infinity, {\it
first} hits the outer ``hull's frontier'' or accessible frontier
${\mathcal F}({\mathcal S})$ at point $w \in {\mathcal F}({\mathcal S})$.
For a given point $w \in {\mathcal F}$, let $B(w,a)$ be the ball (i.e., disk)
of radius $a$ centered at $w$. Then $H({\mathcal F} \cap B(w,a))$ is  the total
harmonic measure of the points of ${\mathcal F}$ inside the ball
$B(w,a)$.

\subsubsection{Potential Theory}
One can also consider potential theory near the same
fractal boundary, now charged. One assumes the absorber to be
perfectly conducting, and introduces the harmonic potential
 ${\mathcal H}\left( z\right) $ at an 
exterior point $z \in {\rm  {\mathbb  C}}$, with Dirichlet boundary
conditions ${\mathcal H}\left({w \in {\mathcal F}}\right)=0$ on the outer
(simply connected)  frontier $ \mathcal F$, and
${\mathcal H}(w)=1$ on a circle ``at $\infty$'', i.e., of a large radius
scaling like the average size $R$ of $ \mathcal F$. As is well-known from a theorem due to Kakutani \cite{kakutani},
${\mathcal H}\left( z\right)$ is identical to the probability that a Brownian path
started at $z$ escapes to ``$\infty$'' without having hit
${\mathcal F}$.
The  harmonic measure $H\left({\mathcal F} \cap B(w,a)\right)$ defined above
 then also appears as the integral of the Laplacian of
${\mathcal H}$ in the disk $B(w,a)$, i.e., the boundary
charge in that disk.

\subsubsection{Multifractal Local Behavior}
The multifractal formalism
  \cite{mandelbrot2,hentschel,frisch,halsey} further involves
characterizing subsets ${\mathcal F}_{\alpha }$ of sites of the
 frontier ${\mathcal F}$ by a H\"{o}lder exponent $\alpha ,$
such that the $H$-measure of the frontier points in the ball
$B(w,a)$ of radius $a$ centered at $w_{\alpha}\in
{\mathcal F}_{\alpha }$ scales as
\begin{equation}
{H}\left({\mathcal F} \cap { B}(w_{\alpha}, a) \right)
\approx \left( a/R\right) ^{\alpha }. \label{ha'}
\end{equation}
The Hausdorff or ``fractal dimension'' $f\left( \alpha \right) $
of the set ${\mathcal F}_{\alpha }$ is such that
\begin{equation}
{\rm Card}\, {\mathcal F}_{\alpha} \approx R^{f(\alpha)}.
\label{ca'}
\end{equation}
Then the local behavior of the
potential near point $w_{\alpha} \in \mathcal F_{\alpha}$,
\begin{equation}
{\mathcal H}(z \to w_{\alpha}) \approx r^{\alpha}, \ r=|z-w_{\alpha}|\ ,
\label{Hal}
\end{equation}
scales with the same $\alpha$-exponent as the harmonic measure (\ref{ha'})
around point $w_{\alpha},$ and
$f(\alpha)={\rm dim}\, \mathcal F_{\alpha}$ thus appears as the Hausdorff
dimension of boundary points inducing the local behavior
(\ref{Hal}).

\subsubsection{Harmonic Moments}
One then considers a covering of $\mathcal F$
by balls $B(w,a)$ of radius $a$, and centered at
points $w$ forming a discrete subset $  {\mathcal F}/a$ of $\mathcal F$.
We are interested in the moments of $H$,
averaged over all realizations of RW's and of ${\mathcal S}$
\begin{equation}
{\mathcal Z}_{n}=\left\langle \sum\limits_{w\in   {\mathcal F}/a}{H}^{n}\left({\mathcal F}\cap {B}(w,
a)\right) \right\rangle , \label{Z}
\end{equation}
 where $n$ is, {\it a priori},
a real number. For very large absorbers ${\mathcal S}$ and 
frontiers ${\mathcal F}\left( {\mathcal S}\right) $ of average
size $R,$ one expects these moments to scale as
\begin{equation}
{\mathcal Z}_{n}\approx \left( a/R\right) ^{\tau \left( n\right)
}, \label{Z2}
\end{equation}
where the radius $a$ serves as a microscopic cut-off, reminiscent
of the lattice structure, and where the multifractal scaling
exponents $\tau \left( n\right) $ encode {\it generalized
dimensions}
\begin{equation}
D\left( n\right) =\frac{\tau\left( n\right)}{n-1} , \label{dn'}
\end{equation}
which vary in a non-linear way with
$n$  \cite{mandelbrot2,hentschel,frisch,halsey}. Several {\it a
priori} results are known. $D(0)$ is the Hausdorff dimension of
the accessible frontier of the fractal. By construction, $H$ is a
normalized probability measure, so that $\tau (1)=0.$ Makarov's
theorem  \cite{makarov}, here applied to the H\"{o}lder regular
curve describing the frontier  \cite{ai2}, gives the so-called
information dimension $\tau ^{\prime }\left( 1\right)
=D\left( 1\right) =1$.

The multifractal spectrum $f\left( \alpha \right)$
appearing in (\ref{ca'})
is given by the symmetric Legendre transform of $\tau \left( n\right)$:
\begin{equation}
\alpha =\frac{d\tau }{dn}\left( n\right) ,\quad \tau \left(
n\right) +f\left( \alpha \right) =\alpha n,\quad
n=\frac{df}{d\alpha }\left( \alpha \right) .  \label{alpha}
\end{equation}
Because of the statistical ensemble average (\ref{Z}), values of $%
f\left( \alpha \right) $ can become negative for some domains of
$\alpha $  \cite{cates et witten}. (The
existence of the harmonic multifractal spectrum $f(\alpha)$ for a
Brownian path has been rigorously established in ref. \cite{lawler97}.)

\begin{figure}[tb]
\begin{center}
\includegraphics[angle=0,width=.7\linewidth]{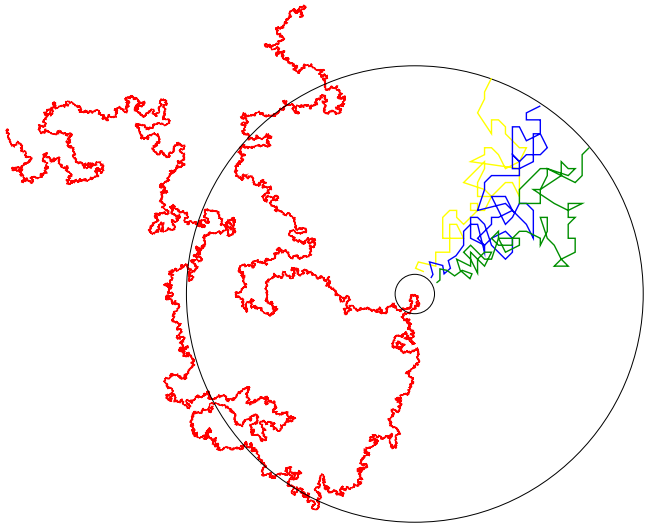}
\end{center}
\caption{Representation of moments (\ref{Z}) by a packets of $n$
independent Brownian paths diffusing away of a SAW, from short distance $a$  to  large distance $R$.}
\label{diffuse}
\end{figure}

\vfill\eject
\subsection{Multifractal Exponents}
\subsubsection{Representation by Random Walks}
By the very definition of the $H$-measure, $n$ independent RW's
diffusing away from the absorber give a geometric representation
of the $n^{th}$ moment ${H}^{n},$ for $n$ {\it integer}, and
convexity arguments give the complete 
continuation to real values (Fig. \ref{diffuse}).

When the absorber is a RW or a SAW of size $%
R,$ the site average of its moments ${H}^{n}$
 is represented by a copolymer star partition function ${\mathcal Z}_{R}\left(
{\mathcal S}  \wedge n\right)$ of the type (\ref{zr}), where we have introduced the
short-hand notation ${\mathcal S}  \wedge n:= {\mathcal
S}\wedge \left(\vee {\mathcal B}\right) ^{n}$ for describing the copolymer star
made by the absorber ${\mathcal S}$ hit by the bunch $\left( \vee {\mathcal B}\right) ^{n}$ at the apex only  \cite{cates et
witten,duplantier8}.\ More precisely, the sum (\ref{Z}) is normalized in such a way that
\begin{equation}
{\mathcal Z}_{n=1}=\left\langle \sum\limits_{w\in  {\mathcal F}/a}{H}\left({\mathcal F}\cap {B}(w,
a)\right) \right\rangle =1, \label{Z1}
\end{equation}
since $H$ is a (first hit) probability. The correct normalization is therefore:
\begin{equation}
\left\langle\sum\nolimits_{w}H^{n}\left( w\right)\right\rangle = {\mathcal Z}_{R}\left(
{\mathcal S}  \wedge n\right) /{\mathcal Z}_{R}\left( {\mathcal S}  \wedge 1\right).
\end{equation}
Because of the scaling (\ref{zrs}), we have
\begin{equation}
\label{ZnZ1}
\left\langle\sum\nolimits_{w}H^{n}\left( w\right)\right\rangle \approx
(a/R)^{x\left( {\mathcal S}  \wedge n\right)-x\left( {\mathcal S}  \wedge 1\right)}.
\end{equation}
The normalizing star partition function ${\mathcal Z}_{R}\left( {\mathcal S}  \wedge 1\right)$ is governed by an exponent
which is identically $x\left( {\mathcal S}  \wedge 1\right)=2$: $n=1$ indeed corresponds
to a single brownian path escaping from the absorber, which represents the potential itself,  and this identity
can be seen as a consequence of Gauss's theorem in two dimensions \cite{cates et witten}. We therefore conclude
that
\begin{equation}
{\mathcal Z}_{n} \approx R^2 {\mathcal Z}_{R}\left( {\mathcal
S}{\wedge }n\right).
\end{equation}
Owing to eqs.(\ref{Z2}), and (\ref{ZnZ1}), we get the scaling relation
\begin{equation}
\label{taunxnx1}
\tau \left( n\right) =x\left( {\mathcal S}  \wedge n\right)
-x\left( {\mathcal S}  \wedge 1\right)=x\left( {\mathcal S}  \wedge n\right) -2.
\end{equation}

\subsubsection{Quantum Gravity Formalism}
The absorber ${\mathcal S}$ near the ball center $w$ (Fig. \ref{diffuse}) is either
a two-RW star ${\mathcal S}={\mathcal B}\vee {\mathcal B}$, where the two strands are independent and mutually-intersecting, or a two-SAW star ${\mathcal S}={\mathcal P} \wedge {\mathcal P},$
made of two non-intersecting
SAW's.

Our formalism (\ref{x}) immediately gives the scaling dimension of the mutually-avoiding set
${\mathcal S}  \wedge n$ as
\begin{equation}
x\left( {\mathcal S}  \wedge n\right) =2V\left(
\tilde{\Delta}\left( {\mathcal S}\right) +U^{-1}\left( n\right)
\right),
\label{xSn}
\end{equation}
 where $\tilde{\Delta}\left( {\mathcal S}%
\right)= U^{-1}\left(\tilde{x}\left( {\mathcal S}%
\right) \right)$ is as above the quantum gravity boundary conformal dimension of
the absorber ${\mathcal S}$ alone. The quantity $U^{-1}\left( n\right)$ represents the
 quantum gravity boundary conformal dimension (\ref{deltatn}) of a packet of $n$ independent paths.
For a RW absorber, we have
\begin{equation*}
\tilde{\Delta}\left( {\mathcal S}\right)=\tilde{\Delta}\left( {\mathcal B} \vee {\mathcal B}\right) =U^{-1}\left( 2\right)
=\frac{3}{2},
\end{equation*}
 while for a SAW (see (\ref{deltaA+deltaB}) and (\ref{numbis}))
\begin{equation*}
\tilde{\Delta}\left( {\mathcal S}\right)=\tilde{\Delta}\left(
{\mathcal P} \wedge {\mathcal P}\right) =2\tilde{\Delta}_{P}(1)=2U^{-1}\left( \frac{5}{8}\right) =%
\frac{3}{2}.
\end{equation*}
The coincidence of these two values gives us the general result:

\noindent {\it In two dimensions the harmonic multifractal exponents $\tau(n)$ and spectra} $f\left(
\alpha \right) $ {\it of a random walk and a self-avoiding walk are
identical.}\\

\subsubsection{Multifractal Spectrum}
\noindent Calculation using (\ref{taunxnx1}-\ref{xSn}) gives  \cite{duplantier8}
\begin{equation}
\tau \left( n\right)=\frac{1}{2}\left( n-1\right)
+\frac{5}{24}\left(
\sqrt{24n+1}-5\right) , \label{tauf}
\end{equation}
\begin{equation}
D\left( n\right) =\frac{1}{2}+\frac{5}{\sqrt{24n+1}+5},\quad n\in \left[ -%
{{\frac{1}{24}}}%
,+\infty \right) .  \label{dna}
\end{equation}
The Legendre transform (\ref{alpha}) gives
\begin{equation}
\alpha =\frac{d\tau }{dn}\left( n\right) =\frac{1}{2}+\frac{5}{2}\frac{1}{%
\sqrt{24n+1}},  \label{alpha1}
\end{equation}
\begin{equation}
f\left( \alpha \right) =\frac{25}{48}\left( 3-\frac{1}{2\alpha -1}\right) -%
\frac{\alpha }{24},\quad \alpha \in \left(
{{\frac{1}{ 2}}}%
,+\infty \right) . \label{mf}
\end{equation}
\begin{figure}[tb]
\begin{center}
\includegraphics[angle=0,width=.7\linewidth]{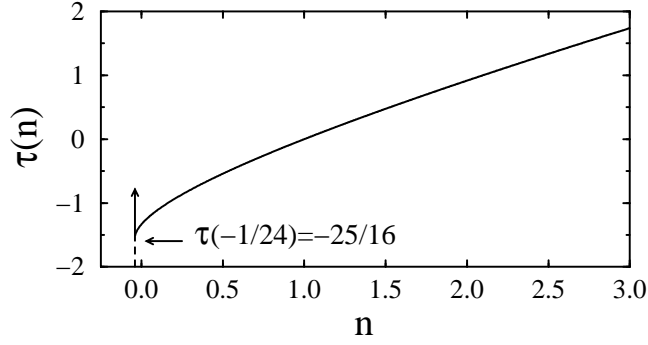}
\end{center}
\caption{Harmonic multifractal dimensions $\tau(n)$ of a two-dimensional RW or SAW.}
\label{Figuredeux}
\end{figure}

\begin{figure}[tb]
\begin{center}
\includegraphics[angle=0,width=.7\linewidth]{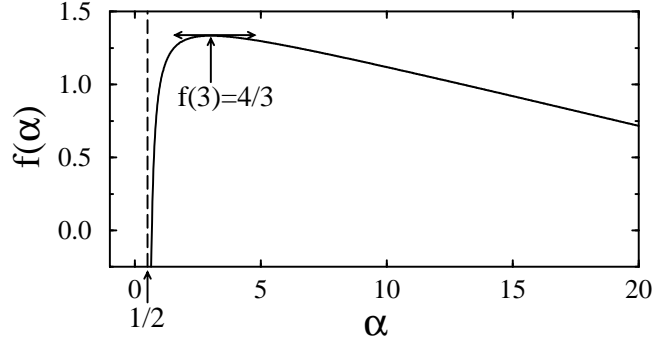}
\end{center}
\caption{Harmonic multifractal spectrum $f(\alpha)$ of
a two-dimensional RW or SAW.}
\label{Figuretrois}
\end{figure}

\subsubsection{Geometrical Properties of Multifractal Curves}
The corresponding universal curves are shown in Figures \ref{Figuredeux} and \ref{Figuretrois}:
$\tau \left( n\right) $ is half a parabola, and $f\left( \alpha
\right) $ a hyperbola. $D\left( 1\right) =\tau ^{\prime }\left(
1\right) =1$ is Makarov's theorem. The singularity at
$\alpha=\frac{1}{2}$ in the multifractal functions $f(\alpha)$
corresponds to points on the fractal boundary $\mathcal F$ where
the latter has the local geometry of a needle. The mathematical
version of this statement is given by Beurling's theorem
 \cite{Beur}, which states that at distance $\epsilon$ from the
boundary, the harmonic measure is bounded above by
\begin{equation}
H \left(z: {\rm inf}_{w \in  {\mathcal F}}|z - w|  \leq
\epsilon)\right) \leq C {\epsilon}^{1/2}, \label{beurling}
\end{equation}
where $C$ is a constant. This insures that the spectrum of
multifractal H{\"o}lder exponents $\alpha$ is bounded below by
$\frac{1}{2}$.
The right branch of the $f\left( \alpha \right)$ curve has a linear asymptote
\begin{equation}
\lim_{\alpha \rightarrow +\infty} \frac{1}{\alpha}f\left( \alpha
\right) = -\frac{1}{24}.
\end{equation}
Its linear shape is quite reminiscent of that of the multifractal
function
 of the growth probability as in the case of a 2D DLA
cluster  \cite {ball}. The domain of large values of $\alpha$
corresponds to the lowest part $n\rightarrow {n^{\ast}}
=-\frac{1}{24}$ of the spectrum of dimensions, which is dominated
by almost inaccessible sites, and the existence of a linear
asymptote to the multifractal function $f$ implies a peculiar
behavior for the number of those sites in a lattice setting.
Indeed define ${\mathcal N}\left( H\right)$ as the  number of
sites having a probability $H$ to be hit:
\begin{equation}
{\mathcal N}\left( H\right)={\rm Card}\left\{w \in {\mathcal F}:
{H}(w)=H \right\}. \label{N(H)}
\end{equation}
Using the MF formalism to change from the variable $H$ to
$%
\alpha $ (at fixed value of $a/R)$, shows that ${\mathcal N}\left(
H\right)$ obeys, for $H\rightarrow 0,$ a power law behavior
\begin{equation}
{\mathcal N}\left( H\right)|_{H\rightarrow 0}\approx
H^{-{\tau}^{\ast}} \label{nh}
\end{equation}
with an exponent
\begin{equation}
\tau ^{\ast }=1+%
\mathrel{\mathop{\lim }\limits_{\alpha \rightarrow +\infty }}%
\frac{1}{\alpha }f\left( \alpha \right)=1+n^{\ast}. \label{to}
\end{equation}
Thus we predict
\begin{equation}
   \tau
^{\ast}=\frac{23}{24}. \label{toc}
\end{equation}

Let us remark that $-\tau \left( 0\right) =\sup_{\alpha }f\left(
\alpha \right) =f\left( 3 \right) =\frac{4}{3}$ is the Hausdorff
dimension of the {\it Brownian frontier} or of a SAW. Thus
Mandelbrot's
classical conjecture identifying the latter two is derived and generalized to the {\it whole} 
$f\left( \alpha \right) $ harmonic spectrum.

\subsection{An Invariance Property of $f(\alpha)$}
\label{subsec.inv}
The expression for $f(\alpha)$ simplifies if one considers the
combination:
\begin{eqnarray}
f\left( \alpha \right)-\alpha&=& \frac{25}{24}
\left[1-\frac{1}{2}\left(2\alpha -1 + \frac{1}{2\alpha
-1}\right)\right]. \label{f-a}
\end{eqnarray}
Thus the multifractal function possesses the invariance symmetry
 \cite{BDH}
\begin{eqnarray}
f\left( \alpha \right)-\alpha=f\left( {\alpha}^\prime
\right)-{\alpha}^{\prime}, \label{inv}
\end{eqnarray}
for $\alpha$ and ${\alpha}^{\prime}$ satisfying the duality
relation:
\begin{eqnarray}
(2\alpha-1)(2{\alpha}^{\prime}-1)=1,
\end{eqnarray}
or, equivalently
\begin{eqnarray}
{\alpha}^{-1}+{{\alpha}^{\prime}}^{-1}=2. \label{aa'}
\end{eqnarray}

When associating an equivalent electrostatic wedge angle $\theta=\pi / \alpha$ to each local
singularity exponent $\alpha$ (see section \ref{sec.conform}), one gets the complementary rule
for angles in the plane  \cite{BDH}
\begin{eqnarray}
\theta+{\theta}^{\prime}=\frac{\pi}{\alpha}+\frac{\pi}{{\alpha}^{\prime}}=2\pi.
\label{tetateta'}
\end{eqnarray}
Notice that by definition of the multifractal dimension $f\left( \alpha \right)$,
$R^{f\left( \alpha \right)-\alpha}$ is the total harmonic measure content of points of type $\alpha$
or equivalent angle $\theta=\pi/\alpha$
along the multifractal frontier. The symmetry (\ref{inv}) thus means that this harmonic content is invariant when
taken at the complementary angle in the plane $2\pi-\theta$. The basic symmetry (\ref{inv}) thus seems to reflect
 that of the frontier itself under the {\it exchange of
interior and exterior domains} (ref. \cite{BDH}).

It is also interesting to note that, owing to the explicit forms
(\ref{alpha1}) of $\alpha$ and (\ref{dna}) of $D(n)$, the
condition (\ref{aa'}) becomes, after a little algebra,
\begin{equation}
D(n)+D(n')=2. \label{DD'}
\end{equation}

\subsection{Higher Multifractality for Brownian or
Self-Avoiding Paths}
\label{subsec.double}
\begin{figure}[tb]
\begin{center}
\includegraphics[angle=0,width=.9\linewidth]{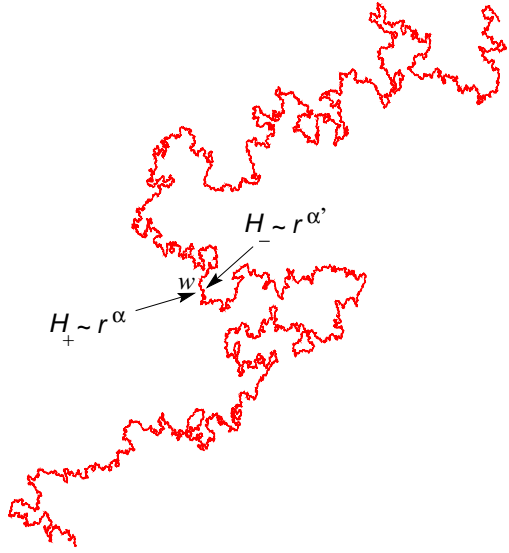}
\end{center}
\caption{Double distribution of harmonic potential ${\mathcal H}$ on both sides of a simple scaling curve
(here a SAW, courtesy of T. G. Kennedy). The local exponents on both sides of point $w=w_{\alpha,\alpha'}$ are
$\alpha$ and $\alpha'$. The Hausdorff dimension of such points along the SAW is $f_2(\alpha,\alpha')$.}
\label{Figure1}
\end{figure}
It is interesting to note that one can define {\it higher
multifractal} spectra as those depending on several $\alpha$
variables  \cite{duplantier10}. A first example is given by the
double moments of the harmonic measure on {\it both} sides of a
random fractal, taken here to be either a Brownian motion or a self-avoiding
walk. (The general case will be further described in section \ref{sec.higher}).

\subsubsection{Double-Sided Potential}
 When it is {\it simple}, i.e., double point free,
a conformally scaling curve $\mathcal F$ can be reached from
both sides. A SAW  is
naturally such a simple curve, therefore accessible from both sides. For a Brownian motion, one can
consider the subset of the {\it pinching} or {\it cut points}, of
Hausdorff dimension $D=2-2\zeta_{2}=3/4$, where the path splits
into two non-intersecting parts.  The Brownian path  is then locally
accessible from
both directions.

Taking Dirichlet boundary conditions on a random curve, one can then
consider  the joint distribution of potential on both sides, namely ${\mathcal H}_{+}$ on one side,
and ${\mathcal H}_{-}$ on the other, such that
\begin{eqnarray}
 {\mathcal H}_{+}(z\to w_{\alpha,\alpha'}) \sim r^{\alpha}, \ {\mathcal H}_{-} (z\to w_{\alpha,\alpha'})\sim r^{\alpha'}\ ,
\end{eqnarray}
 when approaching a point $w_{\alpha,\alpha'}$ of the subset $\mathcal F_{\alpha,\alpha'}$  at
 distance $r=|z-w_{\alpha,\alpha'}|$ (Fig. \ref{Figure1}).
 Then a double-multifractal spectrum $f_2(\alpha, \alpha')={\rm dim}\,\mathcal F_{\alpha,\alpha'}$ yields
 the Hausdorff dimension of the set of points
 of type $(\alpha,\alpha')$.

\subsubsection{Double Harmonic Moments}
As before, instead of considering directly the potential $\mathcal H$, one can consider
equivalently the harmonic measure content of a covering by small balls centered
along the random fractal. Let us now define:
\begin{equation}
{\mathcal Z}_{n,n'}=\left\langle \sum\limits_{w\in {\mathcal F}/a}
\left[{H}_{+}(w,a)\right]^{n}\left[{H}_{-}(w,a)\right]^{n'}\right\rangle, \label{ZZ'}
\end{equation}
where ${H}_{+}(w,a) := {H}_{+}\left({\mathcal F} \cap
B(w,a)\right)$ and ${H}_{-}(w,a):= {H}_{-}\left({\mathcal F} \cap B(w,a)\right)$ are
respectively the harmonic measure contents of the same ball $B(w,a)$ on ``left''or ``right'' sides
of the random fractal. The ball's center $w$ is taken in a discrete set of points $w\in {\mathcal F}/a$.
\begin{figure}[tb]
\begin{center}
\includegraphics[angle=0,width=.9\linewidth]{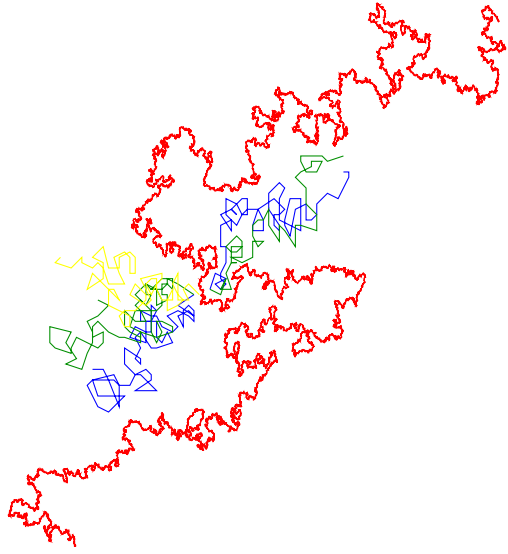}
\end{center}
\caption{Representation of the double moments (\ref{ZZ'}) by two packets of $n$ and $n'$ 
independent Brownian paths diffusing away from a SAW.}
\label{sawbrown}
\end{figure}

They are represented by two packets of $n$ and $n'$
independent Brownian paths
diffusing away from the fractal object (Fig. \ref{sawbrown}).
These moments have a multifractal scaling
behavior
 \begin{equation}
{\mathcal Z}_{n}\approx \left( a/R\right) ^{\tau_2(n,n') },
\label{ZZ2'}
\end{equation}
where the exponents $\tau_2(n,n')$ now depend on two moment orders
$n$ and $n'$.

\subsubsection{Double Legendre Transform}
The generalization of the Legendre transform 
(\ref{alpha}) reads
\begin{eqnarray}
\alpha &=&\frac{\partial\tau_2 }{\partial n}\left( n,n'\right) ,
\quad \alpha' =
\frac{\partial\tau_2 }{\partial n'}\left( n,n'\right), \nonumber \\
f_2\left( \alpha, \alpha' \right) &=&\alpha n+\alpha' n'-\tau_2(n,n'),
 \label{alpha''}\\
n&=&\frac{\partial f_2}{\partial \alpha}\left( \alpha, \alpha'
\right) , \quad n'=\frac{\partial f_2}{\partial \alpha' }\left(
\alpha, \alpha' \right). \nonumber
\end{eqnarray}
It yields the dimension  of the 
subset $\mathcal F_{\alpha,\alpha'}$ of frontier points
$w_{\alpha,\alpha'}$, where the potential $\mathcal H$ scales as
in eqs. (\ref{ZZ'}), or where the harmonic content of a ball $B(w_{\alpha,\alpha'},a)$ scales
as $(a/R)^{\alpha}$ on one side, and $(a/R)^{\alpha'}$ on the other.

\subsubsection{Star Fusion Algebra}
We find the $\tau$ exponents from the star algebra (\ref{x}):
\begin{equation}
\tau_2(n,n')=2V\left( a'+U^{-1}\left( n\right)+U^{-1}\left(
n'\right) \right)-2, \label{taunn'}
\end{equation}
where $a'$ corresponds to the quantum gravity boundary scaling dimension of
the fractal set near where the potential or harmonic measure is evaluated, i.e., the simple SAW curve or
 the Brownian cut-point set.
For Brownian motion near a cut-point, the two-strands appear as two {\it mutually-avoiding} parts  ${\mathcal B} \wedge {\mathcal B}$,
(further separated by the two sets of auxiliary Brownian motions which represent the harmonic measure moments).
Thus we have from (\ref{deltaA+deltaB}) and (\ref{numbis}):
\begin{equation}
a_B'= \tilde{\Delta}\left( {\mathcal B} \wedge {\mathcal B}\right)=2\times \tilde\Delta_{B}
(1)=2U^{-1}\left( 1\right)=2.
\end{equation}
Near a point on a self-avoiding walk, the latter appears by construction as made
of two mutually-avoiding SAW's, and we have from (\ref{deltaA+deltaB}) and (\ref{numbis}):
\begin{equation}
a_P'= \tilde{\Delta}\left( {\mathcal P} \wedge {\mathcal P}\right)=2 \times
\tilde{\Delta}_{P}(1)=2U^{-1}\left( \frac{5}{8}\right)
=\frac{3}{2}.
\end{equation}
 After performing the double Legendre transform and some calculations
 (generalized and detailed in section \ref{sec.higher}), we find
\begin{eqnarray}
f_2\left( \alpha, \alpha'
\right)&=&2+\frac{1}{12}-\frac{1}{3}{a''}^2
{\left[1-\frac{1}{2}\left(\frac{1}{\alpha}+\frac{1}{\alpha'}\right)\right]}^{-1} \nonumber \\
& &-\frac{1}{24}\left(\alpha+\alpha'\right), \label{faa'}
\end{eqnarray}
\begin{equation}
{\alpha}=2
\frac{1}{\sqrt{24n+1}}\left[{a''}+\frac{1}{4}\left(\sqrt{24n+1}+\sqrt{24n'+1}\right)\right],
\end{equation}
and a similar symmetric equation for $\alpha'$. Here $a''$ has the
shifted values:
\begin{eqnarray}
a''&=&a'+\gamma=a'-\frac{1}{2}\\a_B''&=&\frac{3}{2}\ ({\rm RW}),\ {\rm
or}\ a_P''=1\ ({\rm SAW}). \label{a''}
\end{eqnarray}
These doubly multifractal spectra thus are different for RW's and SAW's. The SAW spectrum possesses the required
property $$f_P(\alpha):={\rm sup}_{\alpha'} f_P(\alpha,
\alpha')=f(\alpha),$$ where $f(\alpha)$ is (\ref{mf}) above. For a Brownian path,
the one-sided spectrum $$ f_B(\alpha):={\rm sup}_{\alpha'} f_B(\alpha,
\alpha')=2-\frac{45}{48}-\frac{49}{48}\frac{1}{2\alpha-1}-\frac{\alpha}{24},$$ such that
$f_B(\alpha) < f(\alpha)$,
gives the MF spectrum of
cut-points along the Brownian frontier. This set of Hausdorff dimension $\frac{3}{4} < 1$ is disconnected, and
$f_B(\alpha=1)(= -\frac{49}{48})\neq 1$, in contrast to Makarov's theorem, $f(\alpha=1)= 1 $, for any
connected set in the plane.

\subsubsection{Poly-Multifractality}
The results above can be generalized to a {\it star configuration} made of $m$
random walks or $m$ self-avoiding walks, where one looks at the
simultaneous behavior of the potential in each sector between the mutually-avoiding
arms of the star (see section \ref{sec.higher} below for a  precise description and calculation in
the general case). The quantum gravity boundary dimensions of these stars are respectively
from (\ref{deltaA+deltaB}) and (\ref{numbis}):
\begin{equation}
a'=a_B'(m)= \tilde{\Delta}( \stackrel{m}{\overbrace{{{\mathcal B}\wedge\cdots\wedge {\mathcal B}}}})=m\times \tilde\Delta_{B}
(1)=m\,U^{-1}\left( 1\right)=m,
\end{equation}
for $m$-cut Brownian paths, and
\begin{equation}
a'=a_P'(m)= \tilde{\Delta}(\stackrel{m}{\overbrace{{ {\mathcal P}\wedge\cdots \wedge {\mathcal P}}}})=m \times
\tilde{\Delta}_{P}(1)=m\,U^{-1}\left( \frac{5}{8}\right)
=\frac{3}{4} m,
\end{equation}
for a star made of $m$ self-avoiding walks, all being mutually-avoiding. We give here only these
{\it poly-multifractal} results, which read for Brownian cut-points or self-avoiding paths:
\begin{eqnarray}
f_m\left(\{
{\alpha}_{i=1,...,m}\}\right)&=&2+\frac{1}{12}-\frac{1}{3}{{a''}^2(m)}{\left(1-\frac{1}{2}
\sum_{i=1}^{m}{\alpha}_{i}^{-1}\right)}^{-1} \nonumber \\
& &-\frac{1}{24}\sum_{i=1}^{m}{\alpha}_{i}, \label{fai}
\end{eqnarray}
with
\begin{equation}
{\alpha}_{i}=2
\frac{1}{\sqrt{24n_{i}+1}}\left({a''(m)}+\frac{1}{4}\sum_{j=1}^{m}\sqrt{24n_{j}+1}\right),\label{ai}
\end{equation}
and where
\begin{equation}
a''(m)=a_B''(m)=a_B'(m)-\frac{1}{4}m=\frac{3}{4}m,
\end{equation}
for $m$ {\it random walks pinched} in a mutually-avoiding $m$-star configuration, and
\begin{equation}
a''(m)=a_P''(m)=a_P'(m)-\frac{1}{4}m=\frac{1}{2}m, \label{a''SAW}
\end{equation}
 for $m$ {\it self-avoiding walks} in a mutually-avoiding star configuration. The two-sided case (\ref{faa'}) (\ref{a''})
 above is recovered for $m=2$. The domain of definition of the poly-multifractal function $f$ is given by
\begin{equation}
1-\frac{1}{2}\sum_{i=1}^{m}{\alpha}_{i}^{-1} \geq 0,
\end{equation}
as verified by eq. (\ref{ai}).\\

\section{\sc{Percolation Clusters}}
\label{sec.perco}

\subsection{Cluster Hull and External Perimeter}

\begin{figure}[htbp]
\begin{center}
\includegraphics[angle=0,width=.7\linewidth]{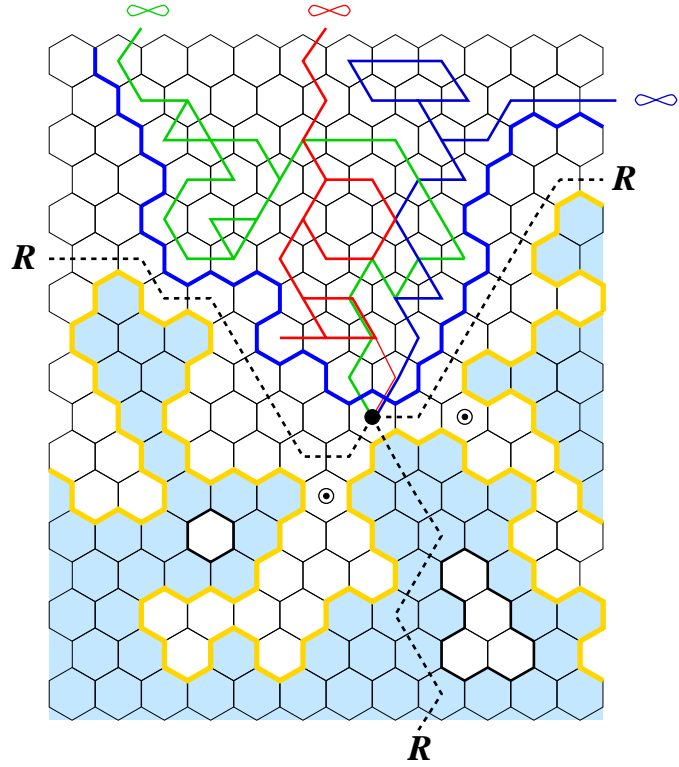}
\end{center}
\caption{An accessible site $(\bullet )$ on the external
perimeter for site percolation on the triangular lattice. It is
defined by the existence, in the {\it scaling limit}, of three non-intersecting,
and connected paths ${\mathcal S}_{3}$ (dotted
lines), one on the incipient cluster, the other two on
the dual empty sites. The entrances of fjords $\odot$ close in the scaling limit.
Point $(\bullet )$ is first reached by three
independent RW's (red, green, blue), contributing to ${H}^3 (\bullet
)$. The hull of the incipient cluster (golden line) avoids the outer
frontier of the RW's (thick blue line). A Riemann map of the
latter onto the real line ${\partial \mathbb H}$ reveals the presence of an underlying
$\ell=3$ path-crossing {\it boundary}
operator, i.e,
a two-cluster boundary operator,
with dimension in the half-plane
$\tilde{x}_{\ell=3 }
=\tilde
{x}^{{\mathcal C}}_{k=2}=2.$ Both accessible hull and Brownian
paths have a frontier dimension $\frac{4}{3}$.}
\label{Figure4}
\end{figure}

Let us consider, for definiteness, site percolation on the
2D triangular lattice. By universality, the results are expected to apply
to other $2D$ (e.g., bond) percolation models in the scaling limit. Consider then
a very large two-dimensional incipient cluster
${\mathcal C}$, at the percolation threshold $p_{c}=1/2$.
Figure \ref{Figure4} depicts such a connected cluster.

\subsubsection{Hull}
The boundary lines of a site
percolation cluster, i.e., of connected sets of occupied hexagons, form
random lines on the
dual hexagonal lattice. (They are actually known to obey the statistics of random loops in the
${ O}\left( N=1\right)$ model,
where $N$ is the loop fugacity, in the so-called ``low-temperature phase'', or
of boundaries of Fortuin-Kasteleyn clusters in the $Q=1$ Potts model  \cite{SD}.)
Each critical connected cluster
thus possesses an external closed boundary, its {\it hull}, the fractal dimension of which is known to be
$D_{\rm H}=7/4$  \cite{SD}.

In the scaling limit, however, the hull, which possesses many pairs of points at relative distances given by a
finite number of lattice meshes $a$, coils onto
itself to become a non-simple curve  \cite{GA}; it thus
develops a smoother outer (accessible) frontier ${\mathcal F}({\mathcal C})$ or {\it external perimeter} (EP).

\subsubsection{External Perimeter and Crossing Paths}
The geometrical nature of this
external perimeter has recently been elucidated and its Hausdorff
dimension found to equal $D_{\rm EP}=4/3$  \cite{ADA}. For a site $w=\left( \bullet \right)$ to belong to the {\it
accessible} part of the hull, it must remain, in the {\it continuous
scaling limit},  the source of at least {\it three
non-intersecting crossing paths} , noted ${\mathcal S}_{3}=
{\mathcal P} \wedge {\bar {\mathcal P_1}} \wedge {\bar {\mathcal P_2}}$,
reaching to a (large) distance $R$ (Fig. \ref{Figure4}). (Recall the notation
$A\wedge B$ for two sets, $A$, $B$, of random
paths, required to
be {\it mutually non-intersecting, }and{\it \ }$A\vee B$ for two {\it independent}, thus
possibly
intersecting, sets.) Each of these  paths is {``\it
monochromatic''}: one path  $\mathcal P$ runs only through occupied sites, which simply means that
$w$ belongs to a particular connected cluster; the other two
{\it dual} lines $ {\bar {\mathcal P}_{i=1,2}}$
run through  empty sites, and doubly
connect the external perimeter site $w$ to
``infinity'' in open space  \cite{ADA}.
The definition of the standard hull requires only the
origination, in the scaling limit, of a {\it ``bichromatic''} pair of lines
${\mathcal S}_2={\mathcal P} \wedge {\bar  {\mathcal P}}$, with one path running on occupied sites, and the dual one on empty ones.
Such hull points lacking a second dual line will not necessarily remain accessible from the outside
after the scaling limit is taken, because
their single exit path becomes a strait pinched by parts of the occupied cluster.
In the scaling limit,
the hull is thus a self-coiling and conformally-invariant (CI) scaling curve
which is not simple, while the external perimeter is a simple CI scaling curve.

The (bichromatic) set ${\mathcal S}_{3 }$ of three
non-intersecting connected paths in the percolation system is governed by a new critical exponent
$x\left( {\mathcal
S}_{3 } \right)(=2/3)$ such that $D_{\rm EP}=2-x\left( {\mathcal
S}_{3 }\right) $, while a  bichromatic pair of non-intersecting paths ${\mathcal
S}_2$ has an exponent $x\left( {\mathcal
S}_{2 }\right)(=1/4) $ such that $D_{\rm H}=2-x\left( {\mathcal
S}_{2}\right) $ (see below).

\subsection{Harmonic Measure of Percolation Frontiers}
Define $H\left( w,a\right):={H}\left({\mathcal F} \cap {B}(w,
a)\right) $
as the probability that a random walker,
launched from infinity, {\it first} hits the outer (accessible)
percolation hull's frontier or external perimeter ${\mathcal F}({\mathcal C})$ in the ball $B(w,a)$ centered at point
$w \in {\mathcal F}({\mathcal C})$. The moments $H^n$ of $H$ are
averaged over all realizations of RW's and ${\mathcal C}$, as in
eq.(\ref{Z}) above:
\begin{equation}
{\mathcal Z}_{n}=\left\langle \sum\limits_{w\in {\mathcal
F}/a}{H}^{n}\left({\mathcal F} \cap {B}(w,
a)\right) \right\rangle . \label{Zpe}
\end{equation}
For very large clusters ${\mathcal C}$ and
frontiers ${\mathcal F}\left( {\mathcal C}\right) $ of average
size $R,$ one expects again these moments to scale as in eq.
(\ref{Z2}): ${\mathcal Z}_{n}\approx \left( a/R\right) ^{\tau
\left( n\right) }$. These exponents $\tau(n)$ have been obtained
recently  \cite{duplantier9}, and we shall see that they are {\it identical} to those
 obtained in the preceding section \ref{sec.harmonic} for Brownian paths and self-avoiding walks.

As before, by the very definition of the $H$-measure, $n$ independent RW's diffusing away or towards
a neighborhood of a EP point $w$, give a geometric representation of the $n^{th}$
moment $H^{n}(w),$ for $n$ {\it integer}. The values so derived for $n\in {\mathbb %
N}$ will be enough, by convexity arguments, to obtain the analytic
continuation for arbitrary $n$'s. Figure \ref{Figure4} depicts such $n$ independent random walks,
in a bunch, {\it
first} hitting the external frontier of a percolation cluster at a
site $w=\left( \bullet \right).$ The packet of independent RW's avoids
the occupied cluster, and defines its own envelope as a set of two
boundary lines separating it from the occupied part of the
lattice. The $n$ independent RW's, or Brownian
paths ${\mathcal B}$ in the scaling limit, in a bunch denoted $\left( \vee
{\mathcal B}\right) ^{n},$ thus  avoid the set ${\mathcal S}_{3 }$ of three {\it
non-intersecting} connected paths in the percolation system, and
this system is governed by a new family of critical exponents
 $x\left( {\mathcal
S}_{3 }\wedge n\right) $ depending on $n.$
The main lines of the derivation of the latter exponents by generalized conformal invariance are as follows.

\subsection{Harmonic and Path Crossing Exponents}

\subsubsection{Generalized Harmonic Crossing Exponents}
The $n$ independent Brownian paths ${\mathcal B}$, in a bunch
 $\left(
\vee {\mathcal B}\right) ^{n},$ avoid a set ${\mathcal S}_{\ell }:= \left(
\wedge
{\mathcal P}\right) ^{\ell }$ of $\ell $  non-intersecting
crossing paths in
the
percolation system. They originate from the same hull site, and each
passes only through occupied sites, or only through empty ({\it dual}) ones
 \cite{ADA}.
The probability that the Brownian and percolation paths altogether traverse the
annulus ${\mathcal %
D}\left( a, R\right) $ from the inner boundary circle of radius $a$ to the outer
one at distance $R$, i.e., are in a ``star'' configuration ${\mathcal S}_{\ell
}\wedge
\left( \vee {\mathcal B}\right) ^{n}$, is expected to scale for $%
a/R\rightarrow 0 $ as
\begin{equation}
{\mathcal P}_{R}\left( {\mathcal S}_{\ell }\wedge n\right) \approx
\left( a/R\right) ^{x\left( {\mathcal S}_{\ell }\wedge n\right) },  \label{xppe}
\end{equation}
where we used ${\mathcal S}_{\ell }\wedge n = {\mathcal S}_{\ell }\wedge \left(
\vee
{\mathcal B}\right) ^{n}$ as a short hand notation, and where $x\left( {\mathcal
S}_{\ell
}\wedge n\right) $ is a new critical exponent
depending on $\ell $ and $n$.  It is convenient to introduce similar boundary
probabilities $\tilde{{\mathcal P}}_{R}\left( {\mathcal S}_{\ell }\wedge
n\right) \approx \left( a/R\right) ^{\tilde{x}\left( {\mathcal S}_{\ell }\wedge
n\right) }$ for the same star configuration of paths, now crossing through the
half-annulus $\tilde{{\mathcal D}}\left( a, R\right) $ in the half-plane $\mathbb H$.

\subsubsection{Bichromatic Path Crossing Exponents}
When $n \to 0$, the probability ${\mathcal P}_{R}\left( {\mathcal S}_{\ell }\right)={\mathcal P}_{R}\left(
{\mathcal S}_{\ell}\wedge 0\right)\approx \left( a/R\right) ^{ x_{\ell}}$
[resp. $\tilde{{\mathcal P}}_{R}\left( {\mathcal S}_{\ell }\right)=\tilde{{\mathcal P}}_{R}\left(
{\mathcal S}_{\ell}\wedge 0\right)\approx \left( a/R\right) ^{\tilde x_{\ell}}$] is the probability
of having $\ell $ simultaneous non-intersecting
path-crossings of the
annulus ${\mathcal %
D}\left( a, R\right) $ in the plane $\mathbb C$ [resp. half-plane $\mathbb H$], with associated exponents $x_{\ell
}:= x\left( {\mathcal S}_{\ell } \wedge
0\right) $ [resp.  $\tilde{x}_{\ell }:= \tilde{x}\left( {\mathcal S}_{\ell } \wedge
0\right)$]. Since these exponents are obtained from the limit $n\to 0$ of the harmonic measure
exponents, at least two paths run on occupied sites or empty sites, and these are
the {\it bichromatic} path crossing exponents \cite{ADA}. The {\it monochromatic} ones are different
in the bulk  \cite{ADA,JK}.

\vfill\eject
\subsection{Quantum Gravity for Percolation}
\subsubsection{$c=0$ KPZ mapping}
Critical percolation is described
by a conformal field theory with the same vanishing central charge $c=0$ as RW's or SAW's
(see, e.g.,  \cite{cardyjapon}). Using again the fundamental mapping of this
conformal field theory (CFT) in the {\it
plane} $%
{\mathbb C}$, to the
CFT on a fluctuating random Riemann surface, i.e., in presence of {\it
quantum gravity}  \cite{KPZ},  the two
universal functions $U$ and $V$ only depend on the central charge $c$ of
the CFT, and  are the same as for RW's, and SAW's:
\begin{eqnarray}
U\left( x\right) &=&\frac{1}{3}x\left( 1+2x\right) , \hskip2mm V\left( x\right)
=\frac{1}{24}\left( 4x^{2}-1\right),  \label{Upe}
\end{eqnarray}
with $V\left( x\right) = U\left(\frac{1}{2}\left(x-\frac{1}{2}\right)\right).$

They suffice to generate all geometrical exponents involving
{\it mutual-avoidance} of random {\it star-shaped} sets of paths of the critical percolation
system. Consider  two arbitrary\ random sets $A,B,$ involving each
a collection of paths in a star configuration, with proper scaling crossing
exponents $x\left( A\right) ,x\left( B\right) ,$ or, in the half-plane, crossing
exponents $\tilde{x}\left( A\right) ,\tilde{x}\left(
B\right) .$ If one fuses the star centers and requires $A$ and $B$ to stay
mutually-avoiding, then the new crossing exponents, $x\left( A\wedge
B\right) $ and $\tilde{x}\left( A\wedge B\right) ,$ obey the same {\it star
fusion algebra} as in (\ref{x}) \cite{duplantier7,duplantier8}
\begin{eqnarray}
x\left( A\wedge B\right) &=&2V\left[ U^{-1}\left( \tilde{x}\left( A\right)
\right) +U^{-1}\left( \tilde{x}\left( B\right) \right) \right]  \nonumber \\
\tilde{x}\left( A\wedge B\right) &=&U\left[ U^{-1}\left( \tilde{x}\left(
A\right) \right) +U^{-1}\left( \tilde{x}\left( B\right) \right) \right] ,
\label{xpe}
\end{eqnarray}
where
$U^{-1}\left( x\right) $ is the inverse function
\begin{equation}
U^{-1}\left( x\right) =\frac{1}{4}\left( \sqrt{24x+1}-1\right) .
\label{u1pe}
\end{equation}

This structure immediately gives both the percolation crossing exponents
$x_{\ell}, \tilde {x}_{\ell}$  \cite{ADA}, and the harmonic crossing
exponents
$x\left( {\mathcal S}_{\ell }\wedge n\right) $ (\ref{xppe}).
\subsubsection{Path Crossing Exponents}
First, for a
set ${\mathcal S}_{\ell }=\left( \wedge {\mathcal P}\right) ^{\ell }$ of $\ell
$ crossing paths, we have from the recurrent use of (\ref{xpe})
\begin{equation}
x_{\ell }=2V\left[ \ell\, U^{-1}\left( \tilde{x}_{1}\right) \right] ,\quad
\tilde{x}_{\ell }=U\left[ \ell\, U^{-1}\left( \tilde{x}_{1}\right) \right] .
\label{xl}
\end{equation}
For percolation, two values of half-plane crossing exponents $\tilde{x}_{\ell }$
are known by
{\it elementary} means: $\tilde{x}_{2}=1,\tilde{x}_{3}=2.$  \cite{ai1,ADA} From (
\ref{xl}) we thus find $U^{-1}\left( \tilde{x}_{1}\right)
=\frac{1}{2}U^{-1}\left(
\tilde{x}_{2}\right) =\frac{1}{3}U^{-1}\left( \tilde{x}_{3}\right) =\frac{1}{
2},$ (thus $ \tilde{x}_{1}=\frac{1}{3}$  \cite{cardy}), which in turn gives
\[
x_{\ell }=2V\left({\frac{1}{ 2}}\ell\right) =\frac{1}{12}\left(
{\ell}^{2}-1\right),
\tilde{x}_{\ell }=U\left({\frac{1}{ 2}} \ell\right)=\frac{\ell }{6}\left( \ell +1\right).
\]
We thus recover the identity  \cite{ADA} $x_{\ell }=x_{L=\ell }^{{O}\left(
N=1\right) }, \tilde{x}_{\ell }=\tilde{x}_{L=\ell +1}^{{O}\left( N=1\right)
}$ with the $L$-line exponents of the associated ${O}\left( N=1\right)$ loop
model, in the ``low-temperature phase''.
For $L$ {\it even}, these exponents also govern the existence of
$k=\frac{1}{2}L$
{\it spanning} clusters, with the identity $x_{k}^{\mathcal C}=x_{\ell =2k}=%
\frac{1}{12}\left( 4k^{2}-1\right) $ in the plane, and $\tilde{x}_{k}^{\mathcal C}=%
\tilde{x}_{\ell =2k-1}=\frac{1}{3}k\left( 2k-1\right) $ in the half-plane
 \cite{SD,D6,D7}.

\subsubsection{Brownian Non-Intersection Exponents}
The non-intersection exponents (\ref{Zeta}) and (\ref{zC2})
of $L$ Brownian paths seen in section \ref{sec.inter} are identical to the percolation
path crossing exponents for
\begin{eqnarray}
2\zeta_L=x_{\ell},\;\; 2\tilde \zeta_L=\tilde{x}_{\ell},\;\;\;{\ell}=2L,
\end{eqnarray}
so we obtain a {\it complete scaling equivalence
between
a Brownian path
and {\it two} percolating crossing paths, in both the plane and half-plane}  \cite{duplantier9}.

\subsubsection{Harmonic Crossing Exponents}
Finally, for the harmonic crossing exponents in (\ref{xppe}), we fuse the two objects
${\mathcal
S}_{\ell}$ and $\left( \vee {\mathcal B}\right) ^{n}$ into a new star ${\mathcal
S}_{\ell}\wedge n $, and use (\ref{xpe}). We just have seen that the
boundary $\ell$-crossing exponent of ${\mathcal S}_{\ell}$, $\tilde{x}_{\ell}$, obeys
$U^{-1}\left( \tilde{x}_{\ell}\right) =\frac{1}{2}\ell.$ The bunch of $n$
independent Brownian paths have their own  half-plane crossing exponent
$\tilde{x}\left( \left( \vee {\mathcal B}\right) ^{n}\right) =n\tilde{x}%
\left( {\mathcal B}\right) =n,$ since the boundary conformal weight of a single Brownian
path
is trivially $\tilde{x}%
\left( {\mathcal B}\right)=1$. Thus we obtain
\begin{equation}
x\left( {\mathcal S}_{\ell}\wedge n\right) =2V\left(
{\textstyle{1}{ 2}}%
\ell
+U^{-1}\left( n\right) \right).  \label{fina}
\end{equation}
Specializing to the case $\ell=3$ finally gives from (\ref{Upe}-\ref{u1pe})
\[
x\left( {\mathcal S}_{3}\wedge n\right) =2+\frac{1}{2}\left( n-1\right) +\frac{5%
}{24}\left( \sqrt{24n+1}-5\right).
\]


\subsection{Multifractality of Percolation Clusters}
\subsubsection{Multifractal Dimensions and Spectrum}
In terms of probability (\ref{xppe}), the harmonic measure moments (\ref{Zpe})
scale simply as ${\mathcal Z}_{n}\approx R^2{\mathcal P}_{R}\left( {\mathcal S}_{\ell =3}\wedge n\right)$
 \cite{cates},
which leads to
\begin{equation}
\tau \left( n\right) =x\left( {\mathcal S}_{3}\wedge n\right) -2.  \label{tt}
\end{equation}
Thus
\begin{equation}
\tau \left( n\right) =\frac{1}{2}\left( n-1\right) +\frac{5%
}{24}\left( \sqrt{24n+1}-5\right)
\end{equation}
is found to be {\it
identical} to (\ref{tauf}) for RW's and SAW's; the generalized dimensions $D\left( n\right)$
are then:
\begin{equation}
D\left( n\right) =\frac{1}{n-1}\tau \left( n\right)=\frac{1}{2}+\frac{5}{\sqrt{24n+1}+5},\quad n\in \left[ -%
{\textstyle{\frac{1}{  24}}}%
,+\infty \right) ,  \label{dn}
\end{equation}
valid for all values of moment order $n,n\geq -\frac{1}{24}.$ The
Legendre transform reads again exactly as in eq. (\ref{mf}):
\begin{equation}
f\left( \alpha \right) =\frac{25}{48}\left( 3-\frac{1}{2\alpha -1}\right) -%
\frac{\alpha }{24},\quad \alpha \in \left(
{\textstyle{\frac{1}{ 2}}}%
,+\infty \right) . \label{f}
\end{equation}

\subsubsection{Comparison to Numerical Results}
\begin{figure}[tb]
\begin{center}
\includegraphics[angle=0,width=.7\linewidth]{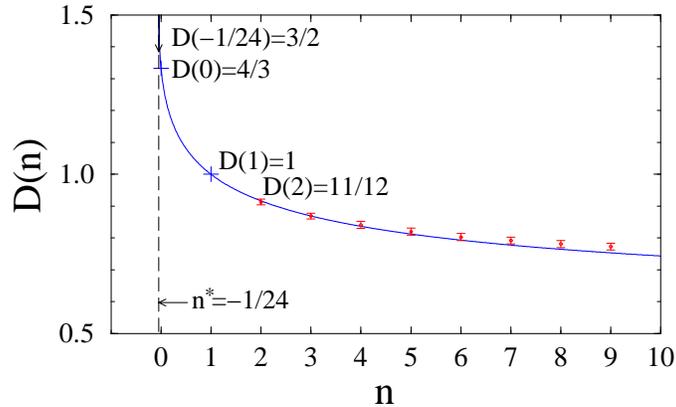}
\end{center}
\caption{Universal generalized dimensions $D(n)$ as a function of $n$, corresponding
to the harmonic measure near a percolation cluster, or to self-avoiding or
random walks, and comparison with 
the numerical data obtained by Meakin et al. (1988) for percolation.}
\label{Figure5}
\end{figure}

Only in the case of percolation has the harmonic measure been
systematically studied numerically, by Meakin et al.
 \cite{meakin}. We show in Figure \ref{Figure5} the exact curve $D\left(
n\right) $ (\ref{dn})
  \cite{duplantier9} together with the
numerical results for $n\in \{2,...,9\} $  \cite{meakin}, showing
fairly good agreement.

The average number ${\mathcal N}(H)$ (\ref{nh}) has been also
determined numerically for percolation clusters in  \cite{MS}, and
for $c=0$, our prediction (\ref{toc}) $\tau ^{\ast
}=\frac{23}{24}=0.95833...$ compares very well with the result
$\tau ^{\ast }=0.951\pm 0.030$, obtained for $%
10^{-5}\leq H\leq 10^{-4}$.

The dimension of the measure's support $D\left(
0\right)=\frac{4}{3} \neq D_{{\rm H}},$ where $D_{{\rm %
H}}=\frac{7}{4}$ is the Hausdorff dimension of the standard hull,
i.e., the outer boundary of critical percolating clusters
 \cite{SD}. The value $ D(0)=\frac{4}{3}$ corresponds to the dimension
of the {\it accessible external perimeter}. A direct derivation of
its exact value is given in  \cite{ADA}. The complement of the
accessible perimeter in the hull is made of deep fjords, which do
close in the scaling limit and are not probed by the harmonic
measure. This is in agreement with the instability phenomenon
observed on a lattice for the hull dimension  \cite{GA}. A striking
fact is the complete identity of the multifractal spectrum  for
percolation to the corresponding results,
eqs.(\ref{tauf}-\ref{mf}), {\it both} for random walks and
self-avoiding walks. Seen from outside, these three fractal simple
curves are not distinguished by the harmonic measure. In fact they
are the same, and one of the main conclusions of this study is
that {\it the external frontiers of a planar Brownian motion, or
of a critical percolation cluster are, in the scaling limit, identical to a critical
self-avoiding walk, with Hausdorff dimension $D=\frac{4}{3}$.}
As we have seen, this fact is linked to the presence of a single
universal conformal field theory (with a vanishing central charge
$c=0$), and to the underlying presence of quantum gravity, which
organizes the associated conformal dimensions. Note that in a
recent work, Smirnov  \cite{smirnov1} proved that critical site
percolation on the triangular lattice has a conformally-invariant
scaling limit, and that the discrete cluster interfaces (hulls)
converge to the same stochastic L\"owner evolution process as the
one involved for Brownian paths, opening the way to a rigorous
derivation of percolation exponents  \cite{lawler6,smirnov2},
previously derived in the physics literature \cite{dennijs,cardy,SD,ADA}.\\

\subsubsection{Double Layer Impedance}
Let us finally consider the different, but related, problem of the {\it
double layer impedance}
of a {\it rough} electrode. In some range of frequencies $\omega $, the
impedance contains an anomalous ``constant
phase angle'' (CPA) term $\left( i\omega
\right) ^{-\beta }$, where $\beta <1$.
From a
natural
RW representation of the impedance, a scaling law was recently
proposed:
$\beta =\frac{D\left( 2\right) }{D\left( 0\right) }$
\label{beta}
(here in 2D), where $D\left( 2\right) $ and $D\left( 0\right)$ are the multifractal dimensions of the
$H$-measure on the rough electrode \cite{halsey7}. In the case of a
2D porous percolative electrode, our
results (\ref{dn}) give $D\left( 2\right)
= \frac{11}{12},$ $D\left( 0\right)=\frac{4%
}{3}$, whence $\beta =\frac{11}{16}=0.6875.$ This
compares very well with a numerical RW algorithm result \cite{MS}, which yields
an
effective CPA exponent $\beta \simeq 0.69,$
nicely vindicating the multifractal description \cite{halsey7}.\\

\section{\sc{Conformally Invariant Frontiers and Quantum Gravity}}
\label{sec.conform}

In the next sections, we present a universal description of
multifractal functions for arbitrary conformally-invariant curves. They are derived
from conformal field theory and
quantum gravity. The geometrical findings are described in
detail, including cases like Ising clusters, or $Q=4$ Potts
Fortuin-Kasteleyn clusters, which are of particular interest.  We also make explicit the relation between
 a conformally-invariant scaling curve with CFT central charge $c$  \cite{duplantier11}, and the
 stochastic L\"owner process ${\rm SLE}_{\kappa}$  \cite{schramm1}. A fundamental geometric duality property
 for the external boundaries in $O(N)$ and Potts models, and SLE is obtained. For several simple paths we also define and calculate
 higher multifractal spectra.

\subsection{Harmonic Measure and Potential near a Fractal Frontier}
\subsubsection{Local Behavior of the Potential}
Consider a single (conformally-invariant) critical random cluster,
generically called ${\mathcal C}$. Let ${\mathcal H}\left( z\right) $ be the
potential at an exterior point $z \in {\rm {\mathbb C }}$, with
Dirichlet boundary conditions ${\mathcal H}\left({w \in \partial
\mathcal
C}\right)=0$ on the outer (simply connected) boundary $\partial
\mathcal C$ of $\mathcal C$, (or frontier  $ {\mathcal F} :=
\partial \mathcal C$) , and ${\mathcal H}(w)=1$ on a circle ``at $\infty$'',
i.e., of a large radius scaling like the average size $R$ of $
\mathcal C$. As is well-known
 \cite{kakutani}, ${\mathcal H}\left( z\right)$ is identical to the {\it
harmonic measure} of the circle ``at $\infty$'' seen from $z$, i.e, the probability that a random walker (more
precisely, a Brownian motion) launched from $z$, escapes to
$\infty$ without having hit ${\mathcal C}$.

The multifractal
formalism  \cite{mandelbrot2,hentschel,frisch,halsey} characterizes
subsets ${\partial\mathcal C}_{\alpha }$ of boundary sites by a
H\"{o}lder exponent $\alpha ,$ and a Hausdorff dimension $f\left(
\alpha \right) ={\rm dim}\left({\partial\mathcal C}_{\alpha
}\right)$, such that their potential locally scales as
\begin{equation}
{\mathcal H}\left( z \to w\in {\partial\mathcal C}_{\alpha }\right) \approx
\left( |z-w|/R\right) ^{\alpha }, \label{ha''}
\end{equation}
in the scaling limit $a \ll r=|z-w| \ll R$  (with $a$ the
underlying lattice constant if one starts from a lattice description before
taking the scaling limit $a \to 0$).

\subsubsection{Equivalent Wedge Angle}
In  2D the {\it complex} potential
$\varphi(z)$ (such that the electrostatic potential ${\mathcal H}(z)=\Im
\varphi(z)$ and the field's modulus $|{\bf E}(z)|=|\varphi'(z)|$) for a
{\it wedge} of angle $\theta$, centered at $w$, is
\begin{equation}
\varphi(z) = (z-w)^{{\pi}/{\theta}}.
\end{equation}
 By eq. (\ref{ha''}) a
H\"older exponent $\alpha$ thus defines a local equivalent
``electrostatic'' angle $\theta={\pi}/{\alpha},$ and the MF
dimension $\hat f(\theta)$ of the boundary subset with such
$\theta$ is
\begin{equation}
\hat f(\theta) = f(\alpha={\pi}/{\theta}). \label{fchapeau}
\end{equation}

\subsubsection{Moments}
It is convenient to define the harmonic measure $H(w,r)=H(\partial \mathcal C \cap B(w,r))$ in a ball  $B(w,r)$
of radius $r$ centered ar $w \in \partial \mathcal C$, as the
probability that a Brownian path started at infinity first hits the frontier 
$\mathcal F=\partial\mathcal C$
inside the ball $B(w,r)$. It is the integral of the Laplacian of
potential ${\mathcal H}$ in the ball $B(w,r)$, i.e., the boundary
charge in that ball. It scales as $r^{\alpha}$ with the same exponent as in (\ref{ha''}).

Of special interest are the moments of ${\mathcal H}$, averaged over all
realizations of ${\mathcal C}$, and defined as
\begin{equation}
{\mathcal Z}_{n}=\left\langle \sum\limits_{z\in {\partial
{\mathcal C}/r}}{H}^{n}\left(w,r \right) \right\rangle , \label{Za}
\end{equation}
 where points $w \in \partial {\mathcal C}/r$
 are the centers of a covering of the frontier $\partial \mathcal C$ by balls of radius $r$, and
 form a discrete subset $\partial {\mathcal C}/r \subset \partial {\mathcal C}$. The moment order
 $n$ can be
 a real number. In the scaling limit, one expects these moments to scale as
\begin{equation}
{\mathcal Z}_{n}\approx \left( r/R\right) ^{\tau \left( n\right)
}, \label{Z2a'}
\end{equation}
where the multifractal scaling exponents $\tau \left( n\right)$
vary in a non-linear way with
$n$ \cite{mandelbrot2,hentschel,frisch,halsey}. As above, they obey the
symmetric Legendre transform $\tau \left( n\right) +f\left( \alpha
\right) =\alpha n,$ with $n=f'\left( \alpha \right), \alpha
=\tau'\left( n\right)$. By normalization:
$\tau (1)=0.$ As noted above, because of the ensemble average
(\ref{Za}), values of $f\left( \alpha \right)$ can become negative
for some domains of $\alpha $  \cite{cates,cates et witten}.

\subsection{Calculation of Exponents from Quantum Gravity}
Let us now give the main lines of the derivation of exponents $\tau\left(
n\right) $, hence $f(\alpha)$,  via generalized {\it conformal invariance}.

\subsubsection{Brownian Representation of Moments}
As above, $n$ {\it independent}
RW's, or Brownian paths ${\mathcal B}$ in the scaling limit,
started at the same point a distance $r$ away from the cluster's
hull's frontier $\partial \mathcal C$, and diffusing without
hitting $\partial \mathcal C$, give a geometric representation of
the $n^{th}$ moment$, H^{n},$ in eq.(\ref{Z}) for $n$ {\it
integer}. Convexity yields the analytic continuation to arbitrary
$n$'s. Let us recall the notation $A\wedge B$ for two random
sets required to traverse, {\it without mutual intersection},
the
annulus ${\mathcal%
D}\left( r, R\right) $ from the inner boundary circle of radius
$r$ to the outer one at distance $R$, and{\it \ }$A\vee B$ for two
{\it independent}, thus possibly intersecting, sets
 \cite{duplantier8}. With this notation, one can define, as in eq. (\ref{zrs}), a grand canonical partition function
 which describes the star configuration
of the Brownian paths and cluster:
${\partial\mathcal C}\wedge {n}:={\partial\mathcal C}\wedge\left( \vee {\mathcal B}\right) ^{n} $.
At the critical point, it
is expected to scale for $%
r/R\rightarrow 0$ as
\begin{equation}
{\mathcal Z}_{R}\left( {\partial\mathcal C}\wedge n\right) \approx
\left( r/R\right) ^{x\left( n\right)+\cdots }, \label{xp}
\end{equation}
where the scaling exponent
\begin{equation}
\label{xnwedge}
x\left(n\right):=x\left({\partial\mathcal C}\wedge n\right)
\end{equation}
depends on $n$ and is associated
with the conformal operator creating the star vertex ${\partial\mathcal C}\wedge {n}$.
The dots after exponent $x(n)$ express the fact that there may be an
additional contribution to the exponent,
independent of $n$, corresponding to the entropy associated with the extremities of the random frontier
(see, e.g., (\ref{zrs})).

By normalization, this contribution actually does not appear in the multifractal moments.
Since $H$ is a probability measure, the sum (\ref{Za}) is indeed normalized as in (\ref{Z1})
\begin{equation}
{\mathcal Z}_{n=1}=1, \label{Za1}
\end{equation}
or in terms of star partition functions:
\begin{equation}
{\mathcal Z}_{n}={\mathcal Z}_{R}\left( {\partial\mathcal C}\wedge n\right) /{\mathcal Z}_{R}\left(  {\partial\mathcal C}\wedge 1\right).
\end{equation}
The scaling behavior (\ref{xp}) thus gives
\begin{equation}
\label{ZnZ1a}
 {\mathcal Z}_{n} \approx
(r/R)^{x\left( n\right)-x\left(1\right)}.
\end{equation}
The last exponent actually obeys the identity $x(1)=x\left(  {\partial\mathcal C}\wedge 1\right)=2$, which
will be obtained directly, and can also be seen as a consequence of Gauss's theorem in two
dimensions \cite{cates et witten}. Thus we can also write as in (\ref{xppe})
\begin{equation}
{\mathcal Z}_{n} = (R/r)^2\ {\mathcal P}_{R}\left(  {\partial\mathcal C}\wedge n\right),
\end{equation}
where ${\mathcal P}_{R}\left( {\partial\mathcal C}\wedge n\right)$ is a (grand-canonical) excursion measure
from $r$ to $R$ for the random set $ {\partial\mathcal C}\wedge n$,
with proper scaling ${\mathcal P}_{R}\approx
\left( r/R\right) ^{x\left( n\right)}$. The factor $(R/r)^2$
is the area scaling factor of the annulus $\mathcal D (r,R)$.

Owing to eqs. (\ref{Z2a'}) (\ref{ZnZ1a}) we get
\begin{equation}
\tau \left( n\right) =x(n)
-x\left(1\right)=x\left(n\right) -2.
\label{tt'}
\end{equation}

\subsubsection{Quantum Gravity}
To calculate these exponents, we again use the fundamental mapping between
the conformal field theory, describing a critical statistical system in the
plane ${\mathbb  C}$ or half-plane $\mathbb H$, and
the same CFT on a random planar surface, i.e., in
presence of {\it quantum gravity}  \cite{KPZ,DK,david2}. Two
universal functions $U$ and $V$, which now depend on the central
charge $c$ of the CFT, describe the KPZ map between conformal dimensions in bulk or boundary QG and those in
the standard plane or half-plane:
\begin{eqnarray}
U\left( x\right)=U_{\gamma}\left( x\right):=x\frac{x-\gamma}{1-\gamma} , \hskip2mm V\left(
x\right)= V_{\gamma}\left( x\right)=\frac{1}{4}\frac{x^{2}-\gamma^2}{1-\gamma},  \label{Ua}
\end{eqnarray}
with
\begin{equation}
\label{UVa}
V_{\gamma}\left( x\right):= U_{\gamma}\left(\frac{1}{2}\left(x+\gamma
\right) \right).
\end{equation}
The parameter $\gamma$ is the {\it string susceptibility exponent}
of the random 2D surface (of genus zero), bearing the CFT of
central charge $c$ \cite{KPZ}; $\gamma$ is the solution of
\begin{equation}
c=1-6{\gamma}^2(1-\gamma)^{-1}, \gamma \leq 0. \label{cgamma}
\end{equation}
In order to simplify the notation, we shall hereafter in this section
drop the subscript $\gamma$ from functions $U$ and $V$.

The function $U$ maps quantum gravity conformal weights, whether in the bulk or on a boundary,
into their counterparts in $\mathbb C$ or $\mathbb H$, as in (\ref{KPZg})(\ref{KPZgb}). The function $V$ has been
tailored to map quantum gravity
boundary dimensions to the corresponding conformal dimensions in the full plane $\mathbb C$,
as in (\ref{Zetal}) (\ref{ZetaL}).
The {\it positive} inverse
function of $U$, $U^{-1}$, is
\begin{equation}
U^{-1}\left( x\right)
=\frac{1}{2}\left(\sqrt{4(1-\gamma)x+\gamma^2}+\gamma\right),
\label{U1a}
\end{equation}
and transforms conformal weights of a conformal operator in $\mathbb C$ or $\mathbb H$ into
the conformal weights of the same operator in quantum gravity, in the bulk or on the boundary.
Note the shift relation
\begin{equation}
U^{-1}\left( x\right) =\frac{1}{2}V^{-1}\left(
x\right)+\frac{1}{2}\gamma\ , \label{shift}
\end{equation}
where the inverse of $V$,
\begin{equation}
V^{-1}\left( x\right) = \sqrt{4(1-\gamma)x+\gamma^2}, \label{V1}
\end{equation}
transforms the bulk conformal weight in $\mathbb C$ of a given conformal operator into the
boundary conformal weight of the corresponding boundary operator in quantum gravity (see appendix \ref{BBapp}).

\subsubsection{Boundary Additivity Rule}
Consider two arbitrary  random sets $A,B,$ with boundary scaling
exponents $\tilde{x}\left( A\right)
,\tilde{x}\left( B\right)$ in the {\it half-plane} $\mathbb H$ with Dirichlet boundary conditions.
When these two sets are mutually-avoiding, the scaling exponent $x\left( A\wedge
B\right)$ in $\mathbb C$, as in (\ref{xnwedge}), or $\tilde x\left( A\wedge
B\right)$ in $\mathbb H$ have the universal structure
 \cite{duplantier8,duplantier9,duplantier11}
\begin{eqnarray}
x\left( A\wedge B\right) &=&2V\left[ U^{-1}\left( \tilde{x}\left(
A\right)
\right) +U^{-1}\left( \tilde{x}\left( B\right) \right) \right],  
\label{xa}
\end{eqnarray}
\begin{eqnarray}
\tilde x\left( A\wedge B\right) &=&U\left[ U^{-1}\left( \tilde{x}\left(
A\right)
\right) +U^{-1}\left( \tilde{x}\left( B\right) \right) \right].  
\label{xb}
\end{eqnarray}
We have seen these fundamental relations in the $c=0$ case above; they are established
for the general case in appendix \ref{BBapp}. $U^{-1}\left( \tilde{x} \right)$ is, on the random disk
with Dirichlet boundary conditions, the boundary scaling dimension corresponding to
$\tilde{x}$ in the half-plane  ${\mathbb  H}$, and  in eqs. (\ref{xa}) (\ref{xb})
\begin{eqnarray}
U^{-1}\left(\tilde x\left( A\wedge B\right)\right) &=&U^{-1}\left( \tilde{x}\left(
A\right)
\right) +U^{-1}\left( \tilde{x}\left( B\right) \right)  
\label{xc}
\end{eqnarray}
is a
{\it linear} boundary exponent corresponding to the fusion of two ``boundary
operators'' on the random disk, under the Dirichlet mutual avoidance condition $A \wedge B$.
This quantum boundary conformal dimension is mapped back by $V$ to the scaling
dimension in $\mathbb C$, or by $U$ to the boundary scaling dimension in $\mathbb H$ \cite{duplantier11} (see appendix \ref{BBapp}).

\subsubsection{Exponent Construction}
\begin{figure}[tb]
\begin{center}
\includegraphics[angle=0,width=.7\linewidth]{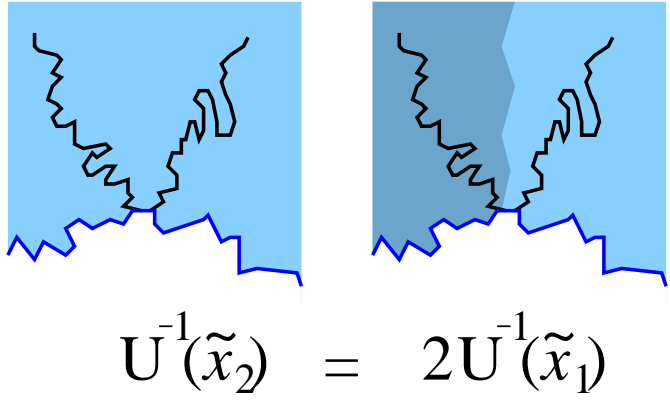}
\end{center}
\caption{Illustration of the additivity rule (\ref{xtilde}): each of the two non-intersecting strands of
a simple random path
lives in its own sector of the random disk near the Dirichlet boundary.}
\label{split}
\end{figure}
\begin{figure}[tb]
\begin{center}
\includegraphics[angle=0,width=.7\linewidth]{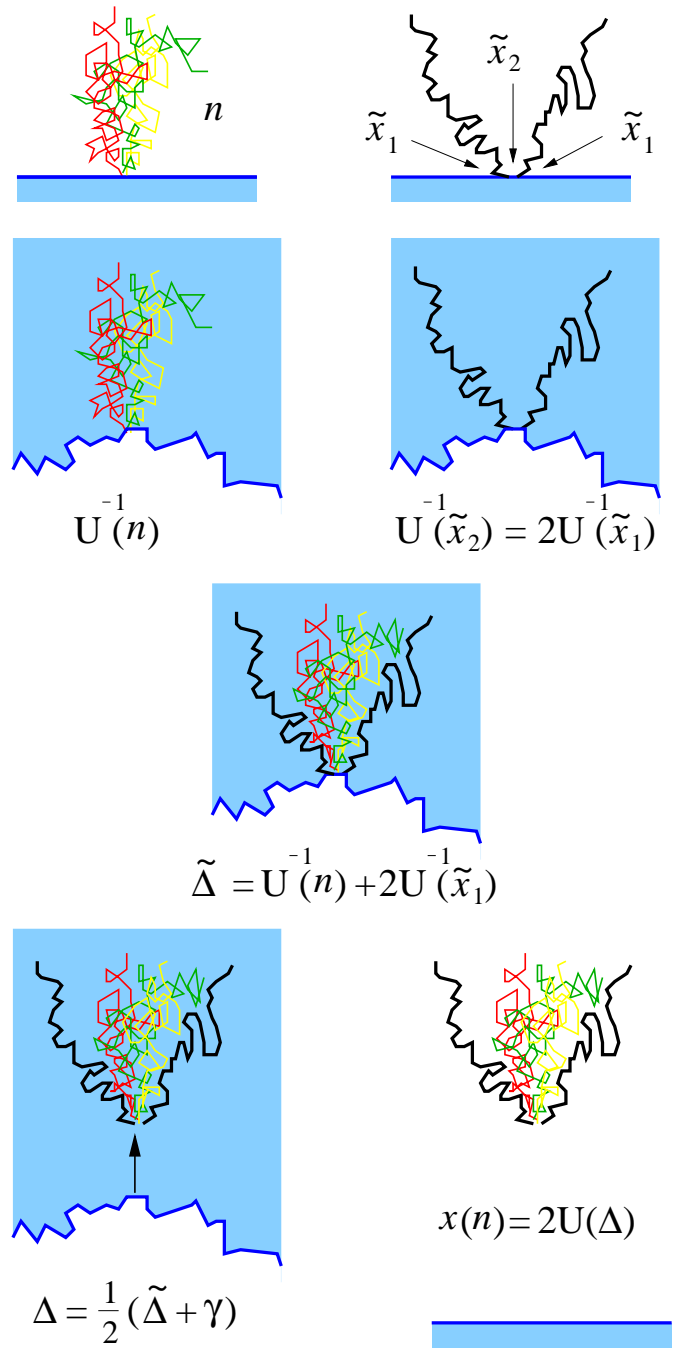}
\end{center}
\caption{The quantum gravity construction (\ref{xtilde}) (\ref{deltatildea}) of exponents (\ref{finab}).}
\label{sumsmall}
\end{figure}

For determining the harmonic exponents $x(n)$ (\ref{xnwedge}),
we use (\ref{xa}) for $A={\partial \mathcal C}$ and $B=\left( \vee {\mathcal B}\right) ^{n}$.

\noindent $\bullet$ We first need the {\it boundary} (conformal) scaling dimension (b.s.d.)
$\tilde x_2:=\tilde{x}\left({\partial \mathcal C}\right)$ associated with the presence of the random frontier
near the Dirichlet boundary $\mathbb H$. Since
this frontier is simple, it can be seen as made of two non-intersecting semi-infinite strands (Fig. \ref{split}).
Its b.s.d. in quantum gravity thus obeys
(\ref{xc})
\begin{equation}
U^{-1}\left( \tilde{x_2}\right) =2U^{-1}\left( \tilde{x_1}\right),
\label{xtilde}
\end{equation}
where $\tilde x_1$ is the boundary scaling dimension of a semi-infinite frontier path
 originating at the boundary $\mathbb H$.

\noindent $\bullet$ The packet of $n$ independent Brownian paths has 
$\tilde{x}\left( \left( \vee {\mathcal B}\right) ^{n}\right)=n,$
since $\tilde{x} \left( {\mathcal B}\right)=1$.

\noindent $\bullet$ From (\ref{xc}) the QG boundary dimension of the whole set is (see Fig. \ref{sumsmall}):
\begin{equation}
\tilde \Delta:=U^{-1}\left[\tilde x\left({\partial \mathcal C} \wedge n\right)\right]=2U^{-1}\left( \tilde{x_1}\right)
 +U^{-1}\left( n\right).
\label{deltatildea}
\end{equation}
Its associated QG bulk conformal dimension is therefore
$\Delta=\frac{1}{2}(\tilde \Delta+\gamma)$ (appendix \ref{BBapp}). From eqs. (\ref{UVa}) or (\ref{xa})
we finally find
\begin{eqnarray}
\nonumber
 x\left( n\right) &=&2U(\Delta)=2V(\tilde\Delta)\\
\label{finab}
 &=&2V\left[2U^{-1}\left( \tilde{x_1}\right)+U^{-1}\left( n\right) \right].
\end{eqnarray}
The whole construction is illustrated in Fig. \ref{sumsmall}.

\noindent $\bullet$ The value of the QG b.s.d. of a {\it simple} semi-infinite random path is
\begin{equation}
\label{tildex1}
U^{-1}\left( \tilde{x_1}\right) =\frac{1}{2}(1-\gamma).
\end{equation}
It is derived in section \ref{sec.SLEKPZ} below and in appendix \ref{ONapp}
from the exponents of the $O(N)$ model, or of the SLE. It can be directly derived from Makarov's theorem:
\begin{equation}
\alpha(n=1)=\tau'(n=1)=\frac{ dx}{ d n}(n=1)=1,
\end{equation}
which, applied to (\ref{finab}), leads to the same result. We thus finally get
\begin{equation}
x\left( n\right) =2V\left(1-\gamma
 +U^{-1}\left( n\right) \right)=2U\left( \frac{1}{2}+\frac{1}{2}
 U^{-1}\left( n\right) \right).  \label{finaa}
\end{equation}
This result satisfies the identity: $x(1)=2U(1)=2$, which is related to Gauss's theorem, as mentioned above.\\

\subsubsection{Multifractal Exponents}

\noindent $\bullet$ The multifractal exponents $\tau(n)$ (\ref{tt'}) are obtained from
(\ref{Ua}-\ref{U1a}) as \cite{duplantier11}
\begin{eqnarray}
\nonumber
\tau\left( n\right)&=&x(n)-2\\
&=&\frac{1}{2}(n-1)+\frac{1}{4}\frac{2-\gamma}{1-\gamma}
[\sqrt{4(1-\gamma)n+{\gamma}^2}-(2-\gamma)]\ .
\label{tauoriginal}
\end{eqnarray}
Similar exponents, but
associated with moments taken at the tip,
later appeared in the context of the ${\rm SLE}$ process
(see II in  \cite{lawler4}, and  \cite{lawleresi}; see also  \cite{hastings} for Laplacian random walks.)
The  whole family will be given in section \ref{sec.multifSLE}.

\noindent $\bullet$ The Legendre transform is easily performed to yield:
\begin{eqnarray}
\alpha &=&\frac{d{\tau} }{dn}\left( n\right)=\frac{1}{2}
+\frac{1}{2} \frac{2-\gamma}{\sqrt{4(1-\gamma)n+{\gamma}^2}};
\label{a'}
\\
\nonumber
\\
f\left( \alpha \right)&=&
\frac{1}{8}\frac{(2-\gamma)^2}{1-\gamma}\left(3- \frac{1}{2\alpha
-1}\right) -\frac{1}{4}\frac{\gamma^2}{1-\gamma}\alpha,
\label{foriginal}
\\
\quad \alpha &\in& \left(
{\textstyle{ \frac{1}{ 2}}}%
,+\infty \right) . \nonumber
\end{eqnarray}

\noindent It is convenient to express the results in terms of the central charge $c$ with the help of:
\begin{eqnarray}
\label{c,gamma}
\frac{1}{4}\frac{(2-\gamma)^2}{1-\gamma}=\frac{25-c}{24};\;\;\;\;
\frac{1}{4}\frac{\gamma^2}{1-\gamma}=\frac{1-c}{24}.
\end{eqnarray}
We finally find the\\
\noindent $\bullet$ {\it Multifractal Exponents}
\begin{eqnarray}
\label{taunc}
\tau\left( n\right) &=&\frac{1}{2}(n-1)+\frac{25-c}{24}
\left(\sqrt{\frac{24n+1-c}{25-c}}-1\right),
\\
\nonumber
\\
D\left( n\right) &=&\frac{\tau\left( n\right)}{n-1}=\frac{1}{2}+
{\left(\sqrt{\frac{24n+1-c}{25-c}}+1\right)}^{-1},
\label{D''}\\
\quad n&\in& \left[ n^{\ast}= -\frac{1-c}{24} ,+\infty \right)\ ;
\nonumber
\end{eqnarray}
$\bullet$ {\it Multifractal Spectrum}
\begin{eqnarray}
\alpha &=&\frac{d{\tau} }{dn}\left( n\right)=\frac{1}{2}
+\frac{1}{2} \sqrt{\frac{25-c}{24n+1-c}}; \label{a'a}
\\
\nonumber
\\
\label{foriginalbis} f\left( \alpha \right)&=& \frac{25-c}{48}\left(3-
\frac{1}{2\alpha -1}\right) -\frac{1-c}{24}\alpha,
\\
\quad \alpha &\in& \left(
\frac{1}{2}
,+\infty \right) . \nonumber
\end{eqnarray}

\subsubsection{Other Multifractal Exponents}
This formalism immediately allows generalizations. For instance,
in place of a packet of $n$ independent random walks, one can consider a packet of $n$ {\it
independent self-avoiding walks} $\mathcal P$, which avoid the fractal boundary.
The associated multifractal exponents $ x\left( {\partial\mathcal
C}\wedge \left( \vee {\mathcal P}\right)^{n} \right)$ are given by
(\ref{finaa}), with the argument $n$ in $U^{-1}(n)$ simply
replaced by  ${\tilde x}\left( \left( \vee {\mathcal P}\right)
^{n}\right) =n{\tilde x} \left( {\mathcal P}\right) =\frac{5}{8}n
$  \cite{duplantier8}. These exponents govern the universal
multifractal behavior of the moments of the probability that a SAW
escapes from $\mathcal C$. One then gets a spectrum $\bar f\left(\bar\alpha\right)$ such
that $${\bar f}\left(\bar\alpha=\tilde{x} \left( {\mathcal
P}\right)\alpha \right) =
f\left(\alpha=\pi/\theta\right)={\hat f}(\theta),$$ thus unveiling
the  same invariant
underlying wedge distribution as the harmonic measure (see also \cite{cardy2}).\\

\subsection{{Geometrical Analysis of Multifractal Spectra}}
\label{subsec.geometry}

\subsubsection{Makarov's Theorem}
The generalized dimensions $D(n)$ satisfy, for any
$c$, $\tau'(n=1)=D(n=1)=1$, or equivalently $f(\alpha=1)=1$, i.e.,
{\it Makarov's theorem}  \cite{makarov}, valid for any simply
connected boundary curve. From (\ref{D''}), (\ref{a'a}) we also
note a fundamental relation, independent of $c$:
\begin{equation}
3-2D(n)=1/\alpha=\theta/\pi. \label{Dtheta}
\end{equation}
We also have the {\it superuniversal} bounds: $\forall c, \forall
n,\frac{1}{2}=D(\infty) \leq D(n) \leq D(n^{\ast})=\frac{3}{2}$,
corresponding to $0 \leq \theta\leq 2\pi$.

\subsubsection{Symmetries}
It is interesting to note that the general multifractal function
(\ref{foriginalbis}) can also be written as
\begin{eqnarray}
f\left( \alpha \right)-\alpha&=& \frac{25-c}{24}
\left[1-\frac{1}{2}\left(2\alpha -1 + \frac{1}{2\alpha
-1}\right)\right]. \label{f-aa}
\end{eqnarray}
The multifractal functions $f\left( \alpha \right)-\alpha
=\hat f(\theta)-\frac{\pi}{\theta}$ thus possess the invariance property (\ref{inv})
upon substitution of primed variables
given by
\begin{equation}
2\pi=\theta+{\theta}^{\prime}=\frac{\pi}{\alpha}+\frac{\pi}{{\alpha}^{\prime}};
\label{tetateta''}
\end{equation}
this corresponds to the complementary domain of the wedge
$\theta$. This condition reads also $D(n)+D(n')=2.$ This basic
symmetry, first observed  \cite{BDH} for the $c=0$
result of  \cite{duplantier8} (see section \ref{subsec.inv}), is valid for {\it any} conformally
invariant boundary.

\subsubsection{Equivalent Wedge Distribution}
The geometrical
multifractal distribution of wedges $\theta$ along the boundary takes the form:
\begin{eqnarray}
\hat
f(\theta)=f\left(\frac{\pi}{\theta}\right)=\frac{\pi}{\theta}-\frac{25-c}{12}
 \frac{(\pi-\theta)^2}{\theta (2\pi -\theta)}\ .
\label{fchap}
\end{eqnarray}
Remarkably enough, the second term also describes the contribution
by a wedge to the density of electromagnetic modes in a cavity
 \cite{BD}. The simple shift in (\ref{fchap}), $25 \to 25 -c$, from
the $c=0$ case to general values of $c$, can then be related to
results of conformal invariance in a wedge  \cite{DuCa}. The
partition function for the two sides of a wedge of angle $\theta$
and size $R$, in a CFT of central charge $c$, indeed scales as
 \cite{Ca}
\begin{equation}
 \hat {\mathcal Z} (\theta,c) \approx R^{-
 {c(\pi-\theta)^2}/12\,{\theta (2\pi -\theta)}}\ .
\label{hatZ}
\end{equation}
Thus, one can view the $c$ dependance of result (\ref{fchap}) as follows: the
number of sites, $R^{\hat f(\theta,c)}$,
 with local wedge angle $\theta$ along a random path with central charge $c$, is the same
 as the number of sites, $R^{\hat f(\theta,c=0)}$, with wedge angle $\theta$
along a {\it self-avoiding walk} ($c=0$), renormalized  by the partition function $\hat
{\mathcal Z}(\theta)$ representing the presence of a
$c$-CFT along such wedges:
$$R^{\hat f(\theta,c)} \propto R^{\hat f(\theta,c=0)}/\hat {\mathcal Z} (\theta,c).$$

\subsubsection{Hausdorff Dimension of the External Perimeter}
The maximum of $f(\alpha)$ corresponds
to $n=0$, and gives the Hausdorff dimension $D_{\rm EP}$ of
 the support of the measure, i.e.,
the {\it accessible} or {\it external perimeter} as:
\begin{eqnarray}
D_{\rm
EP}&=&{\sup}_{\alpha}f(\alpha)=f(\alpha(n=0))\\&=&D(0)=\frac{3-2\gamma}{2(1-\gamma)}
=\frac{3}{2}-\frac{1}{24}\sqrt{1-c}\left(\sqrt{25-c}-\sqrt{1-c}\right).
\label{D(c)}
\end{eqnarray}
This corresponds to a {\it typical} sigularity exponent
\begin{equation}
\hat\alpha={\alpha(0)}=1-\frac{1}{\gamma}=\left(\frac{1}{12}\sqrt{1-c}\left(\sqrt{25-
c}-\sqrt{1-c}\right)\right)^{-1}=(3-2D_{\rm EP})^{-1}\ ,
\label{halpha}
\end{equation}
and to a  typical wedge angle
\begin{equation}
\hat\theta={\pi}/{\hat \alpha}=\pi(3-2D_{\rm EP})\ .
\label{htheta}
\end{equation}

\subsubsection{Probability Densities}
The
probability $P(\alpha)$ to find a singularity exponent $\alpha$
or, equivalently, $\hat P (\theta)$ to find an equivalent opening
angle $\theta$ along the frontier is
\begin{equation}
P(\alpha)=\hat P(\theta)\propto R^{f(\alpha)- f(\hat\alpha)} \ .
\end{equation}
Using the values found above, one can recast this probability as
(see also  \cite{cardy2})
\begin{equation}
P(\alpha)=\hat P(\theta)\propto \exp\left[-\frac{1}{24}\ln
R\left(\sqrt{1-c}\sqrt{\omega}-\frac{\sqrt{25- c}}{2 \sqrt \omega}
\right)^2\right]\ , \label{prob}
\end{equation}
where  $$\omega=\alpha-\frac{1}{2}=\frac{\pi}{\theta}-\frac{1}{2}\
.$$

\subsubsection{Universal Multifractal Data}
\begin{figure}[tb]
\begin{center}
\includegraphics[angle=0,width=.7\linewidth]{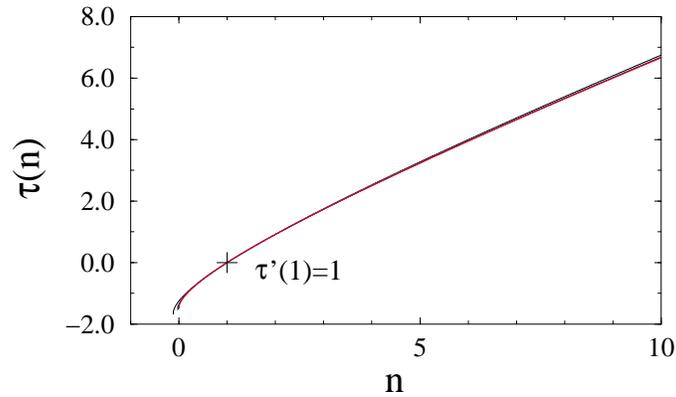}
\end{center}
\caption{Universal multifractal exponents
$\tau(n)$. The curves are indexed by the central charge $c$, and correspond
 to the same colors as in Figure \ref{Figure8} below: (black: 2D spanning trees ($c=-2$); green:
 self-avoiding or random walks, and
percolation ($c=0$); blue:
Ising clusters or $Q=2$ Potts clusters ($c=\frac{1}{2}$); red: $N=2$ loops, or $Q=4$
Potts clusters
($c=1$). The curves are almost
indistinguishable at the scale shown.}
\label{Figure6}
\end{figure}

\begin{figure}[tb]
\begin{center}
\includegraphics[angle=0,width=.7\linewidth]{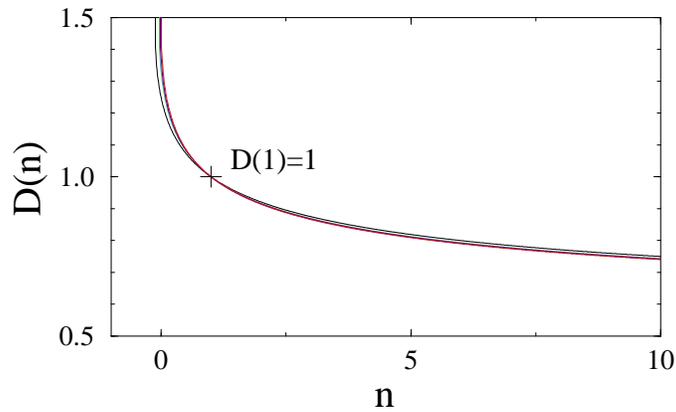}
\end{center}
\caption{Universal generalized dimensions
$D(n)$. The curves are indexed by the same colors as in Figure \ref{Figure8} below but are almost
indistinguishable at the scale shown.}
\label{Figure7}
\end{figure}

\begin{figure}[tb]
\begin{center}
\includegraphics[angle=0,width=.7\linewidth]{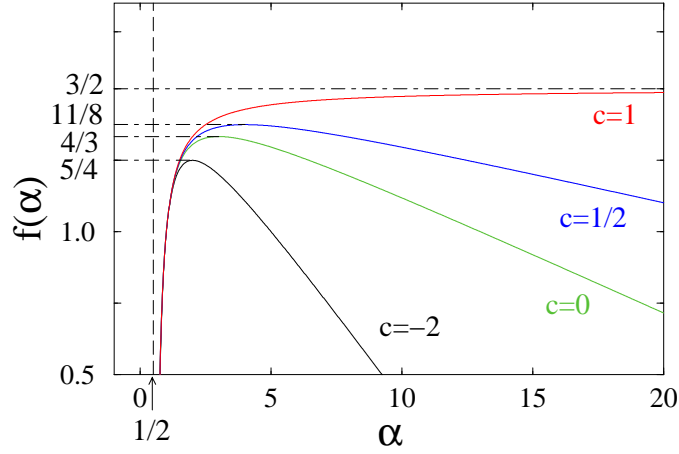}
\end{center}
\caption{Universal harmonic multifractal spectra
$f(\alpha)$. The curves are indexed by the central charge $c$, and correspond
 to: 2D spanning trees ($c=-2$); self-avoiding or random walks, and
percolation ($c=0$);
Ising clusters or $Q=2$ Potts clusters ($c=\frac{1}{2}$); $N=2$ loops, or $Q=4$
Potts clusters
($c=1$). The maximal dimensions are those of the {\it
accessible} frontiers. The left branches of the various $f(\alpha)$ curves are largely
indistinguishable, while their right branches split for large $\alpha$,
corresponding
 to negative values
of $n$.}
\label{Figure8}
\end{figure}

The multifractal exponents $\tau(n)$ (Fig. \ref{Figure6}) or generalized
dimensions $D(n)$ (Fig. \ref{Figure7}) appear quite similar for various
values of $c$, and a numerical simulation would hardly distinguish
the different universality classes, while the $f(\alpha)$
functions, as we see, do distinguish these classes,
especially for negative $n$, i.e. large $\alpha$. In Figure \ref{Figure8}
 we display the multifractal functions $f$, eq. (\ref{foriginalbis}),
corresponding to various values of $-2 \leq c \leq 1$, or,
equivalently, to a number of components $N \in [0, 2]$, and $Q \in
[0,4]$ in the $O(N)$ or Potts models (see below).

\subsubsection{Needles}
The singularity at $\alpha=\frac{1}{2}$, or $\theta=2\pi$, in the
multifractal functions $f$, or $\hat f$, corresponds to boundary
points with a needle local geometry, and Beurling's theorem
 \cite{Beur} indeed insures that the H{\"o}lder exponents $\alpha$ are
 bounded below by $\frac{1}{2}$. This corresponds to large
values of $n$, where, asymptotically, for {\it any}
 universality class,
\begin{equation}
\forall c, \lim_{n \to \infty} D(n)=\frac{1}{2}. \label{1/2}
\end{equation}

\subsubsection{Fjords}
The right branch of $f\left( \alpha \right) $ has a linear
asymptote
\begin{equation}
\lim_{\alpha \rightarrow \infty} f\left(\alpha \right)/{\alpha} =
n^{\ast}=-(1-c)/24.
\end{equation}
The $\alpha \to \infty$ behavior corresponds to moments of lowest
order $n\rightarrow {n^{\ast}}$, where $D(n)$ reaches its maximal
value: $\forall c, D(n^{\ast})=\frac{3}{2}$, common to {\it all}
simply connected, conformally-invariant, boundaries.

This describes almost inaccessible sites: Define ${\mathcal N}\left(
H\right)$ as the number of boundary sites having a given
probability $H$ to be hit by a RW starting at infinity; the MF
formalism yields, for $H\rightarrow 0,$ a power law behavior
\begin{equation}
{\mathcal N}\left( H\right)|_{H\rightarrow 0}\approx
H^{-(1+{n}^{\ast})} \label{nha}
\end{equation}
with an exponent
\begin{equation}
1+n^{\ast}=\frac{23+c}{24}<1.
\end{equation}

\subsubsection{Ising Clusters}
 A critical  Ising
cluster ($c=\frac{1}{2}$) possesses a multifractal spectrum with respect to the
harmonic measure:
\begin{eqnarray}
\tau \left( n\right)&=&\frac{1}{2}\left( n-1\right)
+\frac{7}{48}\left(
\sqrt{48n+1}-7\right) , \label{taufis}\\
f\left( \alpha \right) &=&\frac{49}{96}\left( 3-\frac{1}{2\alpha -1}\right) -%
\frac{\alpha }{48},\quad \alpha \in \left(
\frac{1}{2}
,+\infty \right), \label{fis}
\end{eqnarray}
with the dimension of the accessible perimeter
\begin{equation}
D_{\rm EP}={\rm sup}_{\alpha}f(\alpha,
c=\frac{1}{2})=\frac{11}{8}.
\end{equation}

\subsubsection{$Q=4$ Potts Clusters, and ``ultimate Norway''}
The {\it limit} multifractal spectrum is obtained for $c=1$,  which is an upper
 or lower bound for all $c$'s, depending on the position of $\alpha$ with respect to $1$:
\begin{eqnarray}
\nonumber
f(\alpha,c<1) &<& f(\alpha,c=1),\;  1 < \alpha,\\
\nonumber
f(\alpha=1,c)&=&1,\; \forall c,\\
\nonumber
f(\alpha,c<1) &>& f(\alpha,c=1),\;  \alpha <1.
\end{eqnarray}
This MF spectrum provides an exact example of a {\it left-sided} MF spectrum,
with an asymptote $f\left(\alpha \to \infty, c=1\right)\to
\frac{3}{2}$ (Fig. \ref{Figure8}). It corresponds to singular boundaries
where ${\hat f}\left(\theta \to 0, c=1\right)=\frac{3}{2}=D_{\rm
EP}$, i.e., where the external perimeter is everywhere dominated
by ``{\it fjords}'', with typical angle $\hat \theta =0$. It is
tempting to call it the ``ultimate Norway''.

The frontier of a $Q=4$ Potts Fortuin-Kasteleyn cluster, or the ${\rm SLE}_{\kappa=4}$ provide such an example for  this {\it
left-handed} multifractal spectrum ($c=1$) (see section \ref{sec.geodual}). The MF data are:
\begin{eqnarray}
\tau \left( n\right)&=&\frac{1}{2}\left( n-1\right) +
\sqrt{n}-1, \label{taufc}\\
f\left( \alpha \right) &=&\frac{1}{2}\left( 3-\frac{1}{2\alpha
-1}\right),\quad \alpha \in \left(
\frac{1}{2}
,+\infty \right), \label{fc}
\end{eqnarray}
with accessible sites forming a set of Hausdorff dimension
\begin{equation}
D_{\rm EP}={\rm sup}_{\alpha}f(\alpha,c=1)=\frac{3}{2},
\end{equation}
which is also the {\it maximal} value common to all multifractal
generalized dimensions $D(n)=\frac{1}{n-1}\tau(n)$. The external perimeter
which bears the electrostatic charge is a non-intersecting
{\it simple} path. We therefore arrive at the
striking conclusion that in the plane, a conformally-invariant
scaling curve which is simple has a Hausdorff dimension at
most equal to $D_{\rm EP}=3/2$ \cite{duplantier11}. The
corresponding $Q=4$ Potts frontier, while still possessing a set
of double points of dimension $0$, actually develops
a logarithmically growing number of double points  \cite{aharony}.\\

\vfill\eject
\section{\sc{Higher Universal Multifractal Spectra}}
\label{sec.higher}
\subsection{Double-Sided Spectrum}
\subsubsection{Simple Random Paths \& Double-Sided Potential}
As in section \ref{subsec.double}, we consider here the specific case where the
fractal set
${\mathcal C}$ is a (conformally-invariant) {\it simple} scaling
curve, that is, it does not contain double points. The frontier
$\partial\mathcal C$ is thus identical with the set itself:
$\partial\mathcal C=\mathcal C$.
Each point of the curve can then be reached from infinity, and we
address the  question of the simultaneous behavior
of the potential on both sides of the curve. Notice, however, that one
could also address the case of non-simple random paths, by concentrating
on the double-sided potential near cut-points, as we did in section \ref{subsec.double}
for cut-points in Brownian paths.

The
potential ${\mathcal H}$ scales as
\begin{equation}
{\mathcal H}_{+}\left( z \to w^{+}\in {\partial\mathcal C}_{\alpha,\alpha'
}\right) \approx |z-w|^{\alpha}, \label{ha+}
\end{equation}
when approaching $w$ on one side of the scaling curve, while scaling as
\begin{equation}
{\mathcal H}_{-}\left( z \to w^{-}\in {\partial\mathcal C}_{\alpha,\alpha'
}\right) \approx |z-w|^{\alpha'}, \label{ha-}
\end{equation}
on the other side. The multifractal
formalism now characterizes subsets ${\mathcal C}_{\alpha,\alpha' }$
of boundary sites $w$ with two such H\"{o}lder exponents, $\alpha
,\alpha'$, by their Hausdorff dimension
$f_2\left( \alpha,\alpha' \right) :={\rm dim}\left({\mathcal
C}_{\alpha,\alpha' }\right)$. The standard one-sided multifractal spectrum
$f(\alpha)$ is then recovered as the supremum:
\begin{equation}
f(\alpha)={\rm sup_{\alpha'}}f_2\left( \alpha,\alpha' \right).
\label{sup}
\end{equation}

\subsubsection{Equivalent Wedges}
As above, one can also define two equivalent ``electrostatic''
angles from singularity exponents $\alpha,\alpha'$, as
$\theta={\pi}/{\alpha},\theta'={\pi}/{\alpha'}$ and the MF
dimension $\hat f_2(\theta,\theta')$ of the boundary subset with
such $\theta,\theta'$ is then
\begin{equation}
\hat f_2(\theta,\theta') :=
f_2(\alpha={\pi}/{\theta},\alpha'={\pi}/{\theta'}).
\label{fchapeau'}
\end{equation}

\subsubsection{Harmonic Moments}
Consider the harmonic measure (as seen from infinity) ${H}\left( w,r\right):={H}({\mathcal C} \cap { B}(w, r))$
of the intersection of $\mathcal C$ and the ball
${B}(w, r)$ centered at point $w \in {\mathcal C}$, i.e., the
probability that a Brownian path, launched from infinity, {\it
first} hits the  frontier ${\mathcal C}$ inside the ball ${B}(w, r)$. Let us consider a
covering of the frontier by such balls centered at points
forming a discrete subset ${\mathcal C}/r$ of $\mathcal C$.

 The double multifractal spectrum will be
computed from the double moments of the harmonic measure on {
both} sides of the random fractal curve:
\begin{equation}
{\mathcal Z}_{n,n'}=\left\langle \sum\limits_{w\in {\mathcal
C}/r}\left[{H}_{+}(w,r)\right]^{n} \left[{H}_{-}(w,r)\right]^{n'}\right\rangle, \label{ZZ''}
\end{equation}
where ${H}_{+}(w,r)$
and ${H}_{-}(w,r)$
are
respectively the harmonic measures on the ``left'' or ``right''
sides of the random fractal. These double moments have a
multifractal scaling behavior
 \begin{equation}
{\mathcal Z}_{n,n'}\approx \left( r/R\right) ^{\tau_2 \left(
n,n'\right) },  \label{ZZ2''}
\end{equation}
where the exponent $\tau_2 \left(n,n'\right)$ now depends on two
moment orders $n,n'$. As in section \ref{subsec.double}, the Hausdorff dimension
is given by the double Legendre transform:
\begin{eqnarray}
\nonumber
\alpha &=&\frac{\partial\tau_2 }{\partial n}\left( n,n'\right) ,
\quad \alpha' =\frac{\partial\tau_2 }{\partial n'}\left( n,n'\right), \\
f_2\left( \alpha, \alpha' \right) &=&\alpha n+\alpha' n'-\tau_2 \left( n,n'\right), \label{alpha'''}\\
\nonumber
n&=&\frac{\partial f_2}{\partial \alpha}\left( \alpha, \alpha'
\right) , \quad n'=\frac{\partial f_2}{\partial \alpha' }\left(
\alpha, \alpha' \right).
\end{eqnarray}
From definition (\ref{ZZ''}) and eq. (\ref{ZZ2''}), we recover for $n'=0$ the
one-sided multifractal exponents
\begin{equation}
\tau \left( n\right)=\tau_2 \left( n,n'=0\right),
\end{equation}
and putting these values in the Legendre transform  (\ref{alpha'''}) yields identity (\ref{sup}),
as it must.\\

\subsection{Higher Multifractality of Path Vertices}

\subsubsection{Definition}
One can consider a star configuration ${\mathcal S}_m$ of a number $m$
 of {\it similar simple scaling paths},
all originating at the same vertex $w$. Higher moments
${\mathcal Z}_{n_1,n_2,...,n_m}$ are then defined as
\begin{equation}
{\mathcal Z}_{n_1,n_2,...,n_m}=\left\langle \sum\limits_{w\in
{\mathcal S}_m}\left[{H}_{1}(w,r)\right]^{n_1} \left[{H}_{2}(w,r)\right]^{n_2}\cdots\left[{H}_{m}(w,r)\right]^{n_m}
\right\rangle, \label{Z_m}
\end{equation}
where $${H}_{i}(w,r):= {H}_{i}\left({\mathcal C} \cap
{B}(w, r)\right)$$ is the harmonic measure (or,
equivalently, local potential at distance $r$) in the $i$th sector
of radius $r$ located between paths $i$ and $i+1$, with
$i=1,\cdots,m$, and by periodicity $m+1 \equiv 1$. These higher
moments have a multifractal scaling behavior
 \begin{equation}
{\mathcal Z}_{n_1,n_2,...,n_m}\approx \left( r/R\right) ^{\tau_m
\left(n_1,n_2,...,n_m \right) }, \label{Z_m'}
\end{equation}
where the exponent $\tau_m \left(n_1,n_2,...,n_m\right)$ now
depends on the set of moment orders $n_1,n_2,...,n_m$. The
generalization of the usual Legendre transform of multifractal
formalism eqs. (\ref{alpha}) (\ref{alpha'''}) now involves a multifractal
function $f_m\left(\alpha_1, \alpha_2,\cdots, \alpha_m \right)$,
 depending on $m$ local exponents $\alpha_i$:
\begin{eqnarray}
\alpha_i &=&\frac{\partial\tau_m }{\partial n_i}\left(
\{n_j\}\right) ,
 \nonumber \\
f_m\left(\{\alpha_i\}\right) &=&\sum_{i=1}^{m}\alpha_i n_i-
\tau_m \left( \{n_j\}\right),  \label{alpha_m}\\
\nonumber
n_{i}&=&\frac{\partial f_m}{\partial
\alpha_i}\left(\{\alpha_j\}\right).
\end{eqnarray}

\subsubsection{Summing over Contact Points}
At this point, a caveat is in order. The reader may wonder about
the meaning of the sum over points $w$ in (\ref{Z_m}), since there is
only one $m$-vertex in a star. This notation is kept
for consistency with the $m=2$ case, and can be understood as follows.
 Along a scaling path, one can consider the subset ${\mathcal S}_m$ of contact points of order $m$,
 where the path folds onto itself several times on large (macroscopic) scales $R$,  and returns to itself at
 a short scale $r$ , thereby forming local stars of order $m$. The sum in moment (\ref{Z_m}) runs over such
 higher contact points along the path, and for $n_i=0, \forall i$, their number in a domain of size $R$ then
 scales for $r/R\to 0$ as
\begin{equation}
{\mathcal Z}_{0,0,...,0}  \approx\left( r/R\right) ^{\tau_m
\left(0,0,...,0 \right) }\sim R^{D_m}, \label{Z_m''}
\end{equation}
so that the formal Hausdorff dimension $D_m$ associated with this set of order $m$ contact points is
\begin{equation}
D_m:=-\tau_m \left(0,0,...,0\right)
={\rm sup}_{\{\alpha_i\}}f_m\left(\{\alpha_i\}\right).
\label{Dmtau0}
\end{equation}
One can equivalently consider the density, or the probability for these
  points to appear: ${\mathcal P}_{{\mathcal S}_m} (r) \approx (r/R)^{x_m
\left(0,0,...,0 \right)}$, such that
\begin{equation}
{\mathcal Z}_{0,0,...,0} \approx (R/r)^2{\mathcal P}_{{\mathcal S}_m} (r), \label{Z_m'''}
\end{equation}
whence $\tau_m
\left(0,0,...,0 \right)=x_m
\left(0,0,...,0 \right)-2$.
For $m$ large
enough, the density vanishes for $r \to 0$  fast enough, so that $x_m
\left(0,0,...,0 \right) \geq 2$, and $D_m \leq 0$ (see below).\\

\subsubsection{Local Moments}
Alternatively, one can define shifted exponents
$$\tilde \tau_m:= \tau_m +D_m=\tau_m
\left(n_1,n_2,...,n_m\right)-\tau_m \left(0,0,...,0\right),$$ which
 describe the scaling
of local averages at a given $m$-vertex
\begin{equation}
\left\langle \left[{H}_{1}(w,r)\right]^{n_1} \left[{H}_{2}(w,r)\right]^{n_2}\cdots\left[{H}_{m}(w,r)\right]^{n_m}
\right\rangle \approx \left( r/R\right) ^{\tilde \tau_m
\left(n_1,n_2,...,n_m \right)}\,. \label{H_m}
\end{equation}
By the Legendre transform (\ref{alpha_m}) these exponents give the
{\it subtracted} spectrum $f_m\left(\{\alpha_i\}\right)-{\rm
sup}_{\{\alpha_i\} }f_m\left(\{\alpha_i\}\right)$ directly. The
latter has a direct physical meaning:
 the probability
$P(\{\alpha_i\})$ to find a set of local singularity exponents
$\{\alpha_i\}$ in the $m$ sectors of an $m$-arm star scales as:
\begin{equation}
P_m(\{\alpha_i\})\propto R^{f_m\left(\{
{\alpha}_{i}\}\right)}/R^{\,{\rm sup} f_m} \, . \label{probastar}
\end{equation}
\\

\subsubsection{Recursion between Spectra}
From definition (\ref{Z_m}) and eq. (\ref{Z_m'}), we get the lower
$(m-1)$-multifractal spectrum as
\begin{equation}
\tau_m^{[m-1]} \left(n_1,n_2,\cdots, n_{m-1}\right):=\tau_m \left(
n_1,n_2,\cdots, n_{m-1}, n_m=0\right).
\end{equation}
In these exponents, the subscript $m$ stays unchanged since it
counts the number of arms of the star, while the potential is
evaluated only at $m-1$ sectors amongst the $m$ possible ones. More
generally, one can define exponents
$$\tau_m^{[p]} \left(n_1,n_2,\cdots, n_{p}\right):=
\tau_m \left( n_1,n_2,\cdots, n_{p}~; n_{p+1}=0,\cdots,
n_m=0\right),$$ where $p$ takes any value in $1\leq p \leq m$.
Note that according to the commutativity of the star algebra for
exponents between mutually-avoiding paths (see eq.(\ref{x}) and
below), the result does not depend on the choice of the $p$
sectors amongst $m$. Putting these values $n_{p+1}=0,\cdots,
n_m=0$ in the Legendre
transform  (\ref{alpha_m}) yields
the identity:\\
\begin{equation}
f_m^{[p]}(\alpha_1,\cdots,\alpha_{p})={\rm sup}_{\alpha_{p+1},\cdots,\alpha_m}
f_m\left( \alpha_1, \alpha_2,\cdots, \alpha_{p},\alpha_{p+1},\cdots,\alpha_m \right).
\label{sup_m}
\end{equation}
Note that the usual one-sided spectrum is in this notation $f(\alpha)=f_2^{[1]}(\alpha)$.\\

\vfill\eject
\subsection{Explicit Higher Multifractal Exponents and Spectra}
\subsubsection{Scaling Dimensions of Multiple-Sided Paths}
In analogy to eqs. (\ref{tt'}), (\ref{finab}), the exponent
$\tau_2(n,n')$ is associated with a scaling dimension $x_2(n,n')$
\begin{eqnarray}
\nonumber
\tau_2(n,n')&=&x_2(n,n')-2 \\
x_2(n,n')&=&2V\left[
1-\gamma +U^{-1}\left( n\right)+U^{-1}\left( n'\right) \right].
\label{finaa'}
\end{eqnarray}
Similarly, the $m$-order case
is given by
\begin{eqnarray}
\nonumber
\tau_m \left(n_1,n_2,...,n_m\right)&=&x_m \left(n_1,n_2,...,n_m\right)-2 \\
\nonumber x_m \left(n_1,n_2,...,n_m\right)&=&2V\left[
\tilde \Delta_m +U^{-1}\left( n_1\right)+U^{-1}\left(
n_2\right)+...+U^{-1}\left( n_m\right) \right].
\end{eqnarray}
Here $\tilde \Delta_m$ is the quantum gravity boundary scaling
dimension of the $m$-star ${\mathcal S}_m$ made of $m$ (simple)
scaling paths. According to the star algebra (\ref{xc}) valid for simple paths, we have:
\begin{equation}
\tilde \Delta_m=m\, U^{-1}\left( \tilde x_1\right)=\frac{m}{2}
U^{-1}\left( \tilde{x}_2\right)=m \frac{1-\gamma}{2}\ ,
\label{Delta_m}
\end{equation}
where $\tilde {x}_2$ is the boundary scaling
dimension of a scaling path, i.e., a $2$-star, already considered
in eq. (\ref{xtilde}), and such that $U^{-1}\left(
\tilde{x}_2\right)=1-\gamma$.\\

\subsubsection{Multiple Legendre Transform}

The calculation of the multiple Legendre transform eq.
(\ref{alpha'''}) is as follows.
We start from a total
(boundary) quantum scaling dimension of the $m$-path $\mathcal S_m$ dressed by Brownian paths
\begin{equation}
\delta:=m \frac{1-\gamma}{2} +\sum_{i=1}^{m}U^{-1}\left(
n_i\right)\ ,
  \label{delta_m}
\end{equation}
such that
\begin{equation}
x_m \left(n_1,n_2,...,n_m\right)=2V\left(\delta\right).
\label{x-delta_m}
\end{equation}
Using the shift identity (\ref{shift})
$$U^{-1}\left( n\right)=\frac{\gamma}{2}+\frac{1}{2}V^{-1}\left( n \right),\;\;
V^{-1}\left( n \right)=\sqrt{4(1-\gamma)n+\gamma^2},$$ gives
\begin{equation}
\delta= \frac{m}{2} +\frac{1}{2}\sum_{i=1}^{m}V^{-1}\left(
n_i\right), \label{delta_m'}
\end{equation}
and
\begin{eqnarray}
\alpha_i &=&\frac{\partial x_m }{\partial n_i}=2V'(\delta)\,\frac{\partial \delta }{\partial
n_i} =V'(\delta)\, \left[V^{-1}\left( n_i\right)\right]'.
\end{eqnarray}
Since
$$V'(\delta)=\frac{1}{2}\frac{\delta}{1-\gamma},$$
we get
\begin{eqnarray}
\alpha_i=\frac{\delta}{\sqrt{4(1-\gamma)n_i+\gamma^2}}=\frac{\delta}{V^{-1}\left(
n_i\right)}, \label{alpha-n_i}
\end{eqnarray}
or, equivalently
\begin{eqnarray}
V^{-1}\left( n_i\right)=\frac{\delta}{\alpha_i},\;\;\;\;n_i=V\left(\frac{\delta}{\alpha_i}\right).
\label{alpha-n_i'}
\end{eqnarray}
One gets from eqs. (\ref{delta_m'}) and (\ref{alpha-n_i'}) the useful identity
\begin{equation}
\delta=\frac{m}{2}\left(1-\frac{1}{2}\sum_{i=1}^m
{\alpha_i}^{-1}\right)^{-1}. \label{identity'}
\end{equation}
This yields the simple expression for $f_m$
\begin{equation}
f_m\left( \{\alpha_i\}\right)=2-V(\delta)+\sum_{i=1}^m\alpha_i
V\left(\frac{\delta}{\alpha_i}\right). \label{f_m}
\end{equation}
Recalling (\ref{Ua}), collecting the $\delta$ terms, and
 using identity (\ref{identity'}) for $\delta$,
finally gives after some calculations the explicit formulae
\begin{eqnarray}
f_m\left(\{
{\alpha}_{i=1,...,m}\}\right)&=&2+\frac{\gamma^2}{2(1-\gamma)}-
\frac{1}{8(1-\gamma)}{m}^2{\left(1-\frac{1}{2}\sum_{i=1}^{m}{\alpha}_{i}^{-1}\right)}^{-1} \nonumber \\
& &-\frac{\gamma^2}{4(1-\gamma)}\sum_{i=1}^{m}{\alpha}_{i},
\label{faig}
\end{eqnarray}
with
\begin{equation}
{\alpha}_{i}=\frac{1}{\sqrt{4(1-\gamma)n_{i}+\gamma^2}}\left({\frac{m}{2}}+
\frac{1}{2}\sum_{j=1}^{m}\sqrt{4(1-\gamma)n_{j}+\gamma^2}\right).\label{aic}
\end{equation}

Substituting expressions (\ref{c,gamma}) gives in terms of $c$
\begin{eqnarray}
f_m\left(\{ {\alpha}_{i=1,...,m}\}\right)&=&\frac{25-c}{12}-
\frac{1}{8(1-\gamma)}{m}^2{\left(1-\frac{1}{2}\sum_{i=1}^{m}{\alpha}_{i}^{-1}\right)}^{-1} \nonumber \\
& &-\frac{1-c}{24}\sum_{i=1}^{m}{\alpha}_{i} \ , \label{faic}
\end{eqnarray}
where the central charge $c$ and the parameter $\gamma$ are
related by eq. (\ref{c,gamma}). The self-avoiding walk case
(\ref{fai}) is recovered for $c=0,\gamma=-1/2$.

The domain of definition of the poly-multifractal function $f$ is
independent of $c$ and given by
\begin{equation}
1-\frac{1}{2}\sum_{i=1}^{m}{\alpha}_{i}^{-1} \geq 0,
\label{domain}
\end{equation}
as verified by eq. (\ref{aic}).
\\

\subsubsection{One and Two-Sided Cases}

Notice that the $m=1$ case,
\begin{eqnarray}
f_{1}(\alpha)=\frac{25-c}{12}-
\frac{1}{8(1-\gamma)}\left(1-\frac{1}{2 \alpha}\right)
-\frac{1-c}{24}{\alpha}, \label{faic1}
\end{eqnarray}
corresponds to the potential in the
vicinity of the tip of a conformally-invariant scaling path, and naturally
differs from the usual $f(\alpha)={\rm sup}_{\alpha'}
f_2(\alpha,\alpha'))$ spectrum, which describes the potential on
one side of the scaling path.

The two-sided case
 is obtained for $m=2$ as
\begin{eqnarray}
f_2\left( \alpha, \alpha'
\right)&=&\frac{25-c}{12}-\frac{1}{2(1-\gamma)}
{\left[1-\frac{1}{2}\left(\frac{1}{\alpha}+\frac{1}{\alpha'}\right)\right]}^{-1} \nonumber \\
& &-\frac{1-c}{24}\left(\alpha+\alpha'\right), \label{f_2c}
\end{eqnarray}
\begin{equation}
{\alpha}=\frac{1}{\sqrt{4(1-\gamma)n+\gamma^2}}\left[1
+\frac{1}{2}\left(\sqrt{4(1-\gamma)n+\gamma^2}+\sqrt{4(1-\gamma)n'+\gamma^2}\right)\right],
\label{alphann'c}
\end{equation}
This doubly multifractal spectrum
possesses the desired properties, like $${\rm sup}_{\alpha'}
f_2(\alpha, \alpha')=f(\alpha),$$ where $f(\alpha)$ is
(\ref{foriginalbis}) above.\\

\subsubsection{Local Wedge Description}
We can also substitute equivalent ``electrostatic'' angles
$\theta_i=\pi/\alpha_i$ for the variables $\alpha_i$. This gives a new
distribution:
\begin{eqnarray}
{\hat f}_m\left(\{ {\theta}_{i=1,...,m}\}\right)= f_m\left(\{
{\alpha}_{i=1,...,m}\}\right)
&=&2+\frac{\gamma^2}{2(1-\gamma)}-\frac{\gamma^2}{4(1-\gamma)}\sum_{i=1}^{m}\frac{\pi}{\theta_{i}} \nonumber \\
& &-
\frac{1}{8(1-\gamma)}{m}^2{\left(1-\frac{1}{2\pi}\sum_{i=1}^{m}{\theta}_{i}\right)}^{-1}. \label{fthetaig}
\end{eqnarray}
The domain of definition of distribution $\hat f_m$ is the image
of domain (\ref{domain}) in $\theta$-variables:
 \begin{equation}
\sum_{i=1}^{m}{\theta}_{i} \leq 2\pi. \label{domaint}
\end{equation}
The total {\it electrostatic} angle is thus less than $2\pi$,
which simply accounts for the electrostatic screening of local
wedges by fractal randomness, as expected.
\\

\subsubsection{Maxima and Global Hausdorff Dimension}
The maxima of $f_m$ or ${\hat f}_m$ are by construction obtained
for $n_i=0,\ \forall i=1,...,m.$ Eq. (\ref{aic}) gives the values
of {\it typical} singularity exponents $\hat \alpha_i$ at the maximum of $f_m$:
\begin{equation}
\hat
\alpha_i=\frac{\pi}{\hat\theta_i}=\frac{m}{2}\left(1-\frac{1}{\gamma}\right)^{-1},
\ \forall i=1,...,m, \label{hataic}
\end{equation}
corresponding to a maximal value of $f_m$ or $\hat f_m$:
\begin{eqnarray}
\nonumber D_m&=&{\rm sup}f_m = f_m\left(\{ {\hat
\alpha}_{i=1,...,m}\}\right) ={\hat f}_m\left(\{ {\hat
\theta}_{i=1,...,m}\}\right)\nonumber \\ &=& 2-2V(\tilde
\Delta_m)=\frac{(2-\gamma)^2}{2(1-\gamma)}-
\frac{1}{8}(1-\gamma){m}^2. \nonumber
\end{eqnarray}
As anticipated, for $m$ large enough, i.e., $m \geq 2\frac{2-\gamma}{1-\gamma}$, this
Hausdorff dimension $D_m$ (\ref{Dmtau0}) formally becomes negative. Since $-\infty \leq \gamma \leq 0$, and $m$ integer,
this already happens for $m\geq 3$ (3-stars), except for $\gamma =0$, which gives the condition $m\geq 4$ and again
corresponds to the $c=1$ ``ultimate Norway'' fractal boundary.
\\

\subsubsection{Local Probabilities and Shifted Spectrum}
As mentioned above, the following interpretation of the poly-multifractal spectrum holds.
The probability $P(\{\alpha_i\})\equiv\hat
P(\{\theta_i\})$ to find a set of local singularity exponents
$\{\alpha_i\}$ or equivalent angles $\{\theta_i\}$ in the $m$
sectors of an $m$-arm star is given by the ratio
\begin{equation}
P_m(\{\alpha_i\})\propto R^{f_m\left(\{
{\alpha}_{i}\}\right)}/R^{\,{\rm sup} f_m} \label{proba_m}
\end{equation}
of the respective number of configurations to the total one. We
therefore arrive at a probability, here written in terms of the
equivalent electrostatic angles:
\begin{eqnarray}
\hat P_m(\{\theta_i\})&\propto&R^{{\hat f}_m\left(\{ {\theta}_{i}\}\right)-{\hat f}_m(\{ {\hat \theta}_{i}\})},\\
\hat f_m\left(\{ {\theta}_{i}\}\right)- \hat f_m(\{ \hat
{\theta}_{i}\})&=&
-\frac{1}{8(1-\gamma)}{m}^2{\left(\frac{2\pi}{\sum_{i=1}^{m}{\theta}_{i}}-1\right)}^{-1} \nonumber \\
&
&-\frac{\gamma^2}{4(1-\gamma)}\sum_{i=1}^{m}\frac{\pi}{\theta}_{i}. \label{ftheta-supf}
\end{eqnarray}
For a large scaling star, the typical set of singularity
exponents
$\{\hat \alpha_i\}$, or wedge angles $\{\hat \theta_i\}$, is thus given by the symmetric set of values (\ref{hataic}).\\

\section{\sc{Winding of Conformally Invariant Curves}}
\label{sec.winding}

\begin{figure}[htb]
\epsfxsize=8.5truecm{\centerline{\epsfbox{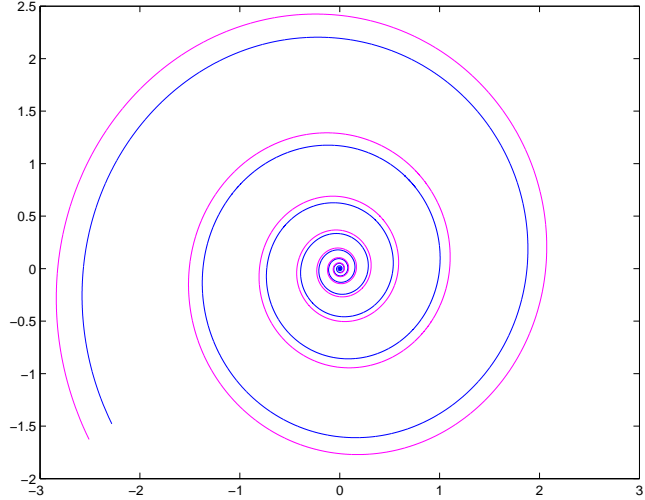}}}
\caption{A double logarithmic spiral mimicking the local geometry of the two strands of the conformally-invariant frontier.}
\label{spiral2}
\end{figure}

Another important question arises concerning the {\it   geometry
 of the  equipotential lines} near a random (CI)
fractal curve. These lines are expected to  rotate wildly, or wind, in a spiralling motion that closely follows
the boundary itself. The key geometrical object is here the {\it  logarithmic spiral},
which is conformally invariant (Fig. \ref{spiral2}).
The MF description should generalize to a
{\it   mixed} multifractal spectrum, accounting for {\it   both scaling
and winding} of the equipotentials  \cite{binder}.

In this
section, we describe the exact solution to this mixed MF spectrum
for any random CI curve \cite{DB}. In particular, it is shown
to be related by a scaling law to the usual 
harmonic MF spectrum. We use the same conformal tools as before, fusing quantum
gravity and Coulomb gas methods, which allow the
description of Brownian paths interacting and winding with CI curves,
thereby providing a probabilistic description of the potential map near any CI random curve.
\\

\subsection{Harmonic Measure and Rotations}

Consider a single
(CI) critical random cluster, generically
called ${\mathcal C}$. Let ${\mathcal H}\left( z\right) $ be the potential at an 
exterior point $z \in {\rm  {\mathbb  C}}$, with Dirichlet boundary
conditions ${\mathcal H}\left({w \in \partial \mathcal C}\right)=0$ on the outer
(simply connected) boundary $\partial \mathcal C$ of $\mathcal C$, and
${\mathcal H}(w)=1$ on a circle ``at $\infty$'', i.e., of a large radius
scaling like the average size $R$ of $ \mathcal C$. The potential
${\mathcal H}\left( z\right)$ is identical to the probability that a Brownian path
started at $z$ escapes to ``$\infty$'' without having hit
${\mathcal C}$. 

Let us now consider the {\it  degree with which the
curves wind in the complex plane about point} $w$ and call
$\varphi(z)={\rm  arg}\,(z-w)$. In the scaling limit, the multifractal formalism, here generalized to take into
account rotations  \cite{binder}, characterizes subsets
${\partial\mathcal C}_{\alpha,\lambda}$ of boundary sites by a
H\"{o}lder exponent $\alpha$, and a rotation rate $\lambda$,
such that their potential lines respectively scale and {\it logarithmically spiral} as
\begin{eqnarray}
\nonumber
{\mathcal H}\left( z \to w\in {\partial\mathcal C}_{\alpha,\lambda }\right) &\approx& r ^{\alpha },\\
\varphi\left( z \to w\in {\partial\mathcal C}_{\alpha,\lambda
}\right) &\approx& \lambda \ln\, r\ , 
\label{ha}
\end{eqnarray}
in the limit  $r=|z-w| \to 0$. The Hausdorff dimension
${\rm  dim}\left({\partial\mathcal C}_{\alpha,\lambda }\right)=f\left(
\alpha, \lambda\right)$ defines
 the mixed MF spectrum, which is CI since {\it under a conformal map
 both $\alpha$ and $\lambda$ are locally invariant}.

\begin{figure}[htb]
\begin{center}
\includegraphics[angle=0,width=.5\linewidth]{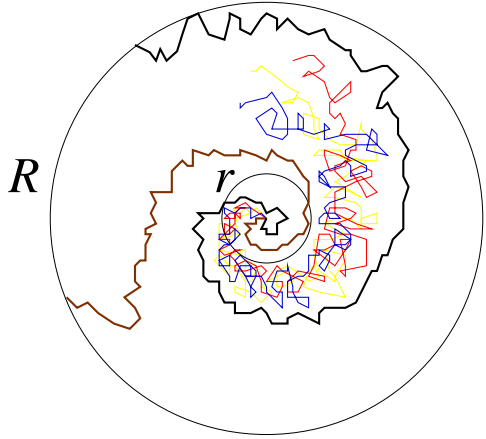}
\end{center}
\caption{Two-sided boundary curve $\partial \mathcal C$ and Brownian $n$-packet winding together from the disk of 
radius $r$ up to distances of order $R$, as measured by the winding angle 
$\varphi (w,r)={\rm  arg}({\partial\mathcal C}\wedge n)$ as in (\ref{ZDB}) and in (\ref{xnp}).}
\label{Fig.escape}
\end{figure}

As above, we 
consider the  harmonic measure $H\left(w,r\right)$, which is the integral of the Laplacian of 
${\mathcal H}$ in a disk $B(w,r)$ of radius $r$ centered at $w \in \partial \mathcal C$, i.e., the boundary 
charge in that disk. It scales as $r^{\alpha}$ with the same exponent as in (\ref{ha}), and 
is also the probability that a Brownian path started at large distance $R$ first hits the boundary at a point inside 
 $B(w,r)$. Let $\varphi (w,r)$ be the associated winding angle of the path down
to distance $r$ from $w$. The {\it   mixed} moments of $H$ and
$e^{\varphi}$, averaged over all realizations of ${\mathcal C}$, are defined as
\begin{equation}
{\mathcal Z}_{n,p}=\left\langle \sum\limits_{w\in {\partial {\mathcal C}}/r} 
H^{n}\left(w,r\right) \exp\, (p\,\varphi (w,r)) \right\rangle
\approx \left( r/R\right) ^{\tau \left(
n,p\right) }, \label{ZDB}
\end{equation}
where the sum runs over the centers of a covering of the boundary by disks of radius $r$, and where $n$ and $p$ are 
real numbers. As before, the $n^{\rm th}$ moment of $H\left(w,r\right)$ is the probability that $n$ independent Brownian 
paths diffuse 
along the boundary and all first hit it at points inside the disk $B(w,r)$. The angle $\varphi (w,r)$ is then their common winding angle down to 
distance $r$ (Fig. \ref{Fig.escape})
 
The scaling limit in (\ref{ZDB}) involves 
multifractal scaling exponents $\tau \left( n,p\right)$
which vary in a non-linear way with $n$ and $p$.
 They give the multifractal spectrum $f\left(
\alpha, \lambda\right)$ via a symmetric double Legendre transform:
\begin{eqnarray}
\nonumber
\alpha &=&\frac{\partial\tau }{\partial n}\left( n,p\right) ,
\quad \lambda =\frac{\partial\tau }{\partial p}\left( n,p\right), \\ \nonumber
f\left( \alpha, \lambda \right)&=&\alpha n+\lambda p-\tau \left( n,p\right) ,\\
n&=&\frac{\partial f}{\partial \alpha}\left( \alpha, \lambda
\right) , \quad p=\frac{\partial f}{\partial \lambda }\left(
\alpha, \lambda \right).
\label{legendre}
\end{eqnarray}
Because of the ensemble average (\ref{ZDB}), $f\left(
\alpha,\lambda \right)$ can become negative for some
$\alpha,\lambda$.

\subsection{Exact Mixed Multifractal Spectra}

 The 2D conformally
invariant random statistical system
 is labelled by its {\it   central charge} $c$, $c\leq 1$  \cite{BPZ}. 
 The main result is the following exact scaling law  \cite{DB}:
\begin{eqnarray}
\label{scalinglaw}
 f(\alpha,\lambda)&=&(1+\lambda^2) f\left(\frac{\alpha}{1+\lambda^2}\right)-b \lambda^2\ ,\\
\nonumber
 b&:=&\frac{25-c}{12}\geq 2\ ,
\end{eqnarray}
where $f\left({\alpha}\right)=
f\left({\alpha},\lambda=0\right)$ is the usual harmonic
MF spectrum in the absence of prescribed winding, first
obtained in  \cite{duplantier11}, and described in section \ref{sec.conform},  eq. (\ref{foriginalbis}). It can be recast as:
\begin{eqnarray}
 f(\alpha)&=&\alpha+b-\frac{b\alpha^2}{2\alpha-1},\\ \nonumber b&=&\frac{25-c}{12}.
\label{falpha}
\end{eqnarray}
We thus arrive at the very simple formula for the mixed spectrum:
\begin{eqnarray}
 f(\alpha,\lambda)=\alpha+b-\frac{b\alpha^2}{2\alpha-1-\lambda^2}\ .
\label{falphalambda}
\end{eqnarray}
Notice that by conformal symmetry $${\sup}_{\lambda}f(\alpha,\lambda)=f(\alpha,\lambda=0),$$
 i.e., 
the most likely situation in the absence of prescribed rotation 
is the same as $\lambda=0$, i.e. {\it  winding-free}.
The domain of definition of the usual $f(\alpha)$ (\ref{falpha}) is $1/2 \leq \alpha 
$  \cite{duplantier11,Beur}, thus for $\lambda$-spiralling points eq. (\ref{scalinglaw}) gives
\begin{eqnarray}
  \frac{1}{2}({1+\lambda^2}) \leq {\alpha}\ ,
\label{alpha'}
\end{eqnarray}
in agreement with a theorem by Beurling  \cite{Beur,binder}.

We have seen in section \ref{subsec.geometry} the geometrical meaning to the exponent $\alpha$:  
For an angle with opening $\theta$, $\alpha={\pi}/{\theta}$, the quantity ${\pi}/{\alpha}$ can be regarded as
 a local generalized angle with respect to the harmonic measure. The geometrical MF spectrum
of the boundary subset with such opening angle $\theta$ and spiralling rate
$\lambda$ reads from (\ref{falphalambda})
\begin{eqnarray}
\hat f(\theta,\lambda)\equiv f(\alpha=\frac{\pi}{\theta},\lambda)=\frac{\pi}{\theta}+b-b\frac{\pi}{2}
\left(\frac{1}{\theta}+
 \frac{1}{\frac{2\pi}{1+\lambda^2} -\theta}\right).
 \nonumber
\label{fthetalambda}
\end{eqnarray}
As in (\ref{alpha'}), the domain of definition in the
$\theta$ variable is
\begin{eqnarray} 
\nonumber
0 \le \theta \le \theta(\lambda),\;\;\;
\theta(\lambda)={2\pi}/({1+\lambda^2}).
\end{eqnarray} 
The maximum is reached when the two frontier strands about point $w$ locally collapse into a single 
$\lambda$-spiral, whose inner opening angle is $\theta(\lambda)$  \cite{Beur}.

In the absence of prescribed winding ($\lambda=0$), the maximum
$D_{\rm  EP}:= D_{\rm  EP}(0)={\sup}_{\alpha}f(\alpha,\lambda=0)$
gives the dimension of the {\it   external perimeter} of the fractal
cluster, which is a {\it simple} curve without double points, and may differ from the full hull  \cite{duplantier11,ADA}.
Its dimension (\ref{D(c)}) reads in this notation
$$D_{\rm  EP}=\frac{1}{2}(1+b)-\frac{1}{2}\sqrt{b(b-2)},\;\;\;b=\frac{25-c}{12}.$$
It corresponds to typical values $\hat
\alpha=\alpha(n=0,p=0)$ and $\hat\theta={\pi}/{\hat \alpha}=\pi(3-2D_{\rm  EP}).$

For spirals, the maximum value
$D_{\rm  EP}(\lambda)={\sup}_{\alpha}f(\alpha,\lambda)$ still
corresponds  in the Legendre transform (\ref{legendre}) to $n=0$,
and gives the dimension of the {\it   subset of the  external
perimeter made of logarithmic spirals of type $\lambda$}. Owing to (\ref{scalinglaw})
we immediately get
\begin{eqnarray}
D_{\rm  EP}(\lambda)=(1+\lambda^2)D_{\rm  EP} -b \lambda^2 \ .
\label{supf}
\end{eqnarray}
This corresponds to typical scaled values $$\hat
\alpha(\lambda)=(1+\lambda^2)\hat \alpha, \;\; \hat
\theta(\lambda)=\hat \theta/(1+\lambda^2).$$ 
Since $b \geq 2$ and
$D_{\rm  EP} \leq 3/2$, the EP dimension decreases
with spiralling rate, in a simple parabolic way.
\begin{figure}[t]
\begin{center}
\includegraphics[angle=0,width=1.\linewidth]{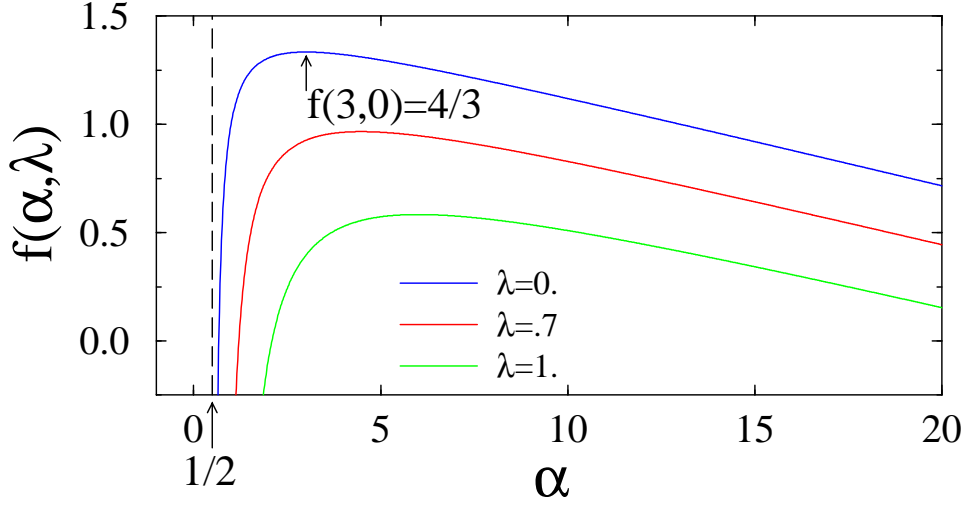}
\end{center}
\caption{Universal multifractal spectrum $f(\alpha,\lambda)$ for
$c=0$ (Brownian frontier, percolation EP and
SAW), and for three different values of the 
spiralling rate $\lambda$. The maximum $f(3,0)=4/3$ is the Hausdorff dimension of the frontier.}
\label{Figure1DB}
\end{figure}
\begin{figure}[t]
\begin{center}
\includegraphics[angle=0,width=1.\linewidth]{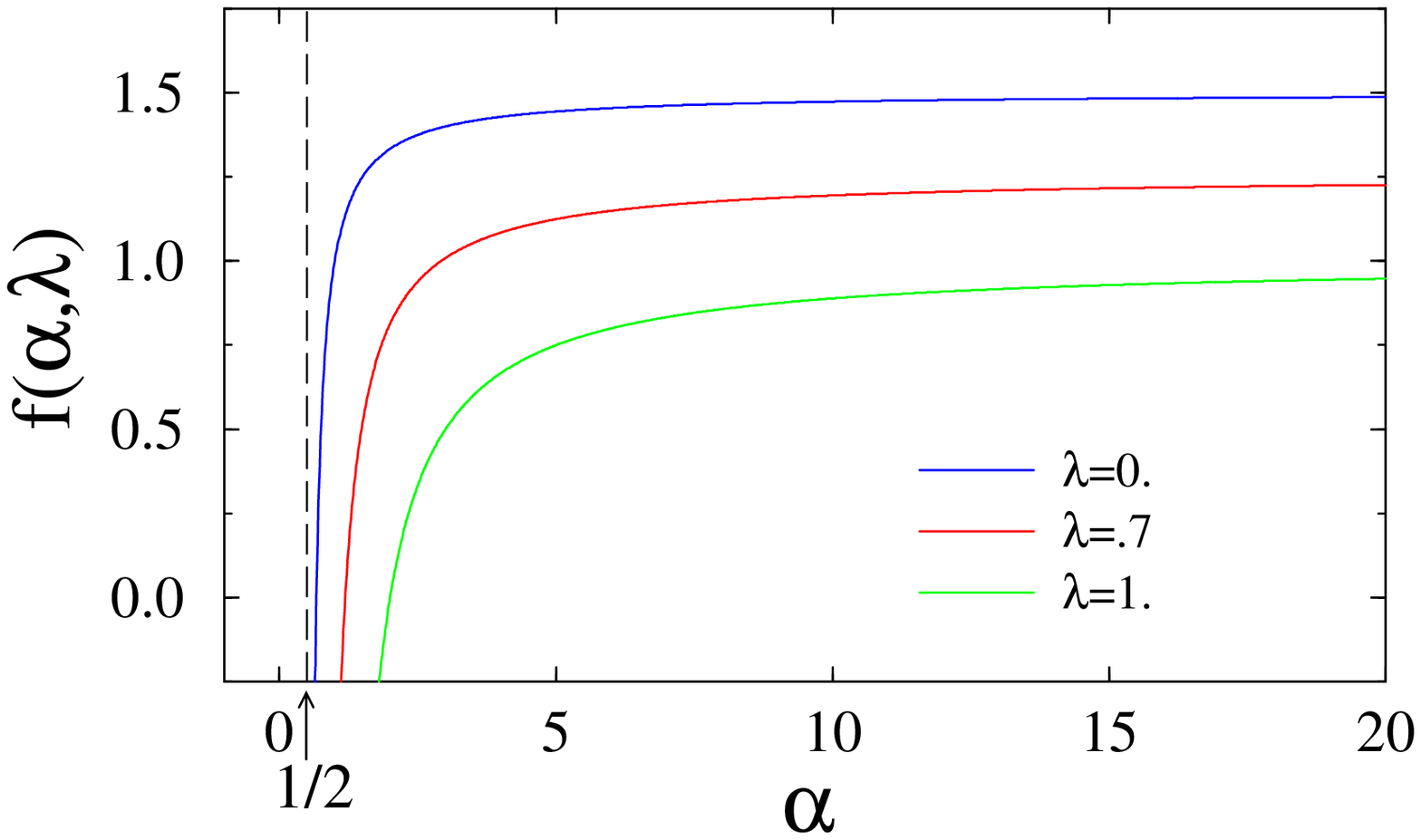}
\end{center}
\caption{Left-sided multifractal spectra $f(\alpha,\lambda)$ for
the limit case $c=1$, the ``ultimate Norway'' (frontier of a $Q=4$ Potts cluster or ${\rm  {\rm SLE}}_{\kappa=4}$).}
\label{Figure2DB}
\end{figure}

Fig. \ref{Figure1DB}  displays typical multifractal functions
$f(\alpha,\lambda;c)$. The example choosen, $c=0$,  
corresponds to the cases of a SAW,
or of a percolation EP, the scaling limits of which both coincide
with the Brownian frontier  \cite{duplantier8,duplantier9,lawler3}. The original
singularity at $\alpha=\frac{1}{2}$ in the
rotation free MF functions $f(\alpha,0)$, which describes boundary points with a needle local
geometry, is shifted for $\lambda \ne 0$ towards the minimal value (\ref{alpha'}). The right branch of
$f\left( \alpha,\lambda\right) $ has a linear asymptote
$\lim_{\alpha \rightarrow +\infty} f\left(\alpha,\lambda
\right)/{\alpha} =-(1-c)/24.$ Thus the $\lambda$-curves all become parallel for $\alpha \to
+\infty$, i.e., $\theta \to 0^{+}$, corresponding to deep fjords
where winding is easiest.

Limit multifractal spectra are
obtained for $c=1$, which exhibit exact examples of {\it  
left-sided} MF spectra, with a horizontal asymptote
$f\left(\alpha\to +\infty,\lambda; c=1\right)=
\frac{3}{2}-\frac{1}{2}\lambda^2$ (Fig. \ref{Figure2DB}). This corresponds to
the frontier of a $Q=4$ Potts cluster (i.e., the
${\rm  {\rm SLE}}_{\kappa=4}$), a universal random scaling curve, with the
maximum value $D_{\rm  EP}=3/2$, and a vanishing typical opening
angle $\hat \theta=0$, i.e., the ``ultimate Norway'' where the EP
is dominated by ``{\it   fjords}'' everywhere
 \cite{duplantier11,BDjsp}.
\begin{figure}[htbp]
\begin{center}
\includegraphics[angle=0,width=1.\linewidth]{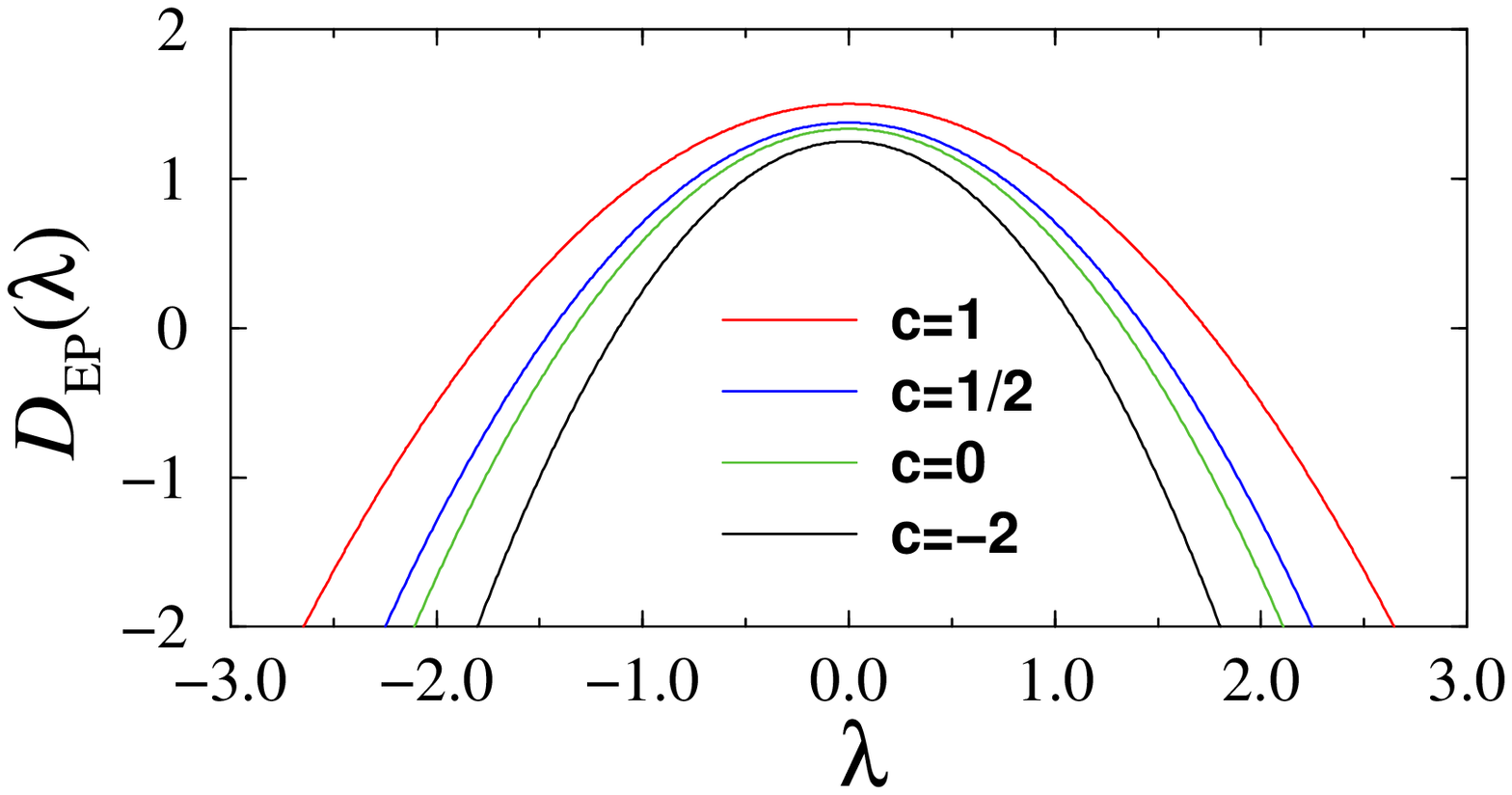}
\end{center}
\caption{Dimensions $D_{\rm  EP}(\lambda)$ of the external frontiers as a function of rotation rate.  The curves
are indexed by the central charge $c$, and correspond
respectively to: loop-erased RW ($c=-2; {\rm {\rm SLE}}_{2}$); Brownian or
percolation external frontiers, and self-avoiding walk ($c=0; {\rm SLE}_{8/3}$);
Ising clusters ($c=\frac{1}{2}; {\rm SLE}_{3}$); $Q=4$ Potts clusters ($c=1; {\rm SLE}_{4}$).}
\label{Figure3DB}
\end{figure}
Fig. \ref{Figure3DB} displays the dimension $D_{\rm
EP}(\lambda)$ as a function of the rotation
rate $\lambda$, for various values of $ c \leq 1$,
corresponding to different statistical systems. Again, the $c=1$
case shows the least decay with $\lambda$, as expected from the
predominence of fjords there.


\subsection{Conformal Invariance and Quantum Gravity} 

We now give the main lines of the derivation of exponents $\tau\left(
n,p\right)$, hence $f(\alpha,\lambda)$  \cite{DB}. As usual, $n$ {\it  
independent} Brownian paths ${\mathcal B}$, starting a small distance $r$ away from
a point $w$ on the frontier $\partial \mathcal C$, and
diffusing without hitting $\partial \mathcal C$, give a geometric
representation of the $n^{th}$ moment, $ H^{n}$, of the harmonic measure in eq.(\ref{ZDB})
for {\it integer} $n$ (Fig. \ref{Fig.escape}). Convexity yields the analytic continuation to
arbitrary $n$'s. Let us introduce an abstract (conformal) field
operator $\Phi_{{\partial\mathcal C}\wedge {n}}$ characterizing the
presence of a vertex where $n$ such Brownian paths and the cluster's
frontier diffuse away from each other in the {\it   mutually-avoiding} configuration 
${\partial\mathcal
C}\wedge {n}$  \cite{duplantier8,duplantier9}; to this operator is associated a scaling
dimension $x(n)$. To measure rotations using the moments (\ref{ZDB}) we have to consider
expectation values with insertion of the mixed operator
\begin{equation}
\Phi_{{\partial\mathcal C}\wedge n} e^{p\,{\rm  arg}({\partial\mathcal
C}\wedge n)} \longrightarrow x\left( n,p\right) , \label{Phinp}
\end{equation}
where ${\rm  arg}({\partial\mathcal C}\wedge n)$ is the winding angle
common to the frontier and to the Brownian paths (see Fig. (\ref{Fig.escape})), and where  $x(n,p)$ is the
{\it scaling dimension} of the operator $\Phi_{{\partial\mathcal C}\wedge n} e^{p\,{\rm  arg}({\partial\mathcal
C}\wedge n)}$. It is directly related to $\tau(n,p)$ by   \cite{duplantier8} 
\begin{equation}
x\left( n,p\right)=\tau \left( n,p\right)+2. \label{xnp}
\end{equation}
For $n=0$, one recovers the previous scaling dimension 
\begin{eqnarray}
\nonumber
x(n,p=0)&=&x(n),\\
\nonumber
\tau(n,p=0) &=&\tau \left( n\right)=x\left(
n\right) -2.
\end{eqnarray}
As in section \ref{sec.conform}, we use the fundamental KPZ mapping of the
CFT in the {\it   plane} ${\mathbb C}$  to the CFT on a
fluctuating abstract random Riemann surface, i.e., in presence of 2D quantum gravity  \cite{KPZ}, and the
 universal functions $U$
and $V$, acting on conformal weights, which describe the map:
\begin{eqnarray}
U\left( x\right) &=&x\frac{x-\gamma}{1-\gamma} , \hskip2mm
V\left( x\right) =\frac{1}{4}\frac{x^{2}-\gamma^2}{1-\gamma}.
\label{UV}
\end{eqnarray}
with $V\left( x\right) = U\left(
\frac{1}{2}
\left( x+\gamma \right) \right)$. As before, the parameter $\gamma$
 is the solution of
$c=1-6{\gamma}^2(1-\gamma)^{-1}, \gamma \leq 0.$

For the purely harmonic exponents $x(n)$,
describing the mutually-avoiding set ${\partial\mathcal C}\wedge n$, we have seen in
eqs. (\ref{finaa}) and (\ref{xtilde}) that
\begin{eqnarray}
x(n)&=&2V\left[2 U^{-1}\left(
\tilde{x_1}\right)
 +U^{-1}\left( n \right) \right],  
\label{xcn}
\end{eqnarray}
where $U^{-1}\left( x\right) $ is the positive inverse of $U$,
\begin{eqnarray}
\nonumber
2 U^{-1}\left( x\right)
=\sqrt{4(1-\gamma)x+\gamma^2}+\gamma\, .
\label{u1DB}
\end{eqnarray}
In (\ref{xcn}), we recall that the arguments $\tilde{x_1}$ and $n$ are respectively the {\it   boundary} scaling dimensions
(b.s.d.) (\ref{xtilde}) of the simple path ${\mathcal S}_1$ representing a
semi-infinite random frontier (such that ${\partial\mathcal C}= {\mathcal
S}_1\wedge{\mathcal S}_1$),
 and of the packet of $n$ Brownian paths, both diffusing into the upper {\it   half-plane} $\mathbb H$.
The function $U^{-1}$ transforms these half-plane b.s.d's into the corresponding b.s.d.'s in quantum
gravity, the {\it linear combination} of which gives, still in QG, the
b.s.d. of the mutually-avoiding set ${\partial\mathcal
C}\wedge n=(\wedge{\mathcal S}_1)^2\wedge n$. The function $V$ finally
maps the latter b.s.d. into the scaling dimension in $\mathbb C$.
The path b.s.d. $\tilde{x_1}$ (\ref{xtilde}) obeys $U^{-1}\left(
\tilde{x_1}\right) =(1-\gamma)/2$.

\begin{figure}[tb]
\begin{center}
\includegraphics[angle=0,width=1.\linewidth]{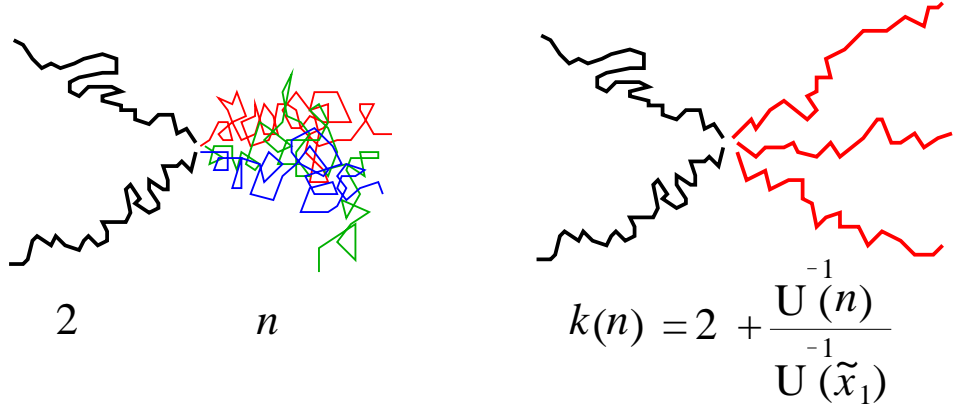}
\end{center}
\caption{Equivalence (\ref{kn}) between two simple paths in a mutually-avoiding configuration 
${\mathcal S}_2={\mathcal S}_1 \wedge {\mathcal S}_1$, further avoided by a packet of $n$ 
independent Brownian motions, and $k(n)$ simple paths in a mutually-avoiding star configuration  ${\mathcal S}_{k(n)}$.}
\label{Fig.equiv}
\end{figure}

It is now useful to consider $k$ semi-infinite
random paths ${\mathcal S}_1$, joined at a single vertex in a
{\it   mutually-avoiding star} configuration ${\mathcal
S}_k=\stackrel{k}{\overbrace{{\mathcal S}_1\wedge{\mathcal
S}_1\wedge\cdots{\mathcal S}_1}}=(\wedge{\mathcal S}_1)^k$.
(In this notation the frontier near any of its points is a two-star
${\partial\mathcal C}={\mathcal S}_2$.)
The scaling dimension of ${\mathcal
S}_k$ can be obtained from the same b.s.d. additivity rule in quantum gravity, as in
(\ref{xa}) or (\ref{xcn})  \cite{duplantier11}
\begin{eqnarray}
x({\mathcal S}_k)&=&2V\left[ k\, U^{-1}\left( \tilde{x_1}\right) \right]\ .  
\label{xk}
\end{eqnarray}
The scaling dimensions (\ref{xcn}) and (\ref{xk}) coincide when
\begin{eqnarray}
\label{equal}
x(n)&=&x({\mathcal S}_{k(n)})\\
\label{kn}
k(n)&=& 2+\frac{U^{-1}\left( n \right)}{U^{-1}\left( \tilde{x_1}\right)}.
\end{eqnarray}
Thus we state the {\it   scaling star-equivalence}
\begin{eqnarray}
{\partial\mathcal C}\wedge n \Longleftrightarrow {\mathcal S}_{k(n)},
\label{equiv}
\end{eqnarray}
{\it   of two mutually-avoiding simple paths ${\partial\mathcal C}={\mathcal S}_2={\mathcal S}_1 \wedge {\mathcal S}_1$, 
further avoiding $n$ Brownian motions, to $k(n)$
simple paths in a mutually-avoiding star configuration ${\mathcal S}_{k(n)}$} (Fig. \ref{Fig.equiv}). This 
equivalence plays an essential role in the computation of the complete rotation
spectrum (\ref{xnp}).

\subsection{Rotation Scaling Exponents}
\label{rotationscaling}
 The Gaussian distribution of
the winding angle about the {\it   extremity} of a scaling path,
like ${\mathcal S}_1$, was derived in  \cite{DSw}, using exact Coulomb
gas methods. The argument can be generalized to the winding angle
of a star ${\mathcal S}_k$ about its center  \cite{BDtopub}, where
one finds that the angular variance is reduced by a factor
$1/k^2$ (see also  \cite{wilson}). The
scaling dimension associated with the rotation scaling operator
$\Phi_{{\mathcal S}_k}e^{p\, {\rm  arg }\left({\mathcal S}_k\right)}$ is found
by analytic continuation of the Fourier transforms evaluated there  \cite{DB}:
\begin{eqnarray}
x({\mathcal S}_k;p)=x({\mathcal
S}_k)-\frac{2}{1-\gamma}\frac{p^2}{k^2}  \ , \nonumber 
\end{eqnarray}
i.e., is given by a quadratic shift in the star scaling exponent. To calculate
the scaling dimension (\ref{xnp}), it is 
sufficient to use the star-equivalence (\ref{kn}) above to
conclude that
\begin{eqnarray}
x(n,p)=x({\mathcal
S}_{k(n)};p)=x(n)-\frac{2}{1-\gamma}\frac{p^2}{k^2(n)}  \ ,
\nonumber
\end{eqnarray}
which is the key to our problem. Using eqs. (\ref{kn}),
(\ref{xcn}), and (\ref{UV}) gives the
useful identity:
\begin{eqnarray}
\frac{1}{8}({1-\gamma})k^2(n)=x(n)-2+b \ ,
\nonumber
\end{eqnarray}
with $b=\frac{1}{2}\frac{(2-\gamma)^2}{1-\gamma}=\frac{25-c}{12}$.
Recalling (\ref{xnp}), we arrive at the multifractal result:
\begin{eqnarray}
\tau(n,p)=\tau(n)-\frac{1}{4}\frac{p^2}{\tau(n)+b}  \ ,
\label{taunp}
\end{eqnarray}
where $\tau(n)=x(n)-2$ corresponds to the purely harmonic
spectrum with no prescribed rotation.

\subsection{Legendre Transform} 

The structure of the full
$\tau$-function (\ref{taunp}) leads by a formal Legendre transform
(\ref{legendre}) directly to the identity
\begin{eqnarray}
 f(\alpha,\lambda)&=&(1+\lambda^2)f(\bar{\alpha})-b\lambda^2 \ ,
 \nonumber
\end{eqnarray}
where $f(\bar{\alpha})\equiv\bar{\alpha}n-\tau(n)$, with $\bar \alpha={d{\tau}(n)
}/{dn}$, is the purely
harmonic MF function. It depends on the natural reduced variable $\bar{\alpha}$
{\it   \`a la} Beurling ($\bar{\alpha} \in \left[\frac{1}{2},+\infty \right)$)
\begin{eqnarray}
\nonumber
\bar{\alpha}&:=&\frac{\alpha}{1+\lambda^2}=\frac{d{x}
}{dn}\left( n\right)=\frac{1}{2} +\frac{1}{2}
\sqrt{\frac{b}{2n+b-2}}\, ,
\end{eqnarray}
whose expression emerges explicitly from (\ref{xcn}). Whence eq.(\ref{scalinglaw}), {\bf QED}.

It is interesting to consider also higher multifractal spectra
 \cite{BDjsp}. For a conformally-invariant scaling curve which is simple,
i.e., without double points, like the external frontier $\partial
\mathcal C$, here taken alone, define the universal function
$f_2(\alpha,\alpha',\lambda)$ which gives the Hausdorff dimension of the
points where the potential varies jointly with distance $r$ as
$r^{\alpha}$ on one side of the curve, and as $r^{\alpha'}$ on
the other, given a winding at rate $\lambda$. This function is
\begin{eqnarray}
f_2\left( \alpha, \alpha'; \lambda
\right)&=&b-\frac{1}{2(1-\gamma)}
{\left(\frac{1}{1+\lambda^2}-\frac{1}{2\alpha}-\frac{1}{2\alpha'}\right)}^{-1} \nonumber \\
& &-\frac{b-2}{2}\left(\alpha+\alpha'\right), \label{f_2twistexp}
\end{eqnarray}
and satisfies the generalization of scaling relation
(\ref{scalinglaw})
\begin{eqnarray}
 {f}_2(\alpha,\alpha';\lambda)&=&(1+\lambda^2){f}_2(\bar{\alpha},{\bar\alpha'};0)-b\lambda^2 \ .
\label{f_2twist}
\end{eqnarray}

This double multifractality can be generalized to higher ones
 \cite{BDjsp}, by considering the distribution of potential
between the arms of a {\it rotating star} ${\mathcal S}_k$, with the
following {\it poly-multifractal} result  \cite{BDBtopub}:
\begin{eqnarray}
f_k\left(\{ {\alpha}_{i}\};\lambda\right)&=&b-
\frac{1}{8(1-\gamma)}{k}^2{\left(\frac{1}{1+\lambda^2}-\sum_{i=1}^{k}\frac{1}{2{\alpha}_{i}}\right)}^{-1} \nonumber \\
& &-\frac{b-2}{2}\sum_{i=1}^{k}{\alpha}_{i} \ . \label{faicrot}
\end{eqnarray}
\\

\section{\sc{Duality for $O(N)$ and Potts Models and the Stochastic L\"owner Evolution}}
\label{sec.geodual}
\subsection{{Geometric Duality in $O(N)$ and Potts Cluster Frontiers}}

\subsubsection{$O(N)$ Model}
The $O(N)$ model partition function is that of a gas $\mathcal G$
of self- and mutually-avoiding {\it loops} on a given lattice,
e.g., the hexagonal lattice  \cite{nien}:
\begin{equation}
{Z}_{O(N)} = \sum_{\mathcal G }K^{{\mathcal N}_{B}}N^{{\mathcal
N}_{P}},
\label{ZON'}
\end{equation}
where $K$ and $N$ are two fugacities, associated respectively with the
total number of occupied bonds ${\mathcal N}_{B}$, and with the
total number of loops ${\mathcal N}_{P}$,
i.e., polygons drawn on the lattice. 
For $N \in [-2,2]$, this model possesses a critical point (CP),
$K_c$, while the whole {\it ``low-temperature''} (low-$T$) phase,
i.e., ${K}_c < K$, has critical universal properties, where the
loops are {\it denser} than those at the critical
point \cite{nien}.\\

\subsubsection{Potts Model}
The partition function of the $Q$-state Potts model on, e.g., the
square lattice, with a second order critical point for $Q \in
[0,4]$, has a Fortuin-Kasteleyn representation {\it at} the CP: $
Z_{\rm Potts}=\sum_{\cup (\mathcal C)}Q^{\frac{1}{2}{\mathcal
N}_{P}},$ where the configurations $\cup (\mathcal C)$ are those
of unions of clusters on
the square lattice, with 
a total number ${\mathcal N}_{P}$ of polygons encircling all
clusters, and filling the medial square lattice of the original
lattice  \cite{nien,dennijs}. Thus the critical Potts model becomes
a {\it dense} loop model, with loop fugacity
$N=Q^{\frac{1}{2}}$, while one can show that its {\it tricritical}
point with site dilution corresponds to the $O(N)$ CP \cite{D6,D7}.\\

\subsubsection{Coulomb Gas}
The $O(N)$ and Potts models thus possess the same ``Coulomb gas''
representations  \cite{nien,dennijs,D6,D7}:
\begin{equation}
N=\sqrt{Q}=-2 \cos \pi g, \nonumber
\end{equation}
with $g \in [1,\frac{3}{2}]$ for the $O(N)$ CP, and $ g \in
[\frac{1}{2},1]$ for the low-$T$ $O(N)$ or critical Potts, xv 
models; the coupling constant $g$ of the Coulomb gas also yields
the central charge:
\begin{equation}
c=1-6{(1-g)^2}/{g}. \label{cg}
\end{equation}
Notice that from the expression (\ref{cgamma}) of $c$ in
terms of $\gamma \leq 0$ one arrives at the simple relation:
\begin{equation}
\gamma=1-g,\ g \geq 1;\ \gamma=1-1/g,\ g\leq 1. \label{ggamma}
\end{equation}
The above representation for $N=\sqrt Q \in [0,2]$ gives a range
of values $- 2 \leq c \leq 1$; our results also apply for $c
\in(-\infty, -2]$, corresponding, e.g., to the $O\left(N\in
[-2,0]\right)$ branch, with a low-$T$ phase for $g \in
[0,\frac{1}{2}]$, and CP for $g \in [\frac{3}{2},2].$\\

\subsubsection{Hausdorff Dimensions of  Hull Subsets}
The fractal dimension $D_{\rm EP}$ of the accessible perimeter,
eq.~(\ref{D(c)}), is, like $c(g)=c(g^{-1})$, a symmetric function of $g$ and $g^{-1}$ once
 rewritten in terms of $g$:
\begin{equation}
D_{\rm EP}=1+ \frac{1}{2}g^{-1}\vartheta(1-g^{-1})+\frac{1}{2}g
\vartheta(1-g), \label{DEP}
\end{equation}
where $\vartheta$ is the Heaviside distribution. Thus $D_{\rm EP}$ is given by two
different analytic expressions on either side of the separatrix
$g=1$. The dimension of the full hull, i.e., the
complete set of outer boundary sites of a cluster, has been
determined for $O(N)$ and Potts clusters  \cite{SD}, and is
\begin{equation}
D_{\rm H}=1+\frac{1}{2}g^{-1}, \label{DH}
\end{equation}
for the {\it entire} range of the coupling constant $g \in
[\frac{1}{2},2]$. Comparing to eq. (\ref{DEP}), we therefore see
that the accessible perimeter and hull Hausdorff dimensions {\it coincide}
for $g\ge 1$, i.e., at the $O(N)$ CP (or for tricritical Potts
clusters), whereas they {\it differ}, namely $D_{\rm EP} < D_H$,
for $g < 1$, i.e., in the $O(N)$ low-$T$ phase, or for critical
Potts clusters. This is the generalization to any Potts model of
the effect originally found in percolation  \cite{GA}. This can be
directly understood in terms of the {\it singly connected} sites
(or bonds) where fjords close in the scaling limit. Their
dimension is given by \cite{SD}
\begin{equation}
D_{\rm SC}=1+\frac{1}{2}g^{-1}-\frac{3}{2}g. \label{DSC}
\end{equation}
For critical $O(N)$ loops, $g \in (1,2]$, so that $D_{\rm SC} <
0,$ hence there exist no closing fjords, thereby explaining the identity:
\begin{equation}
D_{\rm EP} = D_{\rm H}. \label{id}
\end{equation}
In contrast, one has  $g \in [\frac{1}{2},1)$ and $D_{\rm SC} > 0$ for critical Potts
clusters and for the $O(N)$ low-$T$ phase. In this case, pinching points of
positive dimension appear in the scaling limit,
so that $D_{\rm EP} < D_{\rm H}$ (Table 1).\\

\begin{table}[tb]
\begin{tabular}{| c | c | c | c | c | c |}
\hline
$Q$          &     0     &    1     &      2      &      3      &  4          \\
\hline
$c$          &     -2    &    0     &     1/2     &      4/5       &  1        \\
\hline
$D_{\rm EP}$ & ${5}/{4}$ &${4}/{3}$ & ${11}/{8}$  & ${17}/{12}$ & ${3}/{2}$ \\
\hline
$D_{\rm H}$  & $2$       &${7}/{4}$ & ${5}/{3}$   & ${8}/{5}$   & ${3}/{2}$ \\
\hline
$D_{\rm SC}$ & ${5}/{4}$ &${3}/{4}$ &$ {13}/{24}$ & ${7}/{20}$  &   $ 0$        \\
\hline
\end{tabular}
\vskip.5cm
\caption{{Dimensions for the critical $Q$-state Potts
model; $Q=0,1,2$ correspond  to spanning trees,
percolation and Ising clusters, respectively.}}
\end{table}

\subsubsection{Duality}
We then find from eq. (\ref{DEP}), with $g\leq 1$:
\begin{equation}
\left(D_{\rm EP}-1\right) \left( D_{\rm H}-1\right)=\frac{1}{4}.
\label{duali}
\end{equation}
The symmetry point $D_{\rm EP} = D_{\rm H}=\frac{3}{2}$
corresponds to $g=1$, $N=2$, or $Q=4$, where, as expected, the
dimension $D_{\rm SC}=0$ of the pinching points vanishes.

For percolation, described either by $Q=1$, or by the low-$T$
$O(N=1)$ model with $g=\frac{2}{3}$, we recover the result
$D_{\rm EP}=\frac{4}{3}$, recently derived in  \cite{ADA}. For the
Ising model, described either by $Q=2, g=\frac{3}{4}$, or by the
$O(N=1)$ CP with $g'=g^{-1}=\frac{4}{3}$, we observe that the EP
dimension $D_{\rm EP}=\frac{11}{8}$ coincides, as expected, with
that of critical $O(N=1)$ loops, which in fact appear as EP's.
This is a particular case of a further duality relation between
the critical Potts and CP $O(N)$ models:
\begin{equation}
D_{\rm EP}\left(Q(g)\right)= D_{\rm
H}\left[O\left(N(g')\right)\right],\makebox{\rm for}\; g'=g^{-1}, g
\le 1\ .
\end{equation}
In terms of this duality, the central charge takes the simple
expression:
\begin{equation}
c=(3-2g)(3-2g'). \label{cdual}
\end{equation}
\\

\subsection{{Geometric Duality Property of the ${\rm SLE}_{\kappa}$}}
\label{subsec.geoSLE}

\subsubsection{Relation of the ${\rm SLE}_{\kappa}$ trace to $Q$-Potts frontiers or $O(N)$ lines}

  An introduction to the stochastic L\"owner evolution process (${\rm SLE}_{\kappa}$)
 can be found in  \cite{lawleresi}, \cite{stflour}. This process drives a conformally-invariant random path, which
 essentially describes the boundaries of (Potts) clusters or {\it hulls}
 we have introduced above, or the random lines of the $O(N)$ model. The random path can be a simple or non simple path
 with self-contacts.
 The ${\rm SLE}_{\kappa}$ is parametrized by $\kappa$, which describes the rate of an auxiliary Brownian
 motion, which is the source for the process. When $\kappa \in [0,4]$, the random curve is simple,
 while for $\kappa \in (4,8)$, the curve is a self-coiling path.
 For $\kappa \geq 8$ the path is space filling.
 The correspondence to our parameters,
 the central charge $c$, the string susceptibility exponent $\gamma$, or
 the Coulomb gas constant $g$, is as follows.

 In the original work by Schramm  \cite{schramm1}, the variance
  of the Gaussian winding angle $\vartheta$ of the single extremity of a ${\rm SLE}_{\kappa}$ of size $R$ was calculated,
 and found to be $$\langle \vartheta^2\rangle ={\kappa} \,{\rm ln} R.$$ In  \cite{DSw} we found, for instance
 for the extremity of a random line in the $O(N)$ model,
 the corresponding angular variance $$\langle \vartheta^2\rangle=(4/g)\, {\rm ln} R,$$ from which we immediately infer the identity
 \begin{equation}
 \kappa=\frac{4}{g}\ .
 \label{k}
 \end{equation}

The low-temperature branch $g \in [\frac{1}{2},1)$ of the $O(N)$
model, for $N\in [0,2)$, indeed corresponds to $\kappa \in (4,8]$
and describes non simple curves, while $N\in [-2,0], g\in
[0,\frac{1}{2}]$ corresponds to $\kappa \geq 8$. The critical
point branch $g \in [1,\frac{3}{2}], N\in [0,2]$
 gives $\kappa \in [\frac{8}{3},4]$, while $g \in [\frac{3}{2},2], N\in [-2,0]$
 gives $\kappa \in [2,\frac{8}{3}]$.
The range $\kappa \in [0,2)$ probably corresponds to higher
multicritical points with $g>2$. Owing to eq. (\ref{ggamma}) for
$\gamma$, we have
\begin{eqnarray}
\label{gk}
\gamma&=&1-\frac{4}{\kappa},\ \kappa \leq 4\ ;\\
\label{gkd}
\gamma&=&1-\frac{\kappa}{4},\ \kappa \geq 4\ .
\end{eqnarray}

\subsubsection{Duality}
 The central charge (\ref{cgamma}) or (\ref{cg}) is accordingly:
\begin{equation}
c=1-24{\left(\frac{\kappa}{4}-1\right)^2}/{\kappa}\ , \label{ck}
\end{equation}
an expression which of course is symmetric under the {\it duality}
$\kappa/4 \to 4/\kappa=\kappa'$, or
\begin{equation}
\kappa \kappa'=16\ ,
\label{duality}
\end{equation}
 reflecting the symmetry under $gg'=1$  \cite{duplantier11}.
The self-dual form of the central charge is accordingly:
\begin{equation}
c=\frac{1}{4}(6-\kappa)(6-\kappa'). \label{cdualSLE}
\end{equation}
From eqs. (\ref{DH}) and (\ref{DEP}) we respectively find \cite{duplantier11}
\begin{equation}
D_{\rm H}=1+\frac{1}{8}\kappa\ , \label{DHs}
\end{equation}
\begin{equation}
D_{\rm EP}=1+ \frac{2}{\kappa}\vartheta(\kappa-4)+\frac{\kappa}{8}
\vartheta(4-\kappa)\ , \label{DEPs}
\end{equation}
in agreement with some results derived later in probability theory
 \cite{RS,beffara}.

For $\kappa \leq 4$, we have $D_{\rm
EP}(\kappa)=D_{\rm H}(\kappa)$. For $\kappa \geq 4$, the
self-coiling scaling paths obey the duality equation
(\ref{duali}) derived above, recast here in the context of the
${\rm SLE}_{\kappa}$ process:
\begin{equation}
\left[D_{\rm EP}(\kappa)-1\right] \left[ D_{\rm
H}(\kappa)-1\right]=\frac{1}{4},\ \kappa \geq 4\ ,
\label{dualibis}
\end{equation}
where now $$D_{\rm EP}(\kappa)=D_{\rm H}(\kappa'=16/\kappa)\quad
\kappa'\leq 4\ .$$ Thus we predict that the external perimeter of
a self-coiling ${\rm SLE}_{\kappa \geq 4}$ process is, by {\it
duality}, the simple path of the ${\rm SLE}_{(16/{\kappa})=\kappa' \leq
4}$ process.

The symmetric point $\kappa=4$ corresponds to the $O(N=2)$ model,
or $Q=4$ Potts model, with $c=1$. The value $\kappa=8/3, c=0$
corresponds to a self-avoiding walk, which thus appears
 \cite{duplantier9,ADA} as the external frontier of a $\kappa=6$
process, namely that of a percolation hull
 \cite{schramm1,smirnov1}.

 Let us now study more of the ${\rm SLE}$'s random geometry
 using the quantum gravity method described here.

Up to now, we have described general conformally-invariant curves in the plane in terms of the
universal parameters $c$ (central charge) or $\gamma$ (string susceptibility). The multifractal
results described in the
sections above thus apply to the ${\rm SLE}$ after substituting $\kappa$ to $\gamma$ or $c$. Care should be taken,
however, in such a substitution since two dual values of $\kappa$ (\ref{duality}) correspond to a same value of $\gamma$.
The reason is that up to now we have considered geometrical properties of the boundaries which actually were
{\it self-dual}. An exemple is the harmonic multifractal spectrum of the ${\rm SLE}_{\kappa \geq 4}$ frontier, which
is identical to that of the smoother (simple) ${\rm SLE}_{(16/{\kappa})=\kappa' \leq 4}$ path. So we actually saw only
the set of simple SLE traces with $\kappa \leq 4$. When dealing with higher multifractality, we assumed the random
curves to be simple. When dealing with non simple random paths, boundary quantum gravity rules are to be
modified as explained now.\\

\section{\sc{Duality in KPZ}}
\label{sec.duality}
\subsection{Dual Dimensions}

It will be convenient to introduce the following notations. The standard KPZ map reads:
\begin{equation}
x=U_{\gamma}(\Delta)=\Delta \frac{\Delta-\gamma}{1-\gamma}\ ,
\label{KPZS}
\end{equation}
where $x$ is a planar conformal dimension and $\Delta$ its quantum gravity counterpart, and
where we recall that $\gamma$ is the negative root of
\begin{equation}
c=1-6{\gamma}^2(1-\gamma)^{-1}, \gamma \leq 0. \label{cgammaa}
\end{equation}
We introduce the {\it dual quantum dimension} of $\Delta$, $\Delta'$ such that:
\begin{equation}
\Delta' := \frac{\Delta-\gamma}{1-\gamma}\ ,
\label{dualdelta}
\end{equation}
and \begin{equation}
x=U_{\gamma}(\Delta)=\Delta \Delta'\ .
\label{dd'}
\end{equation}
Similarly, let us define the variable $\gamma'$, dual of susceptibility exponent $\gamma$, by:
\begin{equation}
(1-\gamma)(1-\gamma')=1\ ,
\label{dualgamma}
\end{equation}
which is simply the (``non-physical'') positive root of eq. (\ref{cgammaa}):
\begin{equation}
c=1-6{\gamma'}^2(1-\gamma')^{-1}, \gamma' \geq 0. \label{cgammaa'}
\end{equation}
The dual equation of (\ref{dualdelta}) is then:
\begin{equation}
\Delta = \frac{\Delta'-\gamma'}{1-\gamma'}\ ,
\label{dual'delta}
\end{equation}
By construction we have the simultaneous equations:
\begin{equation}
\Delta=U_{\gamma}^{-1}(x),\, \Delta'=\frac{U_{\gamma}^{-1}(x)-\gamma}{1-\gamma},
\label{dualU}
\end{equation}
with the positive solution
\begin{equation}
U_{\gamma}^{-1}\left( x\right)
=\frac{1}{2}\left(\sqrt{4(1-\gamma)x+\gamma^2}+\gamma\right) .
\label{U1aa}
\end{equation}

We define a dual KPZ map $U_{\gamma'}$ by the same equation as (\ref{KPZS}), with ${\gamma'}$ substituted
for ${\gamma}$.
It has the following properties:
\begin{eqnarray}
\label{KPZdual}
x&=&U_{\gamma}(\Delta)=U_{\gamma'}(\Delta')\ ,\\
\label{KPZdualinv}
\Delta'&=&U_{\gamma'}^{-1}(x)=\frac{U_{\gamma}^{-1}(x)-\gamma}{1-\gamma}\ ,\\
\label{KPZdualinv'}
\Delta&=&U_{\gamma}^{-1}(x)=\frac{U_{\gamma'}^{-1}(x)-\gamma'}{1-\gamma'}\ .
\end{eqnarray}

\subsection{Boundary KPZ for non simple paths}
The additivity rules in quantum gravity for the boundary scaling dimensions of mutually-avoiding random paths $A$ and $B$
 are:
\begin{eqnarray}
\label{sa}
\tilde\Delta\left( A\wedge B\right)&=&\tilde \Delta(A)+\tilde\Delta(B)\,\,\,\,\,\,\, {\text{(simple paths),}}\\
\label{nsa}
\tilde\Delta'\left( A\wedge B\right)&=&\tilde\Delta'(A)+\tilde\Delta'(B)\,\,\,\, {\text{(non-simple paths).}}
\end{eqnarray}
For simple paths, like random lines in the $O(N)$ model at its critical point, or the SLE trace for $\kappa \leq 4$ 
the boundary dimensions are additive 
in quantum gravity, a fundamental fact repeatedly used 
above. On the other hand, for non-simple paths, the {\it dual dimensions are additive} in boundary quantum gravity.
 This is the case of random lines in the dense phase of the $O(N)$ model, or, equivalently, of hulls of
 Fortuin-Kasteleyn clusters in the Potts model, or of the ${\rm SLE}_{\kappa \geq 4}$ trace.
 These additivity rules are derived in appendices  \ref{ONapp} and \ref{BBapp} from the consideration of partition functions on a random surface in the dilute or dense phases.
In terms of standard dimensions $\tilde \Delta$ this reads:
\begin{eqnarray}
\label{sa'}
\tilde\Delta\left( A\wedge B\right)&=&\tilde \Delta(A)+\tilde\Delta(B)\,\,\,\,\,\,\,\,\,\,\,\,\,\,\,\,\,\,\,\,\,\,
{\text{ (simple paths),}}\\
\label{nsa'}
\tilde\Delta\left( A\wedge B\right)&=&\tilde\Delta(A)+\tilde\Delta(B)-\gamma\,\,\,\,\,\,\,\,\,\,\,
{\text{ (non-simple paths)}.}
\end{eqnarray}
In the dilute phase the composition rule for boundary dimensions in the upper half-plane $\mathbb H$ reads accordingly: 
\begin{eqnarray}
\label{tildexAB}
\tilde x(A\wedge B)&=&U_{\gamma}\left[ U_{\gamma}^{-1}\left( \tilde{x}\left(A\right)\right)
+U_{\gamma}^{-1}\left( \tilde{x}\left( B\right)\right)  \right].
\end{eqnarray}
In the dense phase, the new rule (\ref{nsa}) in terms of dual variables and functions, gives for
composite boundary dimensions in $\mathbb H$:
\begin{eqnarray}
\label{tildexABdense}
\tilde x(A\wedge B)&=&U_{\gamma}\left[ U_{\gamma}^{-1}\left( \tilde{x}\left(A\right)\right)
+U_{\gamma}^{-1}\left( \tilde{x}\left( B\right)\right)-\gamma  \right],\\
\label{tildexABdual}
&=&U_{\gamma'}\left[ U_{\gamma'}^{-1}\left( \tilde{x}\left(A\right)\right)
+U_{\gamma'}^{-1}\left( \tilde{x}\left( B\right)\right) \right].
\end{eqnarray}
So we see that the composition rules (\ref{tildexABdense}) for non-simple paths are different
from the ones for simple paths, when written in terms of the standard string
susceptibility exponent $\gamma$, but that they are formally identical in terms of the dual exponent $\gamma'$, as shown
by eqs.(\ref{tildexAB}) and (\ref{tildexABdual}).

\subsection{Bulk KPZ for non-simple paths}
For determining the complete set of scaling dimensions, it remains to relate bulk and boundary dimensions.
In the dilute phase, i.e., for simple paths, we have seen the simple relation in a random metric:
\begin{equation}
2\Delta -\gamma=\tilde \Delta \ ,
\label{bulkboundary}
\end{equation}
which is established in appendix \ref{BBapp} (see also appendices  \ref{Brownapp} and \ref{ONapp}. The KPZ map from boundary dimension in quantum gravity to
bulk dimension in the plane reads accordingly
\begin{eqnarray}
x=2 U_{\gamma}(\Delta)=2 U_{\gamma}\left(\frac{1}{2}{(\tilde\Delta+\gamma)}\right)
=2V_{\gamma}(\tilde\Delta),
\label{VD}
\end{eqnarray}
where
\begin{eqnarray}
\label{VV}
V_{\gamma}(x)&=&\frac{1}{4}\frac{x^2-\gamma^2}{1-\gamma}\ ,
\end{eqnarray}
an expression repeatedly used above. When dealing with non-simple paths,
these relations have to be changed to:
\begin{equation}
2\Delta =\tilde \Delta \ ,
\label{bulkboundarydense}
\end{equation}
as shown in appendices  \ref{ONapp} and \ref{BBapp}. At this stage, the reader will
not be surprised that this relation is just identical to the dual of (\ref{bulkboundary})
\begin{equation}
2\Delta' -\gamma'=\tilde \Delta' \ ,
\label{bulkboundarydual}
\end{equation}
when now written in terms of both dual dimensions and susceptibility exponent.
As a consequence, the scaling dimension of a bulk operator in a dense system reads:
\begin{eqnarray}
x=2 U_{\gamma}(\Delta)=2 U_{\gamma}\left(\frac{1}{2}{\tilde\Delta}\right)
=\frac{1}{2}\tilde\Delta\frac{\tilde\Delta -2\gamma}{1-\gamma},
\label{UD}
\end{eqnarray}
which by duality can necessarily be written as:
\begin{eqnarray}
\label{xdensedual}
x&=&2 V_{\gamma'}(\tilde\Delta'),\\
\nonumber
V_{\gamma'}(x)&=&\frac{1}{4}\frac{x^2-{\gamma'}^2}{1-\gamma'}\ ,
\end{eqnarray}
as can be checked easily.

As we have seen before, the composition rule for bulk dimensions of simple paths (in the dilute phase)
in the plane $\mathbb C$ follows from (\ref{VD}), (\ref{VV}), and (\ref{sa}):
\begin{eqnarray}
\label{xAB}
x(A\wedge B)&=&2U_{\gamma}\left[\frac{1}{2}\left( U_{\gamma}^{-1}\left( \tilde{x}\left(A\right)\right)
+U_{\gamma}^{-1}\left( \tilde{x}\left( B\right) \right) +\gamma \right) \right],\\
\label{xABV}
&=&2V_{\gamma}\left[ U_{\gamma}^{-1}\left( \tilde{x}\left(A\right)\right)
+U_{\gamma}^{-1}\left( \tilde{x}\left( B\right)\right) \right].
\end{eqnarray}
The composition rule for bulk dimensions of non-simple paths (dense phase) in the plane $\mathbb C$ differs, 
according to (\ref{nsa'}), (\ref{bulkboundarydense}) and (\ref{UD}): 
\begin{eqnarray}
\label{xABdense}
x(A\wedge B)=2U_{\gamma}\left[\frac{1}{2}\left( U_{\gamma}^{-1}\left( \tilde{x}\left(A\right)\right)
+U_{\gamma}^{-1}\left( \tilde{x}\left( B\right)\right)-\gamma \right) \right],
\end{eqnarray}
which reads also, according to (\ref{xdensedual}) and (\ref{nsa}):
\begin{eqnarray}
x(A\wedge B)=2V_{\gamma'}\left[ U_{\gamma'}^{-1}\left( \tilde{x}\left(A\right)\right)
+U_{\gamma'}^{-1}\left( \tilde{x}\left( B\right)\right) \right].
\label{xABVdense}
\end{eqnarray}
This is formally the same as the rule (\ref{xABV}) in the dilute phase, up to substitution of $\gamma'$ for $\gamma$, and it
 applies to the dense phase of the $O(N)$ model, or to
Potts cluster boundaries, and in particular to the ${\rm SLE}_{\kappa \geq 4}$.
 
In summary, the composition rules for planar scaling dimensions, 
whether on a boundary or in the bulk, take a unique analytic form for both phases (simple or non-simple paths), 
provided one replaces the string susceptibility exponent $\gamma$ in the simple case 
by its dual variable $\gamma'$ in the non-simple case; this can be seen in (\ref{xABV}), eqs. (\ref{tildexAB}), (\ref{tildexABdual})
and  (\ref{xABVdense}).\\

\section{\sc{SLE and KPZ}}
\label{sec.SLEKPZ}

\subsection{Duality for the SLE}
We have seen that the composition rules for dimensions, while they change from the dilute to the dense phase
when dealing with the proper KPZ formalism, scaling dimensions and susceptibility exponent,  stay
invariant when expressed in dual variables.
This duality is perfectly adapted to the parameterization of the ${\rm SLE}_{\kappa}$ process.
Indeed we have from (\ref{gk}) and (\ref{gkd})
\begin{eqnarray}
\label{gg'k}
\gamma&=&1-\frac{4}{\kappa},\ \gamma'=1-\frac{\kappa}{4},\, \kappa \leq 4;\\
\label{gg'kd}
\gamma&=&1-\frac{\kappa}{4},\ \gamma'=1-\frac{4}{\kappa} \ ,\kappa \geq 4,
\end{eqnarray}
so that the analytical forms of $\gamma$ and its dual $\gamma'$ are simply exchanged when passing from simple paths
($\kappa \leq 4$)
to non-simple ones ($\kappa > 4$). Because of the equivalent dual equations (\ref{KPZdual}), by choosing either the 
$\gamma$-solution or the $\gamma'$-solution, depending whether $\kappa \leq 4$ or $\kappa \geq 4$, we can write 
\begin{equation}
x=\left\{\begin{array}{ll}
U_{\gamma(\kappa \leq 4)}(\Delta)={\mathcal U}_{\kappa}(\Delta) & \mbox{$\kappa \leq 4$}\\
U_{\gamma'(\kappa \geq 4)}(\Delta')={\mathcal U}_{\kappa}(\Delta') & \mbox{$\kappa \geq 4$,}
\end{array}
\right.
\label{KPZSLE}
\end{equation}
 with now a single function, valid for all values of parameter $\kappa$
\begin{eqnarray}
{\mathcal U}_{\kappa}(\Delta)=\frac{1}{4}\Delta\left({\kappa}\Delta +4-{\kappa}\right).
\label{USLE}
\end{eqnarray}
Similarly, the inverse KPZ map (\ref{U1aa}) reads, according to (\ref{KPZdualinv}) or (\ref{KPZdualinv'}):
\begin{eqnarray}
\nonumber
\Delta&=&U_{\gamma(\kappa \leq 4)}^{-1}\left( x\right)={\mathcal U}_{\kappa}^{-1}\left( x\right),\,\,\,\kappa \leq 4,\\
\label{KPZSLEinv}
\Delta'&=&U_{\gamma'(\kappa \geq 4)}^{-1}\left( x\right)={\mathcal U}_{\kappa}^{-1}\left( x\right),\,\kappa \geq 4,
\end{eqnarray}
again with a single expression of the inverse function, valid for any $\kappa$
\begin{eqnarray}
{\mathcal U}_{\kappa}^{-1}\left( x\right)=
\frac{1}{2\kappa}\left(\sqrt{16\kappa x+(\kappa-4)^2}+\kappa-4\right).
\label{U-1SLE}
\end{eqnarray}
I emphasize that ${\mathcal U}_{\kappa}$ coincides with the KPZ map for $\kappa \leq 4$, while it
 represents the dual of the latter when $\kappa \geq 4$ and then acts on the dual dimension $\Delta'$. For instance,
  we have the important value at the origin
\begin{equation}
{\mathcal U}_{\kappa}^{-1}\left( 0\right)=\frac{1}{2\kappa}\left[\,|\kappa-4| +\kappa -4\right]
=\left(1-\frac{4}{\kappa}\right) \vartheta (\kappa  -4),
\label{U01}
\end{equation}
which vanishes for simple paths, and is non-trivial for non-simple ones.

 It remains to define
the analogue of the $V$ function (\ref{VV}) or its dual (\ref{xdensedual}):
\begin{eqnarray}
x=\left\{\begin{array}{ll}
2V_{\gamma(\kappa \leq 4)}(\tilde\Delta)=2{\mathcal V}_{\kappa}(\tilde\Delta) & \mbox{$\kappa \leq 4$}\\
2V_{\gamma'(\kappa \geq 4)}(\tilde\Delta')=2{\mathcal V}_{\kappa}(\tilde\Delta') & \mbox{$\kappa \geq 4$,}
\end{array}
\right.
\label{KPZSLEV}
\end{eqnarray}
with again a single function, valid for all values of parameter $\kappa$
\begin{eqnarray}
\nonumber
{\mathcal V}_{\kappa}(\Delta)&=&{\mathcal U}_{\kappa}\left[\frac{1}{2}\left(\Delta+1-\frac{4}{\kappa}\right)\right]\\
&=&\frac{1}{16\kappa}\left[\kappa^2\Delta^2-(\kappa-4)^2\right],
\label{VSLE}
\end{eqnarray}
but acting on the boundary dimension in quantum gravity or on its dual, depending on whether 
$\kappa \leq 4$ or $\kappa \geq 4$.

\subsection{Composition Rules for SLE}
Finally we can conclude with general composition rules for the SLE process. Indeed, the boundary rules in $\mathbb H$ 
(\ref{tildexAB}) or 
its dual (\ref{tildexABdual}), owing to eqs. (\ref{KPZSLE}) and (\ref{KPZSLEinv}), read in a unified way 
in terms of parameter $\kappa$:
\begin{eqnarray}
\label{tildexABk}
\tilde x(A\wedge B)&=&{\mathcal U}_{\kappa}\left[ {\mathcal U}_{\kappa}
^{-1}\left( \tilde{x}\left(A\right)\right)
+{\mathcal U}_{\kappa}^{-1}\left( \tilde{x}\left( B\right)\right)  \right],
\end{eqnarray}
valid for the entire range of $\kappa$. Similarly, the composition rules for SLE's in the plane $\mathbb C$ are found 
from eqs. (\ref{xABV}) or (\ref{xABVdense}), and recast according to (\ref{KPZSLEV}) and (\ref{KPZSLEinv}) 
into a unified formula, valid for any $\kappa$
\begin{eqnarray}
\label{xABk}
x(A\wedge B)=2{\mathcal V}_{\kappa}\left[ {\mathcal U}_{\kappa}^{-1}\left( \tilde{x}\left(A\right)\right)
+{\mathcal U}_{\kappa}^{-1}\left( \tilde{x}\left( B\right)\right) \right].
\end{eqnarray}
Thus we see that by introducing dual equations, we have been able to unify the composition rules for the SLE in a unique way, 
which no longer depends explicitly on the range of $\kappa$.

\subsection{Short Distance Expansion (SDE)}
\subsubsection{Boundary SDE}
Consider the power law governing the behavior of two mutually-avoiding random paths $A$ and $B$ anchored at the Dirichlet boundary line, and 
approaching each other at short distance $r$ along the line. The probability of such an event scales like 
\begin{equation}
\tilde{\mathcal P}_{A, B}(r) \propto r^{\tilde x_{A,B}}, \,\, r \to 0,
\label{SDE}
\end{equation}
where the short-distance  exponent reads  \cite{BPZ, duplantier4}:
\begin{equation}
{\tilde x_{A,B}}= \tilde x(A\wedge B)-\tilde x(A)-\tilde x(B).
\label{SDEexp}
\end{equation}
We simply use the fusion rule (\ref{tildexABk}) and the quadratic map (\ref{USLE}) to immediately get
\begin{equation}
{\tilde x_{A,B}}= \frac{\kappa}{2}{\mathcal U}_{\kappa}^{-1}
\left( \tilde{x}_A\right)\,{\mathcal U}_{\kappa}^{-1}\left(\tilde{x}_B\right),
\label{SDEU}
\end{equation}
where we use $\tilde{x}_A=\tilde{x}\left(A\right)$ as a short-hand notation.
In terms of quantum gravity boundary dimensions, or their dual, this SDE exponent splits into
\begin{eqnarray}
{\tilde x_{A,B}}=\left\{\begin{array}{ll} \frac{\kappa}{2} \tilde\Delta_A\tilde\Delta_B  & \mbox{$\kappa \leq 4$} \\
\frac{\kappa}{2} \tilde\Delta'_A\tilde\Delta'_B & \mbox{$\kappa \geq 4$.}
\end{array}
\right.
\label{SDED}
\end{eqnarray}
So we see that the short-distance expansion along the boundary of $\mathbb H$ is governed by the product of 
the quantum boundary dimensions, or of their duals, 
depending on the phase we are in. In particular, if one chooses the set $B$ to be the chordal SLE trace itself,
its boundary dimension $\tilde x_1=(6-\kappa)/2\kappa$ is such that
$\tilde \Delta_1=U_{\gamma}^{-1}(\tilde x_1)=\frac{1}{2}(1-\gamma)$ in the dilute phase, or $\tilde \Delta_1=U_{\gamma}^{-1}(\tilde x_1)=\frac{1}{2}+\gamma$ 
in the dense phase. That corresponds to the single expression ${\mathcal U}_{\kappa}^{-1}(\tilde x_1)={2}/{\kappa}$, 
 which is $\tilde \Delta_1$ for $\kappa \leq 4$ or  $\tilde \Delta_1^{'}$ for $\kappa \geq 4$. 
 In this case, the expressions (\ref{SDEU}) or (\ref{SDED}) simplify to
\begin{eqnarray}
{\tilde x_{A,1}}&=&{\mathcal U}_{\kappa}^{-1}
\left( \tilde{x}_A\right)=\frac{1}{2\kappa}\left(\sqrt{16\kappa \tilde{x}_A +(\kappa-4)^2}+\kappa-4\right)\\
\nonumber
&=&\left\{\begin{array}{ll}\tilde\Delta_A & \mbox{$\kappa \leq 4$}\\
\tilde\Delta'_A& \mbox{$\kappa \geq 4$.}
\end{array}
\right.
\label{SDESLE}
\end{eqnarray}
This explains the observation made in  \cite{BB} that the boundary SDE of any operator with the SLE trace 
might be seen as exhibiting (boundary) quantum gravity. However, we see that if for $\kappa \leq 4$ 
the SDE exponent (\ref{SDESLE}) is 
indeed the KPZ solution $\tilde \Delta$, for $\kappa \geq 4$ it necessarily transforms to the dual dimension 
$\tilde\Delta'$ introduced above 
in (\ref{dualdelta}) . Moreover, at this stage, this appearance of the quantum gravity dimension
might be seen as a coincidence, since the 
general structure 
of SDE exponent (\ref{SDED}) is clearly still quadratic and given by the product of quantum gravity dimensions or their dual.

\subsubsection{Bulk SDE}
One can also consider the SDE for random paths in the full plane, corresponding to the so-called radial SLE. 
Consider the power law governing the behavior of two mutually-avoiding random paths $A$ and $B$  
approaching each other at short distance $r$ in the plane, with probability  
\begin{equation}
{\mathcal P}_{A, B}(r) \propto r^{x_{A,B}}, \,\, r \to 0,
\label{SDEbulk}
\end{equation}
where the short-distance exponent now reads:
\begin{equation}
{x_{A,B}}= x(A\wedge B)-x(A)-x(B).
\label{SDEbexp}
\end{equation}
We simply use the fusion rule (\ref{xABk}) and the quadratic maps (\ref{USLE}) and (\ref{VSLE}) to get
\begin{equation}
{x_{A,B}}= \frac{\kappa}{4}{\mathcal U}_{\kappa}^{-1}
\left( \tilde{x}_A\right)\,{\mathcal U}_{\kappa}^{-1}\left(\tilde{x}_B \right) + 
\frac{(\kappa -4)^2}{8\kappa}.
\label{SDEV}
\end{equation}
In terms of quantum gravity boundary dimensions, or their dual, this SDE exponent reads
\begin{eqnarray}
{x_{A,B}}=\left\{\begin{array}{ll} \frac{\kappa}{4} \tilde\Delta_A\tilde\Delta_B +\frac{(\kappa -4)^2}{8\kappa} & \mbox{$\kappa \leq 4$}\\
\frac{\kappa}{4} \tilde\Delta'_A\tilde\Delta'_B +\frac{(\kappa -4)^2}{8\kappa} & \mbox{$\kappa \geq 4$.}
\end{array}
\right.
\label{SDEbD}
\end{eqnarray}
So we see that the short-distance expansion  in $\mathbb C$ is again governed by the product of 
the quantum boundary dimensions, or of their duals, plus a shift term. If one chooses in particular the set $B$ to be 
the radial SLE trace itself, taken at a typical medial point,
its boundary scaling dimension $\tilde x_2$ is such that 
$\tilde \Delta_2=U_{\gamma}^{-1}(\tilde x_2)=1-\gamma$ in the dilute phase, 
or $\tilde \Delta_2=U_{\gamma}^{-1}(\tilde x_2)=1+\gamma$ 
in the dense phase. That corresponds to the single expression ${\mathcal U}_{\kappa}^{-1}(\tilde x_2)
=2\,{\mathcal U}_{\kappa}^{-1}(\tilde x_1)={4}/{\kappa}$, 
 which is $\tilde \Delta_2$ for $\kappa \leq 4$ or  $\tilde \Delta_2^{'}$ for $\kappa \geq 4$. 
 In this case the expressions (\ref{SDEV}) or (\ref{SDEbD}) simplify into 
\begin{eqnarray}
\nonumber
{x_{A,2}}&=&{\mathcal U}_{\kappa}^{-1}
\left( \tilde{x}_A\right) +\frac{(\kappa -4)^2}{8\kappa}\\
&=&\left\{\begin{array}{ll} \tilde\Delta_A +\frac{(\kappa -4)^2}{8\kappa} & \mbox{$\kappa \leq 4$}\\
 \tilde\Delta'_A +\frac{(\kappa -4)^2}{8\kappa} & \mbox{$\kappa \geq 4$.}
\end{array}
\right.
\label{SDESLEb}
\end{eqnarray}
So the SDE of the SLE trace with any operator $A$ in the plane again generates the boundary dimension of $A$
in quantum gravity or its dual, modulo a constant shift. Notice that this shift is self-dual 
with respect to $\kappa\kappa'=16$ and reads also
$\frac{(\kappa -4)^2}{8\kappa}=\frac{1-c}{12}$.

\subsubsection{Ka\v {c} Spectrum}

The Ka\v {c} spectrum of conformal weights of a conformal field theory of central charge
$c(\gamma)$ can be written with $\gamma$ as a parameter

\begin{equation}
h_{p,q}^{(\gamma)}=\frac{\left[ (1-\gamma)p-q\right] ^{2}-\gamma^2}{4(1-\gamma)}. \label{Kac}
\end{equation}
Notice that if one substitutes the dual parameter $\gamma'$ to $\gamma$:
\begin{eqnarray}
\nonumber
h_{p,q}^{(\gamma')}&=&\frac{\left[ (1-\gamma')p-q\right] ^{2}-{\gamma'}^2}{4(1-\gamma')}\\
&=& h_{q,p}^{(\gamma)}.
\label{Kac'}
\end{eqnarray}
Hence in metric space the dual dimension is simply that with exchanged indices.
By the inverse KPZ relation the dimension $h_{p,q}^{(\gamma)}$ corresponds to a conformal weight in quantum gravity:
\begin{equation}
\Delta_{p,q}^{(\gamma)}=U_{\gamma}^{-1}(h_{p,q}^{(\gamma)})
=\frac{\left| (1-\gamma)p-q\right|+\gamma}{2}. \label{Kacg}
\end{equation}
The dual dimension (\ref{dualdelta}) reads:
\begin{eqnarray}
\nonumber
\Delta_{p,q}^{{(\gamma)}^{\prime}}&=&\frac{\Delta_{p,q}^{(\gamma)}-\gamma}{1-\gamma}
=\frac{\left| (1-\gamma')q-p\right|+\gamma'}{2}\\
&=&\Delta_{q,p}^{(\gamma')}=U_{\gamma'}^{-1}(h_{q,p}^{(\gamma')}). \label{Kacdual}
\end{eqnarray}
In $O(N)$ model studies (see, e.g.,  \cite{nien,DK,DSd}), it has been observed that
the conformal operator $\Phi_{p,q}$, with conformal weights $h_{p,q}^{(\gamma)}$ (or $\Delta_{p,q}^{(\gamma)}$ in QG),
and describing a given system of random paths, gets its indices $p$ and $q$ interchanged when going
from the dilute phase to the dense phase. In the plane, we see from (\ref{Kac'}) that this corresponds to
keeping the same indices and formally going to the dual string susceptibility, whereas in quantum gravity,
 (\ref{Kacdual}) shows that this corresponds to keeping the same indices, while performing
the double operation of taking the dual dimension together with the dual string susceptibility.

Using (\ref{gk}), (\ref{gkd}), the Ka\v {c} spectrum (\ref{Kac}) can be written in terms of parameter $\kappa$ as:
\begin{eqnarray}
\label{Kackap-}
h_{p,q}^{(\gamma(\kappa\leq 4))}
&=&\frac{\left(4p-\kappa q\right) ^{2}-(\kappa-4)^2}{16\kappa}:=\hbar_{p,q}^{\kappa}, \,\,\,\kappa \leq 4,\\
\label{Kackap+}
h_{p,q}^{(\gamma(\kappa\geq 4))}
&=&\frac{\left(\kappa p-4q\right) ^{2}-(\kappa-4)^2}{16\kappa}=\hbar_{q,p}^{\kappa}, \,\,\,\kappa \geq 4,
\end{eqnarray}
where in this new notation $\hbar_{p,q}^{\kappa}$ coincides with $h_{p,q}^{(\gamma)}$ for $\kappa\leq 4$, and
with $h_{q,p}^{(\gamma)}$ [with {\it interchanged} indices] for $\kappa\geq 4$. It also obeys the duality equation:
$\hbar_{p,q}^{\kappa'=16/\kappa}=\hbar_{q,p}^{\kappa}$. Notice finally that the convention
for placing indices in $\hbar_{p,q}^{\kappa}$ is the reverse of that of ref.  \cite{BB}.

The quantum conformal weight (\ref{Kacg}) reads similarly, depending on the $\kappa$-range:
\begin{eqnarray}
\label{Kacgkap-}
\Delta_{p,q}^{(\gamma(\kappa\leq 4))}
&=&\frac{\left|4p-\kappa q\right| +\kappa-4}{2\kappa}, \,\,\,\kappa \leq 4,\\
\label{Kacgkap+}
\Delta_{p,q}^{(\gamma(\kappa\geq 4))}
&=&\frac{\left|\kappa p-4q\right|+4-\kappa}{8}, \,\,\,\kappa \geq 4.
\end{eqnarray}
Let us also introduce a unified notation for the inverse image of $\hbar_{p,q}^{\kappa}$
(\ref{Kackap-}) by the $\kappa$-dependent map ${\mathcal U}_{\kappa}^{-1}$ (\ref{U-1SLE})
\begin{eqnarray}
\nonumber
\Delta_{p,q}^{\kappa}&:=&{\mathcal U}_{\kappa}^{-1}\left(\hbar_{p,q}^{\kappa}\right)\\
&=&\frac{\left|4p-\kappa q\right| +\kappa-4}{2\kappa}.
\label{KPZSLEinvh}
\end{eqnarray}
We therefore get, depending on the $\kappa$-range:
\begin{eqnarray}
\Delta_{p,q}^{\kappa}=\left\{\begin{array}{ll} \Delta_{p,q}^{(\gamma)} & \mbox{$\kappa \leq 4$}\\
{\Delta_{q,p}^{(\gamma)}}' & \mbox{$\kappa \geq 4$,}
\end{array}
\right.
\label{DD}
\end{eqnarray}
so $\Delta_{p,q}^{\kappa}$ is either a conformal weight ($\kappa \leq 4$), or a dual one ($\kappa \geq 4$).

\subsection{Scaling Dimensions for Multi-Lines in $O(N)$, Potts Models and SLE Process}

We shall need in the following the scaling dimensions associated with several (mutually-avoiding) random paths starting from a same small 
neighborhood, also called star exponents in the above. It is simplest to first give them for the $O(N)$ model, before 
transferring them to the SLE. For completeness, these exponents are also derived explicitly from the random lattice
approach in appendix \ref{ONapp}, in particular in the case in presence of a boundary (see also refs. \cite{DK,kostovgaudin,KK}).

\subsubsection{Boundary and Bulk Quantum Gravity}
Near the boundary of a random surface with Dirichlet conditions, the conformal dimensions read:
\begin{eqnarray}
\label{tdeltaL}
\tilde\Delta_L&=&\frac{L}{2}(1-\gamma)=\Delta_{L+1,1}^{(\gamma)}\, , \\
\label{tdeltaLD}
\tilde\Delta_L^D&=&\frac{L}{2}+\gamma=\Delta_{1,L+1}^{(\gamma)}\, ,
\end{eqnarray}
 where the ``$D$'' superscript stands for the dense phase. The quantum bulk dimensions read similarly
\begin{eqnarray}
\label{deltaLON}
\Delta_L&=&\frac{L}{4}(1-\gamma)+\frac{\gamma}{2}=\Delta_{L/2,0}^{(\gamma)}\, , \\
\label{deltaLD}
\Delta_L^D&=&\frac{L}{4}+\frac{\gamma}{2}=\Delta_{0,L/2}^{(\gamma)}\, .
\end{eqnarray}
In terms of the SLE parameter, the dilute phase corresponds to (\ref{gk}) for $\kappa \leq 4$, while the dense one covers
(\ref{gkd}) with $\kappa \geq 4$:
\begin{eqnarray}
\label{tdeltaLk}
\tilde\Delta_L&=&\frac{2L}{\kappa},\;\;\;\;\;\;\;\;\;\;\;\;\;\; \Delta_L=\frac{1}{2\kappa}(2L+\kappa-4),\;\;  \kappa \leq 4 \\
\label{tdeltaLDk}
\tilde\Delta_L^D&=&\frac{L}{2}+1-\frac{\kappa}{4},\;\; \Delta_L^D=\frac{1}{8}\left(2L+4-\kappa\right),\;\; \kappa \geq 4.
\end{eqnarray}
By using dual dimensions (\ref{dualdelta}) for the dense phase, these results are unified into
\begin{eqnarray}
\label{tdeltaLkk}
\tilde\Delta_L&=&\frac{2L}{\kappa}=\Delta_{L+1,1}^{\kappa}\, ,\;\;\;\;\;\;\;\;\;\;\;\;\;\;\;\;\;\;\;\;\;\;\;\;\;\;\;\;{\kappa \leq 4} \\
\label{deltaLkk}
\Delta_L&=&\frac{1}{2\kappa}(2L+\kappa-4)=\Delta_{L/2,0}^{\kappa}\, ,\;\;\;\;\;\;\;\;\;\kappa \leq 4 \\
\label{tdeltaLDkd}
{{\tilde{\Delta}}_L}^D{}'&=&\frac{2L}{\kappa}=\Delta_{L+1,1}^{\kappa}\, , \;\;\;\;\;\;\;\;\;\;\;\;\;\;\;\;\;\;\;\;\;\;\;\;\;\;\;\;\kappa \geq 4\\
\label{deltaLDkd}
{\Delta}_L^D{}'&=&\frac{1}{2\kappa}(2L+\kappa-4)=\Delta_{L/2,0}^{\kappa}\, ,\;\;\;\;\;\;\;\;\; \kappa \geq 4.
\end{eqnarray}
Hence we again observe that in the dense phase the dual dimensions play the role of the original ones in the dilute phase.

\subsubsection{Scaling Dimensions in $\mathbb H$ and $\mathbb C$}
The scaling dimensions $\tilde x_L$ in the standard complex half-plane $\mathbb H$, or $x_L$ in the complex
plane $\mathbb C$, can now be obtained
from the quantum gravity ones by the KPZ $U$-map (\ref{KPZS}), or, in the SLE formalism, from the $\mathcal U_\kappa$ (\ref{KPZSLE}) or
$\mathcal V_\kappa$ (\ref{KPZSLEV}) adapted KPZ maps. From the last equations (\ref{tdeltaLkk}) to (\ref{deltaLDkd}),
it is clear that by duality the analytic form of the dimensions stays the same in the two phases $\kappa \leq 4$, and
$\kappa \geq 4$. Indeed we get:
\begin{eqnarray}
\label{txLkk}
\tilde x_L&=&\mathcal U_\kappa(\tilde\Delta_L)=\frac{L}{2\kappa}(2L+4-\kappa)=\hbar_{L+1,1}^{\kappa}\, ,\;\;\;\;\;\;\;\;\;\;\;\;\;\;\;\;\;{\kappa \leq 4} \\
\label{xLkk}
 x_L&=&2\mathcal V_\kappa(\tilde\Delta_L)=
 \frac{1}{8\kappa}\left[4L^2-(4-\kappa)^2\right]=\hbar_{L/2,0}^{\kappa}\, ,\;\;\;\;\;\;\;\;\;\kappa \leq 4 \\
\label{txDkd}
\tilde x_L&=&\mathcal U_\kappa({{\tilde{\Delta}}_L}^D{}')=\frac{L}{2\kappa}(2L+4-\kappa)=\hbar_{L+1,1}^{\kappa}\, , \;\;\;\;\;\;\;\;\;\;\;\;\;\;\;\kappa \geq 4\\
\label{xLDkd}
x_L&=&2\mathcal V_\kappa({{\tilde{\Delta}}_L}^D{}')=\frac{1}{8\kappa}\left[4L^2-(4-\kappa)^2\right]=\hbar_{L/2,0}^{\kappa}
\, ,\;\;\;\;\;\;\; \kappa \geq 4.
\end{eqnarray}
We are now in position to determine the multifractal spectrum associated with the harmonic measure near the
$O(N)$ multi-lines, or, equivalently, near the SLE frontier or special points. It also corresponds to the so-called derivative exponents in the SLE
formalism  \cite{lawler4}.\\

\section{\sc{Multifractal Exponents for the SLE}}
\label{sec.multifSLE}

In sections \ref{sec.conform} and \ref{subsec.geometry} above we have
studied in detail the multifractal spectrum associated with the harmonic measure near a
conformally-invariant frontier, generalized to a mixed rotation spectrum in section \ref{sec.winding}.
We also looked at the double-sided distribution of potential near a simple fractal curve. Further generalizations were given to higher
multifractal spectra in between the branches of stars made of several simple paths (section \ref{sec.higher}).
One should note at this stage that we used there the quantum gravity formalism in terms
of the susceptibility exponent $\gamma$, which is valid
for the ``dilute phase'' of critical curves, i.e., for simple CI curves. We have seen in previous
section \ref{sec.SLEKPZ} how to
extend this formalism to the ``dense phase'', namely to non-simple curves, by using duality.
We shall now apply this extended
formalism to the multifractal spectrum of the SLE trace. It would be tedious to repeat all previous calculations,
so we shall rather concentrate on new extended spectra, and on the basic property of duality (\ref{duali})
(\ref{dualibis}) of the SLE trace, which
plays an essential role in the construction (\ref{finaa}) (\ref{foriginal}) of the
standard multifractal spectrum $f(\alpha)$ along any CI random curve.

\subsection{Boundary Multifractal Exponents}
\subsubsection{Definition}
Let us start with the exponents associated with geometrical properties of CI curves at the boundary
of the half-plane $\mathbb H$. In the SLE language, this corresponds to the {\it chordal} case. We look specifically at
the behavior of powers of the harmonic measure, or in SLE terms, to that of powers of the modulus of the derivative
of the Riemann conformal map which maps the SLE trace back to the half-line $\mathbb R= \partial \mathbb H$
 \cite{lawler4,lawleresi,stflour}.

We shall start with the multifractal exponents associated with the $L$-leg boundary operator
${\tilde\Phi}_{{\mathcal S_L}}$ creating a star made of $L$ semi-infinite
random paths $\tilde {\mathcal S}_1$, diffusing in the upper half-plane $\mathbb H$ and started at
a single vertex on the real line $\partial \mathbb H$ in a
{\it   mutually-avoiding star} configuration
$${\mathcal S}_L=\stackrel{L}{\overbrace{{\mathcal S}_1\wedge{\mathcal
S}_1\wedge\cdots{\mathcal S}_1}}=(\wedge{\mathcal S}_1)^L$$ with $L$ lines started at the same
origin, as seen in sections (\ref{sec.winding}) and (\ref{sec.SLEKPZ}). Its boundary scaling dimension $\tilde x_L$ is given by eqs.
(\ref{txLkk}) or (\ref{txDkd}):
\begin{eqnarray}
\label{txL}
\tilde x({\mathcal S}_L)=\tilde x_L=\frac{L}{2\kappa}(2L+4-\kappa)=\hbar_{L+1,1}^{\kappa}\,\, , {\forall \kappa}
\end{eqnarray}
with the inversion formula:
\begin{eqnarray}
\label{uk-1}
\mathcal U_\kappa^{-1}(\tilde x_L)=L\,\mathcal U_\kappa^{-1}(\tilde x_1)=\frac{2L}{\kappa}=\Delta_{L+1,1}^{\kappa}
\, , {\forall \kappa}.
\end{eqnarray}
As explained above in section \ref{sec.SLEKPZ}, the formalism has now been  set up in such a way that the
formulae stay valid in both phases $\kappa \leq 4$ or $\kappa > 4$.

\begin{figure}[tb]
\begin{center}
\includegraphics[angle=0,width=0.7\linewidth]{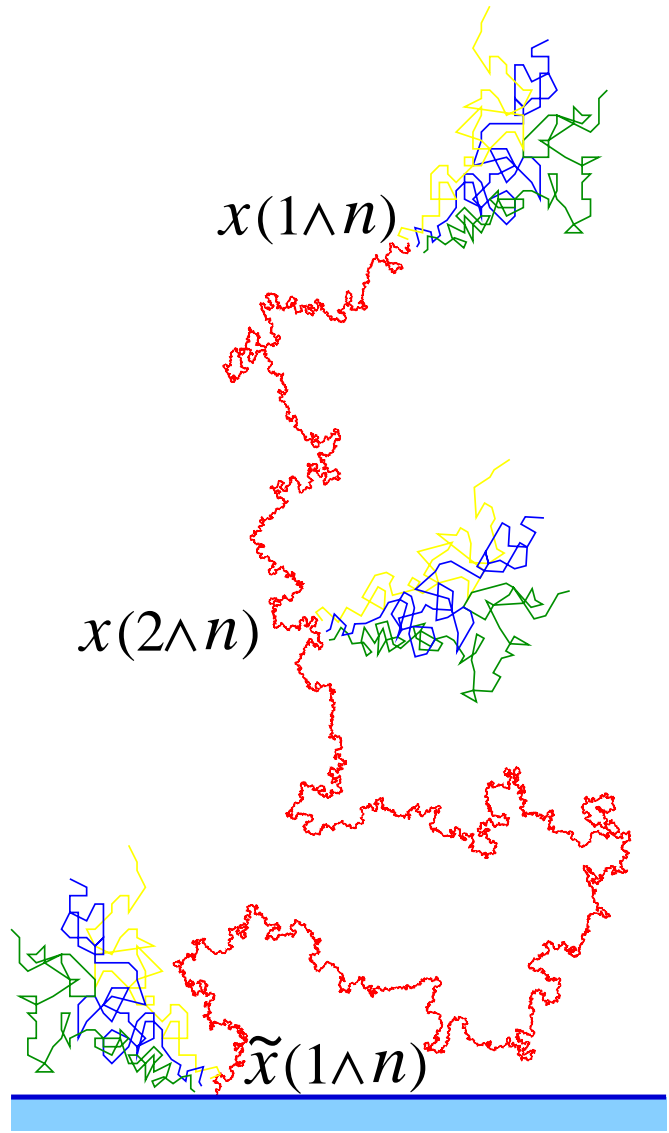}
\end{center}
\caption{Representation of harmonic moments by packets of
independent Brownian paths diffusing away from a single SLE trace, hence a $L=1$ star $\mathcal S_1$.
There are three locations to probe the harmonic measure:
at the SLE origin on the boundary, at the SLE tip in the plane, or along the fractal curve itself.
The corresponding scaling exponents are respectively $\tilde x(1\wedge n)$ (\ref{tildex1nexp}), $x(1\wedge n)$
(\ref{x1nexp}), $x(2\wedge n)$ (\ref{x2nexp}).}
\label{halcbis}
\end{figure}

We now dress this $L$-star ${\mathcal S_L}$ by a packet of $n$ independent Brownian paths diffusing
away from the apex of the star,
located on the boundary, while avoiding the random paths of the star (Fig. \ref{halcbis}).

In our standard notation, this reads:
$${\mathcal S}_L \wedge\{\stackrel{n}{\overbrace{{\mathcal B}\vee{\mathcal B}\vee\cdots{\mathcal B}}}
\} =(\wedge{\mathcal S}_1)^L\wedge (\vee{\mathcal B})^n\equiv L\wedge n.$$
We have, with a slight abuse of notation, introduced the short-hand notation: $L\wedge n:={\mathcal
S}_L \wedge (\vee{\mathcal B})^n$. The corresponding boundary scaling dimension $\tilde x(L\wedge n)$ in $\mathbb H$
is given by repeated application of the boundary KPZ construction (\ref{tildexABk}):
\begin{eqnarray}
\label{tildexLn}
\nonumber
\tilde x(L\wedge n)&=&{\mathcal U}_{\kappa}\left\{ {\mathcal U}_{\kappa}
^{-1}\left( \tilde{x}\left({\mathcal
S}_L\right)\right)
+{\mathcal U}_{\kappa}^{-1}\left[ \tilde{x}\left( (\vee{\mathcal B})^n\right)\right] \right\}\\
\nonumber
&=&{\mathcal U}_{\kappa}\left[ L\,{\mathcal U}_{\kappa}
^{-1}\left( \tilde{x}_1\right)
+{\mathcal U}_{\kappa}^{-1}\left( n\right) \right]\\
&=&{\mathcal U}_{\kappa}\left[ L\,\frac{2}{\kappa}
+{\mathcal U}_{\kappa}^{-1}\left( n\right) \right].
\end{eqnarray}
An explicit calculation with (\ref{USLE}) and (\ref{U-1SLE}) then gives:
\begin{eqnarray}
\label{tildexLnexp}
\tilde x(L\wedge n)=n +\frac{L}{\kappa}\left[L+\frac{1}{2}\sqrt{16\kappa n+(\kappa-4)^2}\right].
\end{eqnarray}
In particular, for the exponent governing the harmonic measure moments near the origin of a single SLE trace
on the boundary
$\partial \mathbb H$ we find:
\begin{eqnarray}
\label{tildex1nexp}
\tilde x(1\wedge n)=n +\frac{1}{\kappa}\left[1+\frac{1}{2}\sqrt{16\kappa n+(\kappa-4)^2}\right].
\end{eqnarray}
\subsubsection{Boundary Derivative Exponents}
It is interesting to isolate in this exponent the contribution $\tilde x_L$ (\ref{txL}) coming from the $L$ random SLE paths, and
which absorbs the non-linearity in $L$, and write:
\begin{eqnarray}
\label{tildexLndec}
\tilde x(L\wedge n)&=&\tilde x_L+n+\frac{L}{2\kappa}\left[\sqrt{16\kappa n+(\kappa-4)^2} +\kappa -4\right]\\
&=&\tilde x_L+n+ L\, {\mathcal U}_{\kappa}^{-1}\left( n\right).
\end{eqnarray}
The structure so obtained is in agreement with the short-distance expansion results (\ref{SDEU}) and (\ref{SDED});
the mutual-avoidance interaction between the random SLE paths and the random Brownian paths
enhances the exponent of independent paths $\tilde x_L+n$ by $L$ times a typical boundary KPZ term.
Let us define the subtracted exponent:
\begin{eqnarray}
\label{tildelambda}
\tilde \lambda_{\kappa}(L\wedge n):=\tilde x(L\wedge n)-\tilde x_L.
\end{eqnarray}
It reads explicitly
\begin{eqnarray}
\label{tildelambdaexp}
\tilde \lambda_{\kappa}(L\wedge n)=n+ L\, {\mathcal U}_{\kappa}^{-1}\left( n\right)=n+ L\, q(\kappa,n),
\end{eqnarray}
with, in the notation of ( \cite{lawler4,lawleresi,stflour}),
\begin{eqnarray}
\label{qkn}
q(\kappa,n)={\mathcal U}_{\kappa}^{-1}\left( n\right)
=\frac{1}{2\kappa}\left[\sqrt{16\kappa n+(\kappa-4)^2} +\kappa -4\right].
\end{eqnarray}

\subsubsection{Boundary Disconnection Exponents}
Notice that for $n=0$ the exponent is not necessarily trivial:
\begin{eqnarray}
\label{tildelambda0}
\nonumber
\tilde \lambda_{\kappa}(L,n=0)&=&\tilde x(L\wedge 0)-\tilde x_L\\
&=& L\, {\mathcal U}_{\kappa}^{-1}\left( 0\right),
\end{eqnarray}
with
\begin{equation}
{\mathcal U}_{\kappa}^{-1}\left( 0\right)=\left(1-\frac{4}{\kappa}\right) \vartheta (\kappa  -4).
\label{U0}
\end{equation}
Hence
\begin{eqnarray}
\tilde \lambda_{\kappa}(L\wedge 0)=\left\{\begin{array}{ll} 0 & \mbox{$\kappa \leq 4$}\\
 L(1-4/\kappa)& \mbox{$\kappa \geq 4$.}
\end{array}
\right.
\label{tildelambdaL0exp}
\end{eqnarray}
So the exponent (\ref{tildelambda0}) takes non-zero values for $\kappa > 4$, i.e. for
self-coiling random CI curves. This is typical of a {\it disconnection exponent}. Consider a point
$z$ located along the boundary $\partial \mathbb H$ at short distance  $r=|z-w|$ from the origin
$w$ where all paths of the SLE star
$\mathcal S_L$ are started. The probability $\tilde{\mathcal P}_{L\wedge 0}$ that point $z$ stays connected to
infinity without being encircled by the collection of
SLE traces scales like
\begin{equation}
\tilde{\mathcal P}_{L\wedge 0}(z) \propto r^{\tilde \lambda_{\kappa}(L\wedge 0)}=r^{\,L(1-{4}/{\kappa})},\; r\to 0, \;\kappa \geq 4.
\label{tildePL0}
\end{equation}
If $\kappa \leq 4$, the probability that the random SLE paths return to the boundary is zero, and any point
$w \ne 0$ stays connected to infinity, hence a vanishing disconnection exponent
$\tilde\lambda_{\kappa \leq 4}(L,0)=0$.

\subsection{Planar Multifractal Exponents}
\subsubsection{Construction from Quantum Gravity}
In this section we deal with exponents similar to the $\tilde x (L\wedge n)$ encountered above,
but now for the harmonic measure near the tip of a collection of $L$ random CI paths in the plane.
In the SLE language, these will give derivative exponents, describing the power law decay of the $n^{\rm th}$ moment
of the modulus of the derivative of the uniformizing Riemann map of the SLE traces to the unit disk $\mathbb U$,
this time for chordal SLEs.

We still use the short-hand $L\wedge n={\mathcal
S}_L \wedge (\vee{\mathcal B})^n$. It then suffices to apply the general composition formalism (\ref{xABk}) in place of (\ref{tildexABk})
 as in (\ref{tildexLn}) above, to get:
\begin{eqnarray}
\label{xLn}
\nonumber
x(L\wedge n)&=&2{\mathcal V}_{\kappa}\left\{ {\mathcal U}_{\kappa}
^{-1}\left( \tilde{x}\left({\mathcal
S}_L\right)\right)
+{\mathcal U}_{\kappa}^{-1}\left[ \tilde{x}\left( (\vee{\mathcal B})^n\right)\right] \right\}\\
\nonumber
&=&2{\mathcal V}_{\kappa}\left[ L\,{\mathcal U}_{\kappa}
^{-1}\left( \tilde{x}_1\right)
+{\mathcal U}_{\kappa}^{-1}\left( n\right) \right]\\
&=&2{\mathcal V}_{\kappa}\left[ L\,\frac{2}{\kappa}
+{\mathcal U}_{\kappa}^{-1}\left( n\right) \right].
\end{eqnarray}

Using (\ref{VSLE}) and (\ref{U-1SLE}), one arrives at the explicit form:
\begin{eqnarray}
\label{xLnexp}
x(L\wedge n)=\frac{n}{2} +B_{\kappa}(L) +A_{\kappa}(L)\frac{1}{\kappa}\sqrt{16\kappa n+(\kappa-4)^2},
\end{eqnarray}
where the various terms read
\begin{eqnarray}
\label{ABL}
B_{\kappa}(L)&=&\frac{L}{4\kappa}(2L+\kappa-4)-\frac{1}{16\kappa}(\kappa-4)^2\\
A_{\kappa}(L)&=&\frac{1}{4}\left(L+ \frac{\kappa}{4}-1 \right).
\end{eqnarray}
Let us specify the first two sets of exponents, corresponding respectively to the tip of the radial SLE ($L=1$), or
to the frontier of the SLE curve ($L=2$):
\begin{eqnarray}
\nonumber
x(1\wedge n)&=&\frac{n}{2} +\frac{\kappa-4+\sqrt{16\kappa n+(\kappa-4)^2}}{16} + \frac{(6-\kappa)(\kappa-2)}{8\kappa},\\
\label{x1nexp}\\
\nonumber
x(2\wedge n)&=&\frac{n}{2} +\left(1+ \frac{\kappa}{4} \right)
\frac{1}{4\kappa}\left[\kappa -4+\sqrt{16\kappa n+(\kappa-4)^2}\right] +1-\frac{\kappa}{8}.\\
\label{x2nexp}
\end{eqnarray}
In each equation, the last term corresponds to the scaling dimension $x_1$ or $x_2$
of the operator $\Phi_{\mathcal S_1}$ or  $\Phi_{\mathcal S_2}$.

\subsubsection{Duality for Multifractal Dimensions}
It is interesting, at this stage, to return at our original developments in terms of
quantum gravity, where the string susceptibility exponent $\gamma$ appeared in a natural way.
Using (\ref{gg'k}), one finds in the dilute phase ($\kappa \leq 4$):
\begin{eqnarray}
\label{xLng}
x(L\wedge n)=\frac{n}{2} +B_{\gamma}(L) +A_{\gamma}(L)\sqrt{4(1-\gamma)n+\gamma^2}
\end{eqnarray}
where the various coefficients read
\begin{eqnarray}
\label{ABLg}
B_{\gamma}(L)&=&\frac{L}{4}\left[\frac{L}{2}(1-\gamma)+\gamma \right] -\frac{1}{4} \frac{\gamma^2}{1-\gamma}\\
A_{\gamma}(L)&=&\frac{1}{4}\left(L+ \frac{\gamma}{1-\gamma} \right).
\end{eqnarray}
In particular for $L=1$ we find:
\begin{equation}
x\left(1\wedge n\right)
=\frac{n}{2}+\frac{1}{8}(1+\gamma)-\frac{1}{4} \frac{\gamma^2}{1-\gamma}+\frac{1}{4}\frac{1}{1-\gamma}
\sqrt{4(1-\gamma)n+{\gamma}^2}\ . \label{x1n}
\end{equation}
For $L=2$ we get:
\begin{equation}
x\left(2\wedge n\right)
=\frac{n}{2}+\frac{1}{2}-\frac{1}{4} \frac{\gamma^2}{1-\gamma}+\frac{1}{4}\frac{2-\gamma}{1-\gamma}
\sqrt{4(1-\gamma)n+{\gamma}^2}\ , \label{x2n}
\end{equation}
which  of course is the same as result (\ref{tauoriginal}) of section 7 above.

For the dense phase, one either uses (\ref{gg'kd}) in (\ref{xLnexp}), or formally substitutes everywhere in (\ref{ABLg})
the dual susceptibility exponent $\gamma'=-\gamma/(1-\gamma)$ (\ref{dualgamma}) to $\gamma$, to get
\begin{eqnarray}
\label{xLng'}
x^{D}(L\wedge n)=\frac{n}{2} +B'_{\gamma}(L) +A'_{\gamma}(L)\sqrt{4(1-\gamma)n+\gamma^2}
\end{eqnarray}
where
\begin{eqnarray}
\label{ABLg'}
B'_{\gamma}(L)&=&B_{\gamma'}(L)=\frac{L}{4}\frac{1}{1-\gamma}\left(\frac{L}{2}-\gamma\right) -\frac{1}{4} \frac{\gamma^2}{1-\gamma}\\
A'_{\gamma}(L)&=&\frac{1}{1-\gamma}A_{\gamma'}(L)=\frac{1}{4}\frac{L-\gamma}{1-\gamma}.
\end{eqnarray}
For the first cases $L=1$ and $L=2$ we respectively find:
\begin{equation}
x^D\left(1\wedge n\right)
=\frac{n}{2}+\frac{1}{4}\frac{1}{1-\gamma}-\frac{1}{4} \frac{\gamma^2}{1-\gamma}
+\frac{1}{4}\sqrt{4(1-\gamma)n+{\gamma}^2}\ , \label{x1nd}
\end{equation}
and
\begin{equation}
x^D\left(2\wedge n\right)
=\frac{n}{2}+\frac{1}{2}-\frac{1}{4} \frac{\gamma^2}{1-\gamma}+\frac{1}{4}\frac{2-\gamma}{1-\gamma}
\sqrt{4(1-\gamma)n+{\gamma}^2}\ .
\label{x2nd}
\end{equation}
We thus observe that the expression for the two-line exponent $x(2\wedge n)$ is invariant between the two phases.
This crucial point explains the fact that when approaching a CI fractal path from outside and converging towards a typical point
associated with the two-line operator $\Phi_{\mathcal S_2}$, one sees only the {\it external frontier},
since all the multifractal exponents (\ref{x2n}) and (\ref{x2nd}) are identical. This identity expresses the screening of
electrostatic interactions by the external perimeter of non-simple paths.

Another way to express this situation is to require the duality identity in the SLE result (\ref{xLnexp})
\begin{eqnarray}
\label{xLnduality}
x_{\kappa}(L\wedge n)=x_{16/\kappa}(L\wedge n)
\end{eqnarray}
which is equivalent to
\begin{eqnarray}
\label{ABLduality}
\nonumber
B_{\kappa}(L)&=&B_{\kappa'=16/\kappa}(L)\\
\nonumber
A_{\kappa}(L)&=&\frac{4}{\kappa}A_{\kappa'=16/\kappa}(L).
\end{eqnarray}
One finds from (\ref{ABL}) a unique solution for $L=2$, which is the geometric duality of frontiers,
already seen in sections (\ref{sec.geodual}) and (\ref{sec.duality}).
 It ascribes the simple path
${\rm SLE}_{\kappa'=16/\kappa}$ as the external frontier of ${\rm SLE}_{\kappa \leq 4}$, which is a non-simple path.

\subsubsection{Planar Derivative Exponents}
As a follow up of eqs. (\ref{x1nexp}) and (\ref{x2nexp}), it is interesting to
separate in (\ref{xLnexp}) the contribution $x_L$ of the star $\mathcal S_L$ itself
\begin{eqnarray}
\label{xL}
x({\mathcal S}_L)= x_L=
 \frac{1}{8\kappa}\left[4L^2-(4-\kappa)^2\right]=\hbar_{L/2,0}^{\kappa}\, ,
\end{eqnarray}
and to write
\begin{eqnarray}
\label{xLndec}
x(L\wedge n)=x_L+\frac{n}{2} +\frac{1}{2}\left(L+ \frac{\kappa}{4}-1 \right)
{\mathcal U}_{\kappa}^{-1}\left( n\right),
\end{eqnarray}
where
\begin{eqnarray}
\label{U-1bis}
{\mathcal U}_{\kappa}^{-1}\left( n\right)=\frac{1}{2\kappa}\left[\sqrt{16\kappa n+(\kappa-4)^2} +\kappa -4\right].
\end{eqnarray}
In analogy to eq. (\ref{tildelambda}), one is thus led to define a subtracted exponent
\begin{eqnarray}
\label{lambda}
\lambda_{\kappa}(L\wedge n):= x(L\wedge n)-x_L.
\end{eqnarray}
Owing to the results above, this set of $\lambda$-exponents reads explicitly:
\begin{eqnarray}
\label{lambdaexp}
\nonumber
\lambda_{\kappa}(L\wedge n)&=&\frac{n}{2}
+\frac{1}{2}\left(L+ \frac{\kappa}{4}-1 \right){\mathcal U}_{\kappa}^{-1}\left( n\right)\\
&=&\frac{n}{2} +\frac{1}{2}\left(L+ \frac{\kappa}{4}-1 \right)
\frac{1}{2\kappa}\left[\sqrt{16\kappa n+(\kappa-4)^2} +\kappa -4\right].
\end{eqnarray}
These exponents, directly related to the harmonic measure moment exponents (\ref{xLn}),
 generalize the so-called {\it derivative exponents} introduced in  \cite{lawler4}. Indeed, for $L=1$, we find:
\begin{eqnarray}
\label{lambda1}
\lambda_{\kappa}(L=1,n)=\frac{n}{2} +\frac{1}{16}\left[\sqrt{16\kappa n+(\kappa-4)^2} +\kappa -4\right],
\end{eqnarray}
in agreement with ( \cite{lawler4,lawleresi,stflour}). This particular exponent describes the conformal behavior of the growth process near the tip of the SLE,
or, equivalently, that of the harmonic measure seen from that tip. The class (\ref{lambdaexp})
generalizes it, for $L=2$, to the case of a typical point along the frontier, and to higher branching points for $L > 2$.

\subsubsection{Planar Disconnection Exponents}
Here again, the geometrical situation will strongly depend on the range of values for $\kappa$,
the fractal sets being directly accessible or not, respectively for $\kappa \leq 4$ and $\kappa \geq 4$. Indeed, let us consider
the values at $n=0$, i.e., the set of disconnection exponents:
\begin{eqnarray}
\label{lambdaL0}
\nonumber
\lambda_{\kappa}(L,n=0)&=&\frac{1}{2}\left(L+ \frac{\kappa}{4}-1 \right)
{\mathcal U}_{\kappa}^{-1}\left( 0\right)\\
&=&\frac{1}{2}\left(L+ \frac{\kappa}{4}-1 \right)\left(1-\frac{4}{\kappa}\right) \vartheta (\kappa  -4).
\end{eqnarray}
This explicitly gives
\begin{eqnarray}
\lambda_{\kappa}(L\wedge 0)=\left\{\begin{array}{ll} 0 & \mbox{$\kappa \leq 4$}\\
 \frac{1}{2}(L-1)(1-\frac{4}{\kappa})+\frac{1}{8}{(\kappa-4)}& \mbox{$\kappa \geq 4$.}
\end{array}
\right.
\label{lambdaL0exp}
\end{eqnarray}
For $L=1$, we find the disconnection exponent associated with the tip of the radial SLE, or,
equivalently, with the tip of a single line in the $O(N)$ model:
\begin{eqnarray}
\lambda_{\kappa}(1,0)=\left\{\begin{array}{ll} 0 & \mbox{$\kappa \leq 4$}\\
 {(\kappa-4)}/{8}& \mbox{$\kappa \geq 4$,}
\end{array}
\right.
\label{lambda10}
\end{eqnarray}
a result also appearing in  \cite{lawler4,RS}.

Consider a point
$z \in \mathbb C$ located  at distance  $r=|z-w|$ from the origin $w$ where all paths of the SLE star
$\mathcal S_L$ begin. The probability $\mathcal P_{L\wedge 0}$ that the point $z$ stays connected to
infinity without being encircled by the collection of
SLE traces scales like
\begin{equation}
\mathcal P_{L\wedge 0}(z) \propto r^{\lambda_{\kappa}(L\wedge 0)}, \; r\to 0.
\label{PL0}
\end{equation}
If $\kappa \leq 4$, the random SLE paths are simple curves which cannot encircle any exterior point
which thus stays connected to infinity, hence a vanishing disconnection exponent
$\lambda_{\kappa \leq 4}(L,0)=0$.

\subsection{Double-Sided Exponents}
\subsubsection{Definition}
For completeness, let us give the expression for the double-sided exponents corresponding to the
double moments of the harmonic measure on both sides of an SLE trace, or, equivalently, to
double-sided derivative exponents  \cite{lawler4}. For the sake of generality, we shall treat them at level $L$,
and specify them for $L=1,2$. We thus have in mind  the configuration where two packets
of $n_1$ and $n_2$ Brownian paths diffuse on both sides of an $\mathcal S_L$ multiple SLE trace, hence
$$\{\stackrel{n_1}{\overbrace{{\mathcal B}\vee{\mathcal B}\vee\cdots{\mathcal B}}}\}\wedge
{\mathcal S}_L \wedge\{\stackrel{n_2}{\overbrace{{\mathcal B}\vee{\mathcal B}\vee\cdots{\mathcal B}}}
\} =(\vee{\mathcal B})^{n_1}\wedge(\wedge{\mathcal S}_1)^L\wedge (\vee{\mathcal B})^{n_2} = n_1\wedge L\wedge n_2.$$
We have again, with a slight abuse of notation, introduced a short-hand notation:
$n_1\wedge L\wedge n_2 :=(\vee{\mathcal B})^{n_1}\wedge{\mathcal
S}_L \wedge (\vee{\mathcal B})^{n_2}$.

\subsubsection{Boundary Exponents}
Let us start with the half-plane configuration, where all paths start at the same origin $w$ on the half-plane boundary
$\partial \mathbb H$. According to the composition rules (\ref{tildexABk}), the associated boundary scaling exponent can
 be constructed immediately as:
\begin{eqnarray}
\label{tildexnLn}
\nonumber
\tilde x(n_1\wedge L\wedge n_2)
\nonumber
&=&{\mathcal U}_{\kappa}\left[{\mathcal U}_{\kappa}^{-1}\left( n_1\right)+ L\,{\mathcal U}_{\kappa}
^{-1}\left( \tilde{x}_1\right)
+{\mathcal U}_{\kappa}^{-1}\left( n_2\right) \right]\\
&=&{\mathcal U}_{\kappa}\left[{\mathcal U}_{\kappa}^{-1}\left( n_1\right)+ \frac{2L}{\kappa }
+{\mathcal U}_{\kappa}^{-1}\left( n_2\right) \right].
\end{eqnarray}
The calculation gives:
\begin{eqnarray}
\label{tildexnLnexp}
\nonumber
&&\tilde x(n_1\wedge L\wedge n_2)  \\
\nonumber
&=&\frac{1}{16\kappa}\left\{\left[4(L-1)+\kappa +
\sqrt{16\kappa n_1+(\kappa-4)^2}
+\sqrt{16\kappa n_2+(\kappa-4)^2}\right]^2-(\kappa -4)^2\right\}
\end{eqnarray}
This is the full boundary  scaling dimension of $L$-SLE paths dressed by $n_1$ and $n_2$ Brownian paths. It is
interesting to compare it to the dimension $\tilde x_L$ (\ref{txLkk}) or (\ref{txDkd}) of the $L$-SLEs alone, and define:
\begin{eqnarray}
\label{tildelamnLn}
\tilde \lambda_{\kappa}(n_1\wedge L\wedge n_2):=\tilde x(n_1\wedge L\wedge n_2)-\tilde x_L.
\end{eqnarray}
We get for this exponent:
\begin{eqnarray}
\label{tildelambdnLnexp}
\nonumber
&&\tilde \lambda_{\kappa}(n_1\wedge L\wedge n_2)  \\
\nonumber
&=&\frac{1}{16\kappa}\left\{\left[4L+\kappa-4 +
\sqrt{16\kappa n_1+(\kappa-4)^2}
+\sqrt{16\kappa n_2+(\kappa-4)^2}\right]^2-(4L+4-\kappa)^2\right\}
\end{eqnarray}
For $L=1$, we find:
\begin{eqnarray}
\label{tildelambn1nexp}
\nonumber
&&\tilde \lambda_{\kappa}(n_1\wedge 1\wedge n_2)  \\
\nonumber
&=&\frac{1}{16\kappa}\left\{\left[\kappa +
\sqrt{16\kappa n_1+(\kappa-4)^2}
+\sqrt{16\kappa n_2+(\kappa-4)^2}\right]^2-(8-\kappa)^2\right\},
\end{eqnarray}
also obtained in  \cite{lawler4}. 

\subsubsection{Double-Sided Boundary Disconnection Exponents}
The values of the generalized disconnection exponents associated with the exponents above are obtained for $n_1=n_2=0$ as:
\begin{eqnarray}
\label{tildex0L0}
\tilde x(0\wedge L\wedge 0)={\mathcal U}_{\kappa}\left[ \frac{2L}{\kappa }
+2\,{\mathcal U}_{\kappa}^{-1}\left( 0\right) \right].
\end{eqnarray}
For the $\lambda$-exponents (\ref{tildelamnLn}) this gives $\tilde \lambda_{\kappa}(0\wedge L\wedge 0)$
 explicitly, with the help of (\ref{U0})
${\mathcal U}_{\kappa}^{-1}\left( 0\right)=\left(1-\frac{4}{\kappa}\right) \vartheta (\kappa  -4)$:
\begin{eqnarray}
\label{tildelamb0L0exp}
\nonumber
\tilde \lambda_{\kappa}(0\wedge L\wedge 0)&=& \tilde x(0\wedge L\wedge 0)-\tilde x_L \\
\nonumber
&=&\frac{1}{2}\left(4L+\kappa-4 \right)\left(1-\frac{4}{\kappa}\right) \vartheta (\kappa-4).
\end{eqnarray}
We therefore find as usual a quite different situation for $\kappa \leq 4$ and $\kappa \geq 4$:
\begin{eqnarray}
\tilde \lambda_{\kappa}(0\wedge L\wedge 0)=\left\{\begin{array}{ll} 0 & \mbox{$\kappa \leq 4$}\\
 \frac{1}{2}\left(4L+\kappa-4 \right)(1-\frac{4}{\kappa}) & \mbox{$\kappa \geq 4$.}
\end{array}
\right.
\label{tildelambda0L0disp}
\end{eqnarray}
Notice that the bulk one-sided disconnection exponent (\ref{lambdaL0}) is related to the double-sided boundary one by 
the identity
\begin{equation}
\lambda_{\kappa}(L\wedge 0)=\frac{1}{4}\tilde \lambda_{\kappa}(0\wedge L\wedge 0).
\end{equation}
\subsubsection{Planar Double-Sided Exponents}
This time we consider the case where the dressed configuration $n_1\wedge n_2$ is located away from the boundary, 
namely in the plane. The associated dimension is obtained from (\ref{xABk}) as:

\begin{eqnarray}
\label{xnLn}
\nonumber
x(n_1\wedge L\wedge n_2)
\nonumber
&=&2{\mathcal V}_{\kappa}\left[{\mathcal U}_{\kappa}^{-1}\left( n_1\right)+ L\,{\mathcal U}_{\kappa}
^{-1}\left( \tilde{x}_1\right)
+{\mathcal U}_{\kappa}^{-1}\left( n_2\right) \right]\\
&=&2{\mathcal V}_{\kappa}\left[{\mathcal U}_{\kappa}^{-1}\left( n_1\right)+ \frac{2L}{\kappa}
+{\mathcal U}_{\kappa}^{-1}\left( n_2\right) \right].
\end{eqnarray}

From (\ref{VSLE}) we find:
\begin{eqnarray}
\label{xnLnexp}
\nonumber
&& x(n_1\wedge L\wedge n_2)  \\
\nonumber
&=&\frac{1}{8\kappa}\left\{\left[2L+\kappa -4+\frac{1}{2}
\sqrt{16\kappa n_1+(\kappa-4)^2}
+\frac{1}{2}\sqrt{16\kappa n_2+(\kappa-4)^2}\right]^2-(\kappa -4)^2\right\}.
\end{eqnarray}
The subtracted $\lambda$-exponents read accordingly
\begin{eqnarray}
\label{lamnLn}
\lambda_{\kappa}(n_1\wedge L\wedge n_2):= x(n_1\wedge L\wedge n_2)- x_L,
\end{eqnarray}
and we get explicitly
\begin{eqnarray}
\label{lambdanLnexp}
\nonumber
&&\lambda_{\kappa}(n_1\wedge L\wedge n_2)  \\
\nonumber
&=&\frac{1}{8\kappa}\left\{\left[2L+\kappa-4 +
\frac{1}{2}\sqrt{16\kappa n_1+(\kappa-4)^2}
+\frac{1}{2}\sqrt{16\kappa n_2+(\kappa-4)^2}\right]^2-4L^2\right\}
\end{eqnarray}

\subsubsection{Double-Sided Bulk Disconnection Exponents}
Generalized disconnection exponents associated with the exponents above are obtained for $n_1=n_2=0$ as:
\begin{eqnarray}
\label{x0L0}
x(0\wedge L\wedge 0)=2{\mathcal V}_{\kappa}\left[ \frac{2L}{\kappa}
+2\,{\mathcal U}_{\kappa}^{-1}\left( 0\right) \right].
\end{eqnarray}
For the $\lambda$-exponent (\ref{lamnLn}) this gives explicitly
\begin{eqnarray}
\label{lamb0L0exp}
\nonumber
\lambda_{\kappa}(0\wedge L\wedge 0)&=&x(0\wedge L\wedge 0)-\tilde x_L \\
\nonumber
&=&\frac{1}{2}\left(2L+\kappa-4 \right)\left(1-\frac{4}{\kappa}\right) \vartheta (\kappa-4)\\
&=&\left\{\begin{array}{ll} 0 & \mbox{$\kappa \leq 4$}\\
 \frac{1}{2}\left(2L+\kappa-4 \right)(1-\frac{4}{\kappa}) & \mbox{$\kappa \geq 4$.}
\end{array}
\right.
\end{eqnarray}
\\
\subsection{Winding Angle Variance of Multiple SLE Strands}
Let us finally return to the winding angle variance at points where $k$ strands come together in a star configuration
$\mathcal S_k$. We have  seen in \S  \ref{sec.winding} that the variance of $k$ paths up to distance $R$ is
reduced by a factor $1/k^2$ with respect to the $k=1$ single path case, namely:
\begin{equation}
\langle \vartheta^2 \rangle_k=\frac{\kappa}{k^2}\, \ln R.
\label{variance-k}
\end{equation}
In the case of {\it non-simple} paths ($\kappa >4$), one can further consider the winding at points where $k$ strands
meet together, amongst which $j$
 adjacent pairs (with $2j\leq k$) are conditioned not to hit each other \cite{wilson}. In each pair the two strands,
 which otherwise would bounce on each other, are disconnected from each other, and that corresponds, in our notations, to
 a star configuration:
\begin{equation}
{\mathcal
S}_{k,j}=\stackrel{k-2j}{\overbrace{{\mathcal S}_1\wedge{\mathcal
S}_1\wedge\cdots{\mathcal S}_1}}\wedge\stackrel{j}{\overbrace{({\mathcal S}_1\wedge 0\wedge{\mathcal S}_1)
\wedge\cdots \wedge({\mathcal S}_1\wedge 0\wedge{\mathcal S}_1)}}.
\label{configuration}
\end{equation}
 Wieland and Wilson  made
 the interesting conjecture that
 in this case the winding angle variance grows like \cite{wilson}
\begin{equation}
\langle \vartheta^2 \rangle_{k,j}=\frac{\kappa}{(k+j\,{\rm max}(0,\kappa/2-2))^2}\, \ln R.
\label{variance-kj}
\end{equation}
This can be derived from the quantum gravity formalism as follows.
A generalization of eq. (\ref{kn}) gives the  number of paths, $k(j)$, which is equivalent to $k$ strands in a star configuration
${\mathcal
S}_{k,j}$ (\ref{configuration}), as
\begin{eqnarray}
\label{kj}
k(j)&=& k+j\frac{{\mathcal U}_{\kappa}^{-1}\left( 0\right)}{{\mathcal U}_{\kappa}^{-1}\left( \tilde{x_1}\right)}.
\end{eqnarray}
Indeed, one simply has to gauge the  extra (quantum gravity) conformal weight $j \times {\mathcal U}_{\kappa}^{-1}\left( 0\right)$,
associated with the $j$ disconnected pairs, by the (QG) boundary conformal weight
${\mathcal U}_{\kappa}^{-1}\left( \tilde{x_1}\right)$ of a single path extremity. This is entirely analogous, for instance,
 to
the way the argument of ${\mathcal V}_{\kappa}$ in the double-sided disconnection exponent, eq. (\ref{x0L0}), was constructed.
Because of the value (\ref{U01})
\begin{equation*}
{\mathcal U}_{\kappa}^{-1}\left( 0\right)
=\left(1-\frac{4}{\kappa}\right) \vartheta (\kappa  -4),
\end{equation*}
and the value (\ref{uk-1})
\begin{equation*}
{\mathcal U}_{\kappa}^{-1}\left( \tilde{x_1}\right)=\frac{2}{\kappa},
\end{equation*}
we find
\begin{eqnarray}
\label{kj1}
k(j)&=& k+j\left(\frac{\kappa}{2}-2\right) \vartheta (\kappa  -4),
\end{eqnarray}
which gives a variance
\begin{equation}
\langle \vartheta^2 \rangle_{k,j}=\langle \vartheta^2 \rangle_{k(j)}=\frac{\kappa}{k^2(j)}\, \ln R,
\label{variance-kjk}
\end{equation}
which is just the conjecture (\ref{variance-kj}), {\bf QED}.
\\

\section*{{Acknowledgements}}

It is a pleasure to thank Michael Aizenman for his collaboration on path-crossing exponents in percolation
(section \ref{sec.perco}) after a seminal discussion together with Bob Langlands, and for
many enjoyable and fruitful discussions over time;
Amnon Aharony for the same shared collaboration;
Ilia A.  Binder for his collaboration on the mixed multifractal rotation spectrum in section \ref{sec.winding};
 Peter Jones and Beno\^{\i}t Mandelbrot for invitations to Yale and to the Mittag-Leffler Institute,
and many stimulating discussions; Emmanuel Guitter for generously preparing the figures; Jean-Marc Luck for 
extensive help with \LaTeX; 
Vincent Pasquier for friendly and interesting discussions; David A. Kosower for a
careful reading of the manuscript; and last but not least, my friend Thomas C. Halsey for many ``multifractal''
discussions over time, and a
careful reading of the manuscript.
\\

\begin{appendix}
\section{\sc{Brownian Intersection Exponents from Quantum Gravity}}
\label{Brownapp}

\subsection{Non-Intersection Exponents and KPZ}

Consider a number $L$ of independent random
walks (or Brownian paths) $B^{(l)},l=1,\cdots,L$ in $\mathbb Z^{d}$ (or $\mathbb R^{d}),$
started at
fixed neighboring points, and the probability $$P_{L}\left( t\right) =P\left\{
\cup^{L}_{l, l'=1} (B^{(l)}\lbrack 0,t\rbrack \cap B^{(l')}\lbrack 0,t\rbrack)
=\emptyset
\right\}$$ that their paths do not intersect up to time $t$. At
large times and for $d<4,$ one expects this probability to decay as $P_{L}\left( t\right) \sim t^{-\zeta_{L}}$, where $\zeta _{L}\left( d\right) $
is a {\it universal} exponent depending 
only on $L$ and $d$. 

For $L$ walks with Dirichlet boundary conditions in $\mathbb H$,
and started at neighboring points near the boundary, the
non-intersection
probability $\tilde{{P}_{L}}\left( t\right) $ scales as $\tilde{P}
_{L}\left( t\right) \sim t^{-\tilde{\zeta}_{L}}$, with a boundary exponent $\tilde \zeta_L$.
In two dimensions, the exponent values are
\begin{equation}
\zeta _{L}=h_{0, L}^{\left( c=0\right) }=\frac{1}{24}\left( 4L^{2}-1\right),
\label{AZeta}
\end{equation}
and for the half-plane
\begin{equation}
2\tilde{\zeta}_{L}=h_{1, 2L+2}^{\left( c=0\right) }=\frac{1}{3}L\left(
1+2L\right), 
\label{AC2}
\end{equation}
where $h_{p,q}^{(c)}$ denotes the Ka\v {c} conformal weight
\begin{equation}
h_{p,q}^{(c)}=\frac{\left[ (m+1)p-mq\right] ^{2}-1}{4m\left( m+1\right) },
\label{AK}
\end{equation}
of a minimal conformal field theory of central charge $c=1-6/m\left(
m+1\right) ,$ $m\in \mathbb N^{*}$  \cite{friedan}. For Brownian motions $
c=0,$ and $m=2.$

This appendix provides the main lines of a derivation of these exponents. One considers the
random walks on a random lattice with planar geometry, or, in other words,
in presence of two-dimensional {\it quantum gravity}  \cite{KPZ}.
There, the conformal dimensions of non-intersecting walks are obtained
from an exact solution. We then use the non-linear KPZ map which exists
between conformal weights $\Delta$ on a random surface and $\Delta^{\left( 0\right)}$ in the plane (\ref{KPZg}),
\begin{equation}
\label{AKPZg}
\Delta ^{\left( 0\right) }=U_{\gamma}(\Delta)=\Delta \frac{\left(
\Delta -\gamma \right) }{(1-\gamma)},
\end{equation}
where $\gamma$, the {\it string susceptibility exponent}, is related to the central charge:
\begin{equation}
\label{Ac(g)}
c=1-6\gamma^{2}/\left(1-\gamma\right);
\end{equation}
for a minimal model of the series~(\ref{AK}), $\gamma=-1/m$,
and with $\Delta^{\left( 0\right) }\equiv
h_{p,q}^{\left( c\right) }.$ For Brownian motions $\gamma=-1/2$,
and the KPZ relation becomes
\begin{equation}
\Delta ^{\left( 0\right) }=U_{(\gamma=-1/2)}(\Delta)=\frac{1}{3}\Delta \left(
1+2\Delta
\right),
\label{AKPZ}
\end{equation}
which indeed bears a striking resemblance to (\ref{AC2}). Eqs. (\ref{AZeta}) and (\ref{AC2}) then correspond to
\begin{eqnarray}
\label{ADelta}
\left\{\begin{array}{ll} \Delta_L=\frac{1}{2}\left(L-\frac{1}{2}\right), & \mbox{$\zeta_L=U_{(\gamma=-1/2)}(\Delta_L)$}\\
\\
\tilde\Delta_L=L, & \mbox{$2\tilde \zeta_L=U_{(\gamma=-1/2)}(\tilde\Delta_L).$}
\end{array}
\right.
\end{eqnarray}

\subsection{Planar Random Graphs}
Consider the set of planar random graphs $G$, built up with,
e.g., trivalent vertices and with a fixed topology, here that of a sphere $%
\left( {\mathcal S}\right) $ or a disk $\left( \mathcal {D}\right) $.
The partition function is
defined as 
\begin{equation}
Z_{}(\beta,\chi )=\sum _G{\frac{1}{ S(G)}}e^{-\beta
\left| G\right| }, 
\label{AZchi1}
\end{equation}
where $\chi $ denotes the Euler characteristic $\chi=2\left( {\mathcal S}\right)
,1\left( {\mathcal D}\right);\left| G\right|$ is the number of vertices of $G$, 
$S\left( G\right) $ its symmetry factor. The partition sum converges for all values
of the parameter $\beta $ larger than some critical $\beta_c$. At $\beta
\rightarrow \beta_c^{+},$ a singularity appears due to the presence of
infinite graphs in (\ref{AZchi1}) 
\begin{equation}
Z\left( \beta, \chi \right) \sim \left( \beta -\beta_c\right) ^{2-\gamma
_{\rm str}(\chi)},
\label{AZchi2}
\end{equation}
where $\gamma _{\rm str}(\chi)$ is the string susceptibility exponent. For pure
gravity as described in (\ref{AZchi1}), with central charge $c=0,$ one has $\gamma _{\rm
str}(\chi)=2-\frac{5}{4}\chi $ \cite{kostov}.

The two-puncture partition function will play an important role. It is defined as
\begin{eqnarray}
\label{AZsec}
Z\lbrack \hbox to 8.5mm{\hskip 0.5mm
$\vcenter{\epsfysize=.45truecm\epsfbox{fig2.eps}}$
              \hskip 10mm}
 \rbrack:=\frac{\partial^{2}}{\partial \beta ^{2}}
Z(\beta, \chi)=\sum _{G({\chi})}\frac{1}{ S(G)} \left| G\right|^2 e^{-\beta
\left| G\right|}.
\label{AZfig}
\end{eqnarray}
It scales as:
\begin{equation}
\frac{\partial^{2}}{\partial \beta ^{2}}
Z(\beta, \chi) \sim \left( \beta -\beta_c\right) ^{-\gamma_{\rm str}(\chi)}.
\label{AZchi"}
\end{equation}
\begin{figure}[tb]
\begin{center}
\includegraphics[angle=0,width=.4\linewidth]{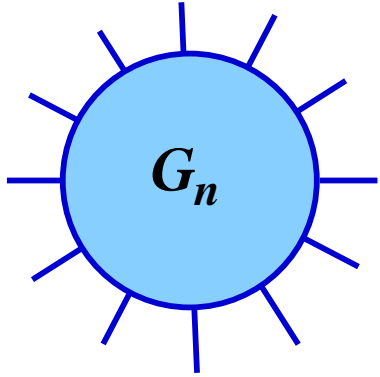}
\end{center}
\caption{A planar random disk with $n$ external legs.}
\label{Figuredisc1}
\end{figure}
The restricted partition function of a planar random
graph with the topology of a disk and a fixed number $n$ of external
vertices (\ref{Figuredisc1}),
\begin{equation}
G_n(\beta ) = \sum _{n-{\rm leg}\ {\rm planar}\ G} e^{-\beta \left| G\right| },
\label{AGn}
\end{equation}
 can be calculated through the large$-{\mathcal N}$ limit of a random ${\mathcal N} \times {\mathcal N}$
matrix integral  \cite{brezin}. It has an integral representation 
\begin{equation}
G_{n}\left( \beta \right) =\int^{b}_{a}d\lambda \, \rho\left( \lambda \right)
{\lambda }^n,
\label{AGnint}
\end{equation}
where $\rho \left( \lambda \right) $ is the spectral eigenvalue density of
the random matrix, for which the explicit expression is known as a function of 
$\lambda ,\beta $ 
 \cite{brezin}. The support $[a, b]$ of the spectral density
depends on $\beta $.

\subsection{Random Walks on Random Graphs}
\subsubsection{Representation by Trees}
Imagine putting a set of $L$ random walks $B^{\left( l\right) },l=1,...,L$
on the {\it random graph} $G$ with the
special constraint that they start at the same vertex $i\in G,$ end at the same
vertex $j \in G$, and have no intersections in between. Consider the set $%
B ^{\left( l\right) }\left[ i,j\right] $ of the points visited on the
random graph by a given walk $B^{\left( l\right) }$ between $i$ and $j$%
, and for each site $k \in B^{\left( l\right) }\left[ i,j\right] $ the
first entry, i.e., the edge of $G$ along which the walk $\left( l\right) $
reached $k$ for the first time. The union of these edges form a tree $%
T^{\left( l\right)}_{i,j}$ spanning all the sites of $B^{\left(
l\right) }\left[ i,j\right]$, called the forward tree. An important
property is that the measure on all the trees spanning a given set of points
visited by a RW is {\it uniform}  \cite{aldous}. This means that we can also
represent the path of a RW by its spanning tree taken with uniform
probability. Furthermore, the non-intersection property of the walks is by
definition equivalent to that of their spanning trees.

\subsubsection{Bulk tree partition function}
\begin{figure}[tb]
\begin{center}
\includegraphics[angle=0,width=.5\linewidth]{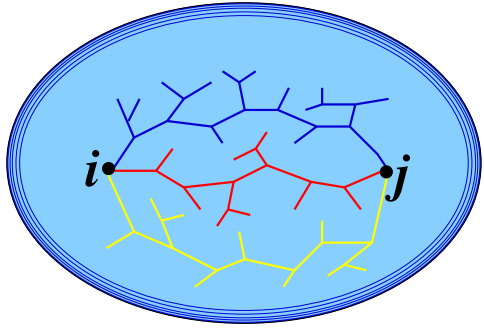}
\end{center}
\caption{$L=3$ mutually-avoiding random trees on a random sphere.}
\label{Fig.wmtree1}
\end{figure}

One
introduces the $L-$tree partition function on the random lattice (Fig. \ref{Fig.wmtree1})
\begin{equation}
Z_L(\beta ,z)=\sum_{{\rm planar}\ G}{\frac{1}{ S(G)}}
e^{-\beta \left| G\right| }\sum _{i,j\in G}
\sum_{\scriptstyle T^{(l)}_{ij}\atop\scriptstyle l=1,...,L}
z^{\left| T\right| },
\label{AZl}
\end{equation}
where $\left\{ T_{ij}^{\left( l\right) }, l=1,\cdots, L \right\}$ is a set of $L$ 
trees,
all constrained to have sites $i$ and $j$ as end-points, and without mutual 
intersections; a fugacity $z$ is in addition associated with the total
number $\left| T\right| =\left| \cup^{L}_{l=1} T^{(l)}\right| $
of vertices of the trees. In principle, the trees spanning the RW
paths can have divalent or trivalent vertices on $G$, but this is immaterial
to the critical behavior, as is the choice of purely trivalent graphs $G$,
so we restrict ourselves here to trivalent trees.

\subsubsection{Boundary Partition Functions}
\begin{figure}[tb]
\begin{center}
\includegraphics[angle=0,width=.5\linewidth]{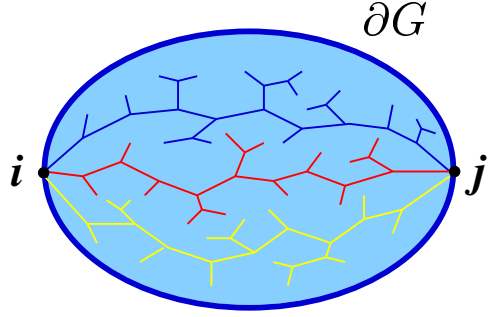}
\end{center}
\caption{$L=3$ mutually-avoiding random trees traversing a random disk.}
\label{Fig.wmtree2}
\end{figure}
We generalize this to the {\it boundary} case where $G$ now has the topology of
a disk and where the trees connect two sites $i$ and $j$ on the
boundary $\partial G$ (Fig. \ref{Fig.wmtree2})%
\begin{equation}
\tilde Z_{L}(\beta , z, \tilde z)=\sum _{ {\rm disk}\ G}
e^{-\beta \left| G\right| }\tilde z^{\left| \partial G\right|}
\sum _{{i,j} \in \partial G}\sum_{\scriptstyle
T^{(l)}_{ij}\atop\scriptstyle l=1,...,L}
z^{\left| T\right| },
\label{AZtilde}
\end{equation}
where $\tilde {z}$ is the fugacity associated with the boundary's length.

The partition function of the disk with two boundary punctures is defined as
\begin{eqnarray}
\label{AZtilde0}
Z( \hbox to 9.5mm{\hskip 0.5mm
$\vcenter{\epsfysize=.45truecm\epsfbox{fig3.eps}}$
              \hskip -80mm}
)&=&\sum _{ {\rm disk}\ G}
e^{-\beta \left| G\right| }\tilde z^{\left| \partial G\right|}\left| \partial G\right|^2\\
\nonumber
&=&\tilde Z_{L=0}(\beta , \tilde z),
\end{eqnarray}
and formally corresponds to the $L=0$ case of the $L$-tree boundary partition functions (\ref{AZtilde}).

\subsubsection{Integral representation}
The partition function (\ref{AZl})
 has been calculated exactly in previous work  \cite{DK}, while (\ref{AZtilde}) was
 first considered in \cite{duplantier7}. The twofold
grand canonical partition function is calculated first by summing over the 
abstract tree
configurations, and then gluing patches of random lattices in between these
trees. A tree generating function is defined as $T(x)
=\sum_{n\geq 1}x^{n}T_{n},$ where $T_{1}\equiv 1
$ and $T_{n}$ is the number of {\it rooted} planar trees with $n$ external
vertices (excluding the root). It reads
 \cite{DK}
\begin{equation}
T\left( x\right) =\frac{1}{2}(1-\sqrt{1-4x}).
\label{AT}
\end{equation}
The result for (\ref{AZl}) is then given by a multiple integral: 
\begin{equation}
Z_{L}(\beta ,z)=\int_{a}^{b}\prod ^{L}_{l=1} 
d\lambda_{l}\, \rho(\lambda_l)
\prod ^{L}_{l=1}{\mathcal T}(z\lambda _l,z\lambda _{l+1}),
\label{AZlint}
\end{equation}
with the cyclic condition  $\lambda _{L+1} \equiv \lambda _{1}$. The geometrical 
interpretation is quite clear (fig. 1).
Each patch $l=1,\cdots,L$ of
random surface between trees $T^{\left( l-1\right) }$, $T^{\left( l\right) }$
contributes as a factor a spectral density $\rho \left( \lambda _{l}\right) $
as in eq.~(\ref{AGnint}), while the backbone of the
each tree $T^{\left( l\right) }$ contributes an inverse ``propagator'' $%
{\mathcal T}\left( z\lambda _{l},z\lambda _{l+1}\right) ,$ which couples the
 eigenvalues $\lambda _{l},\lambda _{l+1}$ associated with the two  patches adjacent to $T^{\left( l\right) }$:
\begin{equation}
{\mathcal T}(x,y):= [1-T(x)-T(y)]^{-1}. 
\label{APropagator}
\end{equation}
\begin{figure}[tb]
\begin{center}
\includegraphics[angle=0,width=.7\linewidth]{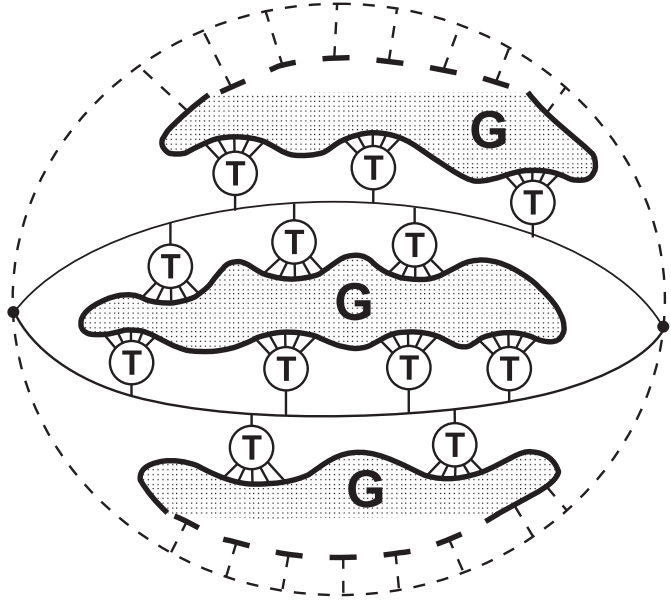}
\end{center}
\caption{Random trees on a random surface. The shaded areas represent portions
of random lattices $G$ with a disk topology (generating function
(\ref{AGn},\ref{AGnint}));
$L=2$ trees connect the end-points, each branch giving a generating
function $T$ (\ref{AT}). Two possible topologies are represented: for the disk,
the dashed lines represent the boundary, whereas for
the sphere the top and bottom dashed lines  should be identified with one another, as should
the upper and lower grey patches.}
\label{FigureA1}
\end{figure}

The integral
representation of the boundary partition function (\ref{AZtilde}) is
\begin{eqnarray}
\tilde Z_{L}(\beta ,z , \tilde z)&=&\int_{a}^{b}%
\prod ^{L+1}_{l=1}d\lambda_{l}\, \rho(\lambda _{l})
\prod ^{L}_{l=1}{\mathcal T}(z\lambda _l,z\lambda
_{l+1}) \nonumber \\
& &\qquad\times (1-\tilde z\lambda_1)^{-1}
(1-\tilde z\lambda _{L+1})^{-1},
\label{AZtildeint}
\end{eqnarray}
with two extra propagators $\mathcal L$ describing the two boundary segments:
\begin{equation}
\label{AL}
\mathcal L(\tilde z\lambda):=(1-z\lambda)^{-1}.
\end{equation}
 This gives for the two-puncture disk partition function
(\ref{AZtilde0})
\begin{equation}
\label{AZtilde0int}
Z( \hbox to 9.5mm{\hskip 0.5mm
$\vcenter{\epsfysize=.45truecm\epsfbox{fig3.eps}}$
              \hskip -80mm}
)
=\int_{a}^{b} d\lambda \, \rho(\lambda )\,
(1-\tilde z\lambda )^{-2}.
\end{equation}

\subsection{Critical Behavior}
\subsubsection{Singularity Analysis}
The critical behavior of the double grand canonical partition function $
Z_{L}\left( \beta ,z\right)$ (\ref{AZlint}) is
obtained by taking the double
scaling limit $\beta \rightarrow \beta_c$ (infinite random surface) and $
z\rightarrow z_c$ (infinite trees or RW's). The latter critical fugacity is obtained as the
smallest $z$ where ${\mathcal T}\left( z\lambda _{l},z\lambda
_{l+1}\right) $ (\ref{APropagator}) diverges. This occurs near the upper edge of
the
support of $\rho\left( \lambda \right)$, i.e., when $\lambda
\to b^{-}$, thus for $4z_{c}b\left( \beta_c\right) =1$ (see
(\ref{AT})).

Hereafter we denote $\beta -\beta_c=:\delta \beta $, and $
z_{c}-z=:\delta z $. In the grand-canonical formalism, their inverses scale like
the mean or typical sizes $|G|$ of the lattice and $| T|$ of the trees:
\begin{equation}
  |G| \sim (\delta \beta)^{-1}, | T| \sim (\delta z)^{-1}.
\end{equation}
Thus we are typically interested in the limits $\delta \beta\to 0$ and $\delta z \to 0$.

For $\lambda \rightarrow b^{-}$ and for $\delta \beta \rightarrow
0$, one knows that $\rho $ has the singular behavior (up to constant
coefficients)  \cite{brezin,DK}
\begin{equation}
\rho(\beta,\lambda)  \sim \left( \delta \beta \right) ^{\frac{1}{2}}\left(
b-\lambda\right) ^{\frac{1}{2}}+\left( b-\lambda \right) ^{\frac{3}{2}}.
\end{equation}
Because of the coupling via propagators $\mathcal T$ of the $\lambda$'s to $z$, hence to $\delta z \to 0$,
and by homogeneity, each integration of a density $\rho $ in (\ref{AZlint}) contributes a singular power
behavior 
\begin{equation}
\label{Arhos}
\int d\lambda \, \rho\sim \left( \delta \beta \right) ^{\frac{1}{2}}\left(
\delta z\right) ^{\frac{3}{2}}+\left( \delta z\right) ^{\frac{5}{2}},
\end{equation}
while each propagator ${\mathcal T}$  (\ref{APropagator}), because of (\ref{AT}),
contributes a square root singularity
\begin{equation}
\label{Ts}
{\mathcal T}\sim \left( \delta z\right) ^{-\frac{1}{2}}.
\end{equation}
We therefore arrive at a formal power behavior
\begin{equation}
\label{AZs}
Z_{L}\sim \left[ \left( \delta \beta \right) ^\frac{1}{2}\left( \delta z\right)
+\left(\delta z\right) ^{2}\right] ^{L}.
\end{equation}

The analysis of this singular behavior is
best performed by using {\it finite-size scaling} (FSS) between the various random sets  \cite{DK}. One
balances the two terms of $Z_{L}$ (\ref{AZs}) against each other so that
\begin{equation}
\delta z\sim \left(\delta
 \beta \right) ^\frac{1}{2}.
\end{equation}
 This corresponds to a {\it dilute phase} where
\begin{equation}
 \left| T\right| \sim \left| G\right|^\frac{1}{2},
\end{equation}
so that the number of sites visited by the random walks grows like the square root of the lattice size.
The partition function (\ref{AZs}) then scales as
\begin{equation}
Z_{L}\left( \beta ,z\right) \sim \left( \beta -\beta_c\right) ^{L}.
\label{AZll}
\end{equation}

\subsubsection{Bulk Quantum Gravity Conformal Weights.}
The partition function $Z_{L}$ (\ref{AZlint})
represents a random surface with two punctures where
two conformal operators of conformal weights $\Delta_{L}$ are
located, here at the
two vertices of $L$ non-intersecting RW's. It scales as
\begin{eqnarray}
\nonumber
Z_{L}&\sim& Z\lbrack \hbox to 8.5mm{\hskip 0.5mm
$\vcenter{\epsfysize=.45truecm\epsfbox{fig2.eps}}$
              \hskip 10mm}
 \rbrack \ \times \left| G\right|
^{-2\Delta_{L}}\
= \frac{\partial^{2}}{\partial \beta ^{2}}
Z_{\chi =2}(\beta)\times
\left| G\right|^{-2\Delta_{L}},\\
&\sim&\left| G\right|^{\gamma _{\rm str}(\chi =2)-2\Delta_{L}},
\label{AZls}
\end{eqnarray}
where the two-puncture partition function (\ref{AZsec}) scales as in
(\ref{AZchi"}). Eq. (\ref{AZll}) immediately gives
\begin{equation}
2\Delta_{L}-\gamma _{\rm str}(\chi =2)=L,
\label{Adeltal}
\end{equation}
where $\gamma _{\rm str}(\chi =2)=\gamma=-1/2$, or
\begin{equation}
\Delta_{L}=\frac{1}{2}\left(L-\frac{1}{2}\right),
\label{Adeltall}
\end{equation}
which is the first of eqs. (\ref{ADelta}), {\bf QED}.

\subsubsection{Boundary Exponents}
For the boundary partition function $\tilde Z_{L}$
(\ref{AZtildeint}) a similar analysis can be performed near the triple critical 
point 
$(\beta _{c},z_{c},\tilde z_{c}=1/b(\beta _{c})),$
where the boundary length also diverges through the singular behavior of its generating function $\mathcal L$ (\ref{AL}).
A {\it triple} finite-size scaling then occurs, with the further equivalence
for $\delta \tilde z=\tilde z_{c}-\tilde z$
\begin{equation}
\delta \tilde z \sim \delta z\sim (\delta \beta)
^\frac{1}{2},
\end{equation}
 obtained by homogeneity since in (\ref{AZtildeint}) $\delta \tilde z$ and $ \delta z$ are
 coupled to the same spectral parameters $\lambda$ in propagators $\mathcal T$ and $\mathcal L$,
 hence $\delta \tilde z \sim \delta z$.
This amounts to the geometrical triple scaling:
\begin{equation}
\label{triples}
\left|\partial
G\right| \sim \left| G\right| ^\frac{1}{2}\sim \left| T\right|,
\end{equation}
with a natural scaling of the boundary length with respect to the area. The latter is characteristic of dilute systems,
and is more complex for dense ones (like ${\rm SLE}_{\kappa \geq 4}$). See appendices \ref{ONapp} and \ref{BBapp},
where a  general
study of boundary quantum gravity
can be found, with applications to the $O(N)$ model, and to the SLE.

The scaling behavior of  $\tilde Z_{L}$ (\ref{AZtildeint}) is obtained using the symbolic notation
\begin{equation}
\label{AtZstar}
\tilde Z_{L} \sim \left(\int \rho d\lambda\right)^{L+1} \star {\mathcal T}^{L} \star \mathcal L^2,
\end{equation}
where the $\star$ operation represents the factorisation of scaling components in terms of
(\ref{Arhos}), (\ref{Ts}) and where the formal powers also represent repeated $\star$ operations. One has
\begin{equation}
\label{Ls}
\mathcal L\sim (\tilde \delta z)^{-1}.
\end{equation}
By analogy to eq. (\ref{AZs}) one therefore arrives at a formal power behavior
\begin{equation}
\label{AtZs}
\tilde Z_{L}\sim \left(\int d\lambda  \rho \star {\mathcal T}\right)^{L} \star \int  d\lambda  \rho \star \mathcal L^2.
\end{equation}
This can be simply recast as
\begin{equation}
\label{AtZs'}
\tilde Z_{L}\sim Z_L \star \int d\lambda \, \rho\star \mathcal L^2.
\end{equation}
Notice that the last two factors precisely correspond to the scaling of the two-puncture boundary partition function
(\ref{AZtilde0int})
\begin{equation}
\label{AtZ0s}
Z( \hbox to 9.5mm{\hskip 0.5mm
$\vcenter{\epsfysize=.45truecm\epsfbox{fig3.eps}}$
              \hskip -80mm}
)=\tilde Z_{0}\sim  \int d\lambda \, \rho\star \mathcal L^2.
\end{equation}

\subsubsection{Boundary Conformal Weights}
Partition function $\tilde Z_{L}$ (\ref{AZtilde})
corresponds to the insertion of two boundary operators of dimensions
${\tilde \Delta}_{L},$ integrated over the boundary $\partial G,$ on a
random surface with the topology of a disk, or in graphical terms:
\begin{equation}
\tilde Z_{L}\sim Z( \hbox to 9.5mm{\hskip 0.5mm
$\vcenter{\epsfysize=.45truecm\epsfbox{fig3.eps}}$
              \hskip -80mm}
) \times \left|\partial G\right| ^{-2\tilde{\Delta }_{L}}.
\label{AZlts}
\end{equation} 
We can use the punctured disk partition function scaling (\ref{AtZ0s})
to divide (\ref{AtZs'}) and get
\begin{equation}
\tilde Z_{L}/Z( \hbox to 9.5mm{\hskip 0.5mm
$\vcenter{\epsfysize=.45truecm\epsfbox{fig3.eps}}$
              \hskip -80mm}
)
\sim Z_{L} , \label{Aratio}
\end{equation} 
where the equivalences hold true in terms of scaling
behavior. Comparing eqs.~(\ref{AZls}) to~(\ref{AZlts}) (\ref{Aratio}),
and using FSS equation (\ref{triples}) gives the general identity between
boundary and bulk exponents: 
\begin{equation}
{\tilde \Delta}_{L}=2\Delta _{L}-\gamma _{{\rm str}}
(\chi =2).
\label{Adeltat}
\end{equation}
From the bulk conformal weight (\ref{Adeltal}), one finds
\begin{equation}
{\tilde \Delta}_{L}=L,
\label{Adeltat'}
\end{equation}
which is the second of eqs. (\ref{ADelta}), {\bf QED}.

Applying  KPZ relation (\ref{AKPZ}) to $\Delta
_{L}$ and ${\tilde \Delta}_{L}$  yields  exponents
$ \zeta_{L}$ (\ref{AZeta}), and $ 2{\tilde
\zeta}_{L}$ (\ref{AC2}), {\bf QED}.

Notice the peculiar relation (\ref{Adeltat}) between boundary and bulk exponents, which is also linear in terms
of the numbers of Dirichlet-like mutually-avoiding components. This is a part of the general results
established in appendices  \ref{ONapp} and \ref{BBapp}.

\subsection{Generalized Non-Intersection Exponents}
\subsubsection{Definitions}
Eqs. (\ref{AZlint}) and (\ref{AZtildeint}) give the key to many generalizations. Indeed the product
of identical
propagators ${\mathcal T}^{L}$ there can be replaced by a product of different propagators
$\prod _{l}{\mathcal T}_{l}$, corresponding to different
geometrical objects, as obvious from the cyclic construction (see figure \ref{FigureA1}).

Consider in particular configurations made of $L$ mutually-avoiding bunches $l=1,\cdots,L$, each of them
made of $n_{l}$ walks {\it
transparent} to each
other, i.e., $n_l$ independent RW's  \cite{werner}. The probability of non-intersection of the $L$ packets
scales as
\begin{equation}
P_{n_{1},\cdots,n_{L}}(t) \sim t^{-\zeta
(n_{1},\cdots,n_{L})},
\label{APbunch}
\end{equation}
and near a Dirichlet boundary
\begin{equation}
\tilde P_{n_{1},\cdots,n_{L}}(t) \sim t^{-\tilde \zeta
(n_{1},\cdots,n_{L})}.
\label{APtbunch}
\end{equation}
The original case of $L$ mutually-avoiding RW's now corresponds to $n_{1}=..=n_{L}=1$.
The generalizations of exponents $\zeta
(n_{1},\cdots,n_{L})=\Delta ^{(0)}
\left\{ n_{l}\right\},$ as well as $2{\tilde \zeta}(n_{1},\cdots,n_{L}) ={\tilde
\Delta}^{(0)}\left\{ n_{l}\right\} ,$ describing these $L$ packets can be calculated from quantum gravity.

\subsubsection{Random Lattice Partition Functions}
On a random lattice, each bunch will contribute a
certain inverse propagator ${\mathcal T}_{n_{l}}$ and yield instead
of~(\ref{AZlint})
\begin{equation}
Z\{n_{1},\cdots,n_{L}\}=\int_{a}^{b}\prod ^{L}_{l=1}
d\lambda_{l}\, \rho(\lambda_l)
\prod ^{L}_{l=1}{\mathcal T}_{n_{l}},
\label{AZlintgen}
\end{equation}
or for~(\ref{AZtildeint})
\begin{eqnarray}
\tilde Z\{n_{1},\cdots,n_{L}\}&=&\int_{a}^{b}%
\prod ^{L+1}_{l=1}d\lambda_{l}\, \rho(\lambda _{l})
\prod ^{L}_{l=1}{\mathcal T}_{n_{l}} \;{\mathcal L}_1\, {\mathcal L}_{L+1}.
\label{AZtildeintgen}
\end{eqnarray}
Recall that the two-puncture boundary partition function reads
\begin{equation}
\label{AtZ0s'}
Z( \hbox to 9.5mm{\hskip 0.5mm
$\vcenter{\epsfysize=.45truecm\epsfbox{fig3.eps}}$
              \hskip -80mm}
)\sim  \int d\lambda \, \rho\star \mathcal L^2.
\end{equation}
In terms of scaling behavior, we can thus rewrite the above integral representations as in
(\ref{AtZstar}) and (\ref{AtZ0s})
\begin{eqnarray}
\label{AZtZ}
Z{\left\{ n_1,\cdots, n_{L}\right\}}
&\sim &\frac{{\tilde Z}{\left\{n_1,\cdots, n_{L} \right\}}}{Z( \hbox to 9.5mm{\hskip 0.5mm
$\vcenter{\epsfysize=.45truecm\epsfbox{fig3.eps}}$
              \hskip -80mm}
)}\\
&\sim& \Bigl (\int d\lambda \, \rho\Bigr )^{L}\star \prod _{l=1}^{L}
{\mathcal T}_{n_{l}}.
\label{AZn}
\end{eqnarray}

\subsubsection{Bulk and Boundary Quantum Gravity Conformal Weights}
From the definition of bulk quantum gravity conformal weights, the bulk partition function
has to be identified, as in(\ref{AZls}), to
\begin{equation}
Z{\left\{n_1,\cdots, n_{L}\right\}} \sim Z\lbrack \hbox to 8.5mm{\hskip 0.5mm
$\vcenter{\epsfysize=.45truecm\epsfbox{fig2.eps}}$
              \hskip 10mm}
 \rbrack \ \times \left| G\right|
^{-2\Delta{\left\{ n_{l}\right\}}} \sim  \left| G\right| ^{\gamma_{\rm str}(\chi=2)-2{ \Delta}%
\left\{ n_{l}\right\}},
\label{AZn1}
\end{equation}
while the normalized boundary partition function scales, as in (\ref{AZlts}), as
\begin{equation}
 \frac{{\tilde Z}{\left\{n_1,\cdots, n_{L}\right\}}}{Z( \hbox to 9.5mm{\hskip 0.5mm
$\vcenter{\epsfysize=.45truecm\epsfbox{fig3.eps}}$
              \hskip -80mm}
)} \sim \left|\partial G\right| ^{-2{\tilde \Delta} \left\{ n_{l}\right\}}  .
\label{AZn2}
\end{equation}
The relative scaling of the boundary length with respect to the area, $|\partial G| \sim |G|^{1/2}$,
together with (\ref{AZtZ}), (\ref{AZn1}), and (\ref{AZn2}), immediately gives the relation
between bulk and boundary conformal weights:
\begin{eqnarray}
\nonumber
{\tilde \Delta}
\left\{n_{1},\cdots,n_{L}\right\}&=&2\Delta \left\{n_{1},\cdots,n_{L}\right\}-\gamma_{\rm str}(\chi=2)\\
&=&2\Delta \left\{n_{1},\cdots,n_{L}\right\}+\frac{1}{2},
\label{bulkbond}
\end{eqnarray}
where we have used the value $\gamma_{\rm str}(\chi=2)=\gamma=-1/2$ for pure quantum gravity ($c=0$).

\subsubsection{Single Packet Partition Functions}
Introduce now the partition functions corresponding to the existence of only one packet of $n_l$ independent
random walks on the random lattice, namely $Z(n_l):=
Z\{n_{1}=0,\cdots,n_l,\cdots,n_{L}=0\}$ and $\tilde Z(n_l):=\tilde Z\{n_{1}=0,\cdots,n_l,\cdots,n_{L}=0\}$.
Using the same notation as in (\ref{AZlintgen}) and (\ref{AZtildeintgen}), we write them as
\begin{eqnarray}
\label{AZlintgenl}
Z(n_l)&=&\int_{a}^{b}
d\lambda \, \rho(\lambda)
{\mathcal T}_{n_{l}},\\
\tilde Z(n_l)&=&\int_{a}^{b}
\prod ^{2}_{l=1}d\lambda_{l}\, \rho(\lambda _{l})
{\mathcal T}_{n_{l}} \;{\mathcal L}_1\, {\mathcal L}_{2}.
\label{AZtildeintgenl}
\end{eqnarray}
As before, these equations imply in terms of scaling behavior
\begin{equation}
Z{\left(n_l\right)} \sim
 \frac{{\tilde Z}{\left(n_l\right)}}{Z( \hbox to 9.5mm{\hskip 0.5mm
$\vcenter{\epsfysize=.45truecm\epsfbox{fig3.eps}}$
              \hskip -80mm}
)} \sim \int d\lambda \, \rho\star \mathcal T.
\label{Aratiol}
\end{equation}

\subsubsection{Conformal Weights}
As usual, the definition of the conformal weights is expressed as:
\begin{equation}
Z{\left(n_l\right)} \sim Z\lbrack \hbox to 8.5mm{\hskip 0.5mm
$\vcenter{\epsfysize=.45truecm\epsfbox{fig2.eps}}$
              \hskip 10mm}
 \rbrack \ \times \left| G\right|
^{-2\Delta{\left( n_{l}\right)}} \sim  \left| G\right| ^{\gamma_{\rm str}(\chi=2)-2{ \Delta}%
\left( n_{l}\right)},
\label{AZnl}
\end{equation}
\begin{equation}
 \frac{{\tilde Z}{\left(n_l\right)}}{Z( \hbox to 9.5mm{\hskip 0.5mm
$\vcenter{\epsfysize=.45truecm\epsfbox{fig3.eps}}$
              \hskip -80mm}
)} \sim \left|\partial G\right| ^{-2{\tilde \Delta} \left( n_{l}\right)}  ,
\label{AtZnl}
\end{equation}
where now ${\Delta}(n)$ and ${\tilde \Delta}(n)$ are the bulk and boundary quantum conformal weights of
a {\it single} bunch of $n$ transparent walks on the random surface. From (\ref{Aratiol}) we
conclude as in (\ref{bulkbond}) that:
\begin{eqnarray}
\nonumber
\tilde \Delta \left( n_{l}\right)&=&2\Delta (n_l)-\gamma_{\rm str}(\chi=2)\\
&=&2\Delta (n_l)+\frac{1}{2}.
\label{bulkbondl}
\end{eqnarray}

\subsubsection{Factorization Properties}
The factorization property (\ref{AZn}), together with (\ref{Aratiol}), immediately gives in terms
of scaling behavior as represented by the $\star$ operation:
\begin{eqnarray}
\label{AZpiZ}
Z\left\{ n_{1},\cdots, n_{L}\right\} &\sim& \prod _{l=1}^{L} \star \left\{Z\left( n_{l}\right)\right\}\\
 &\sim& \prod _{l=1}^{L} \star \left\{\frac{{\tilde Z}\left( n_{l}\right)}{Z(
\hbox to 9.5mm{\hskip 0.5mm
$\vcenter{\epsfysize=.45truecm\epsfbox{fig3.eps}}$
              \hskip -80mm}
)}\right\} \sim \frac{{\tilde Z}\left\{ n_{1},\cdots, n_{L}\right\}}{
Z( \hbox to 9.5mm{\hskip 0.5mm
$\vcenter{\epsfysize=.45truecm\epsfbox{fig3.eps}}$
              \hskip -80mm}
)}.
\label{AtZpiZ}
\end{eqnarray}
Running these scaling relations backwards,
with the help of definitions (\ref{AZn1}), (\ref{AZn2}), (\ref{AZnl}) and (\ref{AtZnl}), immediately gives
the  {\it basic additivity of boundary conformal dimensions, or shifted bulk ones, in presence of
gravity:}
\begin{eqnarray}
\nonumber
{\tilde \Delta}{\left\{n_{1},\cdots,n_{L}\right\}} &=&\sum ^{L}_{l=1}
{\tilde \Delta}(n_{l})\\
&=&\sum ^{L}_{l=1}
{2\Delta}(n_{l})+\frac{1}{2}=2{\Delta}{\left\{n_{1},\cdots,n_{L}\right\}}+\frac{1}{2}.
\label{Adeltan}
\end{eqnarray}

\subsubsection{Exponents ${\tilde \Delta}(n)$}
It does not seem to be easy to calculate the random lattice partition functions (\ref{AZlintgenl}) or (\ref{AZtildeintgenl}) corresponding to a
packet of $n$ transparent walks for $n \neq 1$, since
the walks, as the forward trees spanning the visited sites, will overlap strongly and enclose arbitrarily many patches of random lattice.
In other words, the explicit form of
propagator
$\mathcal T_{n}$ is not easily accessible for $n\geq 2$. For $n=1$ it is given by the expression (\ref{APropagator}) in terms of
the tree generating function (\ref{AT}).

We know, however, the exact value of the  Dirichlet boundary exponent
${\tilde \Delta}(n)$ in presence of quantum gravity corresponding to such a bunch of
$n$ transparent random walks. Indeed, in the standard half-plane $\mathbb H$, it must corresponds to a
 boundary conformal weight
\begin{equation}
{\tilde \Delta}^{(0)}(n)=U\left({\tilde \Delta}(n)\right),
\label{AtD0n}
\end{equation}
which is its image by the KPZ map (\ref{AKPZ}):
\begin{equation}
U(\Delta)=U_{(\gamma=-1/2)}(\Delta)=\frac{1}{3}\Delta \left(
1+2\Delta
\right).
\label{AKPZ'}
\end{equation}
In the half-plane with Dirichlet boundary conditions,
for $n$  {\it independent} Brownian paths, one has by simple additivity
\begin{equation}
{\tilde \Delta}^{(0)}(n)=n{\tilde \Delta}^{(0)}(1)=n,
\label{AtD0n'}
\end{equation}
since for a single Brownian path one has by elementary means: ${\tilde \Delta}^{(0)}(1)=1$.
The inverse of KPZ map (\ref{AKPZ}) then gives the result
\begin{equation}
{\tilde \Delta}(n)=U^{-1}(n)=\frac{1}{4}(\sqrt{24n+1}-1).
\label{Adeltatn}
\end{equation}
This quantum gravity boundary conformal weight is highly non-trivial since the random walks
(in the scaling limit, the Brownian paths), while independent in a fixed metric,  are
 strongly coupled by the fluctuations of the metric in quantum gravity.

\subsubsection{Back to the Complex (Half-) Plane}
Using once again the KPZ relation (\ref{AKPZ'})
for $\tilde \Delta \left\{
n_{l}\right\} $ and ${ \Delta}\left\{ n_{l}\right\} $ gives the general results
in the standard complex  half-plane $\mathbb H$ or plane $\mathbb C$
\begin{eqnarray}
\nonumber
2{\tilde
\zeta}(n_{1},\cdots,n_{L})&=&\tilde \Delta^{(0)}\{n_{1},\cdots,n_{L}\}=U\left(\tilde \Delta\left\{
n_{l}\right\}\right)\\
\nonumber
\zeta (n_{1},\cdots,n_{L})  &=&\Delta^{(0)}\{n_{1},\cdots,n_{L}\}=U\left(\Delta\left\{
n_{l}\right\}\right).
\label{AZetaL}
\end{eqnarray}
Using eqs. (\ref{Adeltan}) and (\ref{Adeltatn}) finally gives
\begin{eqnarray}
\label{ADelta'}
\left\{\begin{array}{ll} 2{\tilde
\zeta}(n_{1},\cdots,n_{L})=U\left(\tilde \Delta\{n_{1},\cdots,n_{L}\}\right)\\
\\
\zeta(n_{1},\cdots,n_{L})=V\left(\tilde \Delta\{n_{1},\cdots,n_{L}\}\right)\\
\\
\tilde \Delta\{n_{1},\cdots,n_{L}\}=\sum_{l=1}^{L} U^{-1}(n_l)
=\sum_{l=1}^{L}\frac{1}{4}(\sqrt{24n_l+1}-1)
\end{array}
\right.
\end{eqnarray}
with
\begin{eqnarray}
\label{AUV}
\left\{\begin{array}{ll}
U(\Delta)=U_{(\gamma=-1/2)}(\Delta)=\frac{1}{3}\Delta \left(
1+2\Delta\right)\\
\\
V(\Delta)=U\left[\frac{1}{2}\left(\Delta-\frac{1}{2}\right)\right]
=\frac{1}{24}(4\Delta^2-1),
\end{array}
\right.
\end{eqnarray}
which is the quantum gravity geometric structure annonced in section \ref{sec.inter}, {\bf QED}.\\

\section{\sc{$O(N)$ Model Multi-Line Exponents from Quantum Gravity}}
\label{ONapp}

Let us derive here the $O(N)$ multi-line exponents (\ref{tdeltaL}), (\ref{tdeltaLD}), (\ref{deltaLON}), and (\ref{deltaLD})
from their study on a random lattice. We shall in particular focus on the relationship between boundary and bulk exponents.
The approach will be similar to the one followed in appendix \ref{Brownapp} for multiple tree exponents. (See also refs.
\cite{DK,kostovgaudin,KK}.)

\subsection{Random Lattice Partition \& Two-Point Functions}

\subsubsection{Partition Function}
Consider again the set of planar random graphs $G$, built up with,
e.g., trivalent vertices and with a fixed topology.
The partition function of the $O(N)$-loop model is
defined as
\begin{eqnarray}
\label{BZchi1}
Z_{O(N)}(\beta, K, \chi)&:=&\sum _{G({\chi})}\frac{1}{ S(G)}e^{-\beta
\left| G\right|}W_{O(N)}(G)\\
W_{O(N)}(G)&:=&\sum _{\mathcal L \in G} K^{|\mathcal L|}N^{{\rm Card}\mathcal L},
\label{WON}
\end{eqnarray}
where $\chi $ denotes the fixed Euler characteristic of the lattice $G=G(\chi)$; $\left| G\right|$ is the number of vertices of $G$,
$S\left(
G\right) $ its symmetry factor. In the $O(N)$ weight $W_{O(N)}(G)$, the sum runs also over all self-avoiding loop
configurations $\mathcal L$ on $|G|$,
with a fugacity $K$ for the total
occupied length $|\mathcal L|$, and a fugacity $N$ for the total number of loops ${\rm Card}\mathcal L$.
Notice that the effective fugacity associated with bonds occupied by $O(N)$ loops is
$$z_{\bullet}:= e^{-\beta}K.$$
The
Euler characteristic is
\begin{equation}
\chi=2-2H-B,
\label{euler}
\end{equation}
where $H$  and $B$ respectively are the numbers of handles and boundaries of $G$. In the following we shall
consider explicitly the cases of the spherical topology ($\chi=2$), and of the disk ($\chi=1$).

The partition function of the random lattice with two punctures will play an important role later. It is defined
 by twice differentiating the
one above with respect to the ``chemical potential'', or ``cosmological constant'' $\beta$,

\begin{eqnarray}
\label{BZsec}
\frac{\partial^{2}}{\partial \beta ^{2}}
Z_{O(N)}(\beta, K, \chi)&=&\sum _{G({\chi})}\frac{1}{ S(G)} \left| G\right|^2 e^{-\beta
\left| G\right|}W_{O(N)}(G)\\
 &=:& Z_{O(N)}\lbrack \hbox to 8.5mm{\hskip 0.5mm
$\vcenter{\epsfysize=.45truecm\epsfbox{fig2.eps}}$
              \hskip 10mm}
 \rbrack \,
\label{BZfig}
\end{eqnarray}
with an intuitively obvious graphical representation for this two-puncture partition function.

For a given $\beta$, the partition sum $Z_{O(N)}$ converges for all values
of the parameter $K$ smaller than some critical $K_c(\beta)$. For $K \to K_c^{-}(\beta)$ the loops fill the lattice and
force it to become infinite: this is the dense phase. There is also a critical line
$\beta =\beta_c(K)$, such that at $\beta
\to \beta_c^{+}(K)$ a singularity appears, solely due to the presence of
infinite graphs in (\ref{BZchi1}), and where the loops stay non-critical. This corresponds to pure gravity.
The two
critical regimes meet for $\beta=\beta^*$ such that $\beta^*=\beta_c[K_c(\beta^*)]$. Then
the loops, for $K \to K_c^-(\beta^*)$, and the non-occupied part of the random lattice,
for $\beta \to \beta^*,$  become infinite simultaneously: this is the dilute phase  \cite{DK}.

\subsubsection{Disk Green's Function}
\begin{figure}[tb]
\begin{center}
\includegraphics[angle=0,width=.4\linewidth]{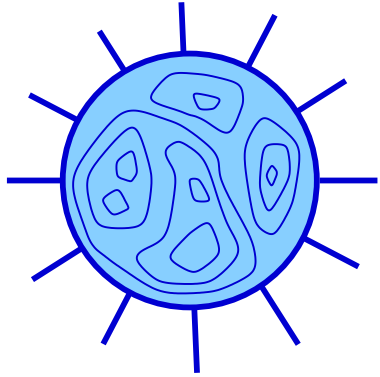}
\end{center}
\caption{Planar random disk with a sea of $O(N)$ loops, and a fixed number $n$ of external
vertices.}
\label{Figuredisc2}
\end{figure}
The {\it disk} partition function, or rather here, Green's function, associated with a planar random
graph with the topology of a disk, and bearing a sea of $O(N)$ loops, with a fixed number $n$ of external
vertices (Fig. \ref{Figuredisc2}), is defined as
\begin{equation}
G_{O(N),n}(\beta,K ) := \sum _{n-{\rm leg}\ {\rm planar}\ G} e^{-\beta \left| G\right|}
\sum _{\mathcal L \in G} K^{|\mathcal L|}N^{{\rm Card}\mathcal L},
\label{BGn}
\end{equation}

\subsubsection{Bulk Two-Point Partition Function}
\begin{figure}[tb]
\begin{center}
\includegraphics[angle=0,width=.5\linewidth]{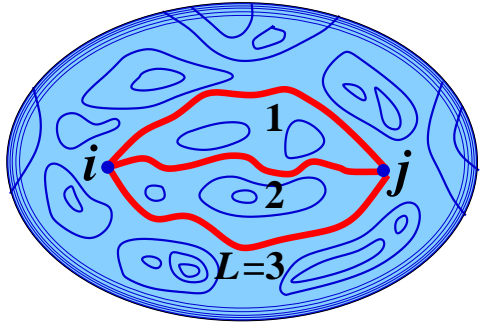}
\end{center}
\caption{The watermelon configuration of $L=3$ random lines among $O(N)$ loops on the random
sphere, building up partition function (\ref{BZl}).}
\label{wm1}
\end{figure}

 Now, let us consider a set of $L$ self- and mutually-avoiding walks
 $\varGamma_{ij}=\{\varGamma_{ij}^{\left( l\right) },l=1,...,L\}$
on the {\it random graph} $G$ with the constraints that they start at the same vertex $i\in G$
(or near it, if $L \geq 4$), end  at (or near) the same
vertex $j \in G$, and have no mutual intersections in between (Fig. \ref{wm1}).  The $L$-walk partition
function on the random lattice with a spherical topology reads  \cite{DK}
\begin{equation}
Z_{O(N),L}(\beta ,z):=\sum _{{\rm planar}\ G}\frac{1}{ S(G)}
e^{-\beta \left| G\right|} W_{O(N)}(G) \sum _{i,j\in G}
\sum_{\scriptstyle \varGamma^{(l)}_{ij}\atop\scriptstyle l=1,...,L}
 z^{\left| \varGamma_{ij}\right| };
\label{BZl}
\end{equation}
a fugacity $z$ is associated with the total
number $\left| \varGamma_{ij}\right| =\left| \cup^{L}_{l=1} \varGamma_{ij}^{(l)}\right| $
of bonds occupied by the $L$ walks. At a later stage, it will be convenient to keep $z$ distinct from the effective fugacity
$z_{\bullet}:= e^{-\beta}K$ associated with the bonds occupied by the $O(N)$ loops.
\subsubsection{Boundary Two-Point Partition Function}
\begin{figure}[tb]
\begin{center}
\includegraphics[angle=0,width=.5\linewidth]{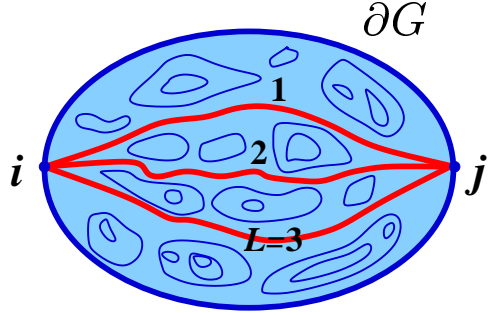}
\end{center}
\caption{The boundary watermelon configuration of $L=3$ random lines among $O(N)$ loops on the random
disk, building up partition function (\ref{BZtilde}).}
\label{wm3}
\end{figure}

We generalize this to the {\it boundary} case where $G$ now has the topology of
a disk and where the $L$ lines connect two sites $i$ and $j$ on the
boundary $\partial G$ (Fig. \ref{wm3}):
\begin{equation}
\tilde Z_{{O(N)},L}(\beta, z, \tilde z):=\sum _{ {\rm disk}\ G}
e^{-\beta \left| G\right|} W_{O(N)}(G)\tilde z^{\left| \partial G\right|}
\sum _{{i,j} \in \partial G}\sum_{\scriptstyle
\varGamma^{(l)}_{ij}\atop\scriptstyle l=1,...,L}
z^{\left| \varGamma_{ij}\right| },
\label{BZtilde}
\end{equation}
where $\tilde {z}$ is the new fugacity associated with the boundary's length.
The partition function of the disk with two boundary punctures will play a important role in the sequel. It corresponds
to eq. (\ref{BZtilde}) in the absence of the $L$ lines, and is defined as
\begin{equation}
\tilde Z_{{O(N)}}( \hbox to 9.5mm{\hskip 0.5mm
$\vcenter{\epsfysize=.45truecm\epsfbox{fig3.eps}}$
              \hskip -80mm}):=\tilde Z_{{O(N)},L=0}(\beta , z, \tilde z)=\sum _{ {\rm disk}\ G}
e^{-\beta \left| G\right|} W_{O(N)}(G)\tilde z^{\left| \partial G\right|}\left| \partial G\right|^2.
\label{BZtilde0}
\end{equation}
\subsection{Random Matrix and Spectral Representation}
\subsubsection{Disk Green's Function}
First, the disk partition function (\ref{BGn}) can be calculated through
the large ${\mathcal N}$ limit of a random $\mathcal N \times \mathcal N $
matrix integral  \cite{kostovgaudin}. It has an integral representation
\begin{equation}
G_{{O(N)},n}\left( \beta,K \right) =\int^{b}_{a}d\lambda\, \rho_{O(N)} \left( \lambda,\beta, K \right)
{\lambda }^n,
\label{BGnint}
\end{equation}
where $\rho_{O(N)} \left( \lambda,\beta,K \right) $ is the spectral eigenvalue density of
the random matrix model  as a function of
$\lambda ,\beta$, $K$ and $N$. The support $[a, b]$ of the spectral density
depends on $\beta$, $K$, and $N$. An explicit expression  for $\rho_{O(N)} \left( \lambda \right)$ is known along the
critical line $K=K_c(\beta)$  \cite{kostovgaudin}.

\subsubsection{Spherical Two-Point Partition Function}
The partition function (\ref{BZl})
 was calculated exactly in previous work \cite{DK}. The two-fold
grand canonical partition function is calculated first by summing over the
$L$-line
configurations, and then by gluing patches of random lattices in between these
lines. In fact, it can also be obtained directly from the tree case of appendix \ref{Brownapp}, by remarking that from each
site belonging to one of the $L$ traversing lines, only a single edge of the random lattice now grows,
instead of a rooted tree. That amounts exactly to replacing the series giving the tree propagator $T(x)$ (\ref{AT}) of appendix \ref{Brownapp}
by its first term, $x$.
The result for (\ref{BZl}) is then given by a multiple integral similar to (\ref{AZlint}):
\begin{equation}
Z_{O(N),L}(\beta ,z)=\int_{a}^{b}\prod ^{L}_{l=1}
d\lambda _{l}\,\rho_{O(N)} (\lambda_l)
\prod ^{L}_{l=1}{\mathcal L}(z\lambda_l,z\lambda_{l+1}),
\label{BZlint}
\end{equation}
with the cyclic condition  $\lambda _{L+1} \equiv \lambda _{1}$.
Each patch $l=1,\cdots,L$ of
random surface between lines $l,\,l+1$
contributes as a factor an $O(N)$ spectral density $\rho_{O(N)} \left( \lambda _{l}\right) $
as in eq.~(\ref{BGnint}), while each line $l$ contributes an inverse ``propagator'' $%
{\mathcal L}\left( z\lambda _{l},z\lambda _{l+1}\right),$ similar to (\ref{APropagator}) which couples the
adjacent eigenvalues $\lambda _{l},\lambda _{l+1}:$%
\begin{equation}
{\mathcal L}(x,y):= (1-x-y)^{-1}.
\label{BPropagator}
\end{equation}

\subsubsection{Boundary Two-Point Partition Function}
The integral
representation of the boundary two-point partition function (\ref{BZtilde}) is similarly
\begin{eqnarray}
\tilde Z_{{O(N)},L}(\beta ,z , \tilde z)&=&\int_{a}^{b}%
\prod ^{L+1}_{l=1}d\lambda _{l}\,\rho_{O(N)} (\lambda _{l})
\prod ^{L}_{l=1}{\mathcal L}(z\lambda _l,z\lambda
_{l+1}) \nonumber \\
& &\qquad\times (1-\tilde z\lambda_1)^{-1}
(1-\tilde z\lambda _{L+1})^{-1}
\label{BZtildeint}
\end{eqnarray}
with two extra propagators $\mathcal L(\tilde z \lambda_{1[{\rm or}\, L+1]})$ describing the
two boundary lines between $i$ and $j$. Hence one sees here that the exterior boundary
lines and the $L$ interior random paths play very similar roles. This is
the case for Dirichlet boundary conditions.

\subsubsection{Disk Partition Function}
The partition function of the disk with two boundary punctures (\ref{BZtilde0}) is given by
the non-trivial case $L=0$ of (\ref{BZtildeint})
\begin{equation}
\label{diskpunc}
\tilde Z_{{O(N)}}( \hbox to 9.5mm{\hskip 0.5mm
$\vcenter{\epsfysize=.45truecm\epsfbox{fig3.eps}}$
              \hskip -80mm})=\tilde Z_{{O(N)},L=0}(\beta ,z , \tilde z)=\int_{a}^{b} d\lambda \, \rho(\lambda )
(1-\tilde z\lambda )^{-2},
\end{equation}
where we again used the graphical representation of the boundary  two-puncture partition function.

\subsection{Scaling Behavior}
\subsubsection{Eigenvalue Density}
The critical behavior of the $O(N)$ model corresponds, for the eigenvalue density $\rho$, as usual to the vicinity of the
 end-point $b$ of the spectrum  \cite{DK,kostovgaudin}. The critical line  $K_c(\beta)$ mentioned above
is given by the critical effective loop fugacity
 $z_{\bullet}=z_{c}(\beta)$ with $$z_{c}(\beta) := e^{-\beta}K_c(\beta)= (2b)^{-1}$$

This can be seen also in the critical behavior of the double-grand canonical partition function
 $Z_{O(N),L}\left( \beta ,z\right)$ (\ref{BZlint})
 associated with $L$ non-intersecting lines on the $O(N)$ random lattice.
The critical point of the $L$ watermelon lines is indeed obtained in the integral representation for the
smallest $z$ where ${\mathcal L}\left( z\lambda _{l},z\lambda_{l+1}\right) $ (\ref{BPropagator}) diverges.
This occurs near the upper edge $b$ of
the
support of $\rho_{O(N)}\left( \lambda \right)$, i.e., when $\lambda
\to b^{-}$, thus again for $z=z_c(\beta)$, with $2z_{c}b=1$ (see
(\ref{BPropagator})). At this critical value of the fugacity, both the internal $O(N)$ loops and the $L$ random lines
become infinite. As explained above,  for reaching the dilute phase a further double scaling is necessary. Ones
 can let $\beta \to {\beta^*}^{+}$ simultaneously, in a finite-size scaling scheme, or right away set $\beta=\beta^*$, and
 afterwards let
$z \to z_c(\beta^*)$.

The eigenvalue density is known exactly along the critical line $K=K_c(\beta)$ or,
equivalently, $z_{\bullet}=z_c(\beta)$  \cite{kostovgaudin}.
We shall only need its behavior for  $\beta- \beta^*$ small and near the upper end of the spectrum, i.e., for
$\lambda \to b^{-}$. There $\rho_{O(N)} $ has the double singular behavior (up to constant
coefficients)  \cite{kostovgaudin,DK}
\begin{equation}
\rho_{O(N)}(\lambda, \beta, K_c(\beta)) \sim \left(  \beta -\beta^*\right) ^{\theta}\left(
b-\lambda\right) ^{1-\theta}+\left( b-\lambda \right) ^{1+\theta},
\label{density}
\end{equation}
where $\theta$ parametrizes the fugacity $N$ in the $O(N)$ model:
\begin{equation}
N=2\cos(\pi \theta),\;0 \leq \theta \leq 1.
\label{parametrisation}
\end{equation}
This behavior (\ref{density}) will be sufficient to determine the critical behavior of the various partition functions.

\subsubsection{Spherical Two-Point Partition Function}
We just have seen the scaling behavior of the eigenvalue density along the critical line $z_{\bullet}=z_c(\beta)$,
and in the vicinity the double scaling critical point $\beta^*$, up to the double critical point $\beta^*$.
In order for the two-point partition function (\ref{BZlint}) to stay finite, it is now
crucial to keep the fugacity $z$
of the watermelon lines in the propagators $\mathcal L$ such that $z < (2b)^{-1}$,
 hence away from the critical value $z_{\bullet}=z_c(\beta)$ of the internal loops. Let us thus introduce the shifted
 quantities
\begin{eqnarray}
\label{dz}
\delta z&:=&z_c(\beta)-z\\
\label{db}
\delta \beta &:=& \beta -\beta^*,
\end{eqnarray}
in terms of which all the scaling analysis, to which we now turn, will be performed.

In multiple integral (\ref{BZlint}), each integration of density $\rho_{O(N)} $
(\ref{density}) yields by homogeneity (power counting) a singular power behavior
\begin{equation}
\label{intrho}
\int \rho\equiv\int d\lambda\, \rho_{O(N)}(\lambda) \sim \left( \delta \beta \right) ^{\theta}\left(
\delta z\right) ^{2-\theta}+\left( \delta z\right) ^{2+\theta}.
\end{equation}
Each propagator ${\mathcal L}$ (\ref{BPropagator}), when integrated over its $\lambda$ variables, brings
in a singularity
\begin{equation}
{\mathcal L}\sim \left( \delta z\right) ^{-1}.
\label{mathLscaling}
\end{equation}
We therefore arrive at a formal power behavior
\begin{eqnarray}
\label{ZLscaling}
Z_{O(N),L}\sim \left(\int \rho\right)^L \left(\delta z\right)^{-L}
= \left(\int \rho \times (\delta z)^{-1}\right)^L,
\end{eqnarray}
where (\ref{intrho}) gives the explicit scaling:
\begin{eqnarray}
Z_{O(N),L}\sim \left[ \left( \delta \beta \right) ^{\theta}\left( \delta z\right)^{1-\theta}
+\left(\delta z\right) ^{1+\theta}\right] ^{L}.
\label{Zscalingexp}
\end{eqnarray}

\subsubsection{Boundary Two-Point Partition Function}
For the boundary partition function $\tilde Z_{O(N),L}$
(\ref{BZtildeint}) a similar analysis is performed near the critical
point
$\tilde z_{c}=1/b,$
where the boundary length also diverges. For Dirichlet boundary conditions, the
scaling limit requires the further equivalence of {\it boundary and bulk loops}, i.e.,
for the shifted boundary fugacity $\tilde z_{c}-\tilde z:= \delta \tilde z$ the equivalence:
\begin{equation}
\delta \tilde z \sim \delta z.
\label{bB}
\end{equation}
The formal scaling of the boundary two-point partition function is therefore:
\begin{eqnarray}
\label{tZ1}
{\tilde Z}_{O(N),L}\sim \left(\int \rho\right)^{L+1} \left(\delta z\right)^{-L} \left( \delta \tilde z\right)^{-2}
\end{eqnarray}
Using equivalence (\ref{bB}) and (\ref{intrho}) yields the explicit scaling behavior
\begin{eqnarray}
\label{tZ2}
{\tilde Z}_{O(N),L}&\sim &\left(\int \rho \times \left(\delta z\right)^{-1}\right)^{L+1}  \left(\delta z\right)^{-1}\\
&\sim&\left[ \left( \delta \beta \right) ^{\theta}\left( \delta z\right)^{1-\theta}
+\left(\delta z\right) ^{1+\theta}\right] ^{L+1} \left(\delta z\right)^{-1}.
\label{tZscaling}
\end{eqnarray}

\subsection{Partition Function Scaling Identities}
In order to simplify the scaling analysis, it is worth noticing several fundamental scaling relations
between partition functions, which are structural and do not require the precise knowledge of the eigenvalue density.
\subsubsection{Bulk-Boundary Relation}
One first notices that one can also rewrite (\ref{tZ1}) as the scaling identity
\begin{equation}
\label{tZ3}
{\tilde Z}_{O(N),L} \sim \left(\int \rho \times \left(\delta z\right)^{-1}\right)^{L} \times
\int \rho \times \left(\delta \tilde z\right)^{-2},
\end{equation}
where, owing to (\ref{ZLscaling}) and (\ref{tZ1}), one respectively recognizes the sphere and the boundary two-point partition functions,
which are thus linked by a fundamental relation, first observed in  \cite{duplantier7}:
\begin{equation}
Z_{O(N),L} \sim \frac{\tilde Z_{O(N),L}}{\tilde Z_{O(N)}( \hbox to 9.5mm{\hskip 0.5mm
$\vcenter{\epsfysize=.45truecm\epsfbox{fig3.eps}}$
              \hskip -80mm}
)}
  . \label{Bratio}
\end{equation}

\subsubsection{Recursion Relation}
We can also write (\ref{tZ1}), together with (\ref{bB}), as a ratio
\begin{equation}
{\tilde Z}_{O(N),L}\sim \frac{\left(\int \rho\right)^{L+2} \left(\delta z\right)^{-(L+1)} \left(\delta \tilde z\right)^{-2}
}{\left(\int \rho\right) \left(\delta \tilde z\right)^{-2}} \times \left(\delta z\right)^{-1}.
\label{tZr}
\end{equation}
Comparing to (\ref{tZ2}) for $L+1$ gives a recursion relation between the two-point boundary partition functions at levels $L$ and $L+1$:
\begin{equation}
\tilde Z_{O(N),L}\sim \frac{\tilde Z_{O(N),L+1}}{\tilde Z_{O(N)}( \hbox to 9.5mm{\hskip 0.5mm
$\vcenter{\epsfysize=.45truecm\epsfbox{fig3.eps}}$
              \hskip -80mm}
)} \times (\delta z)^{-1},
 \label{L+1L}
\end{equation}
a recursion relation which will prove very useful for an easy calculation of scaling exponents.

\subsection{Quantum Gravity Conformal Weights}
\subsubsection{Finite Size Scaling (FSS)}

In the grand-canonical ensemble, and in the critical regime  $\delta \beta \to 0$ for (\ref{db})
the average size $\langle |G|\rangle $ of the lattice $G$ scales as
\begin{eqnarray}
\label{Gsize}
\langle |G|\rangle \sim \left(\delta \beta\right)^{-1},
\end{eqnarray}
Let us denote by $\varGamma$ the set of lattice edges occupied by the $L$ watermelon lines.
Its average size $\langle |\varGamma|\rangle $ in the grand-canonical ensemble scales in the critical regime $\delta z \to 0$ for (\ref{dz}) as
\begin{eqnarray}
\label{Gamsize}
\langle |\varGamma|\rangle\sim \left(\delta z\right)^{-1}.
\end{eqnarray}
Finally, the average size of the boundary line $\partial G$ scales as
\begin{eqnarray}
\label{dGsize}
\langle |\partial G|\rangle\sim \left(\delta \tilde z\right)^{-1}.
\end{eqnarray}
Hereafter we shall keep the simplified notations $|G|$, $|\partial G|$ and $|\varGamma|$ for those averages.
The fugacities $\delta \beta$ and $\delta z$ associated with the two random sets $G$ and $\varGamma$
are expected to have the relative scaling  \cite{DK}
\begin{equation}
(\delta z)^{2\nu} \sim \delta \beta,
\label{dbdz}
\end{equation}
where $\nu$ (denoted $\nu_2 D/2$ in  \cite{DK}) is a critical exponent to be determined.
It depends on the nature of the critical phase. Thus the sizes of the
two random sets $G$ and $\varGamma$ are expected to have the relative scaling  \cite{DK}
\begin{equation}
|\varGamma|^{2\nu} \sim |G|,
\label{GGam}
\end{equation}

For Dirichlet boundary conditions, we have seen that the boundary line and the watermelon lines are
similar, and eqs. (\ref{bB}),  (\ref{Gamsize}), and(\ref{dGsize}) give
\begin{eqnarray}
|\partial G| \sim \left(\delta \tilde z\right)^{-1} \sim \left(\delta z\right)^{-1} \sim|\varGamma|,
\label{dGGam}
\end{eqnarray}
so that we deal with the finite-size scaling regime
\begin{eqnarray}
\left(\delta \tilde z\right)^{2\nu} \sim \left(\delta z\right)^{2\nu} \sim \delta \beta
\label{FSS'}
\end{eqnarray}
or, equivalently
\begin{eqnarray}
|\partial G|^{2\nu}\sim |\varGamma|^{2\nu} \sim |G|.
\label{FSS}
\end{eqnarray}

\subsubsection{String Susceptibility Exponent}
Let us first consider the partition function (\ref{BZchi1}) in the scaling regime. The string susceptibility
exponent $\gamma_{\rm str}(\chi)$ is defined by the scaling behavior of the partition function:
\begin{eqnarray}
\label{Zgammastring}
Z_{O(N)}(\beta, K_c, \chi)&\sim& (\delta\beta)^{2-\gamma_{\rm str}(\chi)},
\end{eqnarray}
where the exponent depends on the topology. Because of the scaling (\ref{Gsize}), one also has
\begin{eqnarray}
\label{Zgammastringbis}
Z_{O(N)}(\beta, K_c, \chi)&\sim& |G|^{\gamma_{\rm str}(\chi)-2},
\end{eqnarray}
The two-puncture partition function (\ref{BZsec}) scales accordingly as:
\begin{eqnarray}
\frac{\partial^{2}}{\partial \beta ^{2}} Z_{O(N)}(\beta, K_c, \chi)\sim (\delta\beta)^{-\gamma_{\rm str}(\chi)}
\sim|G|^{\gamma_{\rm str}(\chi)}.
\label{BZsecgamm}
\end{eqnarray}

The value of $\gamma_{\rm str}$ can be derived from the exact solution of the $O(N)$ model  \cite{kostov,DK}
(actually here also for the simple topologies of the sphere or of the disk from the scaling equations above).
The general formula for $\gamma_{\rm str}$ is also known from the Liouville theory  \cite{david2}.
It reads in terms of the central charge $c$:
\begin{eqnarray}
\label{gammstr}
\gamma_{\rm str}(\chi)=2-\frac{\chi}{24}  \left(25 -c +\sqrt{(1-c)(25-c)}\right).
\end{eqnarray}
A particularly important one is the string susceptibility exponent of the spherical topology
$\gamma\equiv\gamma_{\rm str}(\chi=2)$, in terms of which the KPZ formula is written.
For other topologies we have:
\begin{eqnarray}
\label{gammstr2}
\gamma_{\rm str}(\chi)-2=\frac{\chi}{2} \left[\gamma_{\rm str}(\chi=2)-2\right]
=\frac{\chi}{2} \left(\gamma-2\right).
\end{eqnarray}
In particular, for the disk topology, $\chi=1$, we find:
\begin{eqnarray}
\label{gammstr1}
\gamma_{\rm str}(\chi=1)=1+\frac{1}{2}\gamma.
\end{eqnarray}

\subsubsection{Bulk Conformal Dimensions}
We first consider the planar topology ($\chi=2$).  The two-puncture partition function (\ref{BZfig})
and (\ref{BZsecgamm}) then scales as
\begin{eqnarray}
Z_{O(N)}\lbrack \hbox to 8.5mm{\hskip 0.5mm
$\vcenter{\epsfysize=.45truecm\epsfbox{fig2.eps}}$
              \hskip 10mm}
 \rbrack
 \sim |G|^{\gamma_{\rm str}(\chi=2)}.
\label{BZfiggamm}
\end{eqnarray}
In quantum gravity, the {\it bulk conformal weight} $\Delta_{L}$  is then defined by the scaling of the
two-point function $Z_{O(N),L}$, properly normalized, with respect to the average size of the lattice:
\begin{equation}
Z_{O(N),L}\sim Z_{O(N)}\lbrack \hbox to 8.5mm{\hskip 0.5mm
$\vcenter{\epsfysize=.45truecm\epsfbox{fig2.eps}}$
              \hskip 10mm}
 \rbrack \ \times \left| G\right|
^{-2\Delta_{L}}\ .
\label{BZldelta}
\end{equation}
This definition takes into account the fact that the  order $L$ two-point partition function
$Z_{O(N),L}$ contains, in addition to
the insertion of two operators creating $L$ lines, the two-puncture partition function on the sphere.
Thus we have by definition:
\begin{equation}
Z_{O(N),L}\sim  \left| G\right|
^{\gamma_{\rm str}(\chi=2)-2\Delta_{L}}.
\label{BZldeltag}
\end{equation}

\subsubsection{Boundary Conformal Dimensions}
For measuring length scales along a fluctuating boundary, there are two natural possibilities:
 use the boundary length $|\partial G|$ or the square root $\sqrt{|G|}$ of the fluctuating area.
They differ a priori 
since we have seen that $|G| \sim |\partial G|^{2\nu}$, with $\nu \leq 1$. In boundary quantum gravity, the proper
conformal weights are defined in terms of the effective length $\sqrt {|G|} \sim |\partial G|^{\nu}$.

Let us first consider the disk partition function with two punctures (\ref{diskpunc}). We write its scaling as:
\begin{equation}
\label{diskpuncs}
\tilde Z_{{O(N)},L=0} =\tilde Z_{{O(N)}}( \hbox to 9.5mm{\hskip 0.5mm
$\vcenter{\epsfysize=.45truecm\epsfbox{fig3.eps}}$
              \hskip -80mm})
	      \sim |G|^{\gamma_{\rm str}(\chi=1)-2} \times \left({\sqrt {|G|}}\right)^{2-2 \tilde \Delta_0}.
\end{equation}
The first power law accounts for the scaling of the disk partition function (\ref{Zgammastringbis}) for an
Euler characteristic $\chi =1$, with a susceptibility exponent
$\gamma_{\rm str}(\chi=1)$, while the second term takes into account the integration  and the conformal weight $\tilde \Delta_0$
of the two punctures along the boundary.

For the order $L$ boundary partition function $\tilde Z_{O(N), L}$, we similarly define {\it boundary conformal weights}
$\tilde \Delta_L$, such that:
\begin{equation}
\label{tZltD}
\tilde Z_{{O(N)},L}
	      \sim |G|^{\gamma_{\rm str}(\chi=1)-2} \times \left({\sqrt {|G|}}\right)^{2-2 \tilde \Delta_L}.
\end{equation}
We thus have the simple scaling relation
\begin{equation}
\label{tZL/0}
\frac{\tilde Z_{O(N), L}}{\tilde Z_{{O(N)}}( \hbox to 9.5mm{\hskip 0.5mm
$\vcenter{\epsfysize=.45truecm\epsfbox{fig3.eps}}$
              \hskip -80mm})}=\frac{\tilde Z_{O(N), L}}{\tilde Z_{{O(N)},L=0}}
\sim \left({\sqrt {|G|}}\right)^{2 \tilde \Delta_0- 2 \tilde \Delta_L}.
\end{equation}

Measuring scaling behavior in terms of the boundary length $|\partial G|$ leads to the definition of
another set of boundary quantum gravity dimensions $\tilde \Delta'_L$, such that
\begin{equation}
\label{tZltD'}
\tilde Z_{{O(N)},L}
	      \sim |G|^{\gamma_{\rm str}(\chi=1)-2} \times {|\partial G|}^{2-2 \tilde \Delta'_L},
\end{equation}
or, equivalently
\begin{equation}
\label{tZL/0'}
\frac{\tilde Z_{O(N), L}}{\tilde Z_{{O(N)}}( \hbox to 9.5mm{\hskip 0.5mm
$\vcenter{\epsfysize=.45truecm\epsfbox{fig3.eps}}$
              \hskip -80mm})}=\frac{\tilde Z_{O(N), L}}{\tilde Z_{{O(N)},L=0}}
\sim {|\partial G|}^{2 \tilde \Delta'_0- 2 \tilde \Delta'_L}.
\end{equation}
Because of the FSS relation (\ref{FSS}) we immediately have:
\begin{equation}
\label{d'd}
\tilde \Delta'_L-1= \nu \, (\tilde \Delta_L-1).
\end{equation}

\subsubsection{The Exponent $\nu$}
The scaling exponent $\nu$ can be related to the bulk conformal weight $\Delta_2$ and to the boundary conformal
weight $\tilde \Delta_0$ as follows. The derivative operator ${\partial}/{\partial z}$ inserts
a factor $|\varGamma| \sim |G|^{1/2\nu}$ equal to the total length of the random lines $\varGamma$
in partition functions. But by definition it also geometrically represents
the insertion of a puncture on the watermelon lines, with a local Gibbs weight $|G|^{1-\Delta_2}$, since
$L=2$ random strands originate from this puncture. We therefore conclude that:
\begin{equation}
\label{nudelta2}
\frac{1}{2\nu} = 1-\Delta_2.
\end{equation}
Similarly, differentiating a boundary partition function with respect to the boundary fugacity $\tilde z$
with the operator ${\partial}/{\partial \tilde z}$ gives a factor $|\partial G|\sim |G|^{1/2\nu}$ in that partition function.
But it also represents the insertion of
a puncture operator along the boundary line, with by definition a local Gibbs weight $\left(\sqrt {|G|}\right)^{1-\Delta_0}$.
We therefore arrive at the second identity
\begin{equation}
\label{nutdelta0}
\frac{1}{\nu} = 1-\tilde \Delta_0.
\end{equation}
 A first general identity follows:
\begin{equation}
\label{delta2tdelta0}
2\Delta_2= 1+\tilde \Delta_0,
\end{equation}
which we see as characteristic of Dirichlet boundary conditions in quantum gravity. It means that the fractal dimension of random lines 
in the bulk is the same as that of boundary lines.

\subsubsection{Bulk-Boundary Relation between Conformal Weights}
In the above we observed the scaling relation (\ref{Bratio}) between the two-point function on the sphere and those on the
disk. Relations (\ref{BZldeltag}) and (\ref{tZL/0}) immediately imply the general identity between
boundary and bulk exponents:
\begin{equation}
{\tilde \Delta}_{L}-{\tilde \Delta}_{0}=2\Delta _{L}-\gamma _{{\rm str}}(\chi =2).
\label{Bdeltat}
\end{equation}

\subsubsection{Recursion for Conformal Weights}
We also observed the  scaling recursion relation (\ref{L+1L}) between boundary partition functions, which,
owing to (\ref{dGGam}) and (\ref{FSS}) can be rewritten as
\begin{equation}
\label{L+1L'}
\frac{\tilde Z_{O(N), L+1}}{\tilde Z_{{O(N)}}( \hbox to 9.5mm{\hskip 0.5mm
$\vcenter{\epsfysize=.45truecm\epsfbox{fig3.eps}}$
              \hskip -80mm})}=\tilde Z_{O(N), L} \times |G|^{-{1}/{2\nu}}.
\end{equation}
It is now sufficient to apply  definitions (\ref{tZltD}) and (\ref{tZL/0}) to get the recursion relation
\begin{equation}
{\tilde \Delta}_{0}-{\tilde \Delta}_{L+1}=1-\tilde \Delta_{L}+[\gamma _{{\rm str}}(\chi =1)-2]-\frac{1}{2\nu}.
\label{recur}
\end{equation}
We finally use (\ref{gammstr2}) for $\gamma _{{\rm str}}(\chi =1)-2=\frac{1}{2}[\gamma _{{\rm str}}(\chi =2)-2]$, and the expression 
(\ref{nutdelta0}) of $\nu$ to arrive at the recursion:
\begin{equation}
{\tilde \Delta}_{L+1}=\tilde \Delta_{L}+\frac{1}{2}\left[\tilde \Delta_{0} +1-\gamma _{{\rm str}}(\chi =2)\right],
\label{recur'}
\end{equation}
which finally gives
\begin{equation}
{\tilde \Delta}_{L}=\tilde \Delta_{0}+\frac{L}{2}\left[\tilde \Delta_{0} +1-\gamma _{{\rm str}}(\chi =2)\right].
\label{tdeltafinal}
\end{equation}
The bulk conformal weights  immediately follow from (\ref{Bdeltat}):
\begin{equation}
 \Delta_{L}=\frac{L}{4}\left[\tilde \Delta_{0} +1-\gamma _{{\rm str}}(\chi =2)\right]+\frac{1}{2}\gamma _{{\rm str}}(\chi =2).
\label{deltafinal}
\end{equation}
Notice that for $L=2$, one has identically $\Delta_2=\frac{1}{2}(\Delta_0 +1)$, in agreement with  (\ref{delta2tdelta0}). 

\subsubsection{Dilute and Dense Phases}
At this stage, it is remarkable  that the multi-line exponents of the $O(N)$ model (\ref{tdeltafinal}) and (\ref{deltafinal}) 
have been entirely determined from the existence and structure of the spectral representation of the
two-point functions (\ref{BZlint}), and (\ref{BZtildeint}) implying necessary scaling relations, and without using the precise scaling behavior (\ref{density}) of the eigenvalue density, nor the equivalent
forms (\ref{Zscalingexp}) and (\ref{tZscaling}). 
 
The exponents so found (\ref{tdeltafinal}) and (\ref{deltafinal}) depend on only one unknown, the boundary conformal weight
$\tilde \Delta_0$ associated 
with a boundary puncture on a random surface. In the Euclidean half-plane $\mathbb H$, its scaling dimension is given
by the KPZ equation:
\begin{equation}
\label{tdelta0}
\tilde x_0 =U_{\gamma}(\tilde \Delta_0)=\tilde \Delta_0 \frac{\tilde \Delta_0 -\gamma}{1-\gamma},
\end{equation} 
where $\gamma=\gamma_{\rm str}(\chi=2)$. It corresponds to a point insertion along $\partial \mathbb H$, and therefore
is expected to vanish, so that
$\tilde \Delta_0$ is  solution of $\tilde x_0=U_{\gamma}(\tilde \Delta_0)=0$. One solution is trivially $\tilde \Delta_0=0$,
while the other is negative: $\tilde \Delta_0=\gamma$. We thus find from (\ref{nutdelta0}) the two possibilities:
\begin{eqnarray}
\left\{\begin{array}{ll} {\tilde \Delta_{0}}=0,  & \mbox{$\nu =1\;$ (I)} \\
\\
 {\tilde \Delta_{0}}=\gamma, & \mbox{$\nu =\frac{1}{1-\gamma}\;$ (II).}
\end{array}
\right.
\label{tdelta0cornelien}
\end{eqnarray}
This is where  the nature of the critical phase enters. In case I above, $\nu=1$ gives $|\partial G| \sim \sqrt {|G|}$, i.e.,
the boundary is smooth,
with a fractal dimension half that of the random surface. This is what is expected for the {\it dilute phase}. 
Another way to describe the 
dilute phase   is to balance the two scaling terms in (\ref{BZlint}) or (\ref{BZtildeint}), requiring the same scaling 
for both  \cite{DK}. This gives 
$\delta \beta \sim (\delta z)^2$, independently of $\theta$, thus from (\ref{FSS'}) $\nu =1$, as announced. 
Thus one is led to conclude that case II above must correspond to the {\it dense phase}. 

\subsubsection{Quantum Gravity Conformal Weights}
It is interesting to note that the relation (\ref{Bdeltat}) between bulk and boundary conformal weights then takes two different forms, 
depending on the nature of the critical phase:
\begin{eqnarray}
\left\{\begin{array}{ll}  2\Delta_L-\gamma =\tilde \Delta_L, & \mbox{${\tilde \Delta_{0}}=0,\; {\Delta_{0}}=\frac{1}{2}\gamma,\;\nu =1\;$ (dilute)} \\
\\
 2\Delta_L =\tilde \Delta_L,                          & \mbox{${\tilde \Delta_{0}}=\gamma,\; {\Delta_{0}}=\frac{1}{2}\gamma,\;\nu =\frac{1}{1-\gamma}\;$ (dense).}
\end{array}
\right.
\label{tddcornelien}
\end{eqnarray}
The final expressions of the boundary conformal weights follow from (\ref{tdeltafinal}):
\begin{eqnarray}
{\tilde \Delta}_{L}=\left\{\begin{array}{ll} \frac{L}{2}(1-\gamma)=\Delta^{(\gamma)}_{L+1,1},  & \mbox{${\tilde \Delta_{0}}=0,\;\nu =1\;$ (dilute)} \\
\\
  \gamma+\frac{L}{2}=\Delta^{(\gamma)}_{1,L+1},& \mbox{${\tilde \Delta_{0}}=\gamma,\;\nu =\frac{1}{1-\gamma}$\; (dense),}
\end{array}
\right.
\label{tdeltacornelien}
\end{eqnarray}
while the bulk conformal weights  immediately follow from (\ref{deltafinal}):
\begin{eqnarray}
 {\Delta}_{L}=\left\{\begin{array}{ll} \frac{1}{2}\gamma+\frac{L}{4}(1-\gamma)=\Delta^{(\gamma)}_{L/2,0},  
 & \mbox{${\Delta_{0}}=\frac{\gamma}{2},\;\nu =1\;$ (dilute)} \\
 \\
  \frac{1}{2}\gamma+\frac{L}{4}=\Delta^{(\gamma)}_{0,L/2},& \mbox{${\Delta_{0}}=\frac{\gamma}{2},\;\nu =\frac{1}{1-\gamma}$\; (dense).}
\end{array}
\right.
\label{deltacornelien}
\end{eqnarray}
Hence we have established all the $O(N)$ watermelon conformal weights in boundary and bulk quantum gravity, 
namely  equations (\ref{tdeltaL} - \ref{deltaLD}), {\bf QED}.

\subsubsection{Dual Conformal Weights}
In section 12.2.1, we introduced the {\it dual} $\Delta'$ of a conformal weight $\Delta$ by eq. (\ref{dualdelta})
\begin{equation}
\Delta' = \frac{\Delta-\gamma}{1-\gamma},
\label{dualdelta'}
\end{equation}
such that 
$U_{\gamma}(\Delta)=\Delta \Delta'.$ These dual dimensions are natural in the description of the dense phase. Indeed, while 
we stated in (\ref{sa}) or (\ref{nsa}) that in the dilute phase (or for simple SLE paths) the boundary scaling dimensions are additive, 
their duals retain the additivity property in the dense phase (or for non-simple SLE paths), a fact which will be established 
in all generality in the next appendix. 

The nature of these dual dimensions remained slightly mysterious, however. It is interesting to 
observe their appearance in eq.  (\ref{tZltD'}). Indeed, when 
measuring lengths there
 in terms of the  boundary length $|\partial G|$, instead of the square root $\sqrt { |G|} \sim |\partial G|^{\nu}$ of the area, 
 a new exponent $\tilde \Delta'_L$ appeared, such that
 \begin{equation}
 \label{D'L}
\tilde \Delta'_L=1+\nu (\tilde \Delta_L-1)=\frac{\tilde \Delta_L-\tilde \Delta_0}{1-\tilde \Delta_0}.
\end{equation}
 
${\bullet}$ In the {\it dilute phase}, 
the two boundary lengths  scale in the same way, $\nu =1$, $\tilde \Delta_0=0$, and the boundary conformal weights  are unchanged:
$\tilde \Delta'_L=\tilde \Delta_L.$
   
${\bullet}$ In the {\it dense phase},  we have found $\tilde \Delta_0=\gamma,\; \nu = \frac{1}{1-\gamma} \leq 1$, and 
the new dimension  $\tilde \Delta'_L$ (\ref{D'L}) is just the {\it dual} one (\ref{dualdelta'})
 \begin{equation}
 \label{D''L}
\tilde \Delta'_L=\frac{\tilde \Delta_L-\tilde \Delta_0}{1-\tilde \Delta_0}=\frac{\tilde \Delta_L-\gamma}{1-\gamma}.
\end{equation}
  Furthermore, since the dual boundary puncture dimension $\tilde \Delta'_0=0$ in the dense phase, the cyclicity of eq. (\ref{tZL/0'})  clearly shows the dual exponents 
 $\tilde \Delta'_L$  to be {\it linear} in $L$, in agreement with (\ref{nsa}), {\bf QED}. 
 Indeed eq. (\ref{tdeltacornelien} gives for the dense phase:
 \begin{equation}
 \label{D'''L}
\tilde \Delta'_L=\frac{1}{1-\gamma}\frac{L}{2}={(1-\gamma')}{\frac{L}{2}},
\end{equation}
in agreement with duality eqs. (\ref{dualgamma}), (\ref{Kacdual}), (\ref{tdeltaL}),  and (\ref{tdeltaLD}) 
of section ${\bf 12}$, {\bf QED}.
 
\subsubsection{Coulomb Gas Formulae}
For pedagogical purposes, we have given above a derivation of the exponents based on general principles only, in terms of the 
susceptibilty exponent $\gamma$. We could also have derived these results from expressions (\ref{Zscalingexp}) and 
(\ref{tZscaling}) in terms of parameter $\theta$ (\ref{parametrisation}). This analysis again leads to distinguish the 
two cases of dilute and dense phases, and one respectively finds for bulk and boundary exponents:
\begin{eqnarray}
\left\{\begin{array}{ll}  
\Delta_L=\frac{1}{4}(1+\theta)-\frac{1}{2}\theta & \mbox{$\gamma=-\theta,\;\nu =1\; $ (dilute)} \\
\\
 \Delta_L =\frac{1}{4}L-\frac{1}{2}\frac{\theta}{(1-\theta)}  
            & \mbox{$\gamma=-\frac{\theta}{1-\theta},\;\nu ={1-\theta}\;$ (dense).}
\end{array}
\right.
\label{thetacornelien}
\end{eqnarray}
\begin{eqnarray}
\left\{\begin{array}{ll} \tilde \Delta_L=2\Delta_L-\gamma=\frac{1}{2}L(1+\theta),\;  
 & \mbox{$\gamma=-\theta,\;\nu =1\; $ (dilute)} \\
\\
 \tilde \Delta_L=2\Delta_L=\frac{1}{2}L-\frac{\theta}{(1-\theta)}, \;
   
            & \mbox{$\gamma=-\frac{\theta}{1-\theta},\;\nu ={1-\theta}\;$ (dense).}
\end{array}
\right.
\label{tthetacornelien}
\end{eqnarray}
In terms of the Coulomb gas coupling constant, $g$, parametrizing the $O(N)$ model, one has 
\begin{eqnarray}
N&=&2 \cos \pi \theta,\;\;\;\; 0 \leq \theta \leq 1\\
&=&-2 \cos \pi g \left\{\begin{array}{ll} g\in [1,2] & \mbox{$$ (dilute)} \\
\\   g\in [0,1] & \mbox{$$ (dense).}
\end{array}
\right.
\label{gcornelien}
\end{eqnarray}
We thus have:
\begin{eqnarray}
g=\left\{\begin{array}{ll} 1+\theta \in [1,2],\;\gamma=1-g & \mbox{$$ (dilute)} \\
\\ 1-\theta \in [0,1],\; \gamma=1-g^{-1} & \mbox{$$ (dense).}
\end{array}
\right.
\label{gtcornelien}
\end{eqnarray}
Finally, the dilute and dense exponents can be recast in terms of the Coulomb gas coupling constant. The bulk ones read as  
\begin{eqnarray}
\left\{\begin{array}{ll}  
\Delta_L=\frac{1}{4}gL+\frac{1}{2}(1-g) & \mbox{$\gamma=1-g,\;\nu =1\; $ (dilute)} \\
\\
 \Delta_L =\frac{1}{4}L+\frac{1}{2}(1-g^{-1})  
            & \mbox{$\gamma=1-g^{-1},\;\nu =g\;$ (dense),}
\end{array}
\right.
\label{Lgcornelien}
\end{eqnarray}
and the boundary ones 
\begin{eqnarray}
\left\{\begin{array}{ll} \tilde \Delta_L=2\Delta_L-\gamma=\frac{1}{2}gL,\;  
 & \mbox{$\gamma=1-g,\;\nu =1\; $ (dilute)} \\
\\
 \tilde \Delta_L=2\Delta_L=\frac{1}{2}L+1-g^{-1}, \;
   
            & \mbox{$\gamma=1-g^{-1},\;\nu =g\;$ (dense).}
\end{array}
\right.
\label{tLgcornelien}
\end{eqnarray}

\subsubsection{Conformal Weights for the Stochastic L\"owner Evolution}
The quantum gravity conformal weights for the ${\rm SLE}_{\kappa}$ are obtained either directly from the Coulomb gas ones above, 
by the simple 
substitution $\kappa=4/g$, or by using in eqs. (\ref{dualgamma}), (\ref{Kacdual}), (\ref{tdeltaL}),  and (\ref{tdeltaLD}) 
the parametrization $\gamma=1-4/\kappa$ for the dilute phase, i.e., for simple paths ${\rm SLE}_{\kappa \leq 4}$, and $\gamma=1-\kappa/4$ for 
the dense phase, i.e., for non-simple paths $\rm{SLE}_{\kappa \geq 4}$. We find:
\begin{eqnarray}
{\tilde \Delta}_{L}=\left\{\begin{array}{ll} \frac{2L}{\kappa}=\Delta^{\kappa}_{L+1,1},  
& \mbox{$\nu =1\; (\kappa \leq 4)$} \\
\\
  \frac{L}{2}+1-\frac{\kappa}{4}=\Delta^{\kappa'=16/\kappa}_{1,L+1},
  & \mbox{$\nu =\frac{4}{\kappa}\; (\kappa \geq 4),$}
\end{array}
\right.
\label{tkdeltacornelien}
\end{eqnarray}
while the bulk conformal weights are:
\begin{eqnarray}
 {\Delta}_{L}=\left\{\begin{array}{ll} \frac{1}{2}(1-\frac{4}{\kappa})+\frac{L}{\kappa}=\Delta^{\kappa}_{L/2,0},  
 & \mbox{$\nu =1\; (\kappa \leq 4)$} \\
 \\
  \frac{1}{2}(1-\frac{\kappa}{4})+\frac{L}{4}=\Delta^{\kappa'=16/\kappa}_{0,L/2,},& \mbox{$\nu =\frac{4}{\kappa}\; (\kappa \geq 4),$}
\end{array}
\right.
\label{kdeltacornelien}
\end{eqnarray}
where we recall the definition (\ref{KPZSLEinvh}), $$\Delta^{\kappa}_{p,q}=\frac{|4p-\kappa q|+\kappa-4}{2\kappa}.$$
These are eqs. (\ref{tdeltaLk}) and (\ref{tdeltaLDk}) or, equivalently, (\ref{tdeltaLkk}) to (\ref{deltaLDkd}), {\bf QED}.
\\

\section{\sc{Boundary-Bulk Exponents \& Boundary Fusion Rules in Quantum Gravity}}
\label{BBapp}

\subsection{Structural Conformal Weight Relations}
The aim of this last appendix is to establish two different basic scaling relations in quantum gravity, which we often encountered 
in calculating critical exponents.
 
\subsubsection{$\bullet$ From the Boundary to the Bulk}
 The first one concerns the relation between quantum gravity bulk conformal weights $\Delta$,
and their Dirichlet boundary counterparts $\tilde \Delta$:
\begin{equation} 
2\Delta-\gamma_{\rm str}(\chi=2)=\tilde \Delta -\tilde \Delta_0,
\label{C1}
\end{equation} 
where $\gamma_{\rm str}(\chi=2)=\gamma$ is the string susceptibility exponent of the random surface with the sphere topology, 
bearing a certain conformal field theory of central charge $c(\gamma)$, and where $\tilde \Delta_0$ 
is the conformal weight of the boundary puncture operator.

\subsubsection{$\bullet$ Dirichlet Random Sets and Conformal Weight Additivity}
 The second one is the additivity property of boundary conformal weights, $\tilde \Delta_A$ and $\tilde \Delta_B$,
corresponding to two random sets $A$ and $B$ living on the random disk, and mutually-avoiding each other, namely experiencing 
mutual Dirichlet conditions, in addition to those felt at the disk boundaries. This restriction has been noted $A \wedge B$ 
throughout the paper. This additivity property reads, in its general setting:
\begin{equation} 
\tilde \Delta_{A\wedge B} -\tilde \Delta_0=\tilde \Delta_A -\tilde \Delta_0+\tilde \Delta_B -\tilde \Delta_0.
\label{C2}
\end{equation} 

\subsubsection{KPZ Original and $\kappa$-Modified Maps}
We recall the form of the function involved in the KPZ equation:
\begin{equation}
h=U_{\gamma}(\Delta)=\Delta \frac{\Delta -\gamma}{1-\gamma},
\label{C3}
\end{equation}
which maps the quantum conformal weights $\Delta$ onto (classical) Ka\v c-like conformal weights $h$ in the plane or half-plane. 
 For the SLE process, we introduced a modified KPZ map:
\begin{equation}
{\mathcal U}_{\kappa}(\Delta)=\frac{1}{4} \Delta (\kappa\Delta +4-\kappa),
\label{C4}
\end{equation}
with inverse:
\begin{equation}
{\mathcal U}_{\kappa}^{-1}\left( x\right)=
\frac{1}{2\kappa}\left(\sqrt{16\kappa x+(\kappa-4)^2}+\kappa-4\right).
\label{C6}
\end{equation}

\subsubsection{Conformal Weight of the Boundary Puncture}
We have 
seen in appendix \ref{ONapp} that  
$\tilde \Delta_0$ satisfies 
\begin{equation}
U_{\gamma}(\tilde \Delta_0)=0,
\label{C7}
\end{equation}
and takes one of the two values $(0,\gamma)$, depending whether one is at a dilute critical point, or in a dense phase.

\subsubsection{Additivity Rule for the ${\rm SLE}_{\kappa}$ Process}
As shown in section \ref{sec.SLEKPZ}, when using the ${\mathcal U}_{\kappa}^{-1}$ inverse function, instead of the original
one $U_{\gamma}^{-1}$, the additivity rule (\ref{C2}) can be recast in a unique formula, 
independently of the value of $\tilde \Delta_0$, and of the range of $\kappa$:
\begin{equation}
{\mathcal U}_{\kappa}^{-1}(\tilde x_{A\wedge B})={\mathcal U}_{\kappa}^{-1}(\tilde x_{A})+
{\mathcal U}_{\kappa}^{-1}(\tilde x_{ B}),
\label{C10}
\end{equation}
where the conformal weights $\tilde x$ are those on the boundary of the half-plane $\mathbb H$
\begin{equation}
\tilde x=U_{\gamma}(\tilde \Delta).
\label{C11}
\end{equation}

\subsection{Partition Functions}

\subsubsection{Two-Puncture Partition Functions}
We consider a random lattice $G$ bearing a given statistical system, whose critical properties correspond to a conformal field theory 
with central charge $c$, and string susceptibility exponent $\gamma$. The two-puncture partition 
function in the spherical topology reads
\begin{eqnarray}
\label{CZsec}
Z\lbrack \hbox to 8.5mm{\hskip 0.5mm
$\vcenter{\epsfysize=.45truecm\epsfbox{fig2.eps}}$
              \hskip 10mm}
 \rbrack \ =\frac{\partial^{2}}{\partial \beta ^{2}}Z(\beta)
=\sum _{{\rm planar}\ G}\frac{1}{ S(G)} \left| G\right|^2 e^{-\beta
\left| G\right|}W(G),
\end{eqnarray}
where $W(G)$  is the weight due to the background statistical system bore by the random lattice,
depending on some unspecified fugacities, and where  $Z$ is the partition function of 
the random lattice and statistical system in the absence of punctures.
The two-puncture boundary partition function is, in a similar way:
\begin{equation}
\tilde Z( \hbox to 9.5mm{\hskip 0.5mm
$\vcenter{\epsfysize=.45truecm\epsfbox{fig3.eps}}$
              \hskip -80mm})=\left(\tilde z\frac{\partial}{\partial \tilde z }\right)^2 
              \tilde Z (\beta , \tilde z)=\sum _{ {\rm disk}\ G}
e^{-\beta \left| G\right|} W(G) \left| \partial G\right|^2 \tilde z^{\left| \partial G\right|},
\label{CZtilde}
\end{equation}
where $\tilde Z$ is the disk partition function without punctures.
\subsubsection{General Two-Point Partition Functions}
Imagine a random set $A$ on a random lattice $G$. We have in mind random sets $A$ like
the random trees of appendix \ref{Brownapp}, the random lines
of the $O(N)$ model in appendix \ref{ONapp}, or random paths similar to the SLE process, i.e.,
frontier hulls of Potts or $O(N)$ models.  One can then define two-point partition functions,
which connect two arbitrary points $i$ and $j$ in $G$ (Fig. \ref{Fig.wm2bis}), as the watermelon ones
$Z_{O(N),L}$ (\ref{BZl})
or $\tilde Z_{O(N),L}$ (\ref{BZtilde}) in appendix \ref{ONapp}.

\begin{figure}[tb]
\begin{center}
\includegraphics[angle=0,width=.5\linewidth]{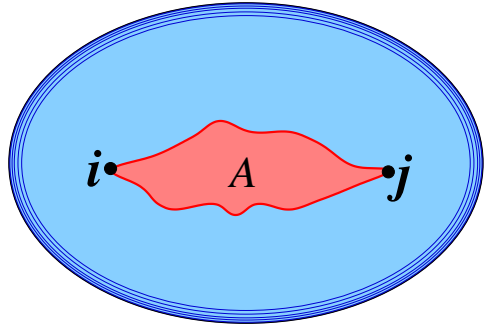}
\end{center}
\caption{Two-point correlator of a random set $A$ on a random sphere,
corresponding to the partition function (\ref{C12}).}
\label{Fig.wm2bis}
\end{figure}
\begin{figure}[tb]
\begin{center}
\includegraphics[angle=0,width=.5\linewidth]{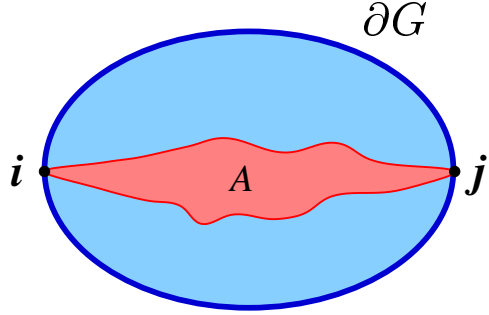}
\end{center}
\caption{Two-point boundary correlator of a random set $A$ on a random disk,
corresponding to the partition function (\ref{CZtildeA}).}
\label{Fig.wm4bis}
\end{figure}
The existence of the random set $A$ at a given point $i\in G$ is
associated with a conformal field operator $\Phi_A(i)$ creating
the process at $i$, and belonging to the conformal field theory borne by $G$. We write the two-point
partition function in a symbolic way
\begin{eqnarray}
\nonumber
Z_A&:=&\langle \sum_{i,j \in G} \Phi_A(i)\Phi_A(j)\rangle\\
&:=&\sum _{{\rm planar}\ G}\frac{1}{ S(G)}
e^{-\beta \left| G\right|} W(G) \sum _{i,j\in G}
\sum_{\scriptstyle \{A_{ij}\}}
 z^{\left| A_{ij}\right| };
\label{C12}
\end{eqnarray}
where 
the average $\langle \cdots \rangle$ is calculated with grand-canonical Gibbs weights;
$W(G)$ again is the weight of the random lattice bearing the background statistical system,
depending on some associated fugacities, and $z$ is a fugacity associated with the number of sites occupied by random set 
$A_{ij}$ between
$i$ and $j$.

Similarly, the two-point boundary partition function reads as in (\ref{BZtilde}) (Fig. \ref{Fig.wm4bis}):
\begin{eqnarray}
\nonumber
\tilde Z_{A}&:=&\langle \sum_{i,j \in \partial G}  \tilde \Phi_A(i)\tilde \Phi_A(j)\rangle\\
&:=&\sum _{ {\rm disk}\ G}
e^{-\beta \left| G\right|} W(G)\,\tilde z^{\left| \partial G\right|}
\sum _{{i,j} \in \partial G}\sum_{\scriptstyle
\{A_{ij}\}} z^{\left| A_{ij}\right|},
\label{CZtildeA}
\end{eqnarray}
where $\tilde {z}$ is the new fugacity associated with the boundary's length.

\begin{figure}[tb]
\begin{center}
\includegraphics[angle=0,width=.5\linewidth]{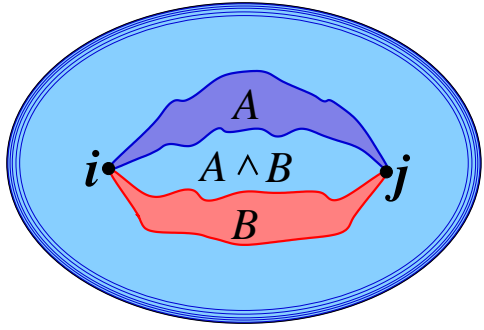}
\end{center}
\caption{Two-point correlation of mutually-avoiding random sets $A\wedge B$ on a random sphere,
as they appear in the partition function (\ref{C12AB}).}
\label{wm2}
\end{figure}

Consider now two sets $A$ and $B$ starting and ending at the same vertices $i$ and $j$ on $G$, and in a mutually-avoiding 
configuration $A\wedge B$ (Fig. \ref{wm2}). Their two-point partition function is defined as:
\begin{eqnarray}
\nonumber
Z_{A\wedge B}&:=&\langle \sum_{i,j \in G} \Phi_{A\wedge B}(i)\Phi_{A \wedge B}(j)\rangle\\
&:=&\sum _{{\rm planar}\ G}\frac{1}{ S(G)}
e^{-\beta \left| G\right|} W(G) \sum _{i,j\in G}
\sum_{\scriptstyle \{A\wedge B|_{ij}\}, }
 z^{\left| A_{ij}\right|+\left| B_{ij}\right|},
\label{C12AB}
\end{eqnarray}
where a common fugacity is attributed to the total number of sites of $A\wedge B$ between $i$ and $j$, 
$|A\wedge B|_{ij}|=\left| A_{ij}\right|+\left| B_{ij}\right|$. 

\begin{figure}[tb]
\begin{center}
\includegraphics[angle=0,width=.5\linewidth]{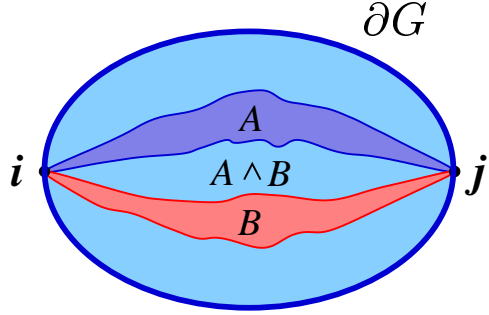}
\end{center}
\caption{Two-point boundary correlation of mutually-avoiding random sets $A\wedge B$ on a random disk,
as they appear in the partition function (\ref{CZtildeAB}).}
\label{wm4}
\end{figure}

The boundary partition function is similarly (Fig. \ref{wm4}):
\begin{eqnarray}
\nonumber
\tilde Z_{A\wedge B}&:=&\langle \sum_{i,j \in \partial G}  \tilde \Phi_{A\wedge B}(i)\tilde \Phi_{A\wedge B}(j)\rangle\\
&:=&\sum _{ {\rm disk}\ G}
e^{-\beta \left| G\right|} W(G)\,\tilde z^{\left| \partial G\right|}
\sum _{{i,j} \in \partial G}\sum_{\scriptstyle
\{A_{ij}\}} z^{\left| A_{ij}\right|+\left| B_{ij}\right|}.
\label{CZtildeAB}
\end{eqnarray}

\subsection{Spectral representation}

\subsubsection{$\bullet$ Partition Function in Spherical Topology:} In the random lattice representations by eigenvalue densities of random matrices
like those used in appendices  \ref{Brownapp} and \ref{ONapp},
the two-point partition function in spherical topology in general can be written as in (\ref{AZlint}) or (\ref{BZlint})
\begin{equation}
Z_{A}=\int
d\lambda \, \rho(\lambda)
\, {\mathcal T}_A(z,\lambda,\lambda),
\label{CZAint}
\end{equation}
where the patch  of
random surface outside set $A$ 
contributes an  spectral density $\rho \left( \lambda\right) $, while $A$ contributes an inverse ``propagator'' $%
{\mathcal T}_A\left( z,\lambda,\lambda\right),$ similar to (\ref{APropagator}) or (\ref{BPropagator}), 
the form of which depends on the nature of set $A$. The variables are the fugacity $z$ of set $A$ and the 
spectral parameter $\lambda$, repeated here since it  characterizes  both sides of the random surface patch
adjacent to $A$.

\subsubsection{$\bullet$ Boundary Two-Point Partition Function (\ref{CZtildeA}):} Its integral
representation  is similarly
\begin{eqnarray}
\label{CZtildeAint}
\tilde Z_{A}&=&\int%
\prod ^{2}_{l=1}d\lambda_{l}\, \rho(\lambda _{l})\,
{\mathcal T}_{A}(z,\lambda_1,\lambda_2) \\
\nonumber
& &\qquad\times (1-\tilde z\lambda_1)^{-1}
(1-\tilde z\lambda_{2})^{-1}
\end{eqnarray}
with two extra propagators $\mathcal L(\tilde z \lambda_{1[{\rm or}\, 2]})$ describing the
two boundary lines between $i$ and $j$:
\begin{eqnarray}
\label{CZtAint}
\tilde Z_{A}&=&\int%
\prod ^{2}_{l=1}d\lambda_{l}\, \rho(\lambda _{l})\,
{\mathcal T}_{A}(z,\lambda_1,\lambda_2)\,{\mathcal L}(\tilde z\lambda_1)\, {\mathcal L}(\tilde z\lambda_2),
\end{eqnarray}
using the simplified notation (see (\ref{BPropagator}))
\begin{equation}
{\mathcal L}(x):={\mathcal L}(x,0)= (1-x)^{-1}.
\label{CPropagator}
\end{equation}

\subsubsection{$\bullet$ Two-Puncture Boundary Partition Function:} (\ref{CZtilde})
can be written as the limit case where $A$ is a
boundary puncture, noted $\bullet$:
\begin{eqnarray}
\nonumber
\tilde Z_{\bullet}=\tilde Z( \hbox to 9.5mm{\hskip 0.5mm
$\vcenter{\epsfysize=.45truecm\epsfbox{fig3.eps}}$
              \hskip -80mm})
              &=&\int%
d\lambda \, \rho(\lambda )\, (1-\tilde z\lambda)^{-2}\\
&=&\int d\lambda \, \rho(\lambda )\, {\mathcal L}^{2}(\tilde z\lambda).
\label{CZtdotint}
\end{eqnarray}

\subsubsection{$\bullet$ Dirichlet Mutually-Avoiding Sets $A\wedge B$:}
The mutual-avoidance $A\wedge B$  requires the random sets $A$ and $B$ to be
separated by a connected piece of random lattice  $G$. Thus the spectral representation associated
with partition functions (\ref{C12AB})
and (\ref{CZtildeAB}) requires to integrate over an extra eigenvalue density $\rho$  in between the two propagators $\mathcal T_A$
and $\mathcal T_B$ associated with traversing random sets $A$ and $B$. Thus partition functions (\ref{C12AB}) and (\ref{CZtildeAB})
can be written as
\begin{equation}
Z_{A\wedge B}=\int
\prod ^{2}_{l=1}d\lambda_{l}\, \rho(\lambda _{l})\,
{\mathcal T}_A(z,\lambda_1,\lambda_2)\, {\mathcal T}_B(z,\lambda_2,\lambda_1),
\label{CZABint}
\end{equation}

\begin{eqnarray}
\nonumber
\tilde Z_{A\wedge B}&=&\int%
\prod ^{3}_{l=1}d\lambda_{l}\, \rho(\lambda _{l})\,
{\mathcal T}_{A}(z,\lambda_1,\lambda_2)\,  {\mathcal T}_{B}(z,\lambda_2,\lambda_3)\\
\nonumber
& &\qquad\times (1-\tilde z\lambda_1)^{-1}
(1-\tilde z\lambda_{3})^{-1}\\
&=&\prod ^{3}_{l=1}d\lambda_{l}\, \rho(\lambda _{l})\,
{\mathcal T}_{A}(z,\lambda_1,\lambda_2)\,  {\mathcal T}_{B}(z,\lambda_2,\lambda_3)\,
{\mathcal L}(\tilde z\lambda_1)\,  {\mathcal L}(\tilde z\lambda_3).
\label{CZtABint}
\end{eqnarray}

\subsection{Quantum Gravity Conformal Weights}
\subsubsection{Bulk Conformal Dimensions}
We first consider the planar topology ($\chi=2$).  The two-puncture partition function (\ref{CZsec}) scales as
\begin{eqnarray}
Z\lbrack \hbox to 8.5mm{\hskip 0.5mm
$\vcenter{\epsfysize=.45truecm\epsfbox{fig2.eps}}$
              \hskip 10mm}
 \rbrack
 \sim |G|^{\gamma_{\rm str}(\chi=2)}.
\label{CZfiggamm}
\end{eqnarray}
In quantum gravity, the {\it bulk conformal weight} $\Delta_{A}$ of operator $\Phi_A$ is then defined by the scaling of the
two-point function $Z_{A}$, properly normalized, with respect to the average size of the lattice:
\begin{equation}
Z_{A}\sim Z\lbrack \hbox to 8.5mm{\hskip 0.5mm
$\vcenter{\epsfysize=.45truecm\epsfbox{fig2.eps}}$
              \hskip 10mm}
 \rbrack \ \times \left| G\right|
^{-2\Delta_{A}}\ .
\label{CZAdelta}
\end{equation}
This definition takes into account the fact that the two-point partition function
$Z_{A}$ contains, in addition to
the insertion of two $\Phi_A$ operators, the two-puncture partition function on the sphere.
Thus we have by definition:
\begin{equation}
Z_{A}\sim  \left| G\right|
^{\gamma_{\rm str}(\chi=2)-2\Delta_{A}}\ .
\label{CZldeltag}
\end{equation}

\subsubsection{Boundary Conformal Dimensions}
Let us first consider the disk partition function with two punctures (\ref{CZtilde}). We write its scaling as:
\begin{equation}
\label{CZtildes}
\tilde Z_{\bullet} =\tilde Z( \hbox to 9.5mm{\hskip 0.5mm
$\vcenter{\epsfysize=.45truecm\epsfbox{fig3.eps}}$
              \hskip -80mm})
	      \sim |G|^{\gamma_{\rm str}(\chi=1)-2} \times \left({\sqrt {|G|}}\right)^{2-2 \tilde \Delta_0}.
\end{equation}
As in eq. (\ref{diskpuncs}) of appendix \ref{ONapp}, the first power law accounts for the scaling of the disk partition
function for an
Euler characteristic $\chi =1$, with a susceptibility exponent
$\gamma_{\rm str}(\chi=1)$, while the second term takes into account the integration  and the conformal weight
$\tilde \Delta_0$
of the two punctures along the boundary.

For the boundary partition function $\tilde Z_{A}$, we similarly define {\it boundary conformal weights}
$\tilde \Delta_A$, such that:
\begin{equation}
\label{CtZltD}
\tilde Z_{A}
	      \sim |G|^{\gamma_{\rm str}(\chi=1)-2} \times \left({\sqrt{ |G|}}\right)^{2-2 \tilde \Delta_A}.
\end{equation}
By using (\ref{CZtildes}) we thus have the simple scaling relation
\begin{equation}
\label{CtZA/0}
\frac{\tilde Z_{A}}{\tilde Z( \hbox to 9.5mm{\hskip 0.5mm
$\vcenter{\epsfysize=.45truecm\epsfbox{fig3.eps}}$
              \hskip -80mm})}
=\frac{\tilde Z_{A}}{\tilde Z_{\bullet}}
\sim \left({\sqrt {|G|}}\right)^{2 \tilde \Delta_0- 2 \tilde \Delta_A}.
\end{equation}

One can also measure the scaling behavior in terms of the boundary length
$|\partial G|=|G|^{1/2\nu}$, where, as shown in appendix \ref{ONapp}, $1/\nu=1-\tilde\Delta_0$. This leads to the definition of
another set of boundary quantum gravity dimensions, $\tilde \Delta'_A$, such that
\begin{equation}
\label{CtZltD'}
\tilde Z_{A}
	      \sim |G|^{\gamma_{\rm str}(\chi=1)-2} \times {|\partial G|}^{2-2 \tilde \Delta'_A},
\end{equation}
or, equivalently,
\begin{equation}
\label{CtZA/0'}
	      \frac{\tilde Z_{ A}}{\tilde Z_{\bullet}}
\sim {|\partial G|}^{2 \tilde \Delta'_0- 2 \tilde \Delta'_A}.
\end{equation}
This gives as in (\ref{d'd}) of appendix \ref{ONapp}, the dimension:
\begin{equation}
\label{Cd'd}
\tilde \Delta'_A-1= \nu (\tilde \Delta_A-1).
\end{equation}
Hence
\begin{equation}
\label{Cd''d}
\tilde \Delta'_A= \frac{\tilde\Delta_A-\tilde\Delta_0}{1-\tilde\Delta_0},
\end{equation}
which is the dual dimension (\ref{dualdelta}) when $\tilde\Delta_0=\gamma$, i.e., in the dense phase.

\subsection{Derivation of Quantum Gravity Scaling Relations}
\subsubsection{Relation between Bulk and Boundary Exponents} We now have to characterize the scaling behavior of
 partition functions, as represented by spectral multiple integrals such as (\ref{CZAint}), (\ref{CZtAint}),
 or (\ref{CZtdotint}). Similarly to the study done in appendix \ref{ONapp}  [see, e.g., eqs. (\ref{intrho}),
 (\ref{mathLscaling}), and (\ref{ZLscaling})], it is useful to introduce a notation characterizing the various scaling
 factors in the multiple integrals. We thus write these partitition functions (\ref{CZAint}), (\ref{CZtAint}),
 and (\ref{CZtdotint}) as
\begin{eqnarray}
\label{CZAs}
Z_{A}&=&\int
 \rho \star {\mathcal T}_A,\\
\label{CZtAs}
\tilde Z_{A}&=&\left(\int \rho \right)^{2}
\star {\mathcal T}_{A} \star {\mathcal L}^{2},\\
\tilde Z( \hbox to 9.5mm{\hskip 0.5mm
$\vcenter{\epsfysize=.45truecm\epsfbox{fig3.eps}}$
              \hskip -80mm})=\tilde Z_{\bullet}
&=&\int  \rho  \star {\mathcal L}^{2},
\label{CZtdots}
\end{eqnarray}
where the $\star$ notation naturally resembles that representing convolutions. Each factor $\mathcal X$ of a
$\star$-product brings in its own contribution to the scaling behavior of the partition function. Ultimately, this scaling
behavior is measured in terms of area $|G|$, and is formally written as
\begin{equation}
\mathcal X \sim |G|^{-[\mathcal X]}
\label{dim}
\end{equation}
 where the notation $[\cdots]$ represents the partial scaling
dimension of factor $\mathcal X$.
All partial scaling behaviors multiply, leading to the simple algebraic rules:
\begin{equation}
[\mathcal X \star \mathcal Y]=[\mathcal X]+[\mathcal Y].
\label{stardim}
\end{equation}
We therefore find:
\begin{eqnarray}
\label{CZAdim}
\left[Z_{A}\right]&=&\left[\int
 \rho \right]+ \left[{\mathcal T}_A\right],\\
\label{CZtAdim}
\left[\tilde Z_{A}\right]&=&2\left[\int%
\rho \right] + \left[{\mathcal T}_{A}\right] + 2 \left[{\mathcal L}\right],\\
\left[\tilde Z( \hbox to 9.5mm{\hskip 0.5mm
$\vcenter{\epsfysize=.45truecm\epsfbox{fig3.eps}}$
              \hskip -80mm})\right]=\left[\tilde Z_{\bullet}\right]
&=&\left[\int  \rho \right] + 2\left[{\mathcal L}\right].
\label{CZtdotdim}
\end{eqnarray}
From definition (\ref{dim}) we also have the simple rule for ratios:
\begin{equation}
\label{CtZA/0dim}
\left[\frac{\tilde Z_{A}}{\tilde Z( \hbox to 9.5mm{\hskip 0.5mm
$\vcenter{\epsfysize=.45truecm\epsfbox{fig3.eps}}$
              \hskip -80mm})}\right]
=\left[\frac{\tilde Z_{A}}{\tilde Z_{\bullet}}\right]=\left[\tilde Z_{A}\right]-\left[\tilde Z_{\bullet}\right].
\end{equation}
Eqs. (\ref{CZAdim}), (\ref{CZtAdim}), and (\ref{CZtdotdim}) then imply
\begin{eqnarray}
\left[\frac{\tilde Z_{A}}{\tilde Z_{\bullet}}\right]=\left[\tilde Z_{A}\right]-\left[\tilde Z_{\bullet}\right]=\left[\int
 \rho \right]+ \left[{\mathcal T}_A\right].
\label{CtZA/0=ZAdim}
\end{eqnarray}
Eq. (\ref{CZAdim}) finally gives
\begin{eqnarray}
\label{dim=dim}
\left[\frac{\tilde Z_{A}}{\tilde Z_{\bullet}}\right]=\left[Z_{A}\right].
\end{eqnarray}
Hence, as expected: \\
{\it In quantum gravity, the two-point boundary partition function,
normalized by the two-puncture boundary one, scales in the same way as
 the two-point bulk partition function}.\\
This can already be seen from (\ref{CZAs}), (\ref{CZtAs}), and (\ref{CZtdots}).
Finally, from the very definitions (\ref{dim}), (\ref{CZldeltag}), and (\ref{CtZA/0}), we have
\begin{eqnarray}
\label{CZADim}
\left[Z_{A}\right]&=&2\Delta_{A}-\gamma_{\rm str}(\chi=2)\\
\label{CtZA/0Dim}
\left[{\tilde Z_{A}}/{\tilde Z_{\bullet}}\right]
&=&{  \tilde \Delta_A}-\tilde \Delta_0.
\end{eqnarray}
From scaling identity (\ref{CtZA/0=ZAdim}) we conclude:
\begin{equation}
2\Delta_{A}-\gamma_{\rm str}(\chi=2)={\tilde \Delta_A}-\tilde \Delta_0,
\end{equation}
which is relation (\ref{C1}), {\bf QED}.

\subsubsection{Boundary Conformal Weights of Dirichlet Sets $A\wedge B$.}
Use is made of same block scaling analysis as above. The spectral representation
(\ref{CZtABint}) of the boundary two-point partition function $Z_{A\wedge B}$ can be symbolically written as
\begin{eqnarray}
\tilde Z_{A\wedge B}=\left(
 \int \rho \right)^3 \star
{\mathcal T}_{A} \star {\mathcal T}_{B}
\star {\mathcal L}^2.
\label{CZtABstar}
\end{eqnarray}
The total scaling dimension (\ref{dim}) of the partition function is therefore:
\begin{eqnarray}
\label{CZtABdim}
\left[\tilde Z_{A\wedge B}\right]=3\left[\int \rho \right] +
\left[{\mathcal T}_{A}\right] + \left[{\mathcal T}_{B}\right]+2 \left[{\mathcal L}\right].
\end{eqnarray}
Considering its ratio to the two-puncture boundary partition function  gives:
\begin{eqnarray}
\label{CtZAB/0dim}
\left[\frac{\tilde Z_{A\wedge B}}{\tilde Z_{\bullet}}\right]
=\left[\tilde Z_{A\wedge B}\right]-\left[\tilde Z_{\bullet}\right]=2\left[\int%
\rho \right] + \left[{\mathcal T}_{A}\right]+\left[{\mathcal T}_{B}\right].
\end{eqnarray}
where use was made of the scaling dimension (\ref{CZtdotdim}).
Comparing to (\ref{CtZA/0=ZAdim}) finally gives
\begin{eqnarray}
\label{dimAB=dimA+dimB}
\left[\frac{\tilde Z_{A\wedge B}}{\tilde Z_{\bullet}}\right]=
\left[\frac{\tilde Z_{A}}{\tilde Z_{\bullet}}\right]+\left[\frac{\tilde Z_{B}}{\tilde Z_{\bullet}}\right].
\end{eqnarray}
From eq. (\ref{CtZA/0Dim}) we conclude:
\begin{equation}
\label{dabdAdB}
\tilde \Delta_{A\wedge B}-\tilde\Delta_0=\tilde \Delta_{A}-\tilde\Delta_0+\tilde \Delta_{B}-\tilde\Delta_0,
\end{equation}
which is eq. (\ref{C2}), {\bf QED}.
\end{appendix}

\bigskip



\end{document}